%% file: SnowmassBook-Energy.tex
\documentclass{tcibook}
\usepackage{fancyhea}
\usepackage{work}
\usepackage{bm}       
\usepackage{graphicx}
\usepackage{multirow}
\usepackage{lineno}
\usepackage{threeparttable}
\usepackage[normalem]{ulem}
\usepackage{color}
\usepackage{hyperref} 
\usepackage{amsmath,amsfonts,amssymb}
\usepackage{placeins} 
\usepackage{array,colortbl}
\usepackage{todonotes}
\setuptodonotes{inline}
\usepackage[caption=false]{subfig}
\usepackage{xspace}
\usepackage[square,numbers,sort&compress]{natbib}

\input workshopsymbols.tex     

\def\authorlist#1#2#3{
    \vskip 0.4in
\begin{center}\begin{large} {{\bf Frontier Conveners:} #1 } \end{large}
    \vskip 0.2in
              {{\bf Topical Group Conveners:} #2}
     \vskip 0.2in
              {{\bf Liaisons:} #3}
     \vskip 0.2in
   \end{center}
}

\begin{document}


\pagenumbering{roman}

\parindent=0pt
\parskip=8pt
\setlength{\evensidemargin}{0pt}
\setlength{\oddsidemargin}{0pt}
\setlength{\marginparsep}{0.0in}
\setlength{\marginparwidth}{0.0in}
\marginparpush=0pt


\pagenumbering{arabic}

\renewcommand{\chapname}{chap:intro_}
\renewcommand{\chapterdir}{.}
\renewcommand{\arraystretch}{1.25}
\addtolength{\arraycolsep}{-3pt}


\begin{center}
{\Huge\bf The Future of US Particle Physics}


\vskip 0.4in
    
{\Huge \bf The Energy Frontier Report}

\vskip 0.4in

{\LARGE \bf  2021  US  Community Study \\ on the Future of Particle Physics}

\smallskip

{\LARGE \bf organized by \\ the APS Division of Particles and Fields}

\vskip 0.8in

{\it{In memory of Meenakshi Narain, \\ 
a friend and a colleague, who passed away on January 1, 2023. \\ This report will be one of her enduring legacies, \\ and a result of her guidance, leadership and vision. \\ We will miss her.}}

\end{center}














\tableofcontents





\input Energy/Energy.tex

\end{document}

%% file: workshopsymbols.tex
\def\beq{\begin{equation}}
\def\eeq#1{\label{#1}\end{equation}}
\def\eeqn{\end{equation}}

\newenvironment{Eqnarray}%
   {\arraycolsep 0.14em\begin{eqnarray}}{\end{eqnarray}}
\def\beqa{\begin{Eqnarray}}
\def\eeqa#1{\label{#1}\end{Eqnarray}}
\def\eeqan{\end{Eqnarray}}

\let\bar=\overbar

\def\lsim{\mathrel{\raise.3ex\hbox{$<$\kern-.75em\lower1ex\hbox{$\sim$}}}}
\def\gsim{\mathrel{\raise.3ex\hbox{$>$\kern-.75em\lower1ex\hbox{$\sim$}}}}

\def\L{{\cal L}}

\def\del{\partial}
\def\Dslash{\not{\hbox{\kern-4pt $D$}}}
\def\dslash{\not{\hbox{\kern-2pt $\del$}}}
\def\pslash{\not{\hbox{\kern-2pt $p$}}}
\def\ETmiss{\not{\hbox{\kern-4pt $E$}}_T}

\def\Dlr{\mathrel{\raise1.5ex\hbox{$\leftrightarrow$\kern-1em\lower1.5ex\hbox{$D$}}}}

\def\eff{{\mbox{\scriptsize eff}}}

\def\ee{e^+e^-}

\def\mz{m_Z}
\def\gz{\Gamma_Z}
\def\mw{m_W}
\def\gw{\Gamma_W}
\def\mt{m_t}

\def\mh{m_H}

\def\alphas{\alpha_s}
\def\MSB{{\bar{M \kern -2pt S}}}
\def\msb{{\bar{\scriptsize M \kern -1pt S}}}

\def\drb{{\bar{\scriptsize D \kern -1pt R}}}



%% file: Energy/Energy.tex
 
\newcommand{\NOTESH}[1]{\textcolor{green}{ \bf[SH: #1 ]}}
\definecolor{raspberry}{rgb}{0.8,0.,0.5}
\newcommand{\NOTEMS}[1]{\textcolor{raspberry}{ \bf[MS: #1 ]}}
\newcommand{\NOTEMB}[1]{\textcolor{purple}{ \bf[MB: #1 ]}}
\newcommand{\NOTEPN}[1]{}
\newcommand{\NOTEAT}[1]{}
\chapter{Energy Frontier} 

\newcommand{\NLO}[1]{N${}^{#1}$LO\xspace}
\newcommand{\NLOone}{NLO\xspace}
\newcommand{\NLOgen}{NLO\xspace}
\newcommand{\NNLOgen}{N$^2$LO\xspace}
\newcommand{\NNNLOgen}{\NLO3}

\authorlist{Meenakshi~Narain$^1$, Laura~Reina$^2$, Alessandro~Tricoli$^3$}
   {Michael~Begel$^3$, Alberto~Belloni$^4$, Tulika~Bose$^5$, Antonio~Boveia$^6$, Sally~Dawson$^3$, Caterina~Doglioni$^7$, Ayres~Freitas$^8$,  James~Hirschauer$^9$, Stefan~Hoeche$^9$,  Yen-Jie~Lee$^{10}$, Huey-Wen~Lin$^{11}$,  Elliot~Lipeles$^{12}$, Zhen~Liu$^{13}$,  Patrick~Meade$^{14}$, Swagato~Mukherjee$^3$, Pavel~Nadolsky$^{15}$,  Isobel~Ojalvo$^{16}$, Simone~Pagan~Griso$^{17}$, Christophe~Royon$^{18}$, Michael~Schmitt$^{19}$, Reinhard~Schwienhorst$^{11}$, Nausheen~Shah$^{20}$, Junping~Tian$^{21}$, Caterina~Vernieri$^{22}$, Doreen~Wackeroth$^{23}$, Lian-Tao~Wang$^{24}$}
{Antonio~Boveia$^6$, Dmitri~Denisov$^3$, Caterina~Doglioni$^7$, Sergey~Gleyzer$^{25}$, Meenakshi Narain$^1$, Peter~Onyisi$^{26}$, Laura~Reina$^2$, Manuel~Franco~Sevilla$^4$, Maksym~Titov$^{27}$, Caterina~Vernieri$^{22}$, Daniel~Whiteson$^{28}$}

\vspace{0.2truecm}
\begin{center}
$^1$ \textit{Physics Department, Brown University, Providence, RI 02912, USA}\\
$^2$ \textit{Physics Department, Florida State University,
Tallahassee, FL 32306-4350, USA}\\
$^3$ \textit{Department of Physics, Brookhaven National Laboratory, Upton, NY, 11973, U.S.A.}\\
$^4$ \textit{Department of Physics, University of Maryland, College Park, MD 20742-4111, USA}\\
$^5$ \textit{Department of Physics, University of Wisconsin, 1150 University Avenue, Madison, WI 53706-1390, USA}\\
$^6$ \textit{Department of Physics and Center for Cosmology \& Astroparticle Physics, The Ohio State University, Columbus, OH 43210, USA}\\
$^7$ \textit{School of Physics \& Astronomy, University of Manchester, Manchester, United Kingdom, and 
Fysiska institutionen, Lund University, Lund, Sweden}\\
$^8$ \textit{Department of Physics \& Astronomy, University of Pittsburgh, Pittsburgh, PA 15260, USA}\\
$^9$ \textit{Fermilab, P.O. Box 500, Batavia, IL 60510, USA}\\
$^{10 }$ \textit{The Massachusetts Institute of Technology, Department of Physics, Cambridge, MA 02139, USA}\\
$^{11}$ \textit{Department of Physics \& Astronomy, Michigan State University, East Lansing, MI 48824, USA}\\
$^{12}$ \textit{Department of Physics \& Astronomy, University of Pennsylvania, Philadelphia, PA 19104, USA}\\
$^{13}$ \textit{School of Physics \& Astronomy, University of Minnesota, Minneapolis, MN 55455, USA}\\
$^{14}$ \textit{C. N. Yang Institute for Theoretical Physics,
Stony Brook University, Stony Brook, NY 11794, USA}\\
$^{15}$ \textit{Department of Physics, Southern Methodist University, Dallas, TX 75275, USA}\\
$^{16}$ \textit{Department of Physics, Princeton University, Princeton, NJ 08554, U.S.A.}\\
$^{17}$ \textit{Physics Division, Lawrence Berkeley National Laboratory, Berkeley, CA 94720, USA}\\
$^{18}$ \textit{Department of Physics \& Astronomy, The University of Kansas, Lawrence, KS 66045, USA}\\
$^{19}$ \textit{Department of Physics \& Astronomy, Northwestern University, Evanston, IL 60208, USA}\\
$^{20}$ \textit{Department of Physics \& Astronomy, Wayne State University, Detroit, MI 48202, USA}\\
$^{21}$ \textit{ICEPP, University of Tokyo, Japan}\\
$^{22}$ \textit{SLAC National Accelerator Laboratory, Stanford, CA 94309, USA}\\
$^{23}$ \textit{Department of Physics, University at Buffalo, Buffalo, NY 14260, USA}\\
$^{24}$ \textit{Physics Department, Enrico Fermi Institute and Kavli Institute for Cosmological Physics, University of Chicago, IL 60637, USA}\\
$^{25}$ \textit{Department of Physics \& Astronomy, University of Alabama, Tuscaloosa, AL 35487-0324, USA}\\
$^{26}$ \textit{Department of Physics, University of Texas, Austin, TX 78712, USA}\\
$^{27}$ \textit{IRFU, CEA, Universit\'e Paris-Saclay, France}\\
$^{28}$ \textit{Department of Physics \& Astronomy, University of California, Irvine, CA  92697-4575, USA}
\end{center}

\section{Executive Summary}
\label{sec:exec-summary}
The Energy Frontier (EF) aims at investigating the fundamental physics of the Universe at the highest energies or – equivalently – the shortest timescales after the Big Bang. We investigate open questions and explore the unknown using various probes to discover and characterize the nature of new physics, through the breadth and multitude of collider physics signatures.  While the naturalness principle suggests new physics to lie at mass scales close to the electroweak scale, in many cases direct searches for specific models have placed strong bounds around 1-2 TeV. Thus, the energy frontier has moved beyond the TeV scale and the exploration of the 10 TeV scale becomes crucial to shed light on physics beyond the Standard Model (SM). 

We need to use both energy reach and precision measurements to push beyond the 1 TeV scale in our exploration. The quest for new physics will be thus conducted in a two-tier approach: 1) looking for indirect evidence of beyond-the-Standard-Model physics (BSM) through precision measurements of the properties of the Higgs boson and other SM particles, and 2)  searching for direct evidence of BSM physics at the energy frontier, reaching multi-TeV scales. 

The EF currently has a top-notch program with the LHC and the High Luminosity LHC (HL-LHC) at CERN, which sets the basis for the EF vision. 
 \textbf{The EF supports continued strong US participation in the success of the LHC, and the HL-LHC construction, operations, computing and software, and most importantly in the physics research programs, including auxiliary experiments.}

The discussions on projects that extend
the reach of the HL-LHC underlined that preparations for the next collider experiments have to start now to maintain and strengthen the vitality and motivation of the community.
Colliders are the ultimate tool to carry out such a program thanks to the broad and complementary set of measurements and searches they enable.
Several projects have been proposed 
such as ILC, CLIC, FCC-ee, CEPC, Cool Copper Collider (C$^3$) or HELEN for e$^+$e$^-$ Higgs Factories, and CLIC at 3 TeV centre-of-mass energy, FCC-hh, SPPC and Muon Collider for multi-TeV colliders. For a detailed discussion of timeline, cost, challenges of those accelerator projects we refer to the Accelerator Frontier Integration Task-Force (ITF) report~\cite{Roser:2022sht} and Appendix~\ref{sec:EFColliderTimelines}. 
Dedicated fora were established across frontiers to bring together diverse expertise in the study of future $e^+e^-$ and $\mu^+\mu^-$ colliders. Results from their studies are available in their reports~\cite{Black:2022cth, Llatas:2022goe} and have informed the studies presented in this report. 

All proposed collider physics experiments need complex detectors as well as software and computing infrastructure, with cutting-edge technologies to meet their ambitious physics goals.
The needs extend beyond generic R\&D.  Experience from R\&D and building previous experiments informs us that it takes about 10 years from CD-0 (Critical Decision - 0) to the end of construction of a collider detector. For e$^+$e$^-$ Higgs factories, immediate investment in targeted detector R\&D is needed at a high priority to explore innovative new technologies and address specific detector challenges in preparation for technical designs. The schedule for multi-TeV colliders and their detectors is technically limited. Therefore for multi-TeV colliders, a long-term program to study the accelerator and detector challenges and reduce overall costs for construction is needed. 

In summary, \textbf{the EF supports a fast start for construction of an e$^+$e$^-$ Higgs factory (linear or circular), and a significant R\&D program for multi-TeV colliders (hadron and muon). The realization of a Higgs factory  will require an immediate, vigorous and targeted detector R\&D program, while the study towards multi-TeV colliders will need significant and long-term investments in a broad spectrum of R\&D programs for accelerators and detectors.} These projects have the potential to be transformative as they will push the boundaries of our knowledge by testing the limits of the SM, and indirectly or directly discovering new physics beyond the SM.

\textbf{The US EF community has also expressed renewed interest and ambition to bring back energy-frontier collider physics to the US soil while maintaining its international collaborative partnerships and
obligations.} 

The EF community proposes several parallel investigations over a time period of ten years or more for pursuing its most prominent scientific goals, namely
1) supporting the full (3 - 4.5 ab$^{-1}$) HL-LHC physics program, 2) proceeding with a Higgs factory, and 3) planning for multi-TeV colliders at the energy frontier.

The proposed plans in five year periods starting 2025 are given below.

{\noindent\bf For the five year period starting in 2025:}
\begin{enumerate} 
\item Prioritize the HL-LHC physics program, including auxiliary experiments,
\vspace{-0.1cm}
\item Establish a targeted $e^+e^-$ Higgs factory detector R\&D program, 
\vspace{-0.1cm}
\item Develop an initial design for a first stage TeV-scale Muon Collider in the US,
\vspace{-0.1cm}
\item Support critical detector R\&D towards EF multi-TeV colliders. 
\end{enumerate}
\vspace{-0.1cm}

{\noindent\bf For the five year period starting in 2030:}
\begin{enumerate}
\item Continue strong support for the HL-LHC physics program, 
\vspace{-0.1cm}
\item Support construction of an $e^+e^-$  Higgs factory, 
\vspace{-0.1cm}
\item Demonstrate principal risk mitigation for a first stage TeV-scale Muon Collider.
\vspace{-0.1cm}
\end{enumerate}

{\noindent\bf  Plan after 2035:}
\begin{enumerate}
\item Continuing support of the HL-LHC physics program to the conclusion of archival measurements,
\vspace{-0.1cm}
\item Support completing construction and establishing the physics program of the Higgs factory,
\item Demonstrate readiness to construct a first-stage TeV-scale Muon Collider,
\vspace{-0.1cm}
\item Ramp up funding support for detector R\&D for energy frontier multi-TeV colliders.
\vspace{-0.1cm}
\end{enumerate}

The EF community recognizes that its success critically depends on the resources obtained by the Accelerator Frontier (AF), as there is a direct linkage between the EF vision and advances in accelerator R\&D. The EF community strongly supports the AF in its proposal to establish an $e^+e^-$ Higgs factory program, and start R\&D for energy frontier multi-TeV colliders with appropriate funding~\cite{Gourlay:2022odf}.
Moreover, the visibly strong interdependence between the EF and the Theory Frontier is key to the success of both frontiers, and EF supports a strong and well funded theory program~\cite{Craig:2022cef}. 
Contributions from Instrumentation Frontier~\cite{Barbeau:2022muf} and the Computational Frontiers~\cite{CompF:sm2021} are key to  the realization of the vision of the EF. In addition, the collaboration with the Community Engagement Frontier~\cite{CEFreport} as well as the cross-fertilization with other Frontiers such as the Rare Processes and Precision Measurements~\cite{Artuso:2022ouk}, Cosmic~\cite{Chou:2022luk}, and Neutrino Physics~\cite{Huber:2022lpm} Frontiers are important.
 Finally, to achieve its goal to advance the current and future scientific endeavors, the EF community has a renewed focus on building and advancing an inclusive workforce, with members from diverse backgrounds, expertise, and career levels.  
 
 The vision of the EF community is presented more extensively in Sec.~\ref{sec:EFvision}, and is motivated by the physics studies that are summarized in Secs.~\ref{sec:EF-bigquestions} to~\ref{sec:BSM}. The Appendices in Sec.~\ref{sec:Appendices} discuss the technological developments that will enable the next generations of EF experiments, as well as opportunities and challenges for the EF community, in terms of collaborations, vibrancy, and diversity as well as needed investments. The estimated timelines and cost for the next generation of large-scale projects in the EF are also provided.

\section{Big Questions in the Energy Frontier}
\label{sec:EF-bigquestions}

\begin{figure}[!ht]
\begin{centering}
\includegraphics[scale=0.5]{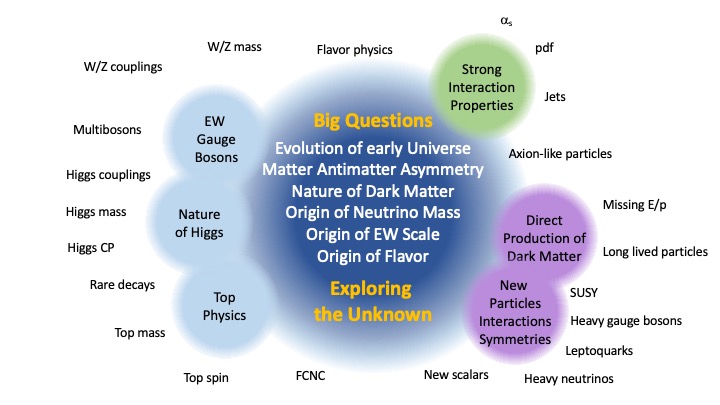} 
\par\end{centering}
\caption{Six categories of Probes and a multiple of Signatures accessible at energy-frontier colliders to address the Big Questions that are at the center of the EF pursuit.}
\label{fig:signatures}
\end{figure}

After decades of pioneering explorations and milestone discoveries,
the Standard Model (SM) of particle physics has been confirmed as the theory that describes electroweak (EW) and strong interactions up to energies of a few hundreds of GeV with great accuracy but it is also leaving some fundamental questions unexplained. In this context, the EF aims at advancing the investigation of some of the still open fundamental questions such as the evolution of the early universe, the matter-antimatter asymmetry of the universe, the nature of dark matter, the origin of neutrino masses, the origin of the EW scale, and the origin of flavor dynamics, with a broad and strongly motivated physics program that will push the exploration of particle physics to the TeV energy scale and beyond. Our sharply focused agenda includes in-depth studies of the SM as well as the exploration of physics beyond the SM to discover new particles and interactions. The vision of the EF must keep its focus on these \textit{big questions}, and must provide opportunities to examine them from as many angles as possible, while also continuing to pursue the \textit{exploration of the unknown}, a leading driver of the EF physics program. This is the core of the EF program as pictorially illustrated in Fig.~\ref{fig:signatures}.

Collider Physics offers a unique opportunity to study a huge number of phenomena and explore the connections between many of the fundamental questions we want to answer. 
The \textit{big questions} outlined at the center of  Fig.~\ref{fig:signatures} unambiguously require new concepts beyond the SM of particle physics. 
A variety of processes can be used as \textit{probes} (depicted as the smaller circles of Fig.~\ref{fig:signatures}) to discover and then characterize the nature of the BSM physics at play and energy-frontier colliders are instrumental to investigating such phenomena via numerous distinct \textit{signatures} (listed as the outermost text in Fig.~\ref{fig:signatures}).

The multi-probe characteristic of high-energy colliders makes them a unique tool in the fundamental quest for answers to the core unknowns of particle physics.
With a combined strategy of precision measurements and high-energy exploration, future lepton colliders starting at energies as low as a few hundreds GeV up to a few TeV can shed substantial light on some of these key questions. Ultimately, it will be crucial to find a way to carry out experiments at higher energies, directly probing new physics at the 10 TeV energy scale and beyond.

Within this big picture, the activities of the Snowmass 2021 EF has been structured around three main \textit{probing} areas, broadly defined as EW Physics (Higgs-boson physics, top-quark and heavy-flavor physics, electroweak gauge bosons physics), Quantum Chromodynamics (QCD) and Strong Interactions (precision QCD, hadronic structure and forward QCD, heavy ions), and BSM physics (model-specific explorations, general explorations, dark matter at colliders), which have focused on three main key questions, as presented below.
\begin{enumerate}
\item What can we learn about the origin of the EW scale and the EW phase transition from an in-depth study of the 
the properties and interactions of SM particles?
\begin{itemize}
\item[] 
The 10$^{\mbox{th}}$ anniversary of the discovery of the Higgs boson~\cite{aad2012observation,chatrchyan2012observation}, a tremendous achievement for our field, is being  celebrated this year, 2022. This discovery provides the last piece of the SM puzzle and at the same time gives a unique connection to new physics beyond the SM that we need to exploit.
The multitude of studies relating to precision Higgs-boson measurements (mass, width, couplings) may help uncover the nature of physics above the EW scale. Together with the full spectrum of EW, top-quark, and flavor-physics precision measurements, these studies will greatly improve the constraining power of global fits of fundamental parameters. The expected large number of Higgs-boson events may lead to measurements of  the shape of the Higgs potential. The study of the Higgs boson, may also give us insight into flavor physics and vice versa.  Last but not least, this study may lead us to understand the implications for naturalness.
\end{itemize}

\item What can we learn about the dynamics of strong interactions in different regimes ?
\begin{itemize}
\item[] 
The fundamental theory of strong interactions, QCD, plays a central role in the SM both as a confining gauge theory that can inspire a deeper understanding of quantum field theories, and at the same time as a ubiquitous crucial component of all measurements at current and future hadron and lepton colliders. Answering the big questions ahead of us intrinsically depends on exploring and understanding the complex phenomenology of strong interactions in different regimes. To harvest the revolutionary progress witnessed in perturbative QCD calculations and Monte Carlo event generators during recent years, residual systematic uncertainties, from parton-shower effects to hadronization models, will have to be understood and reduced.  New ideas to describe the dynamics that drives internal jet structure and their evolution as a function of energy will have to be explored. First-principle calculation of parton densities will have to be benchmarked.
The upcoming era, featuring the HL-LHC, Belle II, the Electron-Ion Collider (EIC), new advances in theory
including in lattice QCD, and potentially a Higgs factory promises to be a golden age for QCD.

\end{itemize}

\item How can we build a complete program of BSM searches which includes both model-specific and model-independent explorations?
\begin{itemize}
\item[] Models connect the high-level unanswered questions in particle physics (dark matter, EW naturalness, CP violation, etc) to specific phenomena, in a self-consistent way. They can be very predictive but model-dependent studies may fail to consider a broad range of new phenomena and search avenues. Many important questions need to be critically addressed.  Which models should be considered and how can model parameter spaces be compared in a consistent way ? Can searches be conducted and interpreted in a model-agnostic way ? How can results from different experiments be compared in a model-independent way to ensure complementarity and avoid gaps in coverage ? Can future colliders chart new regions of model parameter space and also fill in gaps left by existing colliders ? What is the complementarity between precision measurements and direct searches ?
\end{itemize}
\end{enumerate}
A detailed account of the work done by the EF Topical Groups in each of these areas is presented in a series of reports~\cite{Dawson:2022zbb,Agashe:2022plx,Belloni:2022due,Begel:2022kwp,Bose:2022obr}, highlights of which are illustrated 
in Secs.~\ref{sec:EW}-\ref{sec:BSM} of this report.

Finding answers generates more specific questions that will be considered in the studies presented in this report and will be an important factor in building a concrete vision for the future of particle physics explorations at the energy frontier. Among others, 
important aspects that have emerged as most relevant in determining future directions for the EF are the following: the theoretical motivation behind specific future directions, the potential of each future-collider proposal to provide substantial insights in answering the key questions identified as the focus of EF as well as the breadth and complementarity of their physics program, and the identification of what collider and detector developments are necessary to fully pursue the desired physics program of both precision measurements and searches for new physics. 

The collider scenarios discussed in the following are summarized in Table~\ref{tabHiggsFactory} for Higgs factories and in Table~\ref{tabBSMColliders} for multi-TeV colliders.


\begin{table}[!ht]
\parbox{.45\linewidth}{
\begin{center}
 \caption{Benchmark scenarios for Snowmass 2021 Higgs factory studies.}
\begin{tabular}[c]{||l l|c|c|c||}
\hline
 \hline
Collider	&	Type	&	$\sqrt{s}$	&	$\mathcal{P} [\%]$	&	$\mathcal{L}_{\rm int}$	\\
	&		&		&	$e^-/e^+$	&	${\rm ab}^{-1}$	/IP\\ \hline\hline
HL-LHC	&	pp	&	14 TeV	&		&	3 	\\ 
\hline\hline
ILC \& C$^3$	&	ee	&	250 GeV	&	$\pm80/\pm30$	&	2	\\
    &		&	350 GeV	&	$\pm80/\pm30$	&	0.2	\\
	&		&	500 GeV	&	$\pm80/\pm30$	&	4	\\
	&		&	1 TeV	&	$\pm80/\pm20$	&	8	\\
	\hline
CLIC	&	ee	&	380 GeV	&	$\pm80/0$	&	1	\\
\hline
CEPC	&	ee	&	$M_Z$	&		&	50	\\
	&		&	2$M_W$	&		&	3	\\
	&		&	240 GeV	&		&	10	\\ 
	&		&	360 GeV	&		&	0.5 \\
	\hline
FCC-ee	&	ee	&	$M_Z$	&		&	75	\\
	&		&	2$M_W$	&		&	5	\\
	&		&	240 GeV	&		&	2.5	\\
	&		&	2 $M_{top}$	&		&	0.8	\\
\hline\hline
$\mu$-collider	&	$\mu\mu$	&	125 GeV &		&	0.02\\		
\hline \hline
\end{tabular}
\label{tabHiggsFactory}
\end{center}}
%
\hfill
\parbox{.45\linewidth}{
\begin{center}
 \caption{Benchmark scenarios for Snowmass 2021 multi-TeV collider studies.}
\begin{tabular}[c]{||l l|c|c|c||}
\hline
 \hline
Collider	&	Type	&	$\sqrt{s}$	&	$\mathcal{P} [\%]$	& $\mathcal{L}_{\rm int}$	\\
	&		&	(TeV)	& $e^-/e^+$ &  ${\rm ab}^{-1}$/IP	\\ \hline\hline
HE-LHC & 	pp	&	27	&	& 15	\\ 
\hline
FCC-hh	&	pp	&	100	&	&  30	\\
    	\hline
SPPC & pp	&	75-125	&	&  10-20	\\
    	\hline\hline
LHeC	&	ep	&	1.3	&   &  1	\\
FCC-eh	&		&	3.5	&	&	2	\\
\hline\hline
CLIC   &  ee  &	  1.5	&	$\pm 80/0$	&	2.5	\\
	   &	  &   3.0	&	$\pm 80/0$	&	5	\\
\hline \hline
$\mu$-collider	&	$\mu\mu$	&	3	&	&	1	\\
	&		&	10	&	&	10	\\
\hline\hline
\end{tabular}
\label{tabBSMColliders}
\end{center}}
\end{table}

\section{Addressing the Big Questions with EF Colliders}
\label{sec:energyvprecision}

While the existence of BSM physics is well established by observational phenomena, and heavily suggested by theoretical considerations, the energy scale at which it will manifest and its characteristics, e.g. the couplings to known SM particles, are only indirectly constrained.  For example Dark Matter (DM) may be a thermal weakly-interacting massive particle (WIMP) with possible extensions to the multi-TeV range, or it might have an extraordinarily low mass.  Naturalness suggests that the  mass scale of new physics should be close to the EW scale, but in concrete scenarios, such as supersymmetry, it can reach the 10 TeV scale or higher given the measured Higgs-boson mass.  Furthermore it is quite possible that phenomena could show up unexpectedly at unpredicted scales as we have seen historically in our field.

Depending on the mass scale of new physics and the type of collider, the primary method for discovering new physics can vary.  Investigation at the energy frontier allows one to combine direct BSM searches with precision measurements of observables sensitive to scales above the available center-of-mass energy. Furthermore, the type of collider and detectors that are employed will directly influence what signatures can be probed.

The fundamental lessons learned from the LHC thus far are that a Higgs-like particle exists at 125 GeV and there is no other obvious and unambiguous signal of BSM physics. This implies that either there is generically 
a gap to the scale of new physics, or new physics must be weakly coupled to the SM, or hidden in backgrounds at the LHC.  The HL-LHC will either strengthen these conclusions further or potentially point us in a particular direction for discovery.

\begin{figure}[!ht]
\begin{centering}
\includegraphics[scale=0.30]{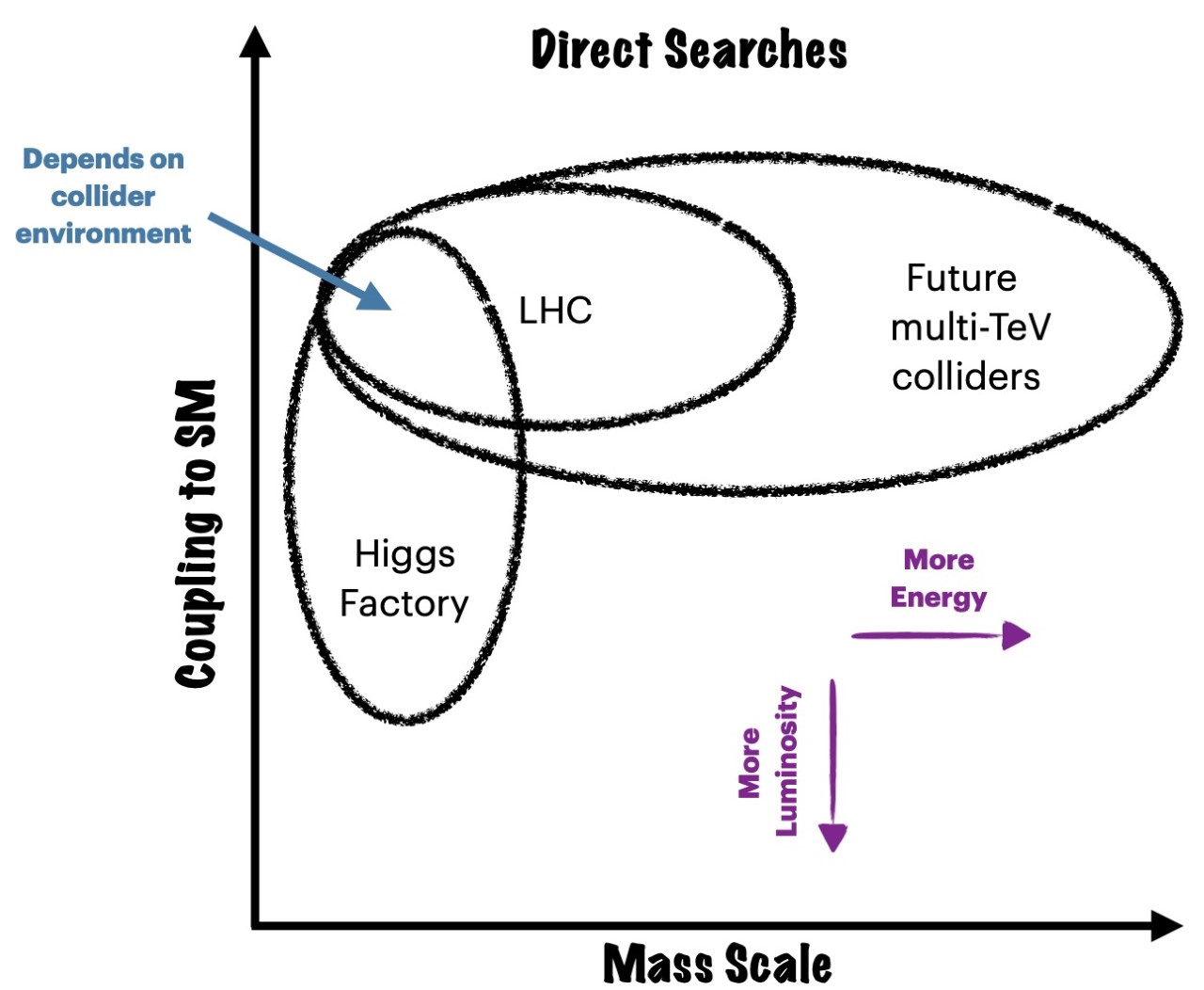} 
\par\end{centering}
\caption{The direct coverage of various colliders in the schematic space of coupling to the SM versus mass scale of BSM physics. ``Higgs factory" and ``multi-TeV colliders" correspond to a generic option among the ones listed in Table~\ref{tabHiggsFactory} and Table~\ref{tabBSMColliders} respectively.}
\label{fig:energyreach}
\end{figure}
To understand how future colliders have complementary potential to unlock the mysteries around these fundamental questions beyond what the LHC and HL-LHC physics program can probe, it is illustrative to use a simplified picture as depicted in Fig.~\ref{fig:energyreach}.  We can imagine that generic new physics lives in a 2D parameter space governed by the coupling of new states to the SM and their mass scale. 
If the center-of-mass energy of a collider is above the one of the LHC, it can {\em directly} search for new states to higher mass scales.  Higgs factories have smaller center-of-mass energies than the LHC, and therefore do not extend the {\em direct} mass reach beyond the LHC.  However, by colliding leptons they offer significantly reduced backgrounds and the ability for triggerless readout, therefore they have the potential to probe new physics that is weakly coupled to the SM.  Additionally, even in the overlap region of Higgs factories with the LHC, the former can be sensitive to new physics that is difficult to discriminate from backgrounds at the LHC. 

\begin{figure}[!ht]
\begin{centering}
\includegraphics[scale=0.3]{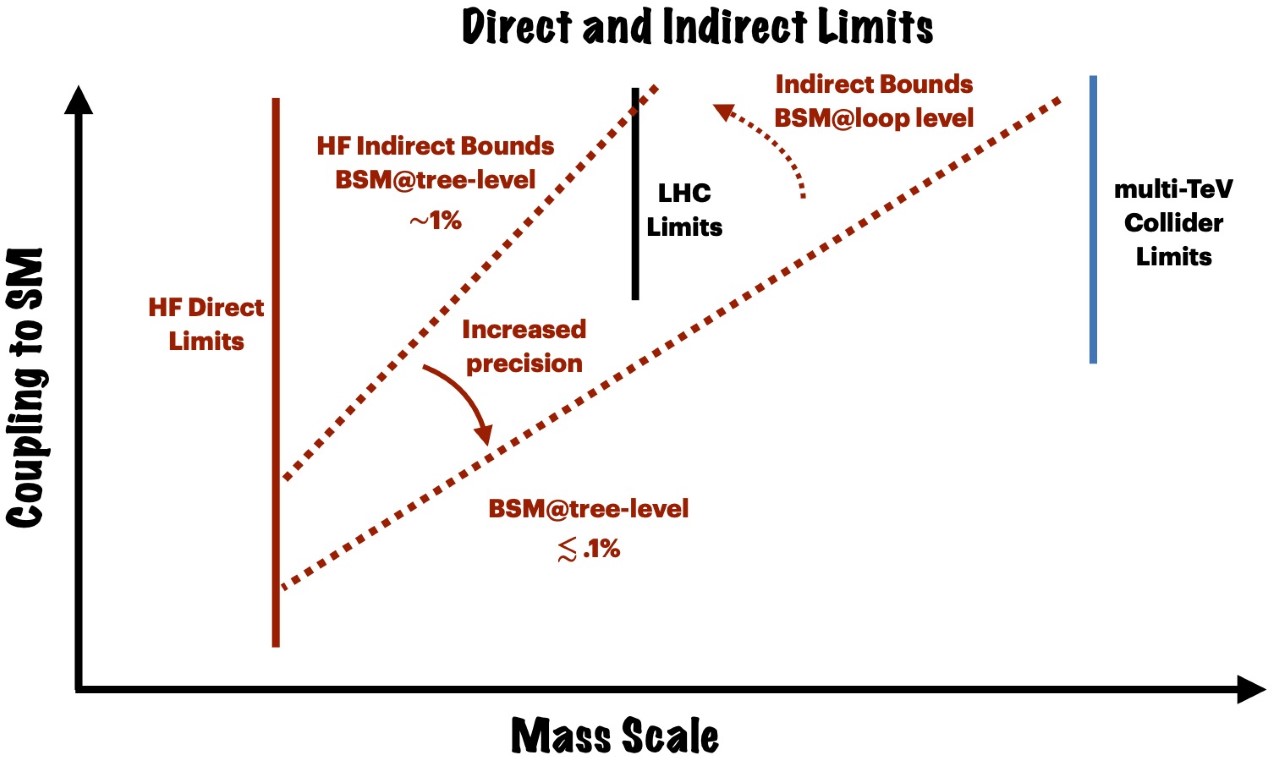} 
\par\end{centering}
\caption{Kinematic reach of direct and indirect searches in the same coupling-mass plane as in Fig.~\ref{fig:energyreach}. The solid lines illustrate direct search limits while the dashed lines represent indirect limits. Higgs Factories (HF) can provide increased reach through indirect searches, benefiting from high-precision measurements.  
A 1\% precision measurement suggests a probed scale of up to a few TeV in perturbative ultra-violet completion models if BSM physics couples at tree-level to the observables of interest. For illustration the potential LHC limit is also depicted.   
If new physics couples at loop-level then the scale probed indirectly is lower.  The ultimate precision, and what scale the multi-TeV colliders can probe is collider specific.}
\label{fig:directorindirect}
\end{figure}

Beyond the direct search for new physics, a key program for the EF is the precision measurement of SM predictions and parameters.  Highly precise measurements allow the probing of scales above the kinematic limit for direct searches at colliders.  This can be captured through Effective Field Theory (EFT) techniques when there is a gap between the probed energy scale and the scale of new physics.  In EF studies, typically this is done by employing specific EFTs, e.g. SMEFT or the more general HEFT formalisms.  
If $M$ is the mass scale of new physics and $g_{BSM}$ is its generic coupling to the SM, then often deviations in SM parameters, $\eta_{SM}$, which occur from integrating heavy particles out at tree-level, scale at the leading order as
\begin{equation}\label{eqn:eft1}
    \delta \eta_{SM}\sim g_{BSM}^2\frac{v^2}{M^2},
\end{equation}
for Higgs related parameters, where $v$ is the vacuum expectation value of the Higgs, or in general as
\begin{equation}\label{eqn:eft2}
    \delta \eta_{SM}\sim g_{BSM}^2\frac{E^2}{M^2},
\end{equation}
where it is assumed that the energy scale $E\ll M$ for the formalism to be applicable.  If new physics only creates loop level deviations in a SM observable, then one can insert a loop factor $\sim 1/16\pi^2$ into Eqns. \ref{eqn:eft1} and \ref{eqn:eft2}. Therefore depending on the precision achievable, as seen in Eqns. \ref{eqn:eft1} and \ref{eqn:eft2}, mass scales larger than the direct reach can be probed.  We can then overlay these types of indirect collider searches, particularly relevant for Higgs factories in Table~\ref{tabHiggsFactory}, on our schematic space of BSM physics shown in Fig.~\ref{fig:energyreach}, as explicitely illustrated in Fig.~\ref{fig:directorindirect}.
As can be seen in Fig.~\ref{fig:directorindirect} the energy versus precision trade-off {\em crucially} depends on the precision attainable.  Suggestively, we have shown a 1\% precision often associated with parameter measurements (except for e.g. the $HZZ$ coupling at Higgs factories), where the scaling typically does not extend beyond the LHC without invoking strong coupling.  However, for quantities that are measured significantly more precisely, e.g. $\lesssim 0.1\%$, at future Higgs-factory programs, such as $M_W$, the reach can extend much further.   This exact scaling in mass reach depends on the type of BSM physics, and both Higgs parameters and EW observables measured at Higgs factories are important for understanding complementary measurements available at future multi-TeV colliders.  The precision that can ultimately be reached and in what types of observables strongly motivates advances in detector technology, increases in luminosity, use of polarization at lepton colliders, and improved theoretical calculations. Moreover depending on the type of collider, for example at a multi-TeV collider, the dichotomy between precision reach and energy reach can potentially be bridged with the availability of large statistics for processes as e.g. Higgs production {\em if} the environment can be fully controlled.

There can be a multitude of phenomena studied at low masses, incompatible with the EFT framework at those energies, that benefit from a reduced background environment at an $e^+e^-$ Higgs factory.  Additionally, even {\em within} one collider, precision measurements and direct searches coexist and offer multiple complementary probes.  To move beyond scaling it is useful to give a few examples of how direct and indirect complementarity play out within the EF, as well as what are potential BSM features at low energy that the LHC may be missing. The remaining of the EF report and the EF Topical Group reports~\cite{Dawson:2022zbb,Agashe:2022plx,Belloni:2022due,Begel:2022kwp,Bose:2022obr} will provide examples and more details, but we list a few illustrative ones here below.

\begin{itemize}
\item Higgs couplings and mass reach: the Higgs boson as the primary target for all future EF colliders provides many examples of the interplay of direct and indirect searches.  For example in Two-Higgs Doublet Models, indirect 
Higgs-boson precision measurements can be complemented by direct resonance searches, EW precision, flavor physics and beyond.  Numerous cases are discussed in Section~\ref{sec:Higgs} and in the corresponding Topical Group report~\cite{Dawson:2022zbb}.
\item Supersymmetry is still a leading example of BSM physics, and while it is a canonical example of direct searches it also can be tested indirectly in numerous ways~\cite{Bose:2022obr}.  However as shown in Section~\ref{sec:susy} for pMSSM parameter scans, indirect searches at Higgs factories do not exceed the typical region covered by the HL-LHC.  Further examples can be found in several of the EF Topical Group reports~\cite{Dawson:2022zbb,Bose:2022obr}.
\item The search for anomalous Trilinear Gauge Couplings (aTGC) and anomalous Quartic Gauge Couplings (aQGC) offer a particularly interesting example of the interplay of several search strategies and measurements~\cite{Belloni:2022due}. At lowest dimension in the SMEFT expansion, deviations are only possible from BSM physics at loop level from new EW charged states.  Therefore in addition to indirect multiboson measurements, direct searches for charged particles, which generate the effects, are powerful probes.  These can include long-lived particle searches in the degenerate mass limit, or more canonical direct searches that 
can also be complemented by precision measurements of Higgs-boson properties when the splitting becomes larger.  Furthermore there are multiple ways to search for vector boson scattering such as in ultraperipheral heavy-ion collisions, as discussed in Section~\ref{sec:ef:qcd:gammagamma} and in Ref.~\cite{Begel:2022kwp}.

\item Higgs-to-invisible decays is an example of where there can be ``holes" in the LHC coverage even at low mass scales.  
A similar consideration can be made for precision measurements of Higgs-boson decays into light quark flavors which are difficult to detect at the LHC despite being copiously produced.  These are both further discussed in Section~\ref{sec:Higgs} and in the corresponding Topical Group report~\cite{Dawson:2022zbb}.
\end{itemize}

In summary, EF colliders are the ultimate tool we possess to clarify the fundamental nature of deviations observed either at collider experiments or at low-energy experiments, thanks to their vast and unique physics programs. 

The current landscape also provides notable examples of deviations in precision measurements of SM parameters that are currently under scrutiny by the particle physics community, and that can provide targets for direct BSM searches. Most notably, anomalies in (semi)leptonic B-hadrons decay, including hints of lepton flavor universality violations, the muon anomalous magnetic moment ($g_\mu - 2$), and the recent $W$-mass measurement might develop into unambiguous BSM signals that call for direct exploration to clarify their nature. Regardless the nature of the present and future anomalies, such occurrences call for a flexible EF program that can develop and deploy a collider able to directly and unambiguously probe energy scales of the order of several TeV. 


\newcommand{\hsm}{h_{\rm SM}}
\section{Electroweak Sector of the Standard Model}
\label{sec:EW}

The investigation of EW interactions has played a crucial role in
defining the SM, predicting all its bosonic and fermionic
constituents, and guiding to their discovery over the last several
decades. The discovery of the Higgs boson has provided the last
piece of evidence to establish the SM as a well-defined quantum field
theory and confirm its predictions, but it has also left unanswered more
fundamental questions on the nature of this particle,  its relation
to defining the EW scale via the spontaneous breaking of the EW
symmetry, as well as its role in the origin of fermion
masses and flavor dynamics. Hence, focused precision studies of the
properties of EW physics still offer a unique handle to
indirectly probe the kind of fundamental new physics that could
explain the origin of the EW scale and relate it to some of the big
open questions we outlined in Sec.~\ref{sec:EF-bigquestions}.

The LHC and HL-LHC will continue exploring EW interactions at the
Energy Frontier with a broad program of measurements of Higgs-boson,
EW gauge-boson ($W,Z$), and heavy-fermions ($t,b$, and also $c$ quark)
properties and interactions. Future colliders, both Higgs Factories
and multi-TeV colliders, will push the precision of such measurements
to indirectly constrain effects of new physics in the 10 TeV range or
above and, at the highest energies, provide direct access to such
physics. In this section we review the current status, future
projections, and overall reach of EW physics studies articulated into
three sections on Higgs-boson physics (Sec.~\ref{sec:Higgs}), top-quark and
heavy-flavor physics (Sec.~\ref{sec:TOPHF}), and EW gauge-boson physics
(Sec.~\ref{sec:EWEFT}). The current and future constraining power of global
fits of EW observables in the SM effective field theory is presented
in Sec.~\ref{sec:gsmeft}.

\subsection{Higgs and BSM physics}
\label{sec:Higgs}

Over the past decade the LHC has fundamentally changed the landscape of high energy particle physics through the discovery of the Higgs boson and the first measurements of many of its properties.  As a result of this, and no other ``discoveries" at the LHC, the questions surrounding the Higgs boson have only become sharper and more pressing for planning the future of particle physics.  

The SM is an extremely successful description of nature, with a basic structure dictated by symmetry.  
Still, symmetry alone is not sufficient to fully describe the microscopic world we explore, and even after specifying the gauge and space-time symmetries, and number of generations, there are still 19 parameters undetermined by the SM (not including neutrino masses).  Out of these parameters 4 are intrinsic to the gauge theory description, the gauge couplings and QCD theta angle.  The other 15 parameters are intrinsic to the Higgs sector or how other SM particles couple to the Higgs, illustrating its paramount importance in the SM.  In particular, the masses of all fundamental  particles, their mixing, CP violation, and the basic vacuum structure are not predicted from symmetry principles of the SM and need to be derived from experimental data to test the consistency of the SM itself. 
However, the centrality of the Higgs boson goes far beyond just dictating the parameters of the SM. 

The Higgs boson is connected to some of our most fundamental questions about the Universe.  Its most basic role in the SM is to provide a source of Electroweak Symmetry Breaking (EWSB). This effect can be formulated via the Higgs potential in the following way:
\begin{equation}
\label{eq:sm-scalar-potential}
    V(H)=-\mu^2 (H^\dagger H)+\lambda (H^\dagger H)^2\,.
\end{equation}
Indeed, if the mass parameter in the potential (the term quadratic in the SM complex scalar doublet $H$)  simply had a positive sign rather than negative, there would be no EWSB and our universe would not exist in its current form.  The explanations of {\it{why}} EWSB occurs and the presence of such a minus sign are outside the realm of the SM Higgs boson.
In other examples of spontaneous symmetry breaking that we have seen manifest in our universe described by Quantum Field Theory (QFT), there has been a dynamical origin.  In principle the Higgs could be a composite of some other strongly coupled dynamics, as we have seen before in history for other particles.  
Alternatively, the Higgs could be a fundamental scalar and EWSB could arise dynamically through its interactions with other BSM fields.  It is even possible that there could be dynamical connections to cosmology or the anthropic principle.   
No matter what the origin of EWSB is, it will leave imprints on the properties of the SM Higgs boson itself.  Moreover, answers to questions such as ``\textit{Is the Higgs boson composite or fundamental?}'' can have ramifications far beyond just the origin of EWSB. 

If the Higgs boson is a fundamental particle, it represents the first fundamental scalar particle discovered in nature. This has profound consequences both theoretically and experimentally.  From our modern understanding of QFT, fundamental scalars should not exist in the low energy spectrum without an UV-sensitive fine tuning if the SM is an EFT of some more fundamental theory. This is known as the naturalness or hierarchy problem. From studying properties of the Higgs boson, one can hope to learn whether there is some larger symmetry principle at work stabilizing the spectrum. For example supersymmetry, neutral naturalness, or if the correct theory is a composite Higgs model, the Higgs could be a pseudo-Goldstone boson.

Experimentally there are also a number of intriguing directions that open up if the Higgs boson is a fundamental particle.  The most straightforward question is whether the Higgs boson is a unique scalar field in our universe, or is it just the first of many.  Additional scalars can always couple to the Higgs at the renormalizable level, and depending on their symmetry properties they can couple to gauge bosons or fermions as well (e.g. the more commonly known Two Higgs Doublet Models). What this implies is that if the Higgs is not unique, there are two complementary methods for investigating this: searching directly for these new scalar states, or measuring their indirect effects on the SM Higgs-boson properties.  Also related to the fact that new scalars can always leave imprints on the Higgs is that a fundamental Higgs particle is special in QFT.  Using only the SM Higgs field, one can construct the lowest-dimension gauge- and Lorentz-invariant operator in the SM.  This means that generically if there are `Hidden" sectors beyond the SM sector (perhaps related to DM), the Higgs boson is the most relevant portal to these sectors, often referred to as the ``Higgs Portal".  Whether these new sectors have a mass scale well below or above the scale of EWSB implies drastically different experimental observables.  For light new sectors of the universe, this can manifests itself as exotic Higgs decays, invisible Higgs decays, and shifts in the Higgs total width, as can be probed well at $e^+e^-$ colliders.  For heavier hidden sectors, the observables are different and the highest energy collider options are needed.

Another aspect of determining if the Higgs is a fundamental scalar particle concerns whether the minimal Higgs potential is correct.  If there are new BSM particles that couple to the Higgs, the potential itself can receive modifications.  This is not solely a question about the potential, because its form has repercussions for both our understanding of the early universe and its ultimate fate.  For the early universe, the SM predicts that the electroweak symmetry should be restored at high temperatures. However, depending on the actual form of the potential the question remains as to whether there even was a phase transition let alone its strength.  Additionally, depending on its form, the Higgs potential can control the future of our universe as our vacuum may be metastable.  Furthermore a strong EW first order phase transition can have implications for Baryogenesis as well. 

Finally, the Higgs boson is connected to some of the most puzzling questions in the universe: flavor, mass, and CP violation. There are effectively two types of interactions in the SM, gauge interactions and Higgs interactions.  Gauge interactions are tightly constrained and do not fundamentally differentiate flavor.   Higgs interactions govern all the important quantities for flavor, mass, and CP violation in the SM.  In particular, all problems connected with flavor and CP -- the origin of fermion masses, the origin of neutrino masses, the origin of the neutrino mixing matrix and CKM angles -- ultimately require knowledge of the fundamental nature of the Higgs sector.  Otherwise, we are just fitting parameters without an understanding.  The full information that we need is only available at high energy by studying the Higgs boson.

\begin{figure}[!ht]
\begin{centering}
\includegraphics[scale=0.28]{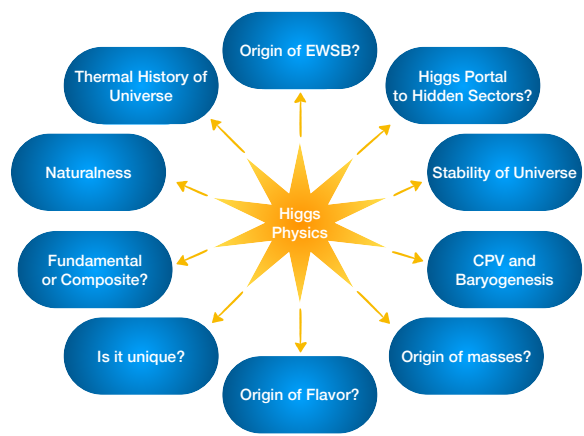} 
\par\end{centering}
\caption{The Higgs boson as the keystone of the Standard Model is connected to numerous fundamental questions that can be investigated by studying it in detail through the many experimental probes illustrated in Fig.~\ref{fig:higgscentral2}.}
\label{fig:higgscentral}
\end{figure}

\begin{figure}[!ht]
\begin{centering}
\includegraphics[scale=0.15]{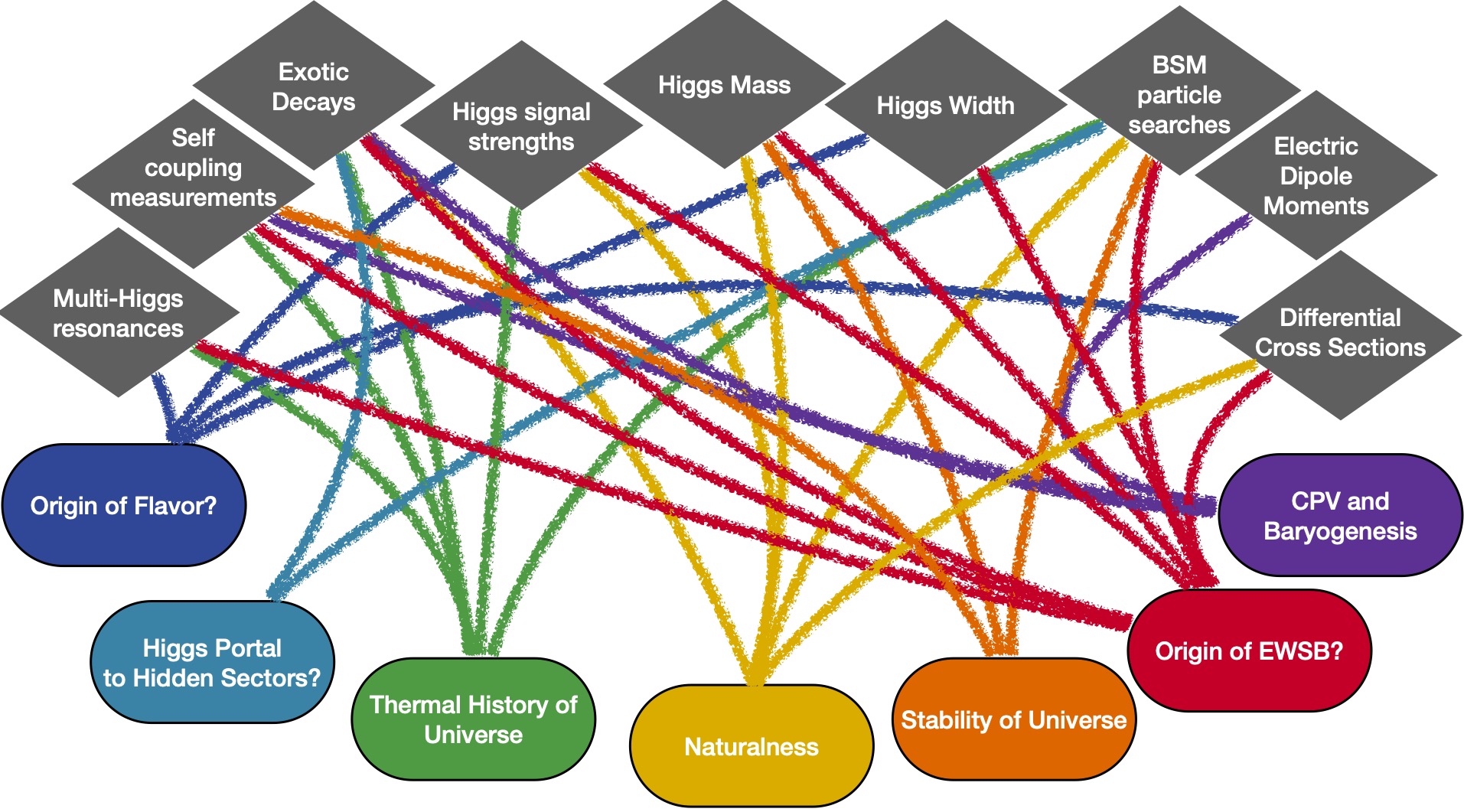} 
\par\end{centering}
\caption{Examples of the interplay between experimental observables and fundamental questions connected to the Higgs boson.}
\label{fig:higgscentral2}
\end{figure}

The fact that understanding the properties of the SM Higgs boson connects to so many fundamental questions illustrates how central it is to the HEP program.  The connections briefly reviewed so far obviously can each be expanded in greater detail, but to collect the various themes in a simple-to-digest manner they are illustrated in Fig.~\ref{fig:higgscentral} and examples of the interplay between experimental observables and fundamental questions are given in Fig.~\ref{fig:higgscentral2}. 
The generality of the concepts and questions posed in Fig.~\ref{fig:higgscentral} could even belie connections to additional fundamental mysteries.  For example, the Higgs portal could specifically connect to DM or other cosmological mysteries.  

%
\subsubsection{Higgs present and future}
The LHC Run 2 with about 140 fb$^{-1}$ of  data analyzed is providing a wealth of new measurements for the Higgs sector. 
The most recent Higgs-boson mass measurements, from CMS and ATLAS, set its value to be 125.38$\pm$0.14 GeV~\cite{CMS:2020xrn} and 124.92$\pm$0.21 GeV~\cite{ATLAS:2020coj} respectively, using both the diphoton and $ZZ$ decay channels. The Higgs-boson mass is a free parameter in the SM and it is now known to per-mille accuracy. We are entering the era of precision Higgs physics, with some of the Higgs-boson couplings measurements approaching $\mathcal{O}$(5-10)\% precision.
All the major production mechanisms of the SM Higgs boson ($h$) have been observed at the LHC: gluon fusion (ggF), vector-boson fusion (VBF), the associated production with a $W$ or $Z$ boson ($Wh$, $Zh$), and the associated production with top quarks ($\mathrm{t\bar{t}h}$, th).  All of these channels are  precisely measured, with the experimental sensitivity of some modes nearing the precision of state-of-the-art theory predictions. Further details of the current LHC measurements at ATLAS and CMS are contained within the Higgs-physics Topical Group report~\cite{Dawson:2022zbb}.  

\begin{figure}[!ht]
\begin{centering}
\includegraphics[scale=0.33]{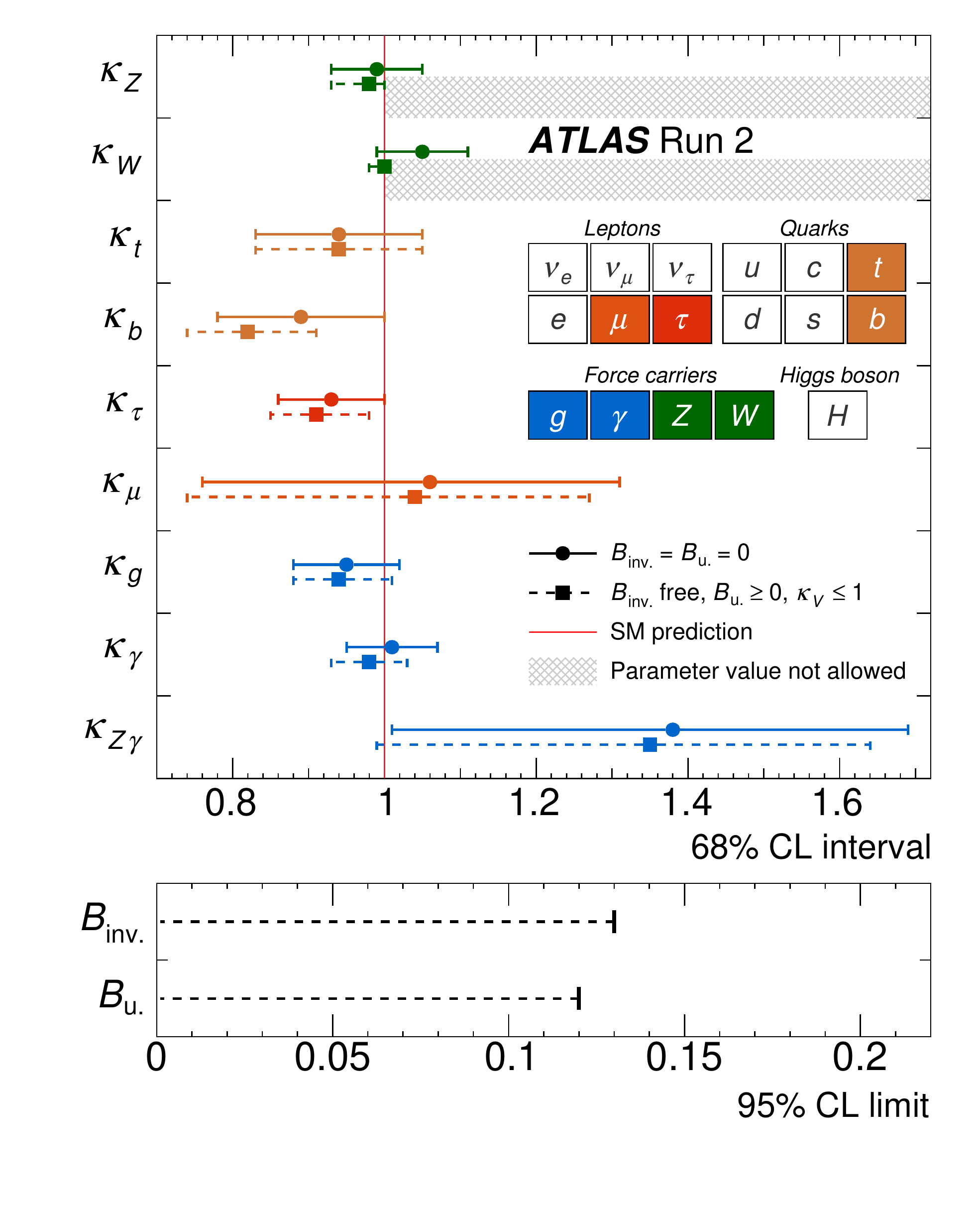} 
\includegraphics[scale=0.40]{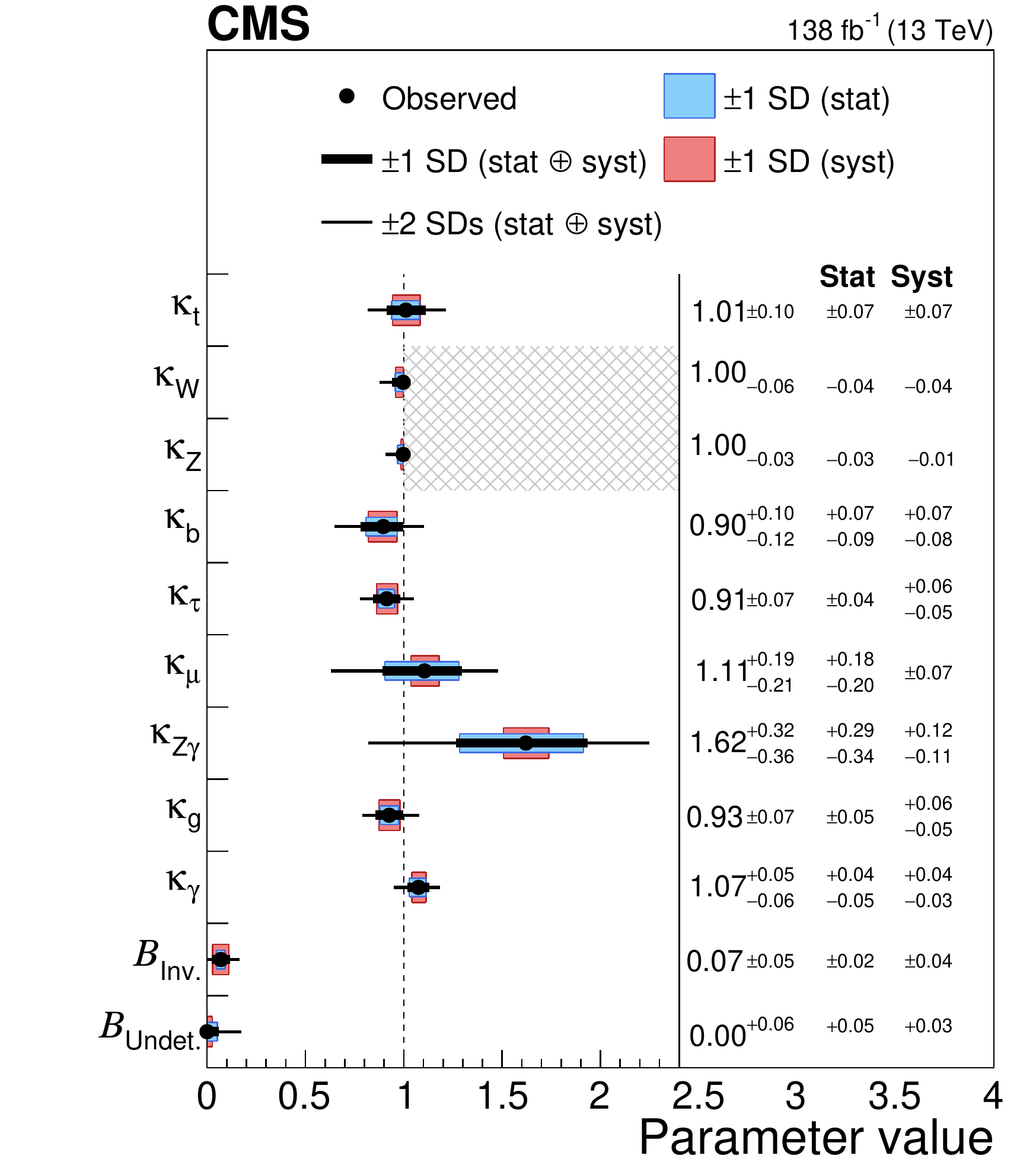} 
\par\end{centering}
\caption{Left, ATLAS best-fit values and uncertainties for Higgs-boson coupling modifiers per particle type with effective photon and gluon couplings, the branching fraction to invisible (B$_i$), and undetected decays (B$_u$) included as free parameters, and the measurement of the Higgs-boson decay rate to invisible final states included in the combination~\cite{ATLAS:2020qdt}. Right, CMS summary of the Higgs-boson couplings modifier best fit. The thick (thin) black lines report the 1$\sigma$ (2$\sigma$) confidence intervals~\cite{CMS:2022dwd}.}
\label{fig:coups_now}
\end{figure}

A simultaneous fit of many individual production rate times branching-fraction measurements is performed to determine the values of the Higgs-boson coupling strength. The $\kappa$-framework defines a set of parameters ($\kappa_X$ for $X=W,Z,\ldots$) that affect the Higgs-boson coupling strengths without altering the shape of any kinematic distributions of a given process. SM values ($\kappa_X\!=\!1$) are assumed for the coupling strength modifiers of first-generation fermions, the other coupling strength modifiers are treated independently. The results are shown in Fig.~\ref{fig:coups_now} for ATLAS and CMS. In this particular fit, the presence of non-SM particles in the loop-induced processes is parameterized by introducing additional modifiers for the effective coupling of the Higgs boson to gluons, photons and $Z\gamma$, instead of propagating modifications of the SM-particle couplings through the loop calculations. In these results, it is also assumed that any potential effect beyond the SM does not substantially affect the kinematic properties of the Higgs boson decay products. The coupling modifiers are probed at a level of uncertainty of 10\%, except for $\kappa_b$ and $\kappa_{\mu}$ ($\approx 20\%$), and $\kappa_{Z\gamma}$ ($\approx 40\%$).  Notably absent are couplings to light fermions, many of whose SM values are out of reach for a LHC measurement (i.e the up, down, strange, charm, and electron couplings) but further details may be found in the Higgs-physics Topical Group report~\cite{Dawson:2022zbb}.  

The scalar potential of the SM 
responsible for the EWSB mechanism, is currently still very far from being probed. After EWSB, the SM scalar potential (see Eq.~\ref{eq:sm-scalar-potential}) gives rise to cubic and quartic terms in the Higgs-boson field, induced by the self-coupling term ($\lambda$). Within the SM, the Higgs-boson self-coupling is fully predicted in terms of the Fermi coupling constant and the Higgs-boson mass, which has been measured at per-mille level accuracy by the ATLAS and CMS experiments~\cite{ATLAS:2020coj, CMS:2020xrn}. The Higgs self-coupling is accessible through Higgs-boson pair production ($hh$) and inferred from radiative corrections to single-Higgs production measurements. Measuring this coupling is essential to shed light on the structure of the Higgs potential, whose exact shape can have deep theoretical consequences. 

The maximum value of the acceptance for the $gg\rightarrow hh$ process is obtained for $\kappa_\lambda \sim 2$, where the cross section is at a minimum. Here $\kappa_\lambda$ refers to the ratio of the measured value to the predicted SM value of the Higgs self coupling and must be unity if the SM is a complete theory. 
Measuring $\kappa_\lambda\neq 1$ would unambiguously imply that there is some new physics beyond the SM. 
The corresponding intervals where  $\kappa_\lambda$ is observed (expected) to be constrained at 95\% CL are listed in Table ~\ref{tab:comb:summary} for the main channels.

\begin{table}[ht!]
\begin{center}
{
\begin{tabular}{|l|c|c|c|}
\hline
\hline
Final state & Collaboration & \multicolumn{2}{c|}{allowed $\kappa_{\lambda}$ interval at 95\% CL} \\
               &             & observed & expected \\
\hline
\multirow{2}{*}{$b\bar{b}b\bar{b}$} & ATLAS & -3.5 -- 11.3 & -5.4 -- 11.4 \\
                        & CMS   & -2.3 -- 9.4  & -5.0 -- 12.0  \\
\hline
\multirow{2}{*}{$b\bar{b}\tau\tau$} & ATLAS & -2.4 -- 9.2 & -2.0 -- 9.0 \\
                              & CMS   & -1.7 -- 8.7  & -2.9 -- 9.8  \\
\hline
\multirow{2}{*}{$b\bar{b}\gamma\gamma$} & ATLAS & -1.6 -- 6.7 & -2.4 -- 7.7 \\
                                  & CMS   & -3.3 -- 8.5 & -2.5 -- 8.2 \\

\hline
\multirow{2}{*}{comb} & ATLAS & -0.6 -- 6.6 & -1.0 -- 7.1 \\
                                  & CMS   & -1.2 -- 6.8 & -0.9 -- 7.1 \\
 
\hline
\hline
\end{tabular}
}
\end{center}
\vspace*{-0.2cm}
\caption{\label{tab:comb:summary} The observed and expected 95\% CL intervals on $\kappa_{\lambda}$ for the most sensitive individual final states analyzed for non-resonant $hh$ production at 13 TeV with about 126-139~fb$^{-1}$ under the assumption of no $hh$ production. All other Higgs boson couplings are set to their SM values. Constraints derived under different statistical assumptions are also available in~\cite{CMS:2020tkr,CMS:hig-21-002,CMS:2022cpr,ATLAS:2022kbf,ATLAS-CONF-2022-035,CMS:2022dwd,ATLAS:2022jtk}. }
\end{table}

The planned HL-LHC, starting in 2029\footnote{This refers to the updated schedule presented in January 2022~\cite{LHCschedule}} will extend the LHC dataset by a factor of $\mathcal{O}(10)$, and produce about 170 million Higgs bosons and 120 thousand Higgs-boson pairs. This would allow an increase in the precision for most of the Higgs-boson couplings measurements. 
The HL-LHC will dramatically expand the physics reach for Higgs physics. Current projections are based on Run 2 results and some basic assumptions that some of the systematic uncertainties will scale with luminosity and that improved reconstruction and analysis techniques will be able to mitigate pileup effects.  The studies also assume that the theory uncertainty is reduced by a factor of 2 relative to current values.  Studies based on the 3000 fb$^{-1}$ HL-LHC dataset estimate that we could achieve $\mathcal{O}(2-4\%)$ precision on the couplings to $W$, $Z$, and third generation fermions. But the couplings to $u$, $d$, and $s$ quarks will still not be accessible at the LHC directly, while the charm-quark Yukawa is projected to be directly constrained to $\kappa_c< 1.75$ at the 95\% CL~\cite{ATLAS:2022hsp}. The Higgs-boson self coupling is a prime target of the HL-LHC and current rough projections claim the trilinear self-coupling will be probed with $\mathcal{O}$(50\%) precision. We will be able to exclude the hypothesis corresponding to the absence of self-coupling at the 95\% CL in these projections for HL-LHC, but not to test the SM prediction~\cite{ATLAS:2022hsp}.  

Future colliders are charged with the challenging tasks of testing the SM predictions of the Higgs-boson Yukawa couplings to light-flavor quarks, and improving the precision on the LHC Higgs-coupling measurements.  
An $e^+e^-$ Higgs factory or Muon Collider can measure these couplings with smaller uncertainties than the HL-LHC due to a combination of knowing the momentum of the incoming particles more precisely, smaller background environments, and better detector resolutions. Indeed, at an $e^+e^-$ Higgs factory the precision can be enhanced by the availability of precise calculations combined with much more democratic production rates: Higgs production is roughly of the same order as other processes in $e^+e^-$ collisions, whereas the LHC must trigger and select Higgs events among backgrounds that are multiple orders of magnitude larger. Moreover, tagging of charm and strange quarks, as previously demonstrated at SLC/LEP,  gives effective probes for precision measurements of charm- and strange-quark Yukawa couplings.  The cleaner $e^+e^-$ environment aided by beam polarization could become a sensitive probe to reveal more subtle phenomena~\cite{Dasu:2022nux}.  For high-energy Muon Colliders, the primary driver is the cleaner environment plus increased statistics~\cite{Forslund:2022xjq}.  The measurement of  the Higgs self-coupling demands access to high-energy center-of-mass collisions to benefit from the larger dataset of $hh$ pairs and is a major goal of all future colliders.  
\begin{figure}[h]
\begin{centering}
\includegraphics[scale=0.7]{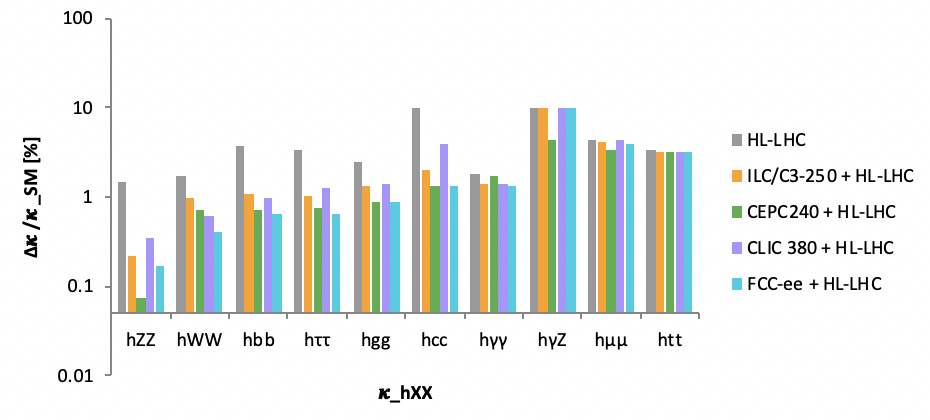} 
\par\end{centering}
\caption{Projected relative Higgs-coupling measurements in $\%$  when combined with HL-LHC results. All values assume no BSM  decay modes. In addition, only the following collider stages are shown: $3~ab^{-1}$ and two interaction points (IPs), ATLAS and CMS, for the HL-LHC at 14~TeV, $2~ab^{-1}$ and 1~IP at 250~GeV for ILC/C$^3$, $20~ab^{-1}$ and 2~IP at 240~GeV for CEPC, $1~ab^{-1}$ and 1~IP at 380~GeV for CLIC, and $5~ab^{-1}$ and 4~IPs at 240~GeV for FCC-ee. Note that the HL-LHC $\kappa_{hcc}$ projection uses only the CMS detector and is an upper bound \cite{ATLAS:2022hsp}. }
\label{fig:higgs_sig_comb_initial_fig}
\end{figure}

While all future colliders give strong contributions to the Higgs precision program, the first stages of $e^+e^-$ Higgs factories are particularly compelling since they can all be constructed in the near future if funding is available, while other collider options require significantly more R\&D. Studies for the five current $e^+e^-$ Higgs factory proposals---ILC~\cite{ILCInternationalDevelopmentTeam:2022izu}, C$^{3}$\cite{Dasu:2022nux}, CEPC~\cite{CEPCPhysicsStudyGroup:2022uwl}, CLIC~\cite{Robson:2018zje}, and FCC-ee~\cite{Bernardi:2022hny}---demonstrate that experiments at these facilities can achieve high precision.  Despite their different strategies, all these proposals lead to very similar projected uncertainties on the Higgs-boson couplings when the colliders are run at the same energies.  The higher luminosity proposed for circular $e^+e^-$ colliders is compensated by the advantages of polarization at linear colliders, yielding very similar projected sensitivity for the precision of Higgs couplings~\cite{Fujii:2018mli,Barklow:2017suo}.  

We show the projected sensitivity for the first stages of proposed $e^+e^-$ colliders combined with HL-LHC projections in Fig.~\ref{fig:higgs_sig_comb_initial_fig}. 
These results are done in the so-called ``kappa-0" framework, which does not allow for BSM decays of the Higgs boson, and are then combined with projections for results from HL-LHC.
It is clear that the dominant improvement from HL-LHC results is in the couplings to $b$'s, $c$'s, and $\tau's$, along with extremely precise measurements of the Higgs interactions with $W$ and  $Z$ bosons.  The future lepton colliders not only can significantly improve on the knowledge of the coupling to the charm quark, but potentially also the coupling to the strange quark, with possible future detector advances, and may even set relevant direct bounds on Higgs couplings to up and down quarks. A dedicated run at the Higgs pole by the FCC-ee has the possibility to measure the coupling of the Higgs to electrons, which would be an important verification of the SM.  Therefore there are subtle differences in the various $e^{+}e^{-}$ Higgs factories and in some cases further study is needed to understand how real the differences are.

Measuring the Higgs couplings can be viewed as part of a global program of fitting to BSM physics in the EFT framework.  In this approach, Higgs interactions are connected to processes without Higgs bosons through the EFT operators, the so-called ``Higgs without Higgs'' events.
The $\kappa$ framework, where the kinematic structure of the Higgs interactions is assumed to be identical to the SM, can be seen as a simplified metric for understanding the capabilities of future colliders for Higgs studies alone, and the only possibility when BSM physics affects Higgs properties at scales not validly described by an EFT approach. 
In these cases a combination of $\kappa$ fits and other observables can be more useful.  The dedicated EFT analysis shown in Section~\ref{sec:gsmeft} combines information from the Higgs sector with information from precision EW measurements, diboson production, and top-quark measurements, including kinematic information, to attempt to gain a deeper understanding of the underlying physics.

Beyond couplings to fermions and gauge bosons, the HL-LHC can constrain the Higgs boson width indirectly from the $ZZ\rightarrow$ 4 lepton channel, with a projected measurement of $\Gamma_{h}=4.1^{+.7}_{-.8}$ MeV, corresponding to roughly a $17\%$ accuracy\cite{ATLAS:2022hsp}. The indirect measurement of the Higgs-boson width can be sensitive to the assumption that there is no new BSM physics contributing to the width.  However, it is more akin to an absolute coupling normalization and can be viewed as part of the larger ``Higgs without Higgs" framework.  BSM physics that invalidates these measurements is not generic, but further complementary information from other colliders is desired.

One distinct advantage of the lepton colliders is the possibility to obtain extremely precise and relatively model-independent measurements of the Higgs boson width. The measurement of the width can confirm the SM prediction given that it can be very sensitive to the scale of new physics.  The fully reconstructed $Z$ boson
in the final state along with the well determined 4-momenta of the initial state leptons in the $Zh$ process allows for a clean determination of the Higgs-boson kinematics regardless of the Higgs decay channel.  The full FCC-ee program (combined with HL-LHC) allows for a $1\%$ measurement of the Higgs-boson width.  Using a SMEFT fit, the ILC finds similar results for the full program, but with just the initial 250 GeV run, a $2\%$ measurement on the total width can be obtained.
A Muon Collider running at $\sqrt{s}=125$ GeV can obtain a model independent measurement of the Higgs-boson width at the 68$\%$ level of $2.7\% ~(1.7\%)$ with $5~ \mbox{ fb}^{-1} (20~ \mbox{fb}^{-1})$  by using a line-shape measurement~\cite{MuonCollider:2022xlm}. 
A high-energy Muon Collider should obtain a similar order of magnitude precision using the indirect methods employed at the LHC with the same theoretical assumptions, and the FCC-hh could in principle also use these methods with further study.

\begin{table}[!ht]
\centering
{\small
\begin{tabular}{ c  c  c  c }
\hline 
collider       &  Indirect-$h$ &  $hh$   & combined   \\
\hline
HL-LHC~\cite{ATL-PHYS-PUB-2022-005}         &  100-200\%    & {50\%}    & 50\%   \\  \hline
ILC$_{250}$/C$^3$-250~\cite{ILCInternationalDevelopmentTeam:2022izu,Dasu:2022nux}    &  49\%    & $-$   &49\%    \\
ILC$_{500}$/C$^3$-550~\cite{ILCInternationalDevelopmentTeam:2022izu,Dasu:2022nux}     &  38\%      & 20\%   &20\%  \\
ILC$_{1000}$/C$^3$-1000~\cite{ILCInternationalDevelopmentTeam:2022izu,Dasu:2022nux}     &  36\%      & 10\%   &10\%  \\
CLIC$_{380}$ ~\cite{Robson:2018zje}  & 50\%     & $-$    &50\%  \\
CLIC$_{1500}$~\cite{Robson:2018zje}   &  49\%   & 36\%  & 29\%  \\
CLIC$_{3000}$~\cite{Robson:2018zje}   &  49\%  & 9\% &  9\% \\
FCC-ee~\cite{Bernardi:2022hny}   &  33\%   & $-$    &33\%   \\
FCC-ee (4 IPs)~\cite{Bernardi:2022hny} &  24\%   & $-$    &24\%  \\  %
FCC-hh  ~\cite{Mangano:2020sao}       &  -  & 3.4-7.8\%  & 3.4-7.8\%  \\
{$\mu$(3 TeV)}~\cite{MuonCollider:2022xlm} &  - & 15-30\% & 15-30\% \\
{$\mu$(10 TeV)}~\cite{MuonCollider:2022xlm} &  - & 4\% &  4\% \\
\hline
\end{tabular}
}
\caption{\label{tab:h3} Sensitivity at 68\% probability on the Higgs cubic self-coupling at the various future colliders. Values for the indirect single Higgs determinations below the first line are taken from \cite{deBlas:2019rxi}.  The values quoted here are combined with an independent determination of the self-coupling with uncertainty 50\% from the HL-LHC. }
\label{tab:future_comp}
\end{table}

A measurement of the Higgs self-coupling is out of reach of Run 3 of the LHC and requires either a larger dataset, or/and a higher collision energy.  The self coupling can be measured by the direct production of $hh$ pairs, or inferred indirectly through the contribution of the Higgs self-coupling to loop corrections to the single-Higgs rate.  However, for the indirect measurement to be relevant, it requires that new physics contributions dominate only the triple Higgs coupling shift.  While this can naively be accounted for in a SMEFT fit, in realistic models this is much more difficult~\cite{McCullough:2013rea}.

The projected sensitivities to the Higg-boson self-coupling at the various future  colliders are presented in Table~\ref{tab:future_comp}. These correspond to projections for a single experiment except for the 'combined' results which for the HL-LHC correspond to the combined projections of ATLAS and CMS experiments.  We see that this is an extremely challenging measurement at all colliders.  Since the measurement is limited by the small number of $hh$ events, the measurement improves with the higher energy colliders.  The indirect measurement improves with the luminosity of the lepton colliders since it is extracted from single-Higgs production. In principle measurements at different center of mass energies can be used to disentangle the indirect effects of shifts in the triple Higgs couplings, however it also depends on the assumptions of what types of other operators can contribute.

The ATLAS and CMS experiments have determined that the observed 125-GeV mass boson has
the Higgs boson quantum numbers $j^{PC}=0^{++}$.  Small violations of CP symmetry in the $hVV$ ($V=W,Z$) and $hf{\overline{f}}$ couplings are still allowed and are an important target of future experimental measurements.  
Hadron colliders provide essentially the full spectrum of possible measurements sensitive to CP violation in the Higgs boson interactions.
Most  processes other than the Higgs gluon interactions could be studied at an $e^+e^-$ collider, especially with the beam energy above the $t {\overline{t}}h$ threshold.  Future $e^+e^-$ colliders are expected to provide comparable sensitivity to HL-LHC in $hf{\overline{f}}$ couplings,  and potentially higher sensitivity in $hZZ$ couplings.  
A Muon Collider operating at the Higgs boson pole allows to measure the CP structure of the $h\mu\mu$ vertex with the beam polarization. 

\begin{figure}[!ht]
\begin{centering}
\includegraphics[scale=0.22]{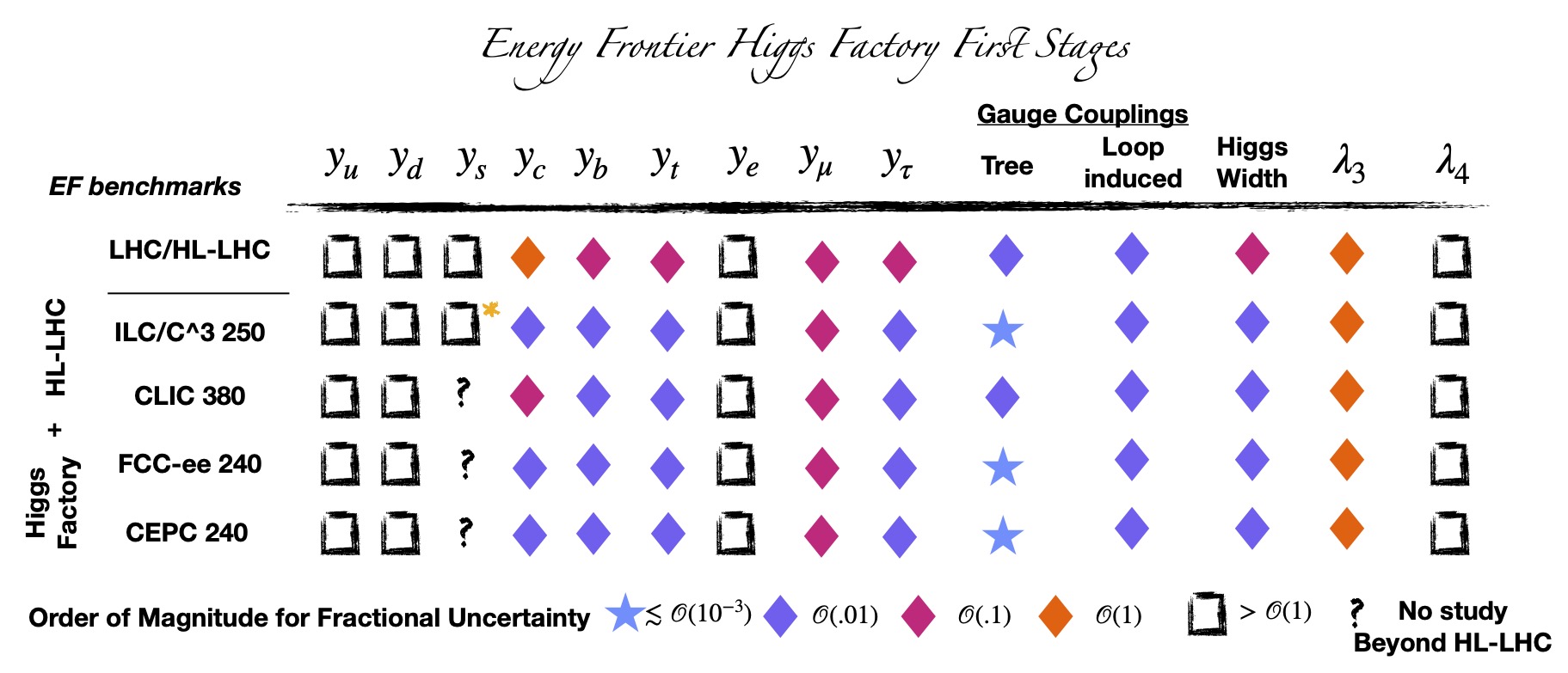} 
\par\end{centering}
\caption{A snapshot of future Higgs precision measurements of SM quantities based on the order of magnitude for the fractional uncertainties with the range defined through the geometric mean.  In this figure the first stages of each $e^+e^-$ Higgs factory are shown in combination with the Hl-LHC, as well as the HL-LHC separately.  The Higgs factories are defined as those listed in Section~\ref{sec:energyvprecision} of the EF Report, excluding the 125 GeV Muon Collider whose timescale is in principle longer term.  The specific precision associated to each coupling can be found in the Higgs-physics Topical Group report~\cite{Dawson:2022zbb} and references therein. A * is put on the ILC measurements for the strange-quark Yukawa coupling to single it out as a new measurement proposed during Snowmass 2021, and is shown in Fig~\ref{fig:2hdmflavor}.  The ? symbol is used in the case where an official study has not yet been performed, for example in the case of strange tagging for CLIC, FCC-ee, and CEPC. This does not mean that they cannot achieve a similar precision, but it is yet to be demonstrated whether based on their detector concepts the measurements is worse or can be improved.
}
\label{fig:higgssummarya}
\end{figure}

\begin{figure}[!ht]
\begin{centering}
\includegraphics[scale=0.42]{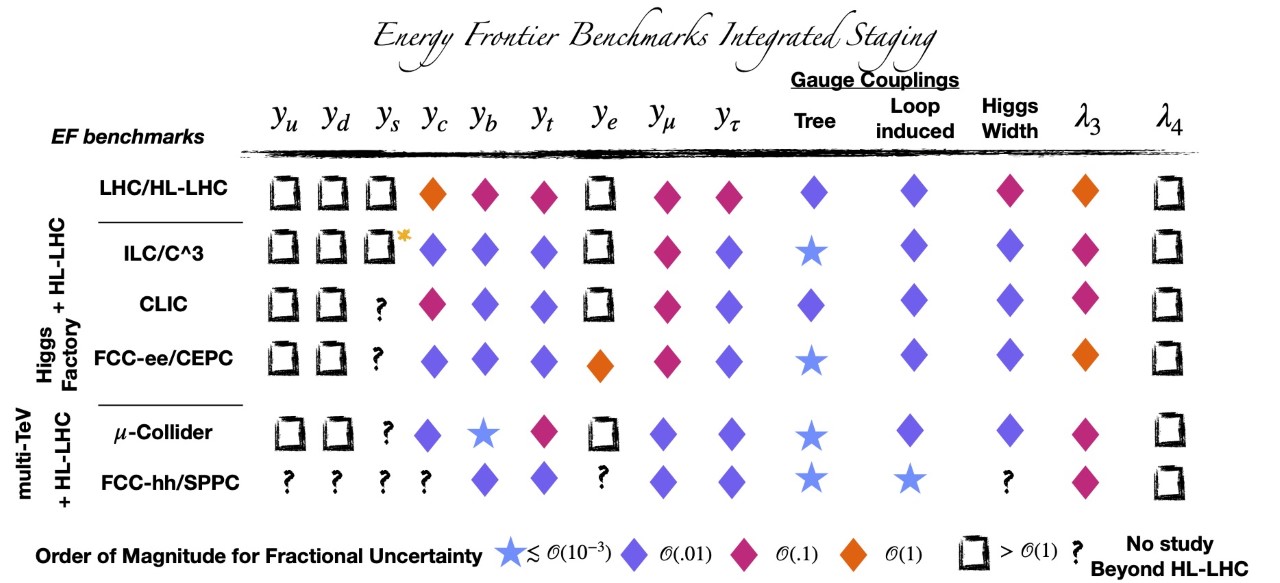} 
\par\end{centering}
\caption{A snapshot of future Higgs precision measurements of SM quantities based on the order of magnitude for the fractional uncertainties with the range defined through the geometric mean.  In this figure the ultimate reach of the final stages of all Higgs factories and multi-TeV colliders are shown in combination with the HL-LHC results, as well as the HL-LHC separately.  All benchmarks and stages are defined in Section~\ref{sec:energyvprecision} of the Energy Frontier Report.  The specific precision associated to each coupling can be found in the Higgs-physics Topical Group report~\cite{Dawson:2022zbb} and references therein. A * is put on the ILC measurements for the strange-quark Yukawa coupling to single it out as a new measurement proposed during Snowmass 2021, and shown in Fig~\ref{fig:2hdmflavor}.  The ? symbol is used in the case where an official study has not yet been performed. It does not connotate that a given collider should be worse than similar ones, but simply that whether it is better or worse based on detector design has not been demonstrated. 
}
\label{fig:higgssummaryb}
\end{figure}

We have presented an overview of the types of precision measurements of Higgs-boson properties that can be done at future colliders,
specific numbers for certain observables at all colliders have been shown, and general coupling fits in the case of the first-stage $e^+e^-$ Higgs factories were shown in Fig.~\ref{fig:higgs_sig_comb_initial_fig}.  However, it is important to understand that there are more observables than discussed here and to consider how high-energy colliders fit into the Higgs precision program. A comprehensive study is presented
in the Higgs-physics Topical Group report~\cite{Dawson:2022zbb}. However, instead of displaying all info as a large table or bar chart, we will conclude this section by displaying a quantitative coarse grained version of all Higgs precision results in Fig.~\ref{fig:higgssummarya} and Fig.~\ref{fig:higgssummaryb}. As can be seen in Fig.~\ref{fig:higgs_sig_comb_initial_fig} and  Fig.~\ref{fig:higgssummarya},  all first-stage Higgs factories have a very similar reach.  However, it also emphasizes that there are still a large number of missing pieces to the Higgs puzzle after only first-stage Higgs factories.  In Fig.~\ref{fig:higgssummaryb}, all stages of future colliders that are discussed in Section~\ref{sec:energyvprecision}, are combined with the HL-LHC.  As we see in this coarse graining, all colliders are compelling, and there is significant progress through Higgs factories and multi-TeV collider proposals compared to the HL-LHC program.  In particular with the additional stages we start to see differences in the various proposed collider programs for Higgs physics.  Linear $e^+e^-$ colliders begin to demonstrate advantages especially in the measurement of the Higgs self coupling compared to circular $e^+e^-$ colliders, whereas the circular colliders can potentially measure the electron Yukawa coupling.  Furthermore, multi-TeV colliders such as the Muon Collider or FCC-hh extend the knowledge of the Higgs boson even further, albeit on different timescales.
However, despite the enormous progress we expect from the currently proposed future colliders in testing the SM Higgs boson sector with high precision, the research for further improvements will have to continue. More specifically, the measurements of certain couplings, e.g. the light-quark Yukawas or the quartic self-coupling of the Higgs boson, that are challenging at the future colliders proposed in the Snowmass 2021 proceedings, motivate a continuing research and development.

\subsubsection{What can we learn about BSM physics from Higgs physics}
\label{sub:BSMPhysics}

The ultimate goal of precision Higgs physics is to learn about new physics at high scales, or to find portals to new physics that could be present at the EW scale or below.  As discussed earlier from an EFT context, the generic scale associated with precision Higgs physics at future colliders typically extends up to a few TeV. 

To go further requires the understanding of the interplay between UV models and Higgs physics. Given that the mapping of fundamental physics questions to Higgs direct and indirect observables is difficult to fully organize comprehensively, the topical report instead focused on specific types of models and observables: Higgs Singlets, Higgs Doublets (including Flavor), Loop-level deviations, and Higgs Exotic Decays.  Fundamental questions of course can be related to all of these types of models and is done so in the Higgs-physics Topical Group report~\cite{Dawson:2022zbb}.  Other connections to fundamental questions are also emphasized in other parts of the EF report, for example whether the Higgs boson is an elementary or composite particle is investigated in Section~\ref{BSM:compositehiggs}.

Given that many of the model-dependent topics have been covered extensively for years, we first wish to highlight some of the results that are new compared to the recent European Strategy Update~\cite{European:2720131}:

\begin{itemize}
    \item The phenomenology of a strong electroweak phase transition is significantly more nuanced than previously envisioned.  It can manifest through shifts in the Higgs cubic coupling, but could still occur without any currently or far future measurable deviation in this coupling~\cite{Meade:2018saz,Baldes:2018nel,Glioti:2018roy}.  Deviations in all types of observables are also possibly correlated with the phase transition, including exotic Higgs decays~\cite{Carena:2022yvx}.
    \item Flavored phenomenology is much richer than previously explored. Flavor-violating decays have now richer possibilities and models~\cite{Altmannshofer:2016zrn,Altmannshofer:2017uvs}. Flavor-preserving deviations in consistent light-quark Yukawa couplings now also exist~\cite{Egana-Ugrinovic:2019dqu,Egana-Ugrinovic:2021uew}, and there are studies for direct probes of this at $e^+e^-$ colliders and related resonance probes from the LHC and other colliders~\cite{Albert:2022mpk}.
    \item  Singlet phenomenology, a canonical example in BSM  Higgs phenomenology, can be quite a bit more varied, including the introduction of scalar resonances decaying to particles with different masses. These searches were further explored~\cite{Adhikari:2022yaa,Robens:2022lkn,Robens:2022erq}.
    \item  There are now viable models of triple-Higgs production at the HL-LHC and beyond~\cite{Robens:2022lkn,Low:2020iua,Egana-Ugrinovic:2021uew,Chen:2022vac}.  The measurement of the quartic coupling should now be considered a standard part of BSM Higgs phenomenology and triple-Higgs and quartic-Higgs measurements should be pursued at future colliders.
\end{itemize}

\begin{figure}
\begin{centering}
\includegraphics[scale=0.3]{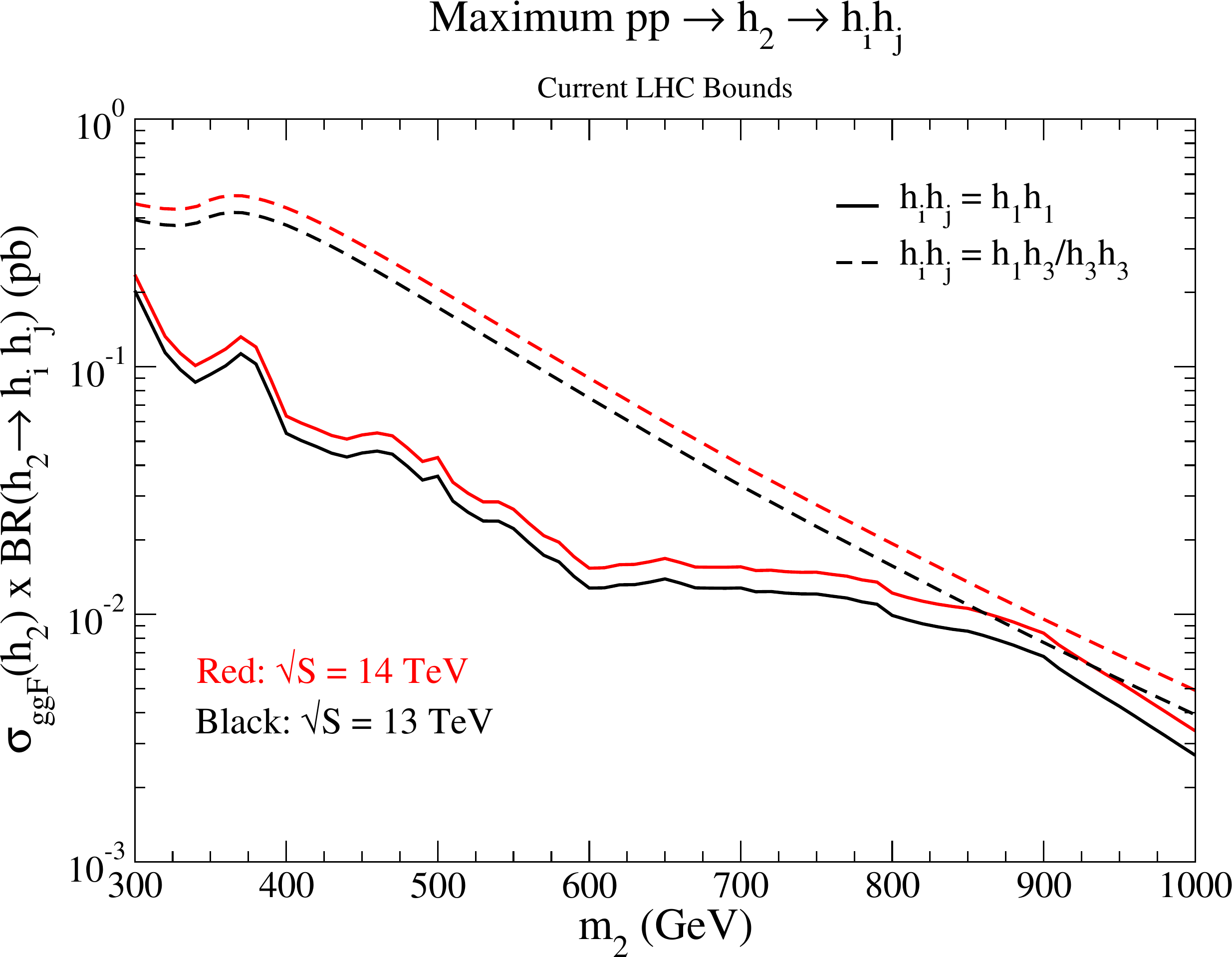} 
\par\end{centering}
\caption{Production of a pair of Higgs bosons in the 
complex singlet model.  In this figure $h_1$ is the SM Higgs boson,
and $h_2$, $h_3$ are new scalars with
$2 m_{h3}< m_{h2}$. 
The maximum rates allowed by current LHC data are shown in black and projections for $\sqrt{s}=14$~TeV in red~\cite{Adhikari:2022yaa}.}
\label{fig:singlet_hh}
\end{figure}

We now give a sampling of results based on UV complete models, starting with the simplest extension of the Higgs sector of the SM with an additional scalar singlet $S$.  Despite the simplicity of this type of models, such results display a wide range of phenomenology and connections to fundamental physics questions.  For example, with a single degree of freedom from a real scalar, one can connect to the EW phase transition and thereby models of baryogenesis.
This Higgs portal can then be connected to dark sectors and DM, or can be viewed as a proxy for models of neutral naturalness.  The existence of a new scalar then also applies to the question of whether or not the Higgs is unique and modifies the Higgs potential.  This can have implications for the stability of our universe.     
Thus, the rich phenomenology of the real singlet scalar is quite extensive.  One can add additional scalar, i.e. a singlet complex scalar, or even more. The phenomenology can be further complicated and projections onto a two-dimensional plane are not sufficient.  In particular, because the masses of the various singlets can be varied, resonance decays have a much wider range of phenomenology.  Figure \ref{fig:singlet_hh} illustrates this possibility, where we see that a rate larger than that for SM di-Higgs production ($h_1h_1$ in Fig.~\ref{fig:singlet_hh}) is possible~\cite{Adhikari:2022yaa}.

Two Higgs Doublet Models (2HDMs) provide the next simplest extension after scalar singlets to the Higgs sector.  They are particularly interesting because they allow for a new state with SM gauge charge that can also acquire a vacuum expectation value while naturally allowing for small EW precision corrections.  The new doublet allows for additional Higgs bosons beyond the observed 125 GeV CP-even neutral scalar $h(125)$, namely, an additional
CP-even neutral scalar $H$, one CP-odd Higgs boson $A$, and a pair of charged Higgs bosons $H^\pm$. Restricting ourselves to the standard types of 2HDM still allows for an enormous range of phenomenology.

The standard parametrization of the physics is done in terms of a ratio of the vacuum expectation values of the 2HDM states, $\tan \beta$, and a mixing angle $\cos (\beta-\alpha)$ as well as the masses of the various scalar eigenstates.

\begin{figure}[!ht]
\begin{centering}
\includegraphics[scale=0.46]{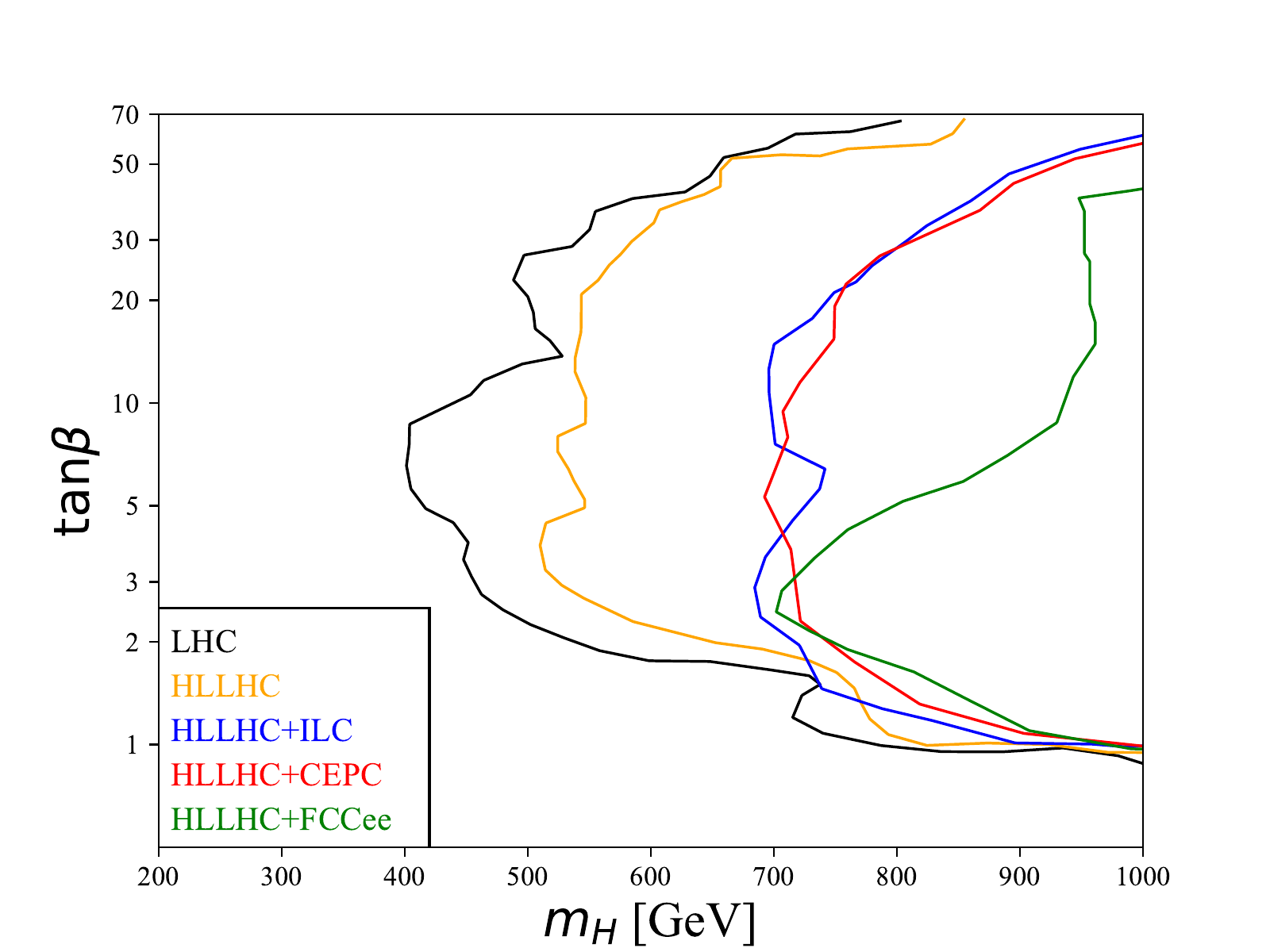} 
\includegraphics[scale=0.275]{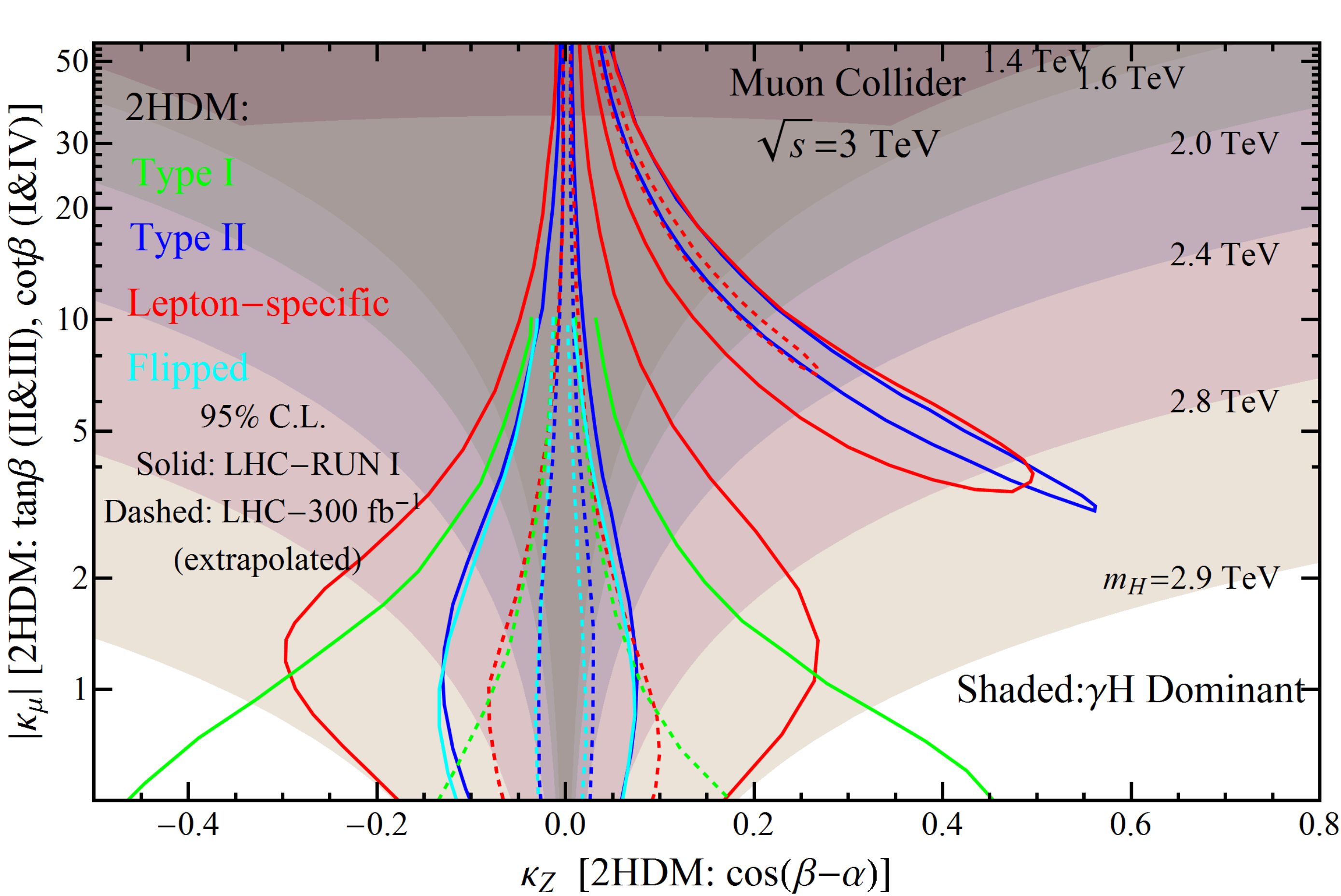} 
\par\end{centering}
\caption{Limits on the parameters of a 2HDM from precision Higgs-coupling measurements. Left: Limits from future $e^+e^-$ colliders combined with limits from the HL-LHC.  The allowed region is to the right of the curves.~\cite{Beniwal:2022kyv}. Right: Limits from a 3 TeV Muon Collider. The TeV scale refers to the mass of the heavier Higgs boson in the 2HDM.}
\label{fig:2hdmfig}
\end{figure}

\begin{figure}[!ht]
\begin{minipage}{.47\textwidth}
\begin{centering}
\includegraphics[scale=0.4]{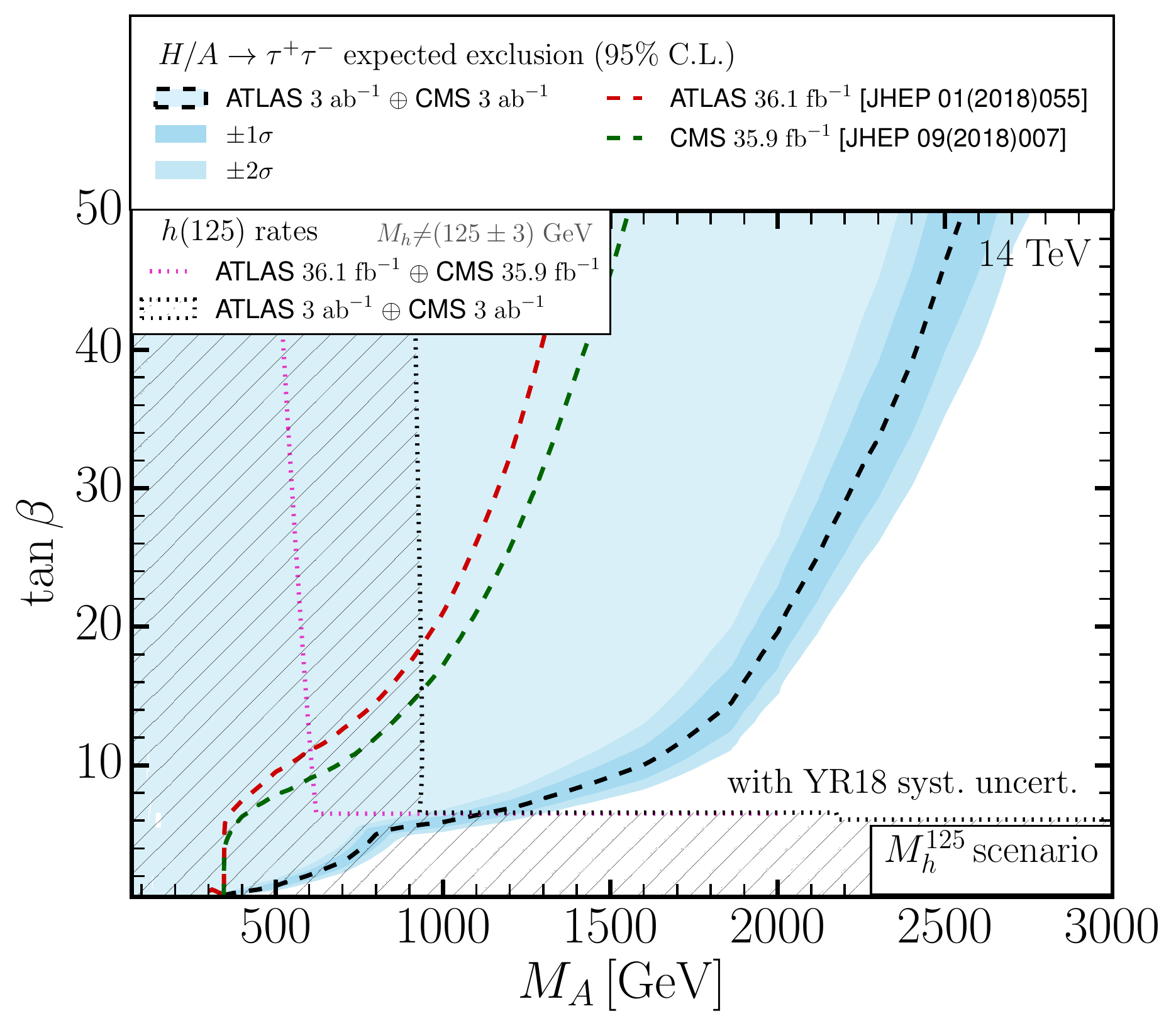}  
\par\end{centering}
\caption{Capability of HL-LHC to probe the scalar sector of the 2HDM. }
\label{fig:2hdmHL}
\end{minipage}%
\begin{minipage}{.47\textwidth}
\begin{centering}
\includegraphics[scale=0.4]{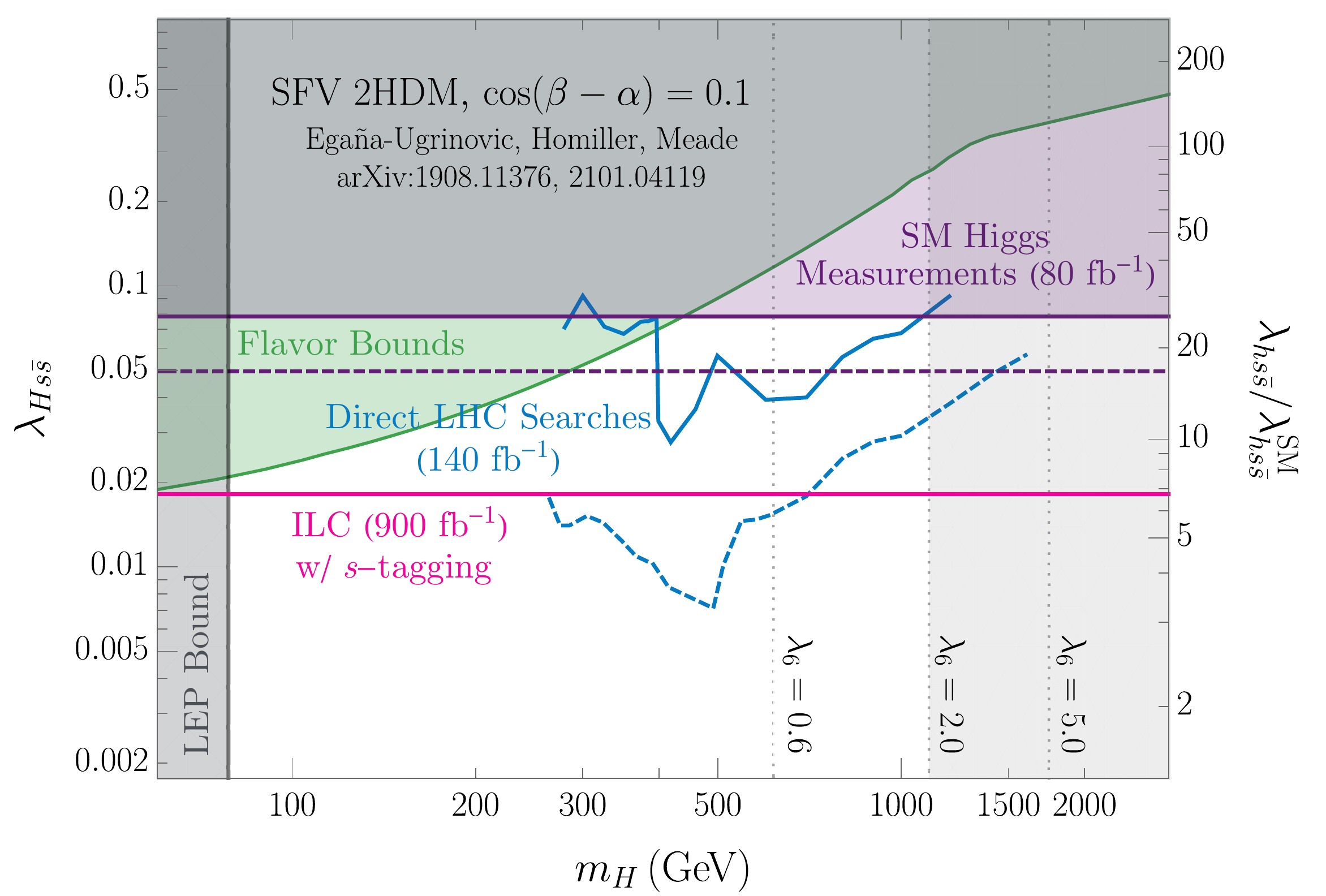}  
\par\end{centering}
\caption{A 2HDM with non standard Yukawa couplings, in the case of an enhanced coupling to the strange quark. 
}
\label{fig:2hdmflavor}
\end{minipage}
\end{figure}

Precision Higgs measurements probe the model parameter space as demonstrated in 
Fig. \ref{fig:2hdmfig}~\cite{Beniwal:2022kyv} and the improvement at lepton colliders for moderate $\tan\beta$ is apparent.  Figure \ref{fig:2hdmfig} (right) demonstrates the ability of a high-energy Muon Collider to probe the parameter space of the 2HDM models.  We note that the region of moderate $\tan\beta$ is best probed by $B$ decays. Limits from direct searches for the heavier Higgs bosons of the 2HDM are from
HL-LHC studies. For high $\tan\beta$, the decay of the heavier Higgs boson to $\tau^+\tau^-$ provides a stringent limit, as seen in Fig. \ref{fig:2hdmHL}~\cite{Bahl_2020}.

A wider range of phenomenology is allowed in the 2HDMs. In particular 2HDMs do not have to be restricted to the usual four types of natural flavor conserving models.  In Fig.~\ref{fig:2hdmflavor} an example of a Spontaneous Flavor Violation (SFV) 2HDM exemplifies the wide range of phenomenology associated to direct and indirect searches, as well as the new techniques proposed at the ILC for tagging strange quarks directly.  These models and measurements can also be further extended into relevant bounds on up and down quark Yukawas as also shown in the ILC white paper~\cite{Albert:2022mpk}.

There are of course numerous connections to deeper questions and additional models covered in the Higgs physics Topical Group report~\cite{Dawson:2022zbb}, but at a general level even
Fig.~\ref{fig:higgscentral2} does provide important lessons.  The first is that many observables map to fundamentally different questions related to the Higgs boson. It is, therefore, non-trivial to connect observables related to Higgs physics with fundamental questions.
This has been referred to as the ``Higgs Inverse Problem", in analogy with the previously coined LHC inverse problem for BSM physics.  Examples of this are given in the Higgs-physics~\cite{Dawson:2022zbb} as well as in the EW physics~\cite{Belloni:2022due} Topical Group reports.  The second important lesson, alluded to in Fig.~\ref{fig:higgscentral2}, is that Higgs related observables do not just fall into the standard $\kappa$ or EFT fits. If there are any deviations in Higgs couplings, or differential measurements etc., there {\em must} be new physics that couples to the Higgs boson which gives origin to it.   How it can be searched for is an ever expanding program and depends on the mass scale of new physics and collider energy.  As mentioned earlier in the context of Higgs width measurements, there is an ever expanding program of ``Higgs without Higgs" measurements and other types of differential probes being discovered.  Suffice it to say, even 10 years after the Higgs discovery, we are still in the earliest stages of fully exploiting the potential connection of the Higgs boson to BSM physics.

\FloatBarrier


\subsection{Heavy-flavor and top-quark physics}
\label{sec:TOPHF}

\begin{figure}[!ht]
    \centering
    \includegraphics[width=0.48\textwidth]{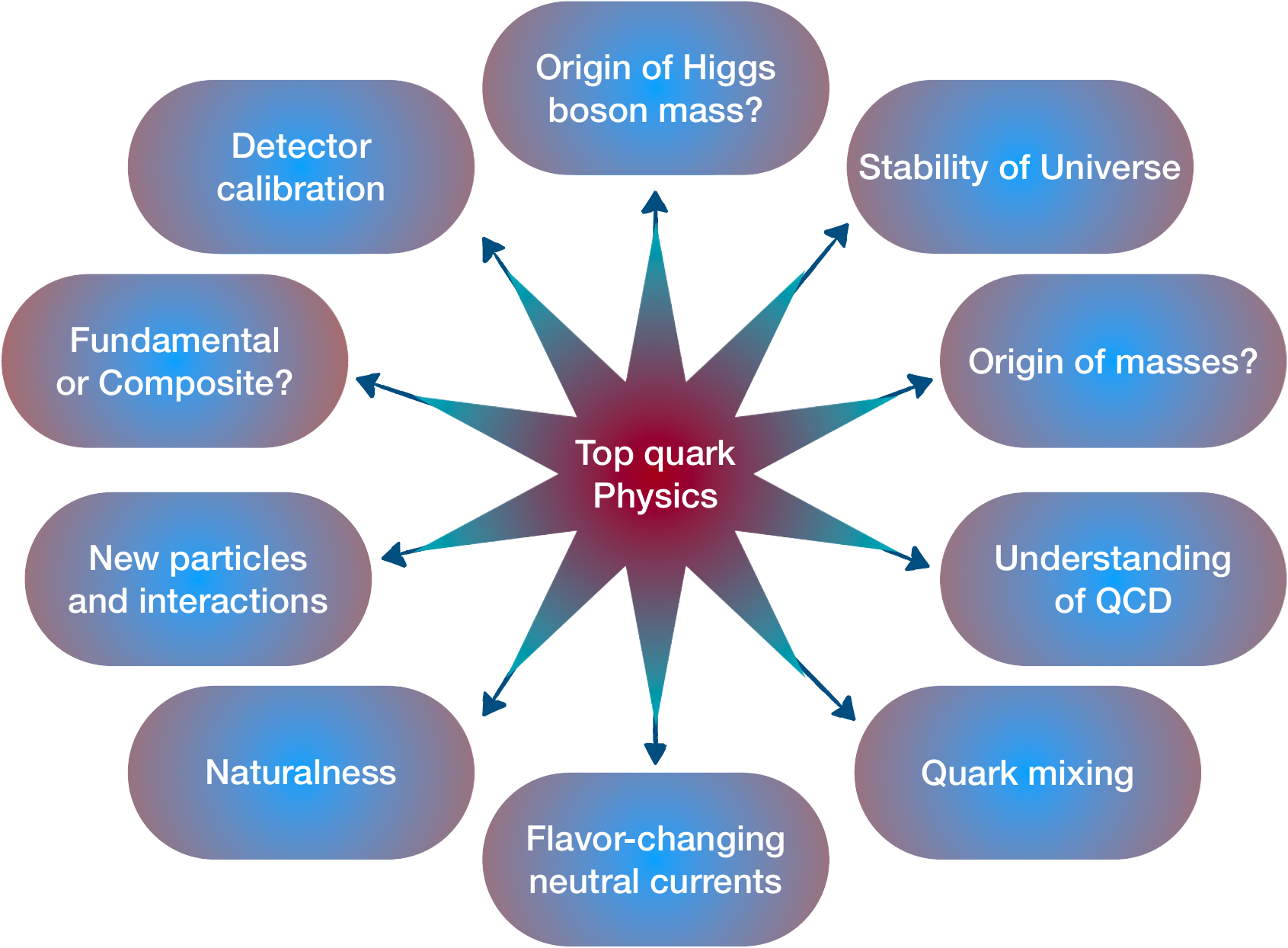}
    \caption{Illustration of the different aspects of the top-quark physics program.}
    \label{fig:top_bubbles}
\end{figure}

The top quark plays a special role in the EW sector of the SM, with a Yukawa coupling ($y_t$) of order unity
($y_t=\sqrt{2} m_t/v \approx 1$, where $m_t$ is the top-quark mass and $v$ is the vacuum expectation value of the Higgs field) introducing large quadratic corrections to the Higgs-boson mass, and affecting the stability of the EW
vacuum~\cite{Degrassi:2012ry}. The top-quark sector is therefore a crucial component of precision EW tests and particularly relevant in searches for BSM physics. Figure~\ref{fig:top_bubbles} illustrates the different topics that are addressed through studying top quarks. 
As the heaviest of all elementary particles, the top quark is relevant for understanding the origin of the Higgs-boson mass and all quark masses. 
Precise measurements of the masses of the top quark (see Section~\ref{sec:TOPHF-mtop}) together with the Higgs boson and the $W$~boson provide a stringent test of the EW sector of the SM. In addition, the top quark decays before it can hadronize, making it the only bare quark that can be studied directly, including at high momenta (see Section~\ref{sec:substructure}). Top-quark production (see Section~\ref{sec:TOPHF-topprod}) and decay kinematic information constrain top-quark EW couplings (see Section~\ref{sec:EWEFT}) and the CKM element $V_{tb}$. Searches for flavor-changing neutral currents (FCNC) and CP violation focus on the top-quark couplings. Direct searches for new particles and interactions look for top-quark partners, SUSY, and high-mass resonances decaying to top quarks (see Section~\ref{sec:BSM}).
Studies of top-quark production at the highest energies (multi-TeV colliders) probe models of compositeness (see Section~\ref{BSM:compositehiggs}). The abundance of top quarks at the LHC makes them ideal for detector calibration of bottom-quark tagging and bottom- and light-quark jet energy calibration.  
Top-quark production processes constitute an important background in many precision measurements and searches. Lepton colliders operating at or above the top-quark production threshold provide significantly improved measurements of the top-quark mass and its couplings. Precision measurements of top-quark and bottom-quark production at lepton colliders are the inputs needed to significantly improve the sensitivity of third generation and global EFT fits, examples of which are presented in Section~\ref{sec:EWEFT}. 
Moreover, running at the $Z$ pole at $e^+e^-$ colliders allows access to high-precision measurements of heavy-quark individual vector and axial-vector couplings. Very precise measurements of the $Z\to b\bar{b}$ rate can be of particular interest for top-quark precision physics given the special relationship of the bottom and top quarks in the SM and BSM models. The prospects for measurements of $Z\to b\bar b$ EW precision observables and of EW couplings $g_{Z,(L,R)}^{bb}$ from a SMEFT global analysis at various future colliders are shown in Table~\ref{tab:efmain:ewpocomp} and Fig.~\ref{fig:efmain:Fit1-HiggsEW}, respectively.

\subsubsection{Top-quark mass}
\label{sec:TOPHF-mtop}

\begin{figure}[!ht]
\includegraphics[width=0.44\linewidth]{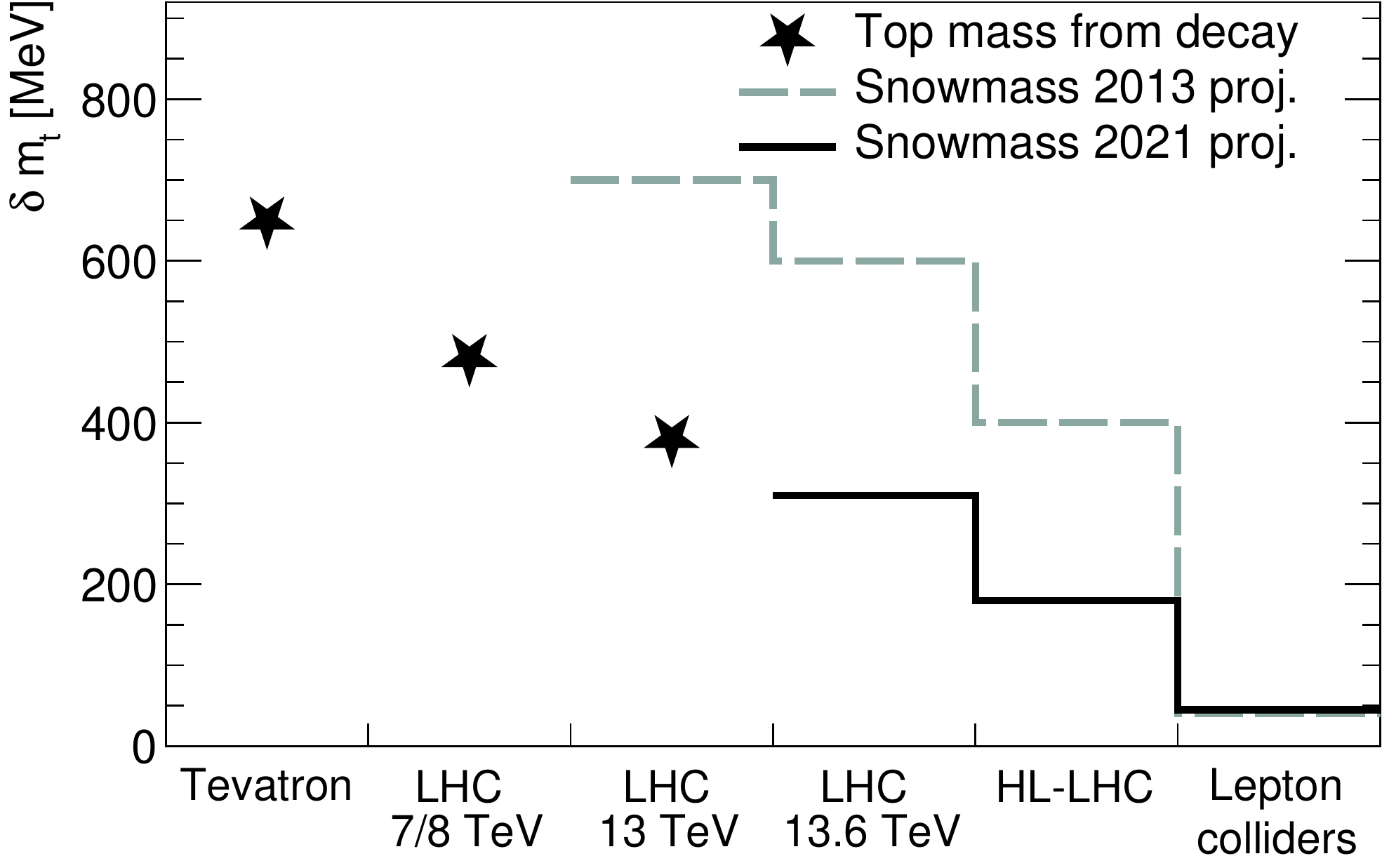}
\includegraphics[width=0.46\linewidth]{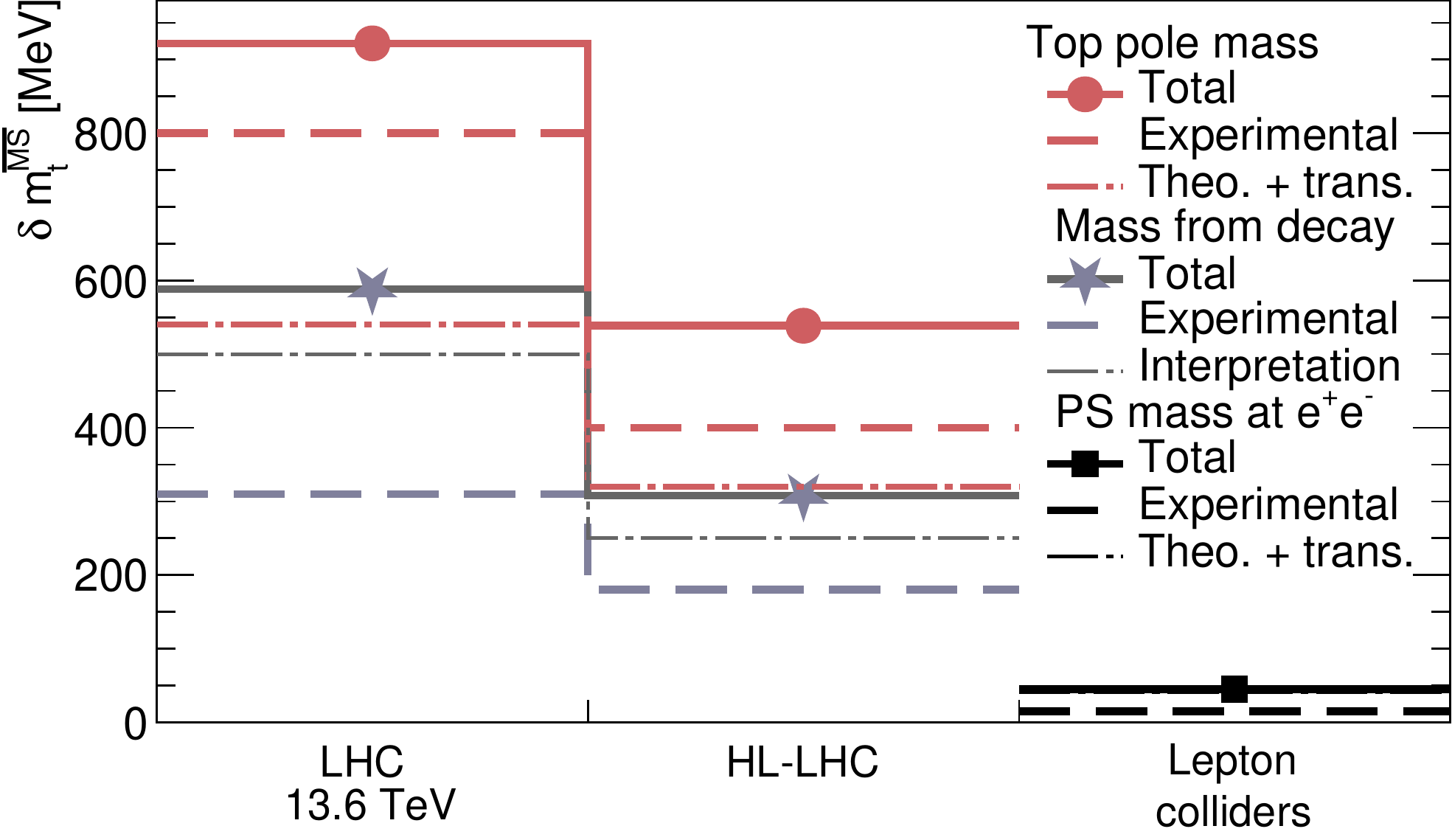}
\caption{(Left) Comparison of top-quark mass measurements from top decay (MC mass) at the Tevatron and the LHC, and projections for future LHC sensitivity and for future mass sensitivity from a top threshold scan at a lepton collider. The interpretation uncertainty for the mass from decay is not included~\cite{Agashe:2022plx}. (Right) 
Comparison of projected $\overline{\rm MS}$ top-quark mass determinations from decay and pole-mass measurements at the LHC and HL-LHC and from a PS mass measurement at a future lepton collider. The dashed-dotted lines show the approximate uncertainty in interpreting the mass from decay as $\overline{\rm MS}$ top-quark mass~\cite{Nason:2017cxd,Hoang:2020iah} (labeled ``\it{Interpretation}") or the combined uncertainty from theory and the conversion to the $\overline{\rm MS}$ scheme (labeled ``\it{Theo.+trans.}").
}
\label{fig:mtop_comp}
\end{figure}

The top-quark mass, $m_t$, is one of the most important parameters of the SM and relevant as an input for precise predictions and for the understanding of SM properties such as the stability of the spontaneously broken vacuum state. Top-quark loop corrections impact the mass of the $W$~boson, e.g. a top-quark mass change of 100~MeV changes the $W$~boson mass by 1~MeV~\cite{Awramik:2003rn}. Thus, given the expected precision on the $W$ mass at the HL-LHC and at future lepton colliders, precisions of better than 500~MeV and 50 MeV are required for top-quark mass measurements at the HL-LHC and at future lepton colliders, respectively, for precision EW fits~\cite{deBlas:2021wap,Haller:2018nnx} (see also Section~\ref{sec:EWprec}).
Since isolated quarks cannot be observed, the top-quark mass is not physical, but a
renormalization-scheme-dependent quantity. This scheme dependence can only be well-defined and controlled
for mass-sensitive observables that are calculable in perturbation theory (at least at the NLO level). The currently most precise top-quark mass determinations at the LHC are obtained from the direct reconstruction of the top-quark decay products (jets and leptons). Analytic perturbative QCD calculations are not available for the kinematic distribution of these objects, therefore the top-quark mass is extracted from comparisons to predictions by Monte Carlo (MC) event generators (so-called \textit{mass from decay} or \textit{MC mass}). It is estimated that interpreting the mass from decay in a well-defined scheme (like the $\bar{\text{MS}}$ scheme) has an uncertainty of about 500~MeV~\cite{Nason:2017cxd,Hoang:2020iah}.
The current and expected precision for measurements of the mass from decay are compared to the projections from Snowmass 2013 in Fig.~\ref{fig:mtop_comp} (left). The recent measurements significantly improved on the projections from 2013. 

Top-quark mass measurements in a well-defined scheme by contrast are based on comparisons of perturbative calculations of differential and total cross sections with measurements that are unfolded to the parton level (so-called \textit{indirect mass} or \textit{pole mass}). Current measurements for a top-quark mass measurement performed in the pole mass scheme have reached a precision of 1.2~GeV~\cite{Czakon:2016teh,ATLAS:2019guf,CMS:2019esx}.
The uncertainty is currently dominated by theoretical uncertainties, in particular due to Parton Distribution Functions (PDFs). For the projection, it is assumed here that the theory uncertainty can be reduced through dedicated PDF fits in $t\bar t$ events and through higher-order calculations of the $t\bar tj$ process. 

The ultimate precision in the top-quark mass will be reached in a scan of the top-quark production threshold at a lepton collider (here denoted as \textit{PS mass}). The corresponding expected precision for the PS mass for different $e^+e^-$ collider options is shown in Table~\ref{tab:TOPHF-PSmtop}. The overall uncertainty is expected to be limited by systematic uncertainties, in particular in the theoretical predictions, including the uncertainty on the strong coupling constant $\alpha_s$. The top-quark width is similarly measured with the highest precision at a lepton collider, in combination with the top-quark mass, in an energy scan of the top-production threshold. The Yukawa coupling of the top quark $y_t$ can be measured in the same scan from the plateau above the top-production threshold~\cite{Bernardi:2203.06520}. 
\begin{table}[!h!tbp]
\begin{center}
\begin{tabular}{l|c|c|c}
$\delta m_t^{{\rm PS}}$ [MeV] & ILC & CLIC  & FCC-ee \\ \hline
$\L [\mbox{fb}^{-1}]$              & 200  & 100 [200] & 200    \\ \hline
Statistical uncertainty &  10  & 20 [13] & 9  \\
Theoretical uncertainty (QCD) &  \multicolumn{3}{c}{40 -- 45}  \\
Parametric uncertainty $\alpha_s$ & 26 & 26 & 3.2 \\
Parametric uncertainty $y_t$ HL-LHC & \multicolumn{3}{c}{5}\\
Non-resonant contributions & \multicolumn{3}{c}{$<40$}\\
Experimental systematic uncertainty  &  \multicolumn{2}{c|}{20 -- 30} & 11 -- 20 \\
\hline
Total uncertainty       &  \multicolumn{3}{c}{40 -- 75} \\ \hline
\end{tabular}
\caption{Anticipated statistical and systematic uncertainties in the measurement of the threshold mass, $m_t^{\mathrm{PS}}$, from a threshold scan around 350 GeV obtained with a one-dimensional fit of the top-quark mass, keeping $\Gamma_t$, y$_t$, and $\alpha_s$ fixed. CLIC assumes a lower integrated luminosity than the other facilities. For comparison, the statistical precision achievable with 200 $\mbox{fb}^{-1}$ for CLIC is also given. It should be noted that the results shown for ILC and FCC-ee assume a 8-point scan with a compressed energy range which improves sensitivity for $m_t^{{\rm PS}}$ at the expense of $y_t$ sensitivity. For the standard 10-point scan assumed for CLIC the statistical uncertainties would be 12 and 10 MeV for ILC and FCC-ee, respectively. The uncertainty due to the current world average for $\alpha_s$ is shown for ILC and CLIC, while for FCC-ee, the run at the $Z$ pole (Tera-$Z$) will reduce this uncertainty significantly. Concrete studies for CEPC are not yet available, but it can be assumed that uncertainties are similar as for FCC-ee.}\label{tab:TOPHF-PSmtop}
\end{center}
\end{table}

To be able to compare the projections for the different top-quark mass measurements, they have been converted to projections for the $\bar{\text{MS}}$ top-quark mass, shown in Fig.~\ref{fig:mtop_comp} (right).  The dash-dotted lines show the approximate uncertainty in interpreting the mass from decay as a $\bar{\text{MS}}$ top-quark mass~\cite{Nason:2017cxd,Hoang:2020iah} or the combined uncertainty from theory and the conversion to the $\bar{\text{MS}}$ scheme (known to ${\cal O}(\alpha_s^4)$ in QCD) in case of the pole and PS mass measurements.

\subsubsection{Top-quark production processes}
\label{sec:TOPHF-topprod}

\begin{figure}[!ht]
    \centering
    \includegraphics[width=0.46\textwidth]{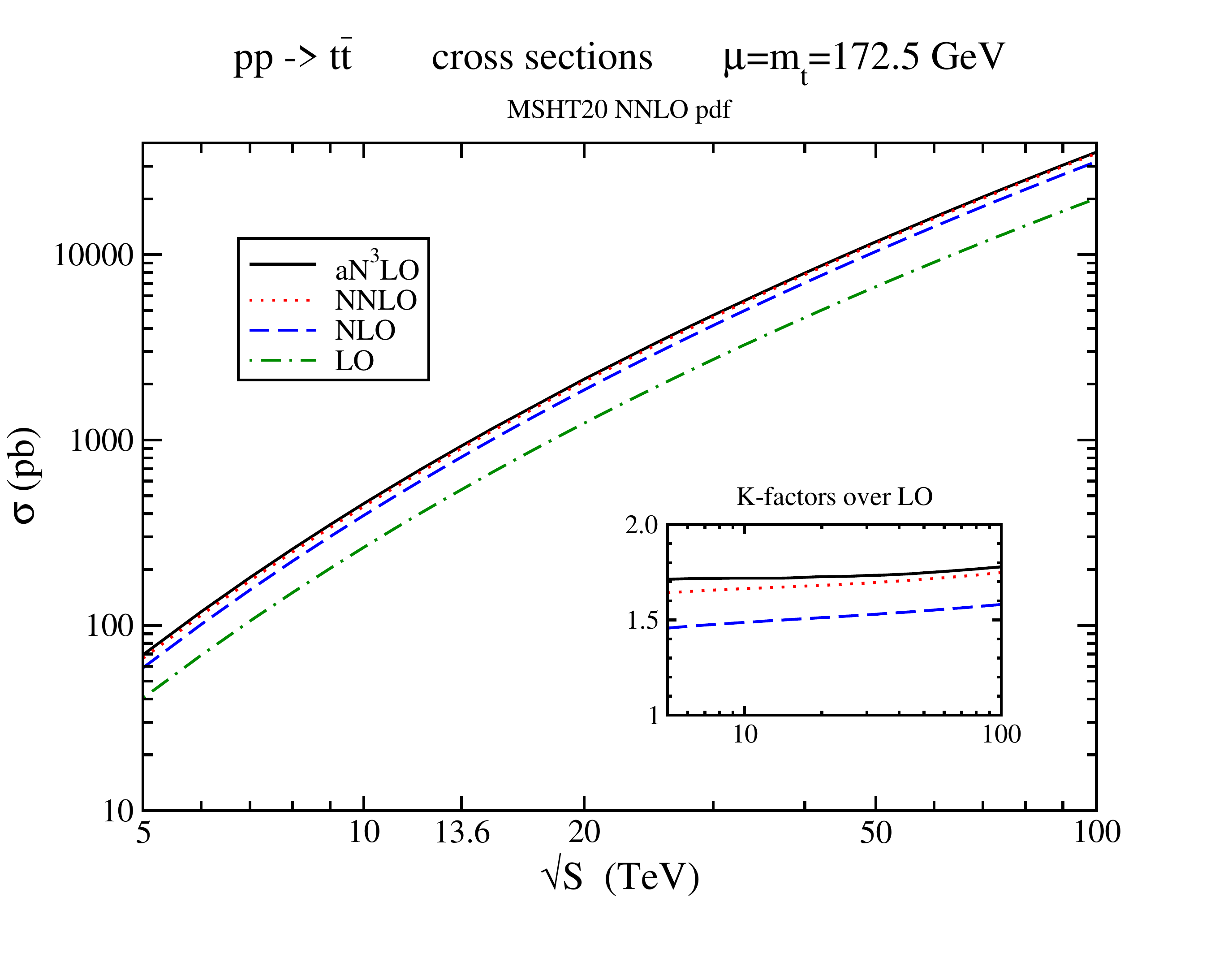}
    \includegraphics[width=0.48\textwidth]{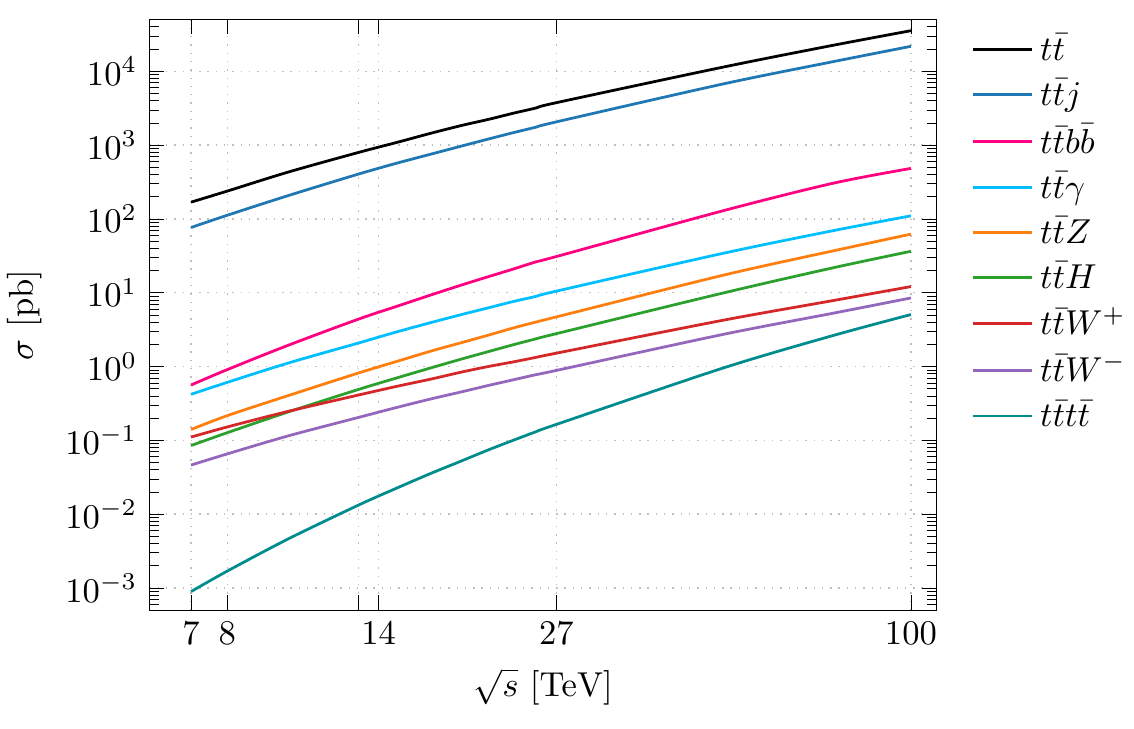}
    \caption{Total cross sections as a function of the 
    center-of-mass energy $\sqrt{s}$ at $pp$ colliders for (left) $t\bar{t}$ production at LO, NLO, \NNLOgen, and approximate N$^3$LO~\cite{Kidonakis:2203.03698} and for (right) various $pp\to t\bar{t}X$ processes at NLO. The $pp\rightarrow t\bar{t}$ cross section is also shown for reference. 
    Light objects are also required to have $p_T > 25$ GeV and
    $|\eta| < 2.5$, while jets are clustered using the anti-$k_T$ algorithm with $R=0.4$. }
    \label{fig:TOPHF-ttX}
\end{figure}

Top quarks are produced copiously at hadron colliders in many different production modes, $t\bar{t}$, single top, and both modes in association with other quarks and bosons. Modeling the different processes requires precision higher-order QCD calculations with heavy quarks. The cross sections for the production of top-quark pairs~\cite{Kidonakis:2203.03698}, as well as top-quark pairs in association with various other particles are shown in Fig.~\ref{fig:TOPHF-ttX}~\cite{Agashe:2022plx}. Measuring all of these processes is possible with high precision at the HL-LHC. Even the process with the lowest production cross section, the production of four top quarks, can be measured with an uncertainty of about 20\% at the HL-LHC, and should be measurable to a few percent at higher-energy hadron colliders (including FCC-hh)~\cite{CMS:2018nqq,ATLAS:2022kld}. This production mode has the highest energy threshold of all top-quark-related SM processes studied at hadron colliders. It is sensitive to $y_t$ and BSM interactions, for example contact interactions~\cite{CMS:2018nqq}. The measurements of $t\bar t H$ production directly probe the top Yukawa coupling.
The measurements of $t\bar t Z$, $t\bar t \gamma$ have intrinsic value, as they form the first constraints on the EW couplings of the top quark.
Precision measurements of the processes shown in Fig.~\ref{fig:TOPHF-ttX} and the corresponding single top-quark processes are utilized in global EFT fits. It should be possible to reach a precision of O(1\%) for $t\overline{t}$ production, larger for other processes. This requires careful calibration, improved modeling of top-quark and other processes, and improved theory calculations; currently the scale uncertainty on $\sigma(t\bar{t})$ is about 3\% at NNLO QCD with NNLL soft-gluon resummation~\cite{Czakon:2013goa}, while the PDF uncertainty is about 3\%~\cite{Butterworth:2015oua}. Extending the differential top-quark measurements to high, multi-TeV transverse momenta gives sensitivity to 4-fermion interactions involving the third generation.

At lepton colliders running at or above the top-quark production threshold allows for high-precision measurements of  $t\overline{t}$ production and of the couplings of the top-quark to the $Z$ boson. These measurements, as well as the corresponding measurements of the production of bottom-quark pairs at similar precision, will allow to significantly extend the sensitivity of global EFT fits, see Fig.~\ref{fig:Fit3-top3}. Producing $t\overline{t}$ in association with the Higgs, $W$ or $Z$ bosons requires significantly higher center-of-mass energies at a lepton collider. 

\subsubsection{Top-quark coupling measurements and EFT fits}
\label{sec:TOPHF-EFT}

\begin{figure}[!h!t]
\centering
\hspace*{1.cm}%
\includegraphics[width=0.65\columnwidth]{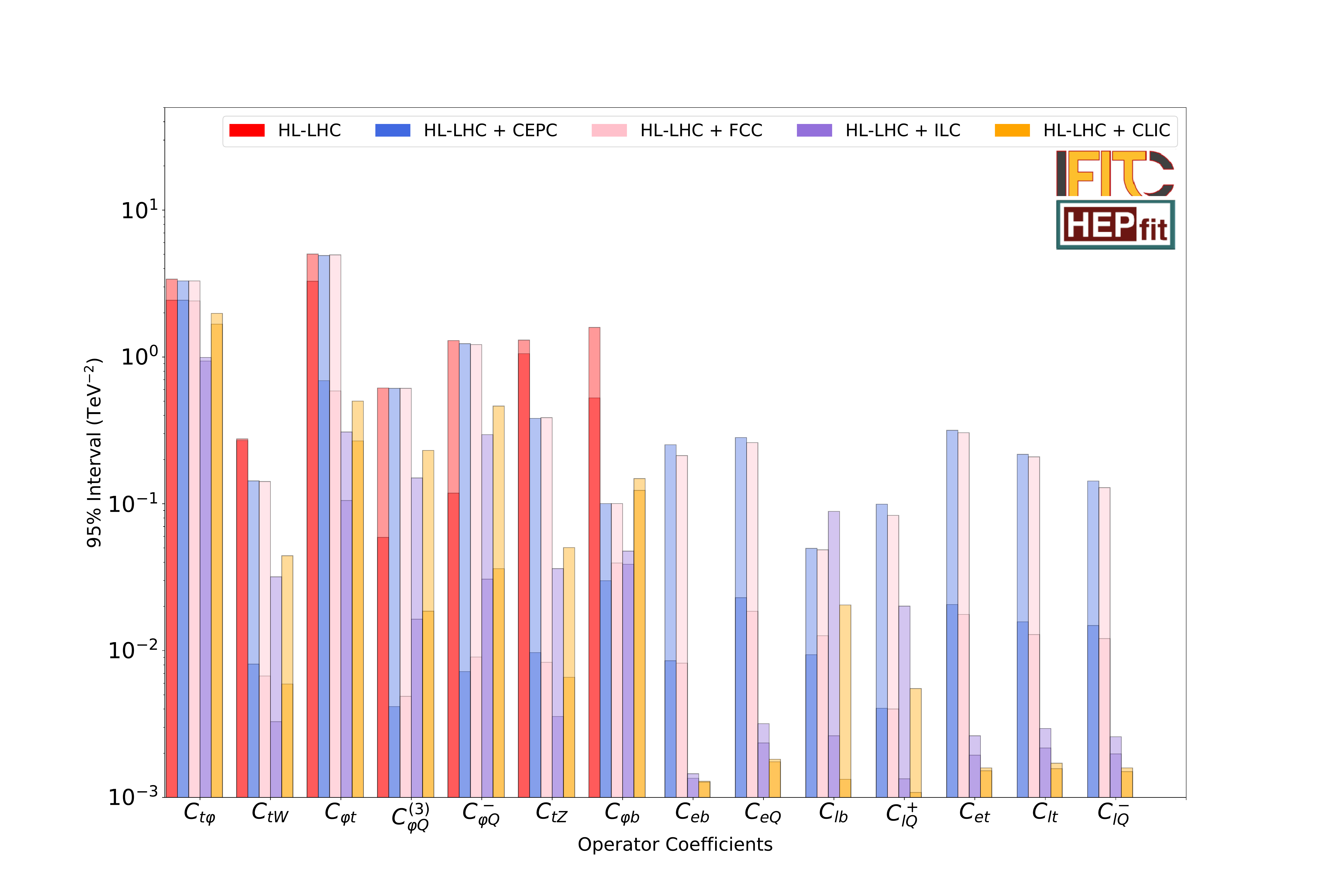}
\caption{Comparison of the constraints expected from a combination of HL-LHC and lepton collider data on Wilson coefficients for EFT operators relevant to top-quark couplings, see Section~\ref{sec:EWEFT}.
The solid bars provide the individual limits of the single-parameter fit and the shaded ones the marginalised limits of the global fit. The details of the fit can be found in Refs.~\cite{GlobalFitWhitePaper,Durieux:2019rbz}.
}
\label{fig:Fit3-top3}
\end{figure}

The measurements of cross sections of top-quark production processes provide important inputs to global EFT fits~\cite{Brivio:2019ius}. Differential production measurements and studies of top-quark decay and top-quark final state correlations provide further constraints~\cite{Durieux:2018tev}. At hadron colliders, these include top-quark pair and single top-quark production processes, and associated production, measured differentially. The precision of the measurements is limited by systematic uncertainties, the largest of which are due to jet energy calibration and QCD modeling of the top-quark final states. 
At lepton colliders, the final state can be fully reconstructed, and most measurements are effectively background-free, in contrast to hadron colliders. 
The lepton collider measurements constrain all EW couplings of the top quark. The gain in precision is particularly pronounced in the couplings to neutral bosons that are tested directly in $e^+e^- \to t\bar{t}$ production. The coupling to the $W$~boson is constrained strongly already in helicity fraction measurements at the Tevatron and the LHC and the improvement compared to HL-LHC is not as dramatic (see Section~\ref{sec:EWEFT}). Figure~\ref{fig:Fit3-top3} shows the reach of the HL-LHC and the improvement that can be expected from adding lepton collider data to a global EFT fit of the Wilson coefficients relevant for top-quark couplings. The fit uses cross sections for various top-quark production processes (Section~\ref{sec:TOPHF-topprod}) and angular correlations at the HL-LHC, and optimal variables at the lepton collider.

\subsubsection{BSM physics from top-quark physics}
\label{sec:TOPHF-fcnc}

\begin{figure}[!h!t]
\begin{minipage}{.5\textwidth}
\hspace{-0.5truecm}
\includegraphics[scale=0.26]{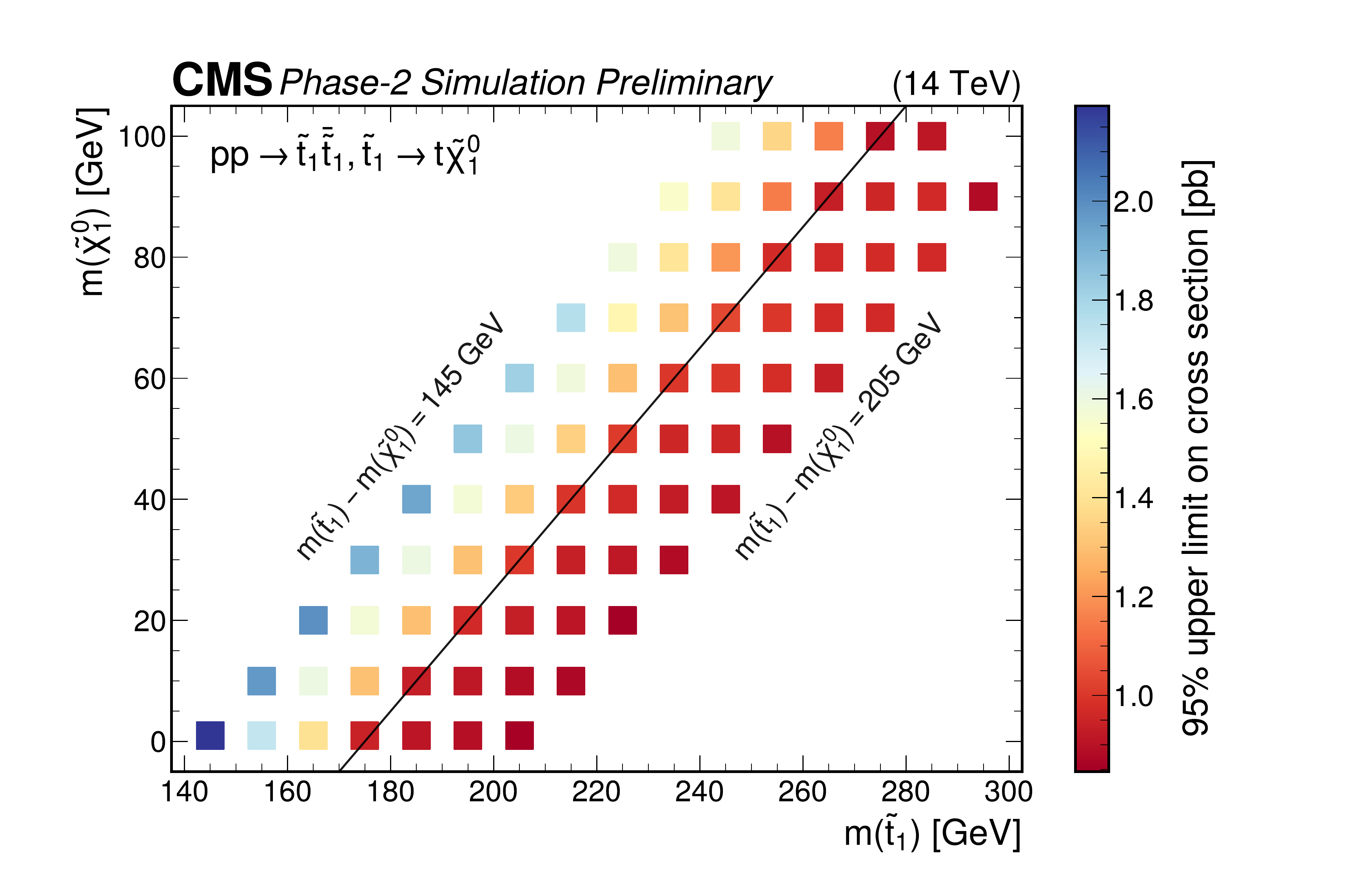}
\caption{Limit on the cross section for SUSY stop production in the compressed region where the stop mass ($m(\tilde{t})$) is close to the neutralino mass ($m(\chi_0)$), $m(\tilde{t}) - m(\chi_0) = 175$~GeV~\cite{CMS:spincor}. Every point in this plot is excluded~\cite{CMS:spincor}. }
    \label{fig:TOPHF-spincor}
\end{minipage}%
\begin{minipage}{.5\textwidth}
\begin{centering}
\includegraphics[scale=0.31]{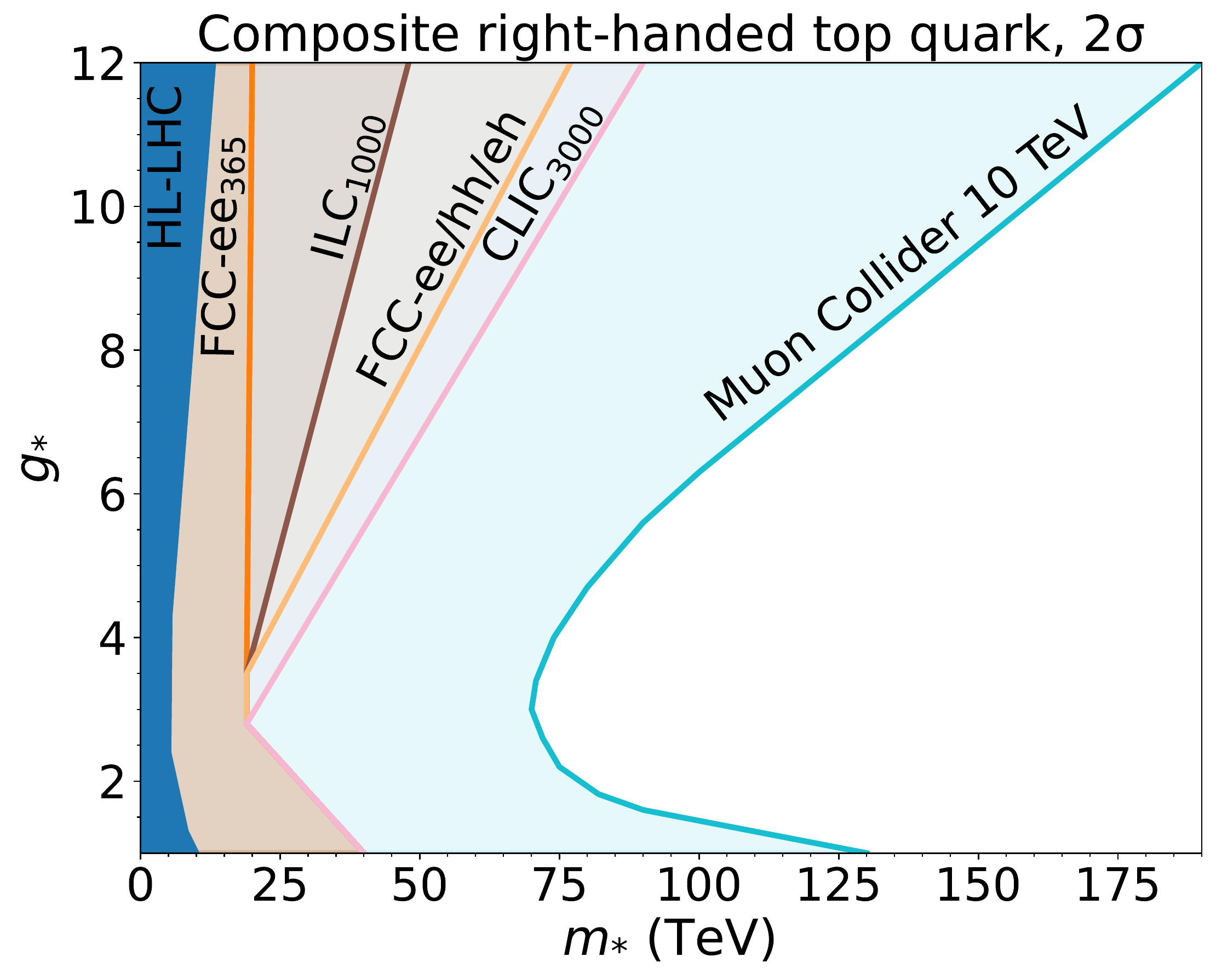} 
\par\end{centering}
\caption{Exclusion (2-$\sigma$) sensitivity projections for compositeness models at future colliders as labeled, for models where both the Higgs boson and the top quark with right-handed couplings are composite. Plot based on Refs.~\cite{Chen:2022msz,Banelli:2020iau}.}
    \label{fig:TOPHF-composite}
\end{minipage}
\end{figure}

Besides indirect searches for BSM physics in the EFT framework, the top quark is also a sensitive probe in direct searches for new physics. BSM models which aim to protect the Higgs boson mass from large quadratic corrections beyond the EW energy scale are often closely connected to the top quark, such as low-energy SUSY. For example, in \cite{CMS:spincor} the correlation of the spins of the two top quarks in $t\bar t$ production at the LHC are used to constrain SUSY models with top squarks with masses close to the top-quark mass (and small neutralino mass). Figure~\ref{fig:TOPHF-spincor} shows the projected limit for a 30~GeV-wide "corridor" in stop mass ($m(\tilde{t})$) and neutralino mass ($m(\chi_0)$) around the top-quark mass ($m(\tilde{t})-m(\chi_0)-m(t)|<30$~GeV)~\cite{CMS:spincor}. The width of the corridor corresponds to the experimental resolution and the region where direct stop searches are not sensitive because of the large $t\bar t$  background. The limits expected for the HL-LHC are a factor two (at low $m(\tilde{t})$) to ten (at high $m(\tilde{t})$) better than the Run~2 limits in this region. The predicted SUSY stop pair production cross section in this region is between 10~pb and 100~pb, meaning the entire area will be excluded.

As the heaviest fermion, it is also expected to play a central role in models of compositeness, together with the Higgs boson~\cite{Chen:2022msz}. Examples of the phenomenology of these type of models in the top-quark sector are the occurrence of new fermionic resonances (top partners), anomalous 4-top quark production, and modified top Yukawa and top-EW couplings. For example, Fig.~\ref{fig:TOPHF-composite} compares the reach of future colliders in the plane of mass scale and coupling $g*$ for a model of total right-handed top-quark compositeness, giving rise to sizeable 4-top Wilson coefficients~\cite{Banelli:2020iau}.

Top-quark interactions are also an excellent probe of FCNC involving the third generation of leptons. Searches for FCNC interactions in single top-quark production at hadron colliders (sensitive to gluon FCNC interactions) and lepton colliders (sensitive to photon and $Z$~boson FCNC interactions) take advantage of needing lower center-of-mass energy to produce one top quark rather than two. The large samples of top quarks collected at the LHC and expected at the HL-LHC allow for searches in the top-quark decay (sensitive to photon, $Z$~boson, and Higgs FCNC interactions). The limits on the top decay branching ratios are around $10^{-4}$ with the Run~2 dataset, these will be improved to around $10^{-5}$ at the HL-LHC. Lepton colliders are sensitive to FCNC couplings of the top quark to the photon and the $Z$~boson, especially at energies below the $t\bar{t}$ production threshold~\cite{Aryshev:2203.07622,Bernardi:2203.06520}. The production of a single top quark together with an up or charm quark provides a unique final state signature. Combining runs at multiple center-of-mass energies provides additional sensitivity, especially at the highest energies reached in $e^+ e^-$ only by CLIC~\cite{deBlas:2018mhx}. This is an area where a Muon Collider might also provide additional sensitivity.

\subsubsection{Heavy-flavor and top-quark physics summary}
Table~\ref{tab:TOPHF-couplings} compares a few top-quark measurements between different future collider options. Each of the measurements can be improved at future colliders beyond the precision at the HL-LHC. Significantly improving the precision of the top-quark Yukawa coupling beyond the 2-4\% uncertainty expected at the HL-LHC~\cite{ATLAS:2022hsp} and measuring the top-Higgs coupling in $t \bar t H$ events requires a high-energy lepton collider at a center-of-mass energy of at least 500~GeV or a hadron collider. The projected top-quark mass uncertainty at lepton colliders is obtained from Table~\ref{tab:TOPHF-PSmtop}, taking the lower end of each uncertainty range.
The precision of the coupling measurements to the SM bosons will all be significantly improved at a lepton collider operating at or above the top-production threshold. The four-top coupling can be probed at hadron colliders, or at lepton colliders running at sufficiently high energies. 
Searches for FCNC via the $Z$ boson or photon are done in top-quark decays at hadron colliders, the sensitivity is significantly extended at lepton colliders operating as a Higgs factory. These projections also rely on a significant theoretical effort to meet the tremendous challenges in providing theory predictions at the required level of precision.
\begin{table}[!h!t]
\begin{center}
\begin{tabular}{l|c|c|c|c}
Parameter & HL-LHC  & ILC 500 &  FCC-ee & FCC-hh \\ \hline
$\sqrt{s}$ [TeV] & 14 & 0.5 & 0.36 & 100  \\ \hline
Yukawa coupling $y_t$ (\%) & 3.4  & 2.8 & 3.1 & 1.0 \\
Top mass $m_t$ (\%) & 0.10 & 0.031 & 0.025 & -- \\
Left-handed top-$W$ coupling $C_{\phi Q}^3$  (TeV$^{-2}$) & 0.08 & 0.02 & 0.006 & -- \\
Right-handed top-$W$ coupling $C_{tW}$ (TeV$^{-2}$) & 0.3 & 0.003 & 0.007 & -- \\
Right-handed top-$Z$ coupling $C_{tZ}$ (TeV$^{-2}$) & 1 & 0.004 & 0.008 & -- \\
Top-Higgs coupling $C_{\phi t}$ (TeV$^{-2}$) & 3 & 0.1 & 0.6 & \\
Four-top coupling $c_{tt}$ (TeV$^{-2}$) & 0.6 & 0.06 & -- & 0.024 \\
\end{tabular}
\caption{Anticipated precision of top-quark Yukawa coupling and mass measurements, and of example EFT Wilson coefficient for the top-quark coupling to $W$, $Z$ and Higgs bosons, as well as a four-top Wilson coefficient. The reaches of the CEPC and SPPC are expected to mirror those of the FCC-ee and the FCC-hh respectively.}
\label{tab:TOPHF-couplings}
\end{center}
\end{table}


\subsection{Electroweak precision physics and new physics constraints}
\label{sec:EWEFT}

The precise measurement of physics observables and the test of their consistency within the SM are an invaluable approach, complemented by direct searches for new particles, to determine the existence of physics beyond the SM.

Indirect searches for new physics, which exploit off-shell and loop contributions of new particles, allow one to explore a much wider range of energy scales than those probed by direct searches in specific BSM scenarios. Such indirect BSM effects are typically inversely proportional to some power of the mass scale of the new degrees of freedom, so that high precision is crucial for probing large energy scales. The achievable precision of an experiment is determined by the statistics of the collected data sample, the experimental and theoretical systematic uncertainties, and their correlations.

\subsubsection*{Electroweak precision physics}
\label{sec:EWprec}

The current precision for a few selected electroweak precision pseudo-observables (EWPOs) is listed in Table~\ref{tab:efmain:ewpolhc}.
The HL-LHC with integrated luminosity of 3000~fb$^{-1}$ can make improved measurements of certain EWPOs. The effective weak mixing angle can be extracted from measurements of the forward-backward asymmetry in Drell-Yan production, $pp \to \ell^+\ell^-$ ($\ell=e,\mu$), while the $W$-boson mass can be determined from measurements of $pp \to \ell\nu$. Both measurements crucially depend on precise knowledge of PDFs and theory input for QCD and EW corrections, where the SM has to be assumed for the latter.

\begin{table}[!ht]
  \begin{center}
  \begin{tabular}{|c|c|c|}
    \hline
    EWPO Uncertainties        &    Current      & HL-LHC     \\
    \hline
    $\Delta\mw$ (MeV)          &    12 / 9.4$^\dag$  &    5      \\
    $\Delta\mt$ (GeV)    &  0.6* &  0.2 \\
    $\Delta\sin\theta^\ell_\eff$ ($\times 10^5$)   &    13 & $<10$     \\ 
    \hline    
  \end{tabular}\\[2ex]
  $^\dag$ \parbox[t]{.6\textwidth}{\footnotesize{The recent $W$ mass measurement from CDF with 9.4~MeV precision \cite{CDF:2022hxs} has not yet been included in the global average \cite{ParticleDataGroup:2020ssz}.}}\\[1ex]
  * \parbox[t]{.6\textwidth}{\footnotesize{This value includes an additional uncertainty due to ambiguities in the top mass definition (see~\cite{Agashe:2022plx} for more details).}}
  \end{center}
\vspace{-1ex}
\caption{The current precision of a few selected EWPOs, based on data from LEP, SLC, Tevatron and LHC \cite{ParticleDataGroup:2020ssz}, and expected improvements from the HL-LHC \cite{ATLAS:2022hsp}. $\Delta$ ($\delta$) stands for the absolute (relative) uncertainty.}
\label{tab:efmain:ewpolhc}
\end{table}

Future high-luminosity $e^+e^-$ colliders can be used to study the masses and interactions of electroweak bosons to much higher precision than before. We here focus on four collider proposals: ILC \cite{Baer:2013cma,Bambade:2019fyw,ILCInternationalDevelopmentTeam:2022izu}, CLIC \cite{Linssen:2012hp,Charles:2018vfv}, FCC-ee \cite{Abada:2019zxq,Bernardi:2022hny}, and CEPC \cite{CEPCStudyGroup:2018ghi,Cheng:2022zyy}. For ILC, CLIC and FCC-ee, we use the run scenarios and integrated luminosities in Table~\ref{tabHiggsFactory}, whereas for CEPC the 50~MW upgrade is assumed, which corresponds to 100~ab$^{-1}$ on the $Z$ pole, 6~ab$^{-1}$ at the $WW$ threshold, and 1~ab$^{-1}$ at the $t\bar{t}$ threshold \cite{Cheng:2022zyy}. For ILC also the Giga-$Z$ option with 100~fb$^{-1}$ on the $Z$ pole is considered. [Note that a $Z$-pole run is also proposed as a possible option for CLIC \cite{Gohil:2687090}.]
Table~\ref{tab:efmain:ewpocomp} summarizes the achievable precision for a range of EWPOs.

\newcommand{\hdashline}{\hline}
\begin{table}[!ht]
  \begin{center}
  \begin{tabular}{|c|c|c|c|c|c|c|}
    \hline
    Quantity       & current  &    ILC250      & ILC-GigaZ   &      FCC-ee            &  CEPC      &  CLIC380     \\
    \hline
    $\Delta\alpha(\mz)^{-1}\;(\times 10^3)$ &    17.8$^*$  & 17.8$^*$  &             &    3.8 (1.2)  & 17.8$^*$ &           \\
    $\Delta\mw$ (MeV)     & 12$^*$     &    0.5 (2.4)  &             &    0.25 (0.3)   & 0.35 (0.3)   &           \\
    $\Delta\mz$ (MeV)    & 2.1$^*$      &    0.7 (0.2)  &  0.2           &    0.004 (0.1)   & 0.005 (0.1)   &    2.1$^*$       \\
    $\Delta\mh$ (MeV)   & 170$^*$       &    14  &             &    2.5 (2)   & 5.9   &    78       \\
    $\Delta\gw$ (MeV)     & 42$^*$ &    2   &             &    1.2 (0.3)   & 1.8 (0.9)   &           \\    
    $\Delta\gz$ (MeV)     & 2.3$^*$ &    1.5 (0.2)  &  0.12    &    0.004 (0.025)   & 0.005 (0.025)   &    2.3$^*$       \\ 
    \hdashline
    $\Delta A_e\;(\times 10^5)$   & 190$^*$ &    14 (4.5)  &   1.5 (8)  &    0.7 (2)   & 1.5 (2)   &    60 (15)       \\ 
    $\Delta A_\mu\;(\times10^5)$  & 1500$^*$ &    82 (4.5)  &   3 (8)  &    2.3 (2.2)   & 3.0 (1.8)   &    390 (14)      \\    
    $\Delta A_\tau\;(\times10^5)$ & 400$^*$ &   86 (4.5)  &   3 (8)  &    0.5 (20)   & 1.2 (20)   &    550 (14)      \\        
    $\Delta A_b\;(\times10^5)$   & 2000$^*$ &   53 (35)  &   9 (50)  &    2.4 (21)   & 3 (21)  &    360     (92)  \\
    $\Delta A_c\;(\times10^5)$  & 2700$^*$  &  140 (25)  &  20 (37)  &    20 (15)   & 6 (30)   &    190     (67)  \\ 
    \hdashline
    $\Delta \sigma_{\rm had}^0$ (pb)  & 37$^*$ &           &          &    0.035 (4)      & 0.05 (2)     &    37$^{*}$       \\     
    $\delta R_e\;(\times10^3)$    & 2.4$^*$ &    0.5 (1.0)    &   0.2 (0.5)  &    0.004 (0.3)   & 0.003 (0.2)   &    2.5 (1.0)       \\
    $\delta R_\mu\;(\times10^3)$  &  1.6$^*$  & 0.5 (1.0)    &   0.2 (0.2)  &    0.003 (0.05)  & 0.003 (0.1)   &    2.5 (1.0)       \\
    $\delta R_\tau\;(\times10^3)$ &  2.2$^*$ &  0.6 (1.0)    &   0.2 (0.4)  &    0.003 (0.1)   & 0.003 (0.1)   &    3.3 (5.0)       \\
    $\delta R_b\;(\times10^3)$    & 3.1$^*$ &   0.4 (1.0)    &   0.04 (0.7)  &   0.0014 ($<0.3$)   & 0.005 (0.2)   &    1.5 (1.0)       \\
    $\delta R_c(\times10^3)$    &  17$^*$ &  0.6 (5.0)    &   0.2 (3.0)  &    0.015 (1.5)   & 0.02 (1)   &    2.4 (5.0)       \\
    \hline    
  \end{tabular}
  \end{center}
\vspace{-1ex}
\caption{EWPOs at future $e^+e^-$ colliders: statistical error (estimated experimental systematic error). $\Delta$ ($\delta$) stands for absolute (relative) uncertainty, while * indicates inputs taken from current data \cite{ParticleDataGroup:2020ssz}. See Refs.~\cite{deBlas:2019rxi,DeBlas:2019qco,LCCPhysicsWorkingGroup:2019fvj,Blondel:2021ema,ILCInternationalDevelopmentTeam:2022izu,Cheng:2022zyy}.}
\label{tab:efmain:ewpocomp}
\end{table}

The table separately lists the expected statistical and experimental systematic uncertainties. Note that the latter are based on assumptions about future performance improvements that cannot be substantiated at this time. Uncertainties due to the physics modeling affect all collider proposals equally. As part of the Snowmass 2021 process, a consistent set of assumptions is being used and applied uniformly. See EF04 Topical Group report for more details~\cite{Belloni:2022due}.

\begin{figure}[t]
    \centering
    \includegraphics[width=0.60\textwidth]{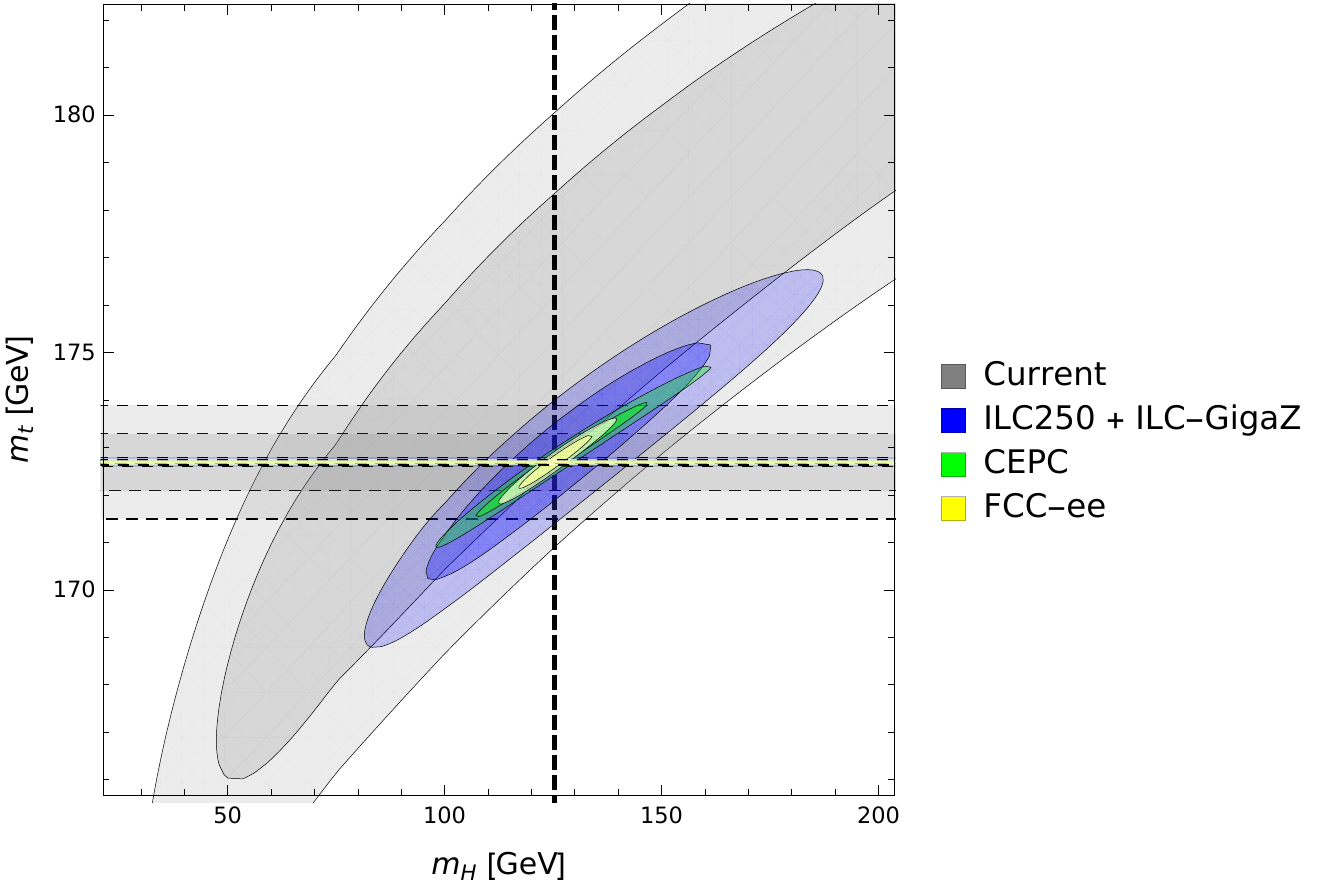}
    \medskip
    \caption{Indirect sensitivity to $\mh$ and $\mt$ for a fit of SM theory predictions to current and projected future data for electroweak precision tests ($W$ mass and $Z$-pole quantities). For comparison, the direct measurement precision is also shown (on the scale of the plot the width of the $\mh$ band is not visible). The light (dark) shaded areas depict 95\% (68\%) confidence level regions. For the future collider scenarios it is assumed that the central values coincide with the SM expectations.}
    \label{fig:efmain:FitSM-EW}
\end{figure}

The impact of these estimated future precision measurements on the indirect determination of the Higgs-boson and top-quark mass is illustrated in Fig.~\ref{fig:efmain:FitSM-EW}. The dependence on $\mh$ and $\mt$ appears in loop corrections to the SM theory predictions for $Z$ coupling parameters and the $W$ mass, and their agreement with direct measurements of these masses is a highly non-trivial test of the SM. 

 For ``canonical'' electroweak precision measurements (Z-pole, $WW$ threshold), circular $e^+e^-$ colliders (FCC-ee, CEPC) have in general a higher sensitivity than linear colliders (ILC, CLIC) due to the high luminosity at center-of-mass energies below 200~GeV. Beam polarization at the linear colliders improves their sensitivity and can help to control systematics. In particular, for a linear collider run on the $Z$ pole, beam polarization would enable measurements of the asymmetry parameters $A_f$ with a precision that is only a factor of a few worse than for circular colliders, in spite of several orders of magnitude larger statistics for $Z$-pole physics at circular colliders.

For many of the most precisely measurable precision observables at linear colliders, the most significant source of experimental systematics stems from the polarization calibration. For the circular colliders, on the other hand, modeling uncertainties for hadronic final states appear to be the dominant systematic error source.

To exploit the full potential of the anticipated precision of any future $e^+e^-$ collider, theory inputs are needed on multiple fronts. Accurate MC tools for the simulation of QED and QCD radiation are crucial for the evaluation of acceptance effects, and theory calculations including higher-order effects are needed for the prediction of irreducible backgrounds. For the interpretation of electroweak precision measurements, one needs to compare the measured values to their expectation within the SM, which requires multi-loop theory computations. For the anticipated experimental precision FCC-ee, CEPC, ILC or CLIC, the current state of the art of theory calculations needs to be extended by at least one order of perturbation theory, i.e. \NNLOgen/NLL contributions for MC tools and backgrounds, and N$^3$LO and partial N$^4$LO contributions for the SM predictions. See EF04 Topical Group report for more details~\cite{Belloni:2022due}.

The SM predictions also rely on other SM parameters as inputs, such as the electromagnetic coupling at the weak scale, $\alpha(\mz)$, the top-quark mass, $\mt$, and the strong coupling, $\alphas$. The current uncertainties for these parameters would severely limit the possibility for future high-precision studies, and thus it is necessary to perform improved measurements of these quantities at the future $e^+e^-$ colliders. See sections~\ref{sec:TOPHF-mtop} and \ref{sec:QCD-alphas} for more information.

\begin{table}[tb]
  \begin{center}
  \begin{tabular}{|c|c|c|c|}
    \hline
    EWPO & Current & \multicolumn{2}{c|}{Projected param.\ error} \\
    \cline{3-4}
    uncertainties        &    param.\ error & Scenario 1 & Scenario 2    \\
    \hline
    $\Delta\mw$ (MeV)          & 5  & 2.8 & 0.6  \\
    $\Delta\gz$ (MeV)     & 0.5 & 0.3 & 0.1 \\ 
    $\Delta\sin^2\theta^\ell_\eff\; (\times 10^{5})$   & 4.2 & 3.7 & 1.1   \\ 
    $\Delta A_\ell\; (\times 10^{5})$   &  30 & 25 & 7.5   \\ 
    $\delta R_\ell\;(\times10^3)$  &  6 &  3.2 & 1.3  \\
    \hline    
  \end{tabular}

\medskip
\begin{tabular}{|l|ccccc|}
\hline
 Input par.\ uncertainties & $\Delta\mt$ [GeV] & $\Delta\mh$ [GeV] & $\Delta\mz$ [MeV] & $\Delta(\Delta\alpha)$ & $\Delta\alphas$ \\
\hline
Current & 0.6 & 0.17 & 2.1 & $10^{-4}$ & $9\times 10^{-4}$ \\
Scenario 1 & 0.3 & 0.02 & 0.8 & $10^{-4}$ & $5\times 10^{-4}$ \\
Scenario 2 & 0.05 & 0.01 & 0.1 & $3 \times 10^{-5}$ & $2\times 10^{-4}$ \\
\hline
\end{tabular}
  \end{center}
\caption{Impact of uncertainties of SM input parameters on the prediction of a few selected EWPOs in top table (see Ref.~\cite{Freitas:2019bre}). Current uncertainties are compared to two future scenarios in the bottom table.}
\label{tab:ewpopar}
\end{table}
The impact of the uncertainties of the SM input parameters on the interpretation of electroweak precision measurements is illustrated in Table~\ref{tab:ewpopar}. In addition to the current measurement precision, two future scenarios are considered, where Scenario 1 assumes improvements from a Higgs factory with moderate luminosity spent on the $Z$ pole and no $t\bar{t}$ running, whereas Scenario 2 displays the full potential of achievable precision at future $e^+e^-$ colliders.
Note that the dependence of the predictions for $\gz$ and $R_\ell$ on $\alphas$ are to a certain extent circular, since these quantities would be used for the extraction of the strong coupling constant at future $e^+e^-$ colliders \cite{dEnterria:2022hzv}.

Experiments at lower-energy $e^+e^-$ colliders, lepton-proton colliders, or neutrino scattering facilities can deliver complementary information about electroweak quantities.
For instance, a beam polarization upgrade of  SuperKEKB/Belle II can enable a measurement of the electroweak mixing angle via tests of vector coupling universality in neutral currents with unprecedented precision. Electron-proton colliders such as EIC, LHeC and FCC-eh could provide information about the running electroweak mixing angle at scales of a few tens of GeV and enable the separate determination of up- and down-quark electroweak couplings with high precision.

A Muon Collider with center-of-mass energy $\sqrt{s} \approx 91$~GeV \cite{Blondel:1997eq} would also be very interesting for electroweak precision measurements, but more studies are needed.

\subsection{EFT and new physics}
\subsubsection{Multi-boson processes}
The SM predicts the existence of multi-boson interactions, which give rise to final states with two or three bosons. Anomalies in the rate and kinematics of these final states can be indicative of new physics not currently described in the SM. Such anomalies can be parametrized through modifications of the strength or form of the SM multi-boson vertices. A newer approach consists in using EFT operators of dimension six or above, where measurements of multi-boson processes can be recast as  determinations of the Wilson coefficients of these operators.

It shall be noted that the sensitivity to BSM effects, or, in other terms, the upper limits to the Wilson coefficients of new operators, scale with a power of the center-of-mass energy, thus making multi-TeV colliders the ideal tools for studying these final states. At this time, the most promising avenues for reaching multi-TeV energies are proton-proton colliders or $\mu^+\mu^-$ colliders.

High-energy ($> 1$ TeV) lepton colliders are effectively boson colliders. The total cross section for many production processes is dominated by vector-boson fusion (VBF) and/or vector-boson scattering (VBS) contributions. However, for studies of BSM effects at very high invariant masses, non-VBF processes become typically more dominant.

At multi-TeV lepton colliders, multiple electroweak gauge-boson production is ubiquitous, and new theoretical tools are needed for calculating and simulating these effects (e.g. theory modeling of EW PDF and fragmentation, initial-state - ISR - and final-state - FSR- radiation).

Plentiful experimental results with multi-boson final states are available.
Both the ATLAS and CMS collaborations have measured di-boson \cite{Aaboud:2019lgy,Aaboud:2019nkz,Aaboud:2019gxl,Aaboud:2018jst,Aaboud:2017rwm,Sirunyan:2019gkh,Sirunyan:2019ksz,Sirunyan:2019bez,Sirunyan:2017zjc}, tri-boson processes \cite{Aad:2019dxu,Aaboud:2017tcq,Sirunyan:2020cjp,Sirunyan:2017lvq}, as well as VBF/VBS processes \cite{Aad:2020zbq,Aad:2019wpb,Aad:2019xxo,Aaboud:2019nmv,Aaboud:2018ddq,CMS:2020zly,Sirunyan:2020gyx,Sirunyan:2020tlu,Sirunyan:2019der,Sirunyan:2017fvv,Sirunyan:2018vkx}, which are characterized by a $VVjj$ final state. Di-boson final states include $W^+W^-$, same-sign $W^\pm W^\pm$, $WZ$, $ZZ$, $Z\gamma$. Tri-boson final states include $W\gamma\gamma$, $Z\gamma\gamma$, $WV\gamma$ (where $V=W,Z$), and $WVV'$ (where $V,V'=W,Z$). 

Bounds on new physics have been determined in the language of anomalous gauge-boson couplings (aGCs) \cite{Aaboud:2019lgy,Aaboud:2018jst,Aaboud:2017rwm,Sirunyan:2019gkh,Sirunyan:2017zjc} and effective operators \cite{Aaboud:2019nkz,Sirunyan:2020gyx,Sirunyan:2020tlu,Sirunyan:2019gkh,Sirunyan:2019der,Sirunyan:2019ksz,Sirunyan:2019bez,Sirunyan:2017fvv,Sirunyan:2017lvq}. The latter is theoretically preferred since it provides a consistent power counting and allows one to implement theoretical consistency constraints. In these studies, only one or two aGCs/operators are allowed to be non-zero at the same time, i.e. no full aGC/SMEFT analysis has been performed.

The most up-to-date limits on gauge-coupling anomalies are available at Refs.~\cite{cmspublic_tgc,atlaspublic_tgc}.
Expected limits at the end of the HL-LHC and potential HE-LHC runs are reported in Ref.~\cite{Azzi:2019yne}.

There are ongoing studies for the determination of aGCs or multi-gauge boson SMEFT operators at future hadron and lepton colliders (see e.g. Ref.~\cite{Abbott:2022jqq,Apyan:2022gis}). Preliminary results indicate that a 100-TeV $pp$ collider or a multi-TeV $\mu^+\mu^-$ collider would have strongly enhanced sensitivity to these BSM effects compared to the LHC, but more work is needed to obtain a comprehensive picture.

\subsubsection{SMEFT global fits} \label{sec:gsmeft}

Assuming new physics scales are significantly higher than the EW scale, EFTs provide a model-independent prescription that allows us to put generic constraints on new physics and to study and combine large sets of experimental data in a systematically improvable QFT approach. 
All new physics effects are represented by a set of higher dimensional operators which consist of only the SM fields and respect the SM gauge symmetries. 

The EFT approach has some features that are of particular interest for studying precision EW physics, for instance: it provides a well-defined theoretical framework that enables the inclusion of radiative corrections for both the SM and BSM parts; 
and the synergies between different precision EW measurements can be explored globally so that a comprehensive picture of the
constraints on new physics can be drawn.
However, the EFT approach also has some practical limitations since it has in principle an infinite number of degrees of freedom, and it is only an adequate description if the new physics scales are larger than the experimentally reachable energies. 
In a realistic global EFT fit, various flavor assumptions and truncations to the lowest order of 
relevant operators often have to be applied.

A model-independent parametrization of the new-physics reach of different colliders is given by the SMEFT framework, where the SM is extended by higher-dimensional operators, with the leading contribution to most observables furnished by dimension-6 operators. Several subsets of such dimension-6 operators have been investigated in a series of global fits across a large number of observables: (a) operators contributing to electroweak gauge-boson interactions; (b) operators contributing to Higgs interactions; (c) operators contrbuting to top-quark interactions; and (d) operators contributing to four-fermion contact interactions.

For Snowmass 2021, the global EFT fit for the European Study Group (ESG)~\cite{deBlas:2019rxi} has been extended in a few directions~\cite{GlobalFitWhitePaper}: the consistent implementation of full EFT treatment in $e^+e^-\to W^+W^-$ using optimal observables;
the new inclusion of a large set of 4-fermion operators; a more complete set of operators that are related to top-quarks. In all the fits, operators for third-generation fermions are treated independently, without assuming flavor universality, and in some cases even universality between the first two generations has been lifted. However, no flavor-changing operators were included in the analysis. The projections of the uncertainties of 
required input observables are provided by EF01 for Higgs related observables~\cite{Dawson:2022zbb}, EF03 for top-quark related observables~\cite{Agashe:2022plx}, EF04 for $W/Z$ related observables~\cite{Belloni:2022due}, and the 
Rare Process and Precision Measurements Frontier for a set of low-energy measurements~\cite{Artuso:2022ouk}. Care has been taken to ensure that the various inputs are consistent and based on similar assumptions, e.g. by using extrapolations to compare inputs from two different $e^+e^-$ colliders.

\begin{figure}[ht]
\begin{center}
    \includegraphics[width=0.8\textwidth]{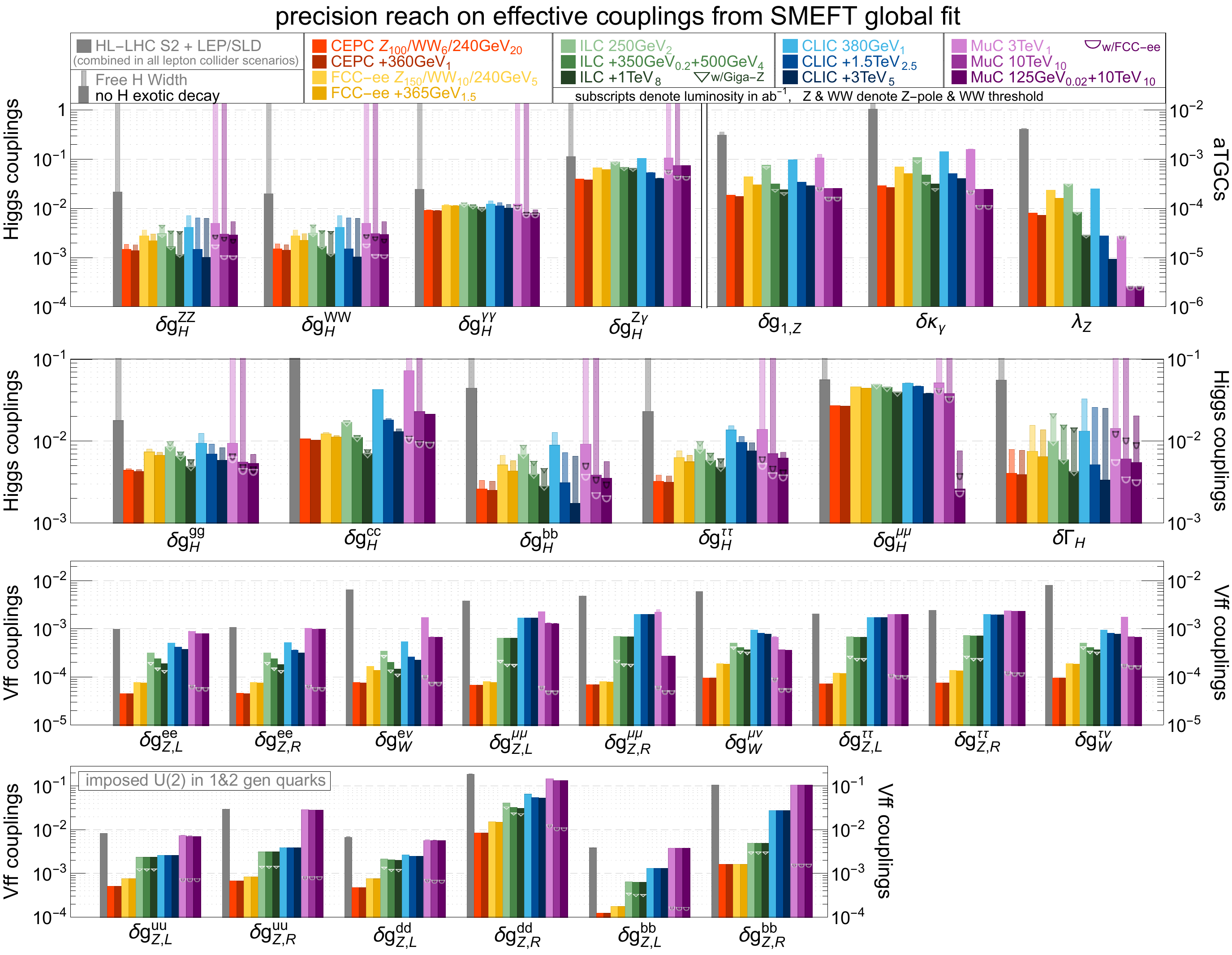}
    \caption{Precision reach on Higgs and electroweak effective couplings from an SMEFT global analysis of the Higgs and EW measurements at various future colliders.  The wide (narrow) bars correspond to the results  
    from the constrained-$\Gamma_H$ (free-$\Gamma_H$) fit. The HL-LHC and LEP/SLD measurements are combined with all future lepton collider scenarios.  For $e^+e^-$ colliders, the high-energy runs are always combined with the low energy ones.  For the ILC, the (upper edge of the) triangle mark shows the results for which a Giga-$Z$ run is also included.  For the Muon Collider, three separate scenarios are considered.  The subscripts in the collider scenarios denote the corresponding integrated luminosity of the run in ${\rm ab}^{-1}$.}
    \label{fig:efmain:Fit1-HiggsEW}
    \end{center}
\end{figure}

Figure~\ref{fig:efmain:Fit1-HiggsEW} displays the result of the global EFT fit for the subset of operators that affect Higgs and EW observables. Instead of showing the projected constraints on the Wilson coefficients of the operators considered, they have been translated into constraints on the effective Higgs and gauge-boson couplings. See EF04 Topical Group report for more details~\cite{Belloni:2022due}.

Generally, future lepton colliders have the best reach for many of the aforementioned operators. Circular $e^+e^-$ colliders have the highest sensitivity to EW operators, due to the large statistical precision of $Z$-pole and $WW$-threshold measurements. All lepton colliders ($e^+e^-$ and $\mu^+\mu^-$) are comparable in their reach for Higgs operators, although a multi-TeV Muon Collider cannot constrain exotic Higgs decays in a model-independent way, and the combination with a run on the $s$-channel Higgs resonance would be required for this purpose. Since some of the same operators contribute to $Z$-pole precision observables, as well as to $HZ$, $WW$, and $ZZ$ pair production cross sections, the operator constraints extracted from the latter can be improved by performing a combined fit with $Z$-pole data. This effect is more significant for circular $e^+e^-$ colliders than for linear $e^+e^-$ colliders, since for the latter beam polarization helps to disentangle the contributions of different operators in $HZ/WW/ZZ$ pair production processes.

\begin{figure}[ht]
\begin{center}
    \includegraphics[width=0.8\textwidth]{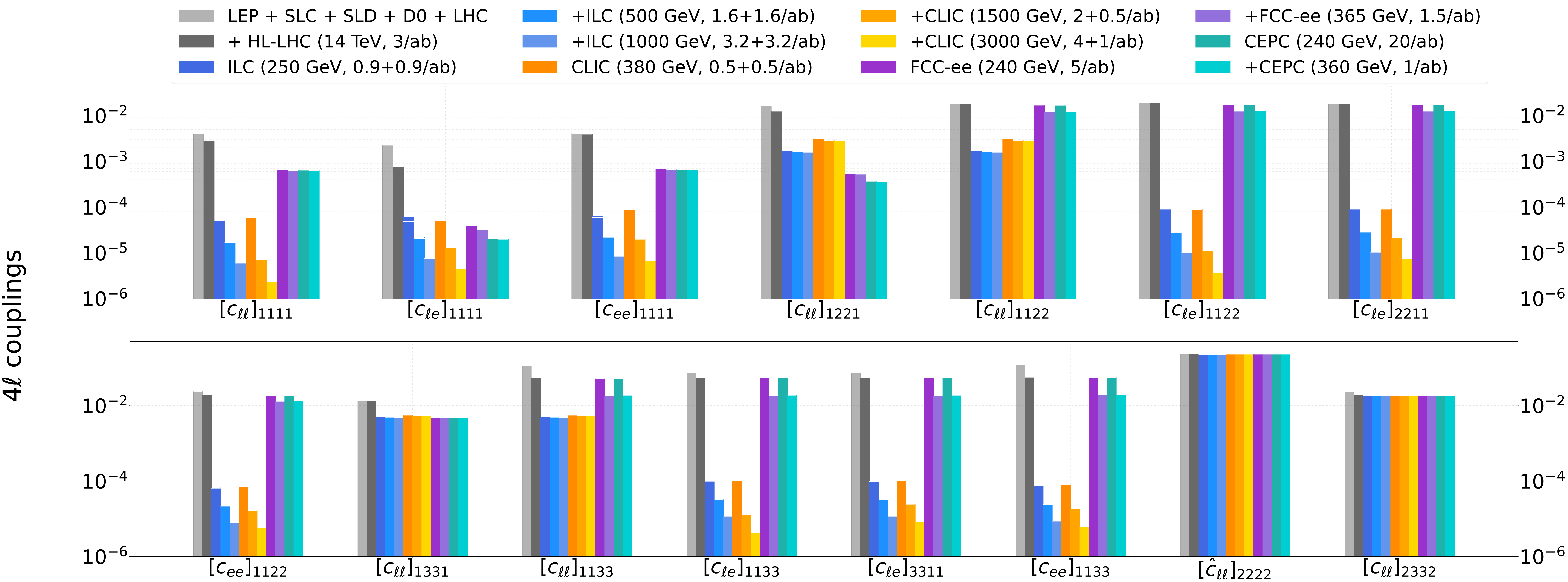}%
    \medskip
    \caption{Precision reach on a subset of 4-fermion operators from a SMEFT global fit at various future lepton colliders. "LEP+SLC+SLD" represents current measurements which are always combined in the future collider scenarios. The horizontal white line for ILC illustrates the global fit results when the pole observables from its Giga-$Z$ option are included.}
    \label{fig:efmain:Fit2-4f1}
    \end{center}
\end{figure}

Figure~\ref{fig:efmain:Fit2-4f1} shows a selection of results for a fit that combines a set of Higgs and EW operators with 4-fermion operators. The latter are better constrained at linear $e^+e^-$ than circular $e^+e^-$ machines, taking advantage of the higher energy reach and beam polarizations\footnote{For now the global fit for 4-fermion operators did not include Muon Colliders.}. A recent analysis of the sensitivity of Muon Colliders to new 4-fermion interactions can be found in Ref.~\cite{Chen:2022msz}. Low-energy measurements (from fixed-target neutrino and electron scattering, tau and meson decays) are needed to close the fit for four-fermion operators. For complete results of the combined fit with 4-fermion operators, see Ref.~\cite{Belloni:2022due}.

Not only four-fermion operators, but also top-quark electroweak operators are best constrained at lepton colliders with $\sqrt{s} \geq 500$~GeV, and measurements at two or more values of $\sqrt{s}$ are crucial for breaking degeneracies. Many constraints on top-quark operators are improved by combining $e^+e^-$ and (HL-)LHC inputs and exploiting synergies between them. See EF04 Topical Group report for more information~\cite{Belloni:2022due}.
\begin{figure}[htb!]
\centering
\includegraphics[trim=0 0 0 20,width=.9\textwidth]{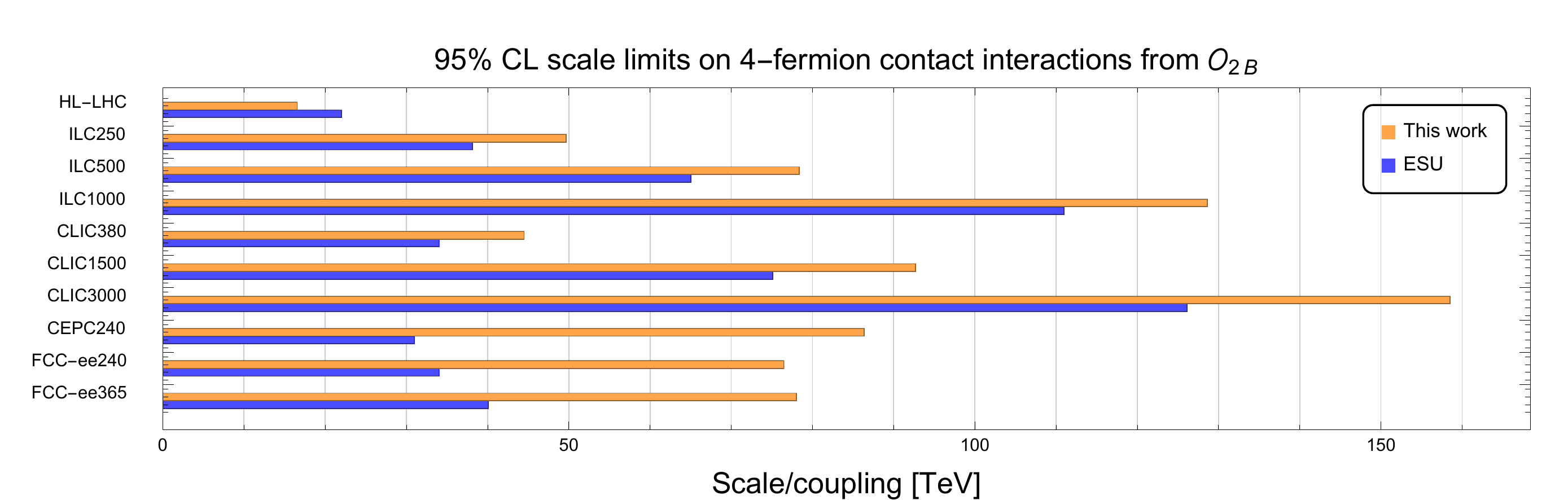}%
\medskip
\caption{95\% CL exclusion reach of different colliders on four-fermion contact interactions from the operator $O_{2B}$ (numbers for ESG are taken from Ref.~\cite{EuropeanStrategyforParticlePhysicsPreparatoryGroup:2019qin}).}
\label{fig:efmain:model-4f}
\end{figure}
\begin{figure}[htb]
\centering
\includegraphics[trim=0 0 0 20,width=.55\textwidth]{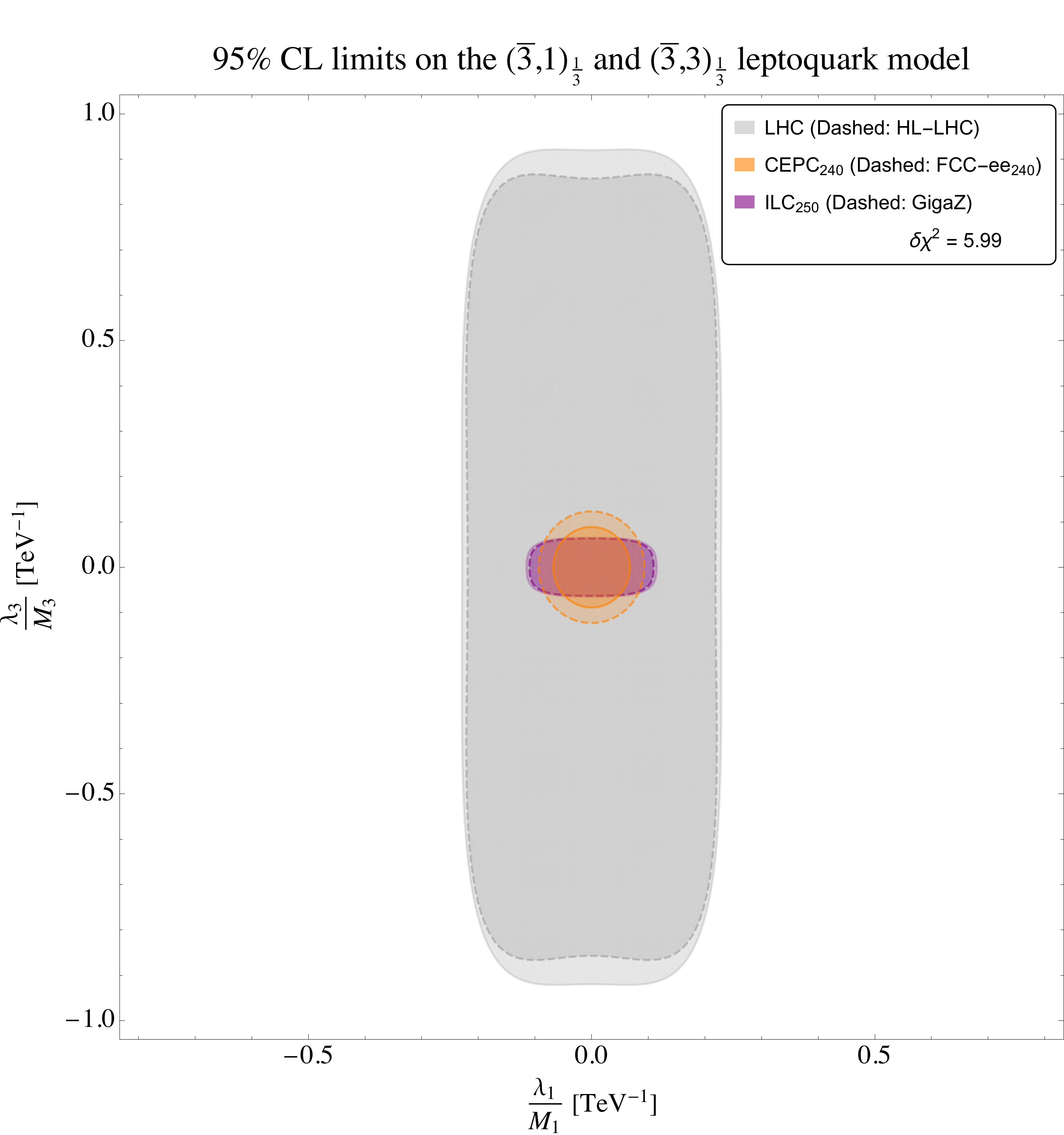}%
\caption{95\% CL exclusion reach of different colliders on the leptoquark model parameters. Only future $e^+e^-$ scenarios with energies below the $t\bar{t}$ threshold have been considered since the analysis did not include any top-quark observables.}
\label{fig:efmain:model-leptoquark}
\end{figure}
The results from global fits can be also interpreted in terms of constraints on simple BSM benchmark models.
Figure~\ref{fig:efmain:model-4f} is one example for a scenario where a flavor-universal 4-fermion contact interaction is introduced, which can be described by the single operator $O_{2B}$ \cite{Giudice:2007fh}. A second example is shown in Fig.~\ref{fig:efmain:model-leptoquark}, which extends the SM by two leptoquark fields, an SU(2) singlet with Yukawa coupling $\lambda_1$ and mass $M_1$, and an SU(2) triplet with Yukawa coupling $\lambda_3$ and mass $M_3$ \cite{Gherardi:2020det,Aebischer:2021uvt}. This model generates several 4-fermion interactions, and the bounds on the latter can be translated to bounds on the model parameters shown in the figure, where the constraining power of future $e^+e^-$ colliders is manifest.
More examples on composite Higgs models and $Z^\prime$ models are also discussed in the BSM Topical Group report~\cite{Bose:2022obr}.


\newcommand{\parbold}[1]{\par{ \bf #1.}}
\section{QCD and Strong Interactions}
\label{sec:QCD}
The fundamental theory of strong interactions, QCD, plays a unique role in the Standard Model. Being a confining gauge theory, it is an interesting quantum field theory to study in its own right. It is also a crucial tool to enable discovery at every high-energy collider. It predicts a rich panoply of phenomena associated with both perturbative and non-perturbative dynamics of the strong interactions. Continued success of the high-energy and nuclear physics research program hinges on an improved understanding of both regimes, as well as the dynamical transition between them.

\NOTEAT{The following par can be removed, as it is basically a table of content and a kind of repetition wrt the 4th par.}\NOTEPN{Shortened this paragraph and trimmed down downstream repetitions. It's good to have this intro to the main research directions.}

Future SM measurements and new physics searches will allow the exploration of new kinematic regions, such as very high 
transverse momentum and very forward rapidities, where large scale hierarchies may induce hitherto unseen QCD effects.
The upcoming era --- featuring the HL-LHC, Belle~II, the EIC, new advances in theory including lattice QCD, and potentially a Higgs factory --- will be a new golden age for QCD, easily rivaling the 1990's when the Tevatron, HERA, and LEP were all operating.

Measurements of jet, heavy-quark, and top-quark cross sections at the HL-LHC will test perturbation theory to an unprecedented level, and constrain PDFs \cite{Amoroso:2022eow} as well as the strong coupling $\alpha_s$~\cite{dEnterria:2022hzv}.  
Accurate prediction of QCD radiative effects will remain a key factor in precision measurements of the $W$-boson and  top-quark masses, and of the weak mixing angle at hadron colliders, as discussed in Sec.~\ref{sec:EWEFT}. 
Versatile techniques are refined at the LHC to analyze the internal particle composition of hadronic jets \cite{Larkoski:2017jix,Marzani:2019hun}. 
These techniques are widely employed to minimize the impact of pileup, probe fundamental and emergent properties 
of the strong force, enhance precision of measurements of highly-Lorentz-boosted particles, and extend the sensitivity
of searches for new particles (cf.\ Sec.~\ref{sec:substructure})~\cite{Nachman:2022emq}.

Detection of the decay products of far-forward hadrons at the proposed Forward Physics Facility (FPF) at the HL-LHC (Sec.~\ref{sec:FPF}) would enable QCD studies and BSM searches in a novel high-energy regime \cite{Anchordoqui:2021ghd,Feng:2022inv}. The FPF would observe neutrinos of all flavors, and possibly new particles, on its own or in coincidence with ATLAS, leading to deeper understanding of small-$x$ dynamics, forward heavy-flavor production, neutrino scattering in the TeV range, and the structure of nuclear matter. 

Due to their QCD-neutral initial state, $e^+e^-$ colliders offer the cleanest environment to study QCD dynamics. 
Belle~II will perform various measurements at low and medium energies~\cite{Accardi:2022oog}, such as the cross section for $e^+e^-\to\text{hadrons}$ (in particular two pions), from which the leading-order hadronic contribution to the anomalous magnetic moment of the muon can be extracted. \NOTEPN{Add a cross reference to the relevant external section ?}
Unprecedented statistical precision combined with low backgrounds will  enable highly impactful studies of factorization, QCD evolution, and multidimensional correlation functions, which will help to constrain MC models of hadronization for the HL-LHC program.
There has been much progress since LEP in understanding hadronic final states at $e^+e^-$ colliders, driven by techniques to analyze jet substructure and energy correlation functions. 
Measurements of event shapes will enable precision extractions of $\alpha_s$~\cite{Abbate:2010xh,Hoang:2015hka}. These colliders offer a particular advantage of producing pure samples of gluon jets via the process $e^+e^- \to hZ$, with $Z$ decaying to leptons, and the Higgs boson decaying to $gg$~\cite{Gao:2019mlt}. 
Better understanding of $b$-quark showering and hadronization, as well as $b$-quark production by secondary gluons, will be important: these are leading sources of systematic uncertainty in the measurement of the $b$-fraction in hadronic decays ($R_b$) and the $b$ forward-backward asymmetry in $Z$ decays~\cite{dEnterria:2020cgt}.

Proposed Muon Colliders offer a physics reach for discoveries that is similar to that of proposed high-energy hadron colliders. While operating at  few-TeV energies, the Muon Colliders maintain appealing experimental aspects of lepton collider environments, e.g. a lack of pileup and underlying event.
Advanced pileup mitigation techniques studied at the LHC could provide versatile handles to remove beam-induced background contamination during reconstruction~\cite{Bertolini:2014bba,Cacciari:2014gra,Berta:2014eza,Komiske:2017ubm}.

The EIC physics program~\cite{AbdulKhalek:2021gbh}, dedicated to exploration of hadronic matter, has strong synergies with the HL-LHC and FPF. The EIC can precisely measure the hadronic structure in deep inelastic scattering (DIS) through both neutral- and charged-current reactions,  to perform spin-dependent three-dimensional tomography of nucleons and various ion species using highly-polarized beams for electrons and light ions. With its variable center-of-mass energy and excellent detection of final hadronic states, the EIC can resolve the flavor composition of unpolarized proton PDFs in the kinematic region of relevance for BSM searches at the HL-LHC, but at QCD scales of only a few (tens of) GeV. The EIC program will stimulate the development of theoretical and numerical tools for QCD at the interface between particle and nuclear physics.

Proposed lepton-hadron colliders operating in the TeV energy range (Muon-Ion Collider \cite{Acosta:2022ejc}, LHeC \cite{LHeCStudyGroup:2012zhm,LHeC:2020van}, FCC-eh \cite{Curtin:2017bxr}) would be both machines for subpercent-level measurements of $\alpha_s$, nucleon structure, EW and Higgs couplings, as well as discovery machines to search for new physics such as compositeness and leptoquarks. Future hadron-hadron colliders, including the FCC-hh \cite{Mangano:2020sao} operating at 100 TeV, would open unprecedented opportunities for precision measurements in perturbative and non-perturbative QCD. Their physics program would require innovative developments both in particle detection and QCD theory (Sec.~\ref{sec:FutureDetectors}), such as parton distributions for electroweak bosons, predictions for boosted final states inside jets, and new types of event generators.

\newcommand{\NLOH}[1]{N${}^{#1}$LO${}_{\rm HTL}$\xspace}
\newcommand{\NLOHone}{NLO${}_{\rm HTL}$\xspace}
\newcommand{\NLOHTL}{NLO${}_{\rm HTL}$\xspace}
\newcommand{\NNLOHTL}{N$^2$LO${}_{\rm HTL}$\xspace}
\newcommand{\NNNLOHTL}{\NLOH3}
\newcommand{\NLOQ}[1]{N${}^{#1}$LO${}_{\rm QCD}$\xspace}
\newcommand{\LOQ}{LO${}_{\rm QCD}$\xspace}
\newcommand{\NLOQone}{NLO${}_{\rm QCD}$\xspace}
\newcommand{\LOQCD}{LO${}_{\rm QCD}$\xspace}
\newcommand{\NLOQCD}{NLO${}_{\rm QCD}$\xspace}
\newcommand{\NNLOQCD}{N$^2$LO${}_{\rm QCD}$\xspace}
\newcommand{\NNNLOQCD}{\NLOQ3}
\newcommand{\NLOE}[1]{N${}^{#1}$LO${}_{\rm EW}$\xspace}
\newcommand{\NLOEone}{NLO${}_{\rm EW}$\xspace}
\newcommand{\LOEW}{LO${}_{\rm EW}$\xspace}
\newcommand{\NLOEW}{NLO${}_{\rm EW}$\xspace}
\newcommand{\NNLOEW}{N$^2$LO${}_{\rm EW}$\xspace}
\newcommand{\NLOD}[1]{N${}^{#1}$LO${}_{\rm QED}$\xspace}
\newcommand{\NLODone}{NLO${}_{\rm QED}$\xspace}
\newcommand{\NLOQED}{NLO${}_{\rm QED}$\xspace}
\newcommand{\NNLOQED}{N$^2$LO${}_{\rm QED}$\xspace}
\newcommand{\NLOSM}{NLO${}_{\rm SM}$\xspace}
\newcommand{\NLOQE}[2]{N${}^{(#1,#2)}$LO${}_{{\rm QCD}\otimes{\rm EW}}$\xspace}
\newcommand{\NLOHE}[2]{N${}^{(#1,#2)}$LO${}^{\rm (HTL)}_{{\rm QCD}\otimes{\rm EW}}$\xspace}
\newcommand{\NLOmixQED}[2]{N${}^{(#1,#2)}$LO${}_{{\rm QCD}\otimes{\rm QED}}$\xspace}
\newcommand{\NLOQmtsix}[1]{N${}^{#1}$LO${}_{\rm QCD}^{(1/{m_t^8})}$\xspace}
\newcommand{\NLOQzzero}[1]{N${}^{#1}$LO${}_{\rm QCD}^{(z\to0)}$\xspace}
\newcommand{\NLOQVBF}[1]{N${}^{#1}$LO${}_{\rm QCD}^{(\rm VBF)}$\xspace}
\newcommand{\NLOQoneVBF}{NLO${}_{\rm QCD}^{(\rm VBF)}$\xspace}
\newcommand{\NLOQCDVBF}{NLO${}_{\rm QCD}^{(\rm VBF)}$\xspace}
\newcommand{\NNLOQCDVBF}{N$^2$LO${}_{\rm QCD}^{(\rm VBF)}$\xspace}
\newcommand{\NLOQoneDIS}{NLO${}_{\rm QCD}^{(\rm DIS)}$\xspace}
\newcommand{\NLOQDIS}[1]{N${}^{#1}$LO${}_{\rm QCD}^{(\rm DIS)}$\xspace}
\newcommand{\NLOEoneVBF}{NLO${}_{\rm EW}^{(\rm VBF)}$\xspace}
\newcommand{\NLOEWVBF}{NLO${}_{\rm EW}^{(\rm VBF)}$\xspace}
\newcommand{\NLOQoneVBFstar}{NLO${}_{\rm QCD}^{(\rm VBF^{*})}$\xspace}
\newcommand{\NLOQVBFstar}[1]{N${}^{#1}$LO${}_{\rm QCD}^{(\rm VBF^{*})}$\xspace}
\newcommand{\NLOQCDVBFstar}{NLO${}_{\rm QCD}^{(\rm VBF^{*})}$\xspace}
\newcommand{\NNLOQCDVBFstar}{N$^2$LO${}_{\rm QCD}^{(\rm VBF^{*})}$\xspace}
\newcommand{\NNNLOQCDVBFstar}{\NLOQVBFstar3}
\newcommand{\NLOEoneVBFstar}{NLO${}_{\rm EW}^{(\rm VBF^{*})}$\xspace}
\newcommand{\NLOggHVtb}[1]{N${}^{#1}$LO${}_{gg\to HZ}^{(t,b)}$\xspace}
\newcommand{\NNLOQCDT}{N$^2$LO${}_{\rm QCD}^{(t)}$\xspace}
\newcommand{\NNLOQCDBC}{N$^2$LO${}_{\rm QCD}^{(b,c)}$\xspace}
\newcommand{\tb}{\bar{t}}
\newcommand{\bb}{\bar{b}}
\newcommand{\qb}{\bar{q}}
\newcommand{\wodecay}{(w/o decay)}
\newcommand{\wdecay}{}
\newcommand{\wodecays}{(w/o decays)}
\newcommand{\wdecays}{}
\newcommand{\wleptdecays}{}
\definecolor{lightgray}{rgb}{0.85,0.85,0.85}
\begin{table}[p!]
\caption{\label{tab:LHWishlist}Summary of the Les Houches precision wish-list for hadron colliders~\cite{LHPrecisionWishlist}.
HTL stands for calculations in heavy top limit, VBF* stands for structure function approximation.}
\small
\renewcommand{\arraystretch}{1.2}
\setlength{\tabcolsep}{5pt}
\centering
  \begin{tabular}{lll}
    \hline\hline
    \multicolumn{1}{c}{process} & \multicolumn{1}{c}{known} &
    \multicolumn{1}{c}{desired} \\
    \hline\hline
\rowcolor{lightgray}
    $pp\to h$ &
    \NNNLOHTL, \NNLOQCDT, \NLOHE11 &
    \NLOH4 (incl.), \NNLOQCDBC\\
    $pp\to h+j$ &
    \NNLOHTL, \NLOQCD, \NLOQE11  &
    \NNLOHTL$\!\otimes\,$\NLOQCD\!+\,\NLOEW \\
\rowcolor{lightgray}
    $pp\to h+2j$ &
    \NLOHone$\!\otimes\,$\LOQCD &
    \NNLOHTL$\!\otimes\,$\NLOQCD\!+\,\NLOEW,
    \\
\rowcolor{lightgray}
    &     \NNNLOQCDVBFstar (incl.),
    \NNLOQCDVBFstar, 
    \NLOEWVBF &
    \NNLOQCDVBF \\
    $pp\to h+3j$ &
    \NLOHone, \NLOQCDVBF &
    \NLOQCD\!+\,\NLOEW \\
\rowcolor{lightgray}
    $pp\to Vh$ &
      \NNLOQCD\!+\,\NLOEW,
      \NLOggHVtb{} & \\
    $pp\to Vh + j$ &
      \NNLOQCD  &
      \NNLOQCD + \NLOEW \\
\rowcolor{lightgray}
    $pp\to hh$ &
    \NNNLOHTL$\!\otimes\,$\NLOQCD &
    \NLOEW  \\
    $pp\to h+t\tb$ &
    \NLOQCD\!+\,\NLOEW,
      \NNLOQCD (off-diag.) &
     \NNLOQCD  \\
\rowcolor{lightgray}
    $pp\to h+t/\tb$ &
      \NLOQCD &
      \NNLOQCD,
      \NLOQCD\!+\,\NLOEW \\
    \hline
    $pp\to V$ &
    \NNNLOQCD,
    \NLOQE11,
    \NLOEW &
      \NNNLOQCD\!+\,\NLOQE11,
      \NLOE2 \\
\rowcolor{lightgray}
    $pp\to VV'$ &
      \NNLOQCD\!+\,\NLOEW{ }\wleptdecays{},
      \!+\,\NLOQCD{ }($gg$) \wleptdecays{} &
      \NLOQCD{ } ($gg$,massive loops)
    \\
\rowcolor{lightgray}
    $pp\to V+j$ &
      \NNLOQCD\!+\,\NLOEW{ }\wleptdecays{} &
      hadronic decays \\
    $pp\to V+2j$ &
      \NLOQCD\!+\,\NLOEW{ }\wleptdecays{},
      \NLOEW{ }\wleptdecays{} &
      \NNLOQCD \wdecays{} \\
\rowcolor{lightgray}
    $pp\to V+b\bar{b}$ &
      \NLOQCD{ }\wleptdecays{} &
      \NNLOQCD \!+\,\NLOEW{ }\wdecays{} \\
    $pp\to VV'+1j$ &
      \NLOQCD\!+\,\NLOEW{ }\wdecays{} &
      \NNLOQCD \\
\rowcolor{lightgray}
    $pp\to VV'+2j$ &
    \NLOQCD \wleptdecays{} (QCD),
    \NLOQCD\!+\,\NLOEW{ }\wleptdecays{}(EW) &
      Full \NLOQCD\!+\,\NLOEW{ }\wdecays{} \\
\rowcolor{lightgray}
    $pp\to W^+W^++2j$ &
      Full \NLOQCD\!+\,\NLOEW{ }\wleptdecays{} &
      \\
\rowcolor{lightgray}
    $pp\to W^+W^-+2j$ &
      \NLOQCD\!+\,\NLOEW{ }\wleptdecays{} (EW component) &
      \\
\rowcolor{lightgray}
    $pp\to W^+Z+2j$ &
      \NLOQCD\!+\,\NLOEW{ }\wleptdecays{} (EW component) &
      \\
\rowcolor{lightgray}
    $pp\to ZZ+2j$ &
      Full \NLOQCD\!+\,\NLOEW{ }\wleptdecays{} &
      \\
   $pp\to VV'V''$ &
      \NLOQCD,
      \NLOEW{ }\wodecays{} &
      \NLOQCD\!+\,\NLOEW \wdecays{} \\
\rowcolor{lightgray}
   $pp\to W^\pm W^+W^-$ &
      \NLOQCD + \NLOEW{ }\wdecays{} &
      \\
    $pp\to \gamma\gamma$ &
      \NNLOQCD\!+\,\NLOEW &
      \NNNLOQCD \\
\rowcolor{lightgray}
    $pp\to \gamma+j$ &
      \NNLOQCD\!+\,\NLOEW &
      \NNNLOQCD \\
    $pp\to \gamma\gamma+j$ &
      \NNLOQCD\!+\,\NLOEW,
      \!+\,\NLOQCD{ }($gg$ channel) & \\
\rowcolor{lightgray}
    $pp\to \gamma\gamma\gamma$ &
      \NNLOQCD &
      \NNLOQCD\!+\,\NLOEW \\
    \hline
    $pp\to 2$\,jets &
      \NNLOQCD, 
      \NLOQCD\!+\,\NLOEW
      &
      \NNNLOQCD\!+\,\NLOEW \\
\rowcolor{lightgray}
    $pp\to 3$\,jets &
      \NNLOQCD\!+\,\NLOEW &
      \\
    \hline
    $pp\to t\tb$ &
    \begin{tabular}{@{}l@{}}
      \NNLOQCD(w/ decays)\!+\,\NLOEW (w/o decays) \\
      \NLOQCD\!+\,\NLOEW{ }(w/ decays, off-shell effects) \\
      \NNLOQCD{ }
    \end{tabular} &
    \begin{tabular}{@{}l@{}}
      \NNNLOQCD
    \end{tabular} \\
\rowcolor{lightgray}
    $pp\to t\tb+j$ &
    \begin{tabular}{@{}l@{}}
      \NLOQCD{ }(w/ decays, off-shell effects) \\
      \NLOEW (w/o decays)
    \end{tabular} &
    \begin{tabular}{@{}l@{}}
      \NNLOQCD\!+\,\NLOEW{ }(w/ decays)
    \end{tabular} \\
    $pp\to t\tb+2j$ &
    \begin{tabular}{@{}l@{}}
      \NLOQCD{ }(w/o decays)
    \end{tabular} &
    \begin{tabular}{@{}l@{}}
      \NLOQCD\!+\,\NLOEW{ }(w/ decays)
    \end{tabular} \\
\rowcolor{lightgray}
    $pp\to t\tb+Z$ &
    \begin{tabular}{@{}l@{}}
      \NLOQCD\!+\,\NLOEW{ }(w/o decays) \\
      \NLOQCD{ }(w/ decays, off-shell effects)
    \end{tabular} &
    \begin{tabular}{@{}l@{}}
      \NNLOQCD\!+\,\NLOEW{ }(w/ decays)
    \end{tabular} \\
    $pp\to t\tb+W$ &
    \begin{tabular}{@{}l@{}}
    \NLOQCD\!+\,\NLOEW{ }(w/ decays, off-shell effects) \\
    \end{tabular} &
    \begin{tabular}{@{}l@{}}
      \NNLOQCD\!+\,\NLOEW{ }(w/ decays)
    \end{tabular} \\
\rowcolor{lightgray}
    $pp\to t/\tb$ &
    \begin{tabular}{@{}l@{}}
      \NNLOQCD{*}(w/ decays) \\
      \NLOEW{ }(w/o decays)
    \end{tabular} &
    \begin{tabular}{@{}l@{}}
      \NNLOQCD\!+\,\NLOEW{ }(w/ decays)
    \end{tabular} \\
    $pp\to tZj$ &
    \begin{tabular}{@{}l@{}}
      \NLOQCD\!+\,\NLOEW{ }(w/ decays)
    \end{tabular} &
    \begin{tabular}{@{}l@{}}
      \NNLOQCD\!+\,\NLOEW{ } (w/o decays)
    \end{tabular} \\
    \hline\hline
  \end{tabular}
\end{table}
\renewcommand{\arraystretch}{1.0}

\subsection{Perturbative QCD}
\vspace{-1\baselineskip} \parbold{Precision calculations}
\label{sec:perturbativePrecisionCalculations}
  Perturbative precision calculations are crucial for measurements of SM parameters
and a key ingredient for the reliable estimation of SM backgrounds to new physics searches. 
They also serve as an input to precision simulations in modern MC event generators for collider physics~\cite{Craig:2022cef}.
There has been significant recent progress in the computation of QCD radiative corrections~\cite{Amoroso:2020lgh,Heinrich:2020ybq,Cordero:2022gsh,LHPrecisionWishlist}. Several groups have used different approaches to achieve the first 
$2\rightarrow3$ \NNLOgen calculations of hadron collider processes. There have also been significant steps forward in the development of improved infrared subtraction schemes including methods to deal with higher-multiplicity processes at \NNLOgen.
The computation of full SM corrections has seen major improvements as well~\cite{LHPrecisionWishlist}.
A summary of the state of the art and targets for future measurements is shown in Table~\ref{tab:LHWishlist}. The listed calculations correspond to the next frontier in cross section calculations for a multitude of processes, and their completion will elevate sensitivity of tests of the Standard Model and new physics searches at the HL-LHC by providing more accurate predictions for signal and background scattering processes. 
This Les Houches precision wish-list has served as a summary and repository for the higher-order QCD and EW calculations 
relevant for high-energy colliders, providing a crucial link between theory and experiment.

\parbold{Strong coupling}
\label{sec:QCD-alphas}
The strong coupling, $\alpha_s$, is a fundamental parameter of the SM and the least well known of its gauge couplings. The uncertainty on $\alpha_s$ will be one of the limiting factors in many measurements including Higgs couplings at the HL-LHC. BSM physics can also impact extractions of $\alpha_s$ in different ways and introduce tensions between their results.
The relative uncertainty in the current world average, assuming no new physics or systematic discrepancies among extractions, is 0.8\% and, within the next decade, can be reduced to $\approx$\,0.4\%  (Table~\ref{tab:alphasprospects})~\cite{dEnterria:2022hzv}. 
This requires completing the necessary pQCD calculations and control of various commensurate factors to the same level.
Many lattice QCD (LQCD) methods have been developed to extract $\alpha_s$, and to provide systematic checks of these LQCD methods~\cite{Aoki:2013ldr,Aoki:2016frl,FlavourLatticeAveragingGroup:2019iem,Aoki:2021kgd}. There are proposals to apply similar checks to phenomenological determinations~\cite{DallaBrida:2020pag,Komijani:2020kst,dEnterria:2022hzv}.
\begin{table}[th]
\caption{Summary of current and projected future (within the decade ahead or, in parentheses, longer time scales) uncertainties in the $\alpha_s(m_Z)$ extractions used today to derive the world average of $\alpha_s$~\cite{dEnterria:2022hzv}. 
\label{tab:alphasprospects}\vspace{0.2cm}}
\small
\centering
\begin{tabular}{lcc} \hline\hline
        & \multicolumn{2}{c}{Relative $\alpha_s(m_Z)$ uncertainty}\\ 
Method  & Current &  Near (long-term) future \\\hline
(1) Lattice &
$0.7\%$ &  $\approx\,0.3\%~(0.1\%)$ \\ 
(2) $\tau$ decays 
 & $1.6\%$ & $<1.\%$ \\
(3) $Q\bar{Q}$ bound states & 
$3.3\%$ &  $\approx\,1.5\%$  \\
(4) DIS \& global PDF fits 
 & $1.7\%$ & $\approx\,1\%$~(0.2\%) \\
(5) $e^+e^-$ jets \& evt shapes 
  & $2.6\%$ & $\approx\,1.5\%$~($<1$\%) \\
(6) Electroweak fits 
 & $2.3\%$  & ($\approx\,0.1\%$) \\
(7) Standalone hadron collider observables
 & 2.4\% & $\approx\,1.5\%$ \\
\hline
World average & $0.8\%$  & $\approx\,0.4\%$~(0.1\%) \\
 \hline\hline
\end{tabular}
\end{table}

The FCC-ee, which combines $3\times10^{12}$ $Z$ bosons decaying hadronically at the $Z$ pole, and a $\sqrt{s}$ calibration to tens of keV accuracy~\cite{Blondel:2021zix}, would provide measurements with unparalleled precision. The FCC-ee extraction of $\alpha_s(m_Z)$ with 0.1\% uncertainty~\cite{dEnterria:2020cpv} will enable searches for small deviations from SM predictions that could signal the presence of new physics contributions. Dedicated high-luminosity $e^+e^-$ runs at the $Z$ pole would also enable further precision tests of QCD through the study of the renormalization group running of the bottom-quark mass~\cite{Aparisi:2022yfn,Fuster:2021ekh}.
Future $ep$ collider experiments would also provide many opportunities for precision determinations of $\alpha_s$. The EIC~\cite{Accardi:2012qut,AbdulKhalek:2021gbh} and EicC~\cite{Anderle:2021wcy} will provide new high-luminosity data that could lead to a few percent uncertainty level~\cite{AbdulKhalek:2021gbh}.
The LHeC~\cite{LHeCStudyGroup:2012zhm,LHeC:2020van} would provide
hadronic final-state observables covering a considerably larger range than was possible at HERA. It could determine $\alpha_s$ from inclusive DIS data alone, something not feasible with HERA data, and an experimental uncertainty possibly reduced to 0.2\%~\cite{LHeC:2020van}.

\parbold{Jet substructure}
\label{sec:substructure}
Quark- and gluon-initiated jets are used in measurements of $\alpha_s$, the extraction of universal objects within factorized QCD, and for tuning parton-shower MC generators.
They are statistically distinguishable due to their different fragmentation processes and
can be separated using new tools such as jet substructure~\cite{Abdesselam:2010pt,Altheimer:2012mn,Altheimer:2013yza,Adams:2015hiv,Larkoski:2017jix,Kogler:2018hem,Marzani:2019hun,Kogler:2021kkw}.
Charm- and bottom-quark jets, such as in the $h\rightarrow b\overline{b}$ and $h\rightarrow c\overline{c}$ final states, can be effectively separated from other jets due to the long lifetime of the heavy quark and the mass of the heavy-flavored hadrons~\cite{CMS:2017wtu,ATLAS:2019bwq,CMS:2021scf,ATLAS:2021cxe,Erdmann:2020ovh,Nakai:2020kuu,Erdmann:2019blf}.
Identifying these types of jets is a standard benchmark for development of new classical and machine learning-based jet taggers and can help enhance certain BSM signals~\cite{Gras:2017jty,Proceedings:2018jsb,Amoroso:2020lgh}.
Tagging has not realized its full potential due to large uncertainties in the modeling of gluon jets~\cite{Gras:2017jty}. High purity samples of gluon jets provided by future lepton colliders through the process $\ell^+\ell^-\to h[\to gg]Z[\to \ell\ell]$ would significantly change this situation~\cite{Gao:2019mlt}.

Jet substructure techniques are usually applied to identify Lorentz-boosted massive particles such as $W$, $Z$, $h$ bosons, top quarks, and BSM particles in complex final states.
Many collider scenarios also result in $h$, $W$, and $Z$ bosons radiating off of very high energy jets (``Weak-strahlung''). There may also be top quarks produced via gluon splitting to $t\overline{t}$ within a jet that originates from light quarks or gluons. 
Identification of these signatures will be crucial for future high-energy colliders~\cite{Nachman:2022emq}.
Unconventional signatures include cases where jets are composed of leptons and hadrons, only leptons, only photons, hadrons and missing transverse energy etc.  In addition to the jet kinematics and substructure, the jet timing information~\cite{Chiu:2021sgs} and other information can be used for classification. Examples include jets containing one or more hard leptons~\cite{Chatterjee:2019brg,Mitra:2016kov,Nemevsek:2018bbt,duPlessis:2021xuc,Dube:2017jgo}, displaced vertices~\cite{Liu:2018wte,Nemevsek:2018bbt,Liu:2020vur}, hard photons~\cite{Wang:2021uyb,Sheff:2020jyw}, or significant missing transverse momentum~\cite{Kar:2020bws,Canelli:2021aps}. Some of these anomalous signatures are already starting to be explored at the LHC~\cite{CMS:2021dzb,ATLAS:2019isd,CMS:2021dzg,ATLAS:2019tkk,CMS:2019qjk}.

The jet substructure program has led to the introduction of techniques that systematically remove low-energy soft radiation~\cite{Abdesselam:2010pt,Altheimer:2012mn,Altheimer:2013yza,Adams:2015hiv,Dasgupta:2013ihk,Larkoski:2014wba} and can significantly reduce the dependence of observables on non-perturbative QCD effects. For a generic infrared and collinear safe observable, one can measure its ``groomed" counterpart, which will be infrared and collinear safe. Although these observables are theoretically cumbersome to compute, they can be very useful, for example for measurements of $\alpha_s$. 

Finer calorimeter granularity~\cite{Yeh:2019xbj,Coleman:2017fiq}, more hermetic coverage of tracking detectors, and precise timing information are expected to improve substructure measurements. In particular, at future Muon Colliders, such detectors in addition to detectors that specifically target the identification of particles originating from beam background, can reduce the impact of beam background on jet substructure observables.

\parbold{New observables}
\label{sec:eventshapes}
Measurements of the flow of radiation,
traditionally studied using event shapes or energy correlation functions,
provide some of the most informative tests of QCD~\cite{Neill:2022lqx}.  Energy correlators exhibit simple structures in perturbation theory~\cite{Chen:2022jhb,Chen:2021gdk,Chen:2020adz,Chen:2019bpb,Chen:2020vvp,Holguin:2022epo}. Their
measurements at future colliders would provide remarkable insights into the dynamics of jets and hadronization~\cite{Kologlu:2019mfz,Neill:2022lqx}.

Modern measurements rely strongly on the use of particle flow and tracking information. However, only observables that are completely inclusive over the spectrum of final states can be computed purely from perturbation theory. The non-perturbative input needed for theoretical predictions of track-based observables is universal and can be parametrized by  so-called ``track functions"~\cite{Chang:2013rca,Chang:2013iba}, 
which describe the fraction of energy carried by charged particles from a fragmenting quark or gluon. Recently it has been shown 
how to compute jet substructure observables at high precision by incorporating track functions~\cite{Jaarsma:2022kdd,Li:2021zcf}. 
Such calculations give promise for precision jet substructure measurements at the HL-LHC. Track functions could be measured precisely at the ILC and other future $e^+e^-$ colliders.

\definecolor{lightgray}{rgb}{0.85,0.85,0.85}
\begin{table}[!ht]
    \caption{Top part: PDF-focused topics explored in Snowmass 2013 \cite{Campbell:2013qaa} and 2021 studies \cite{Amoroso:2022eow}. Bottom part: a selection of new critical tasks to develop a new generation of PDFs that meet the targets of the HL-LHC physics program.}
    \label{table:PDFupdates2021}
\small

    \begin{tabular}{>{\raggedright}p{0.3\textwidth}>{\raggedright}p{0.3\textwidth}>{\raggedright}p{0.3\textwidth}}
\hline 
\textbf{TOPIC} & \textbf{STATUS, Snowmass'2013} & \textbf{STATUS, Snowmass'2021}\tabularnewline
\hline 
\rowcolor{lightgray}Achieved accuracy of PDFs & \NNLOgen for evolution, DIS and vector boson production & \NNLOgen for all key processes; \NNNLOgen for some processes\tabularnewline
PDFs with NLO EW contributions & MSTW'04 QED, NNPDF2.3 QED & LuXQED and other photon PDFs from several groups; PDFs with leptons
and massive bosons\tabularnewline
\rowcolor{lightgray} PDFs with resummations & Small x (in progress) & Small-x and threshold resummations implemented in several PDF sets\tabularnewline
Available LHC processes to determine nucleon PDFs & $W/Z$, single-incl. jet, high-$p_{T}$ $Z,$ $t\overline{t}$, $W+c$
production at 7 and 8 TeV & + $t\overline{t}$, single-top, dijet, $\gamma/W/Z+$jet, low-Q Drell
Yan pairs, \ldots{} at 7, 8, 13 TeV\tabularnewline
\rowcolor{lightgray} Current, planned \& proposed experiments to probe PDFs & LHC Run-2\\DIS: LHeC & LHC Run-3, HL-LHC \\DIS: EIC, LHeC, MuIC, \ldots\tabularnewline
Benchmarking of PDFs for the LHC & PDF4LHC'2015 recommendation in preparation & PDF4LHC'21 recommendation issued\tabularnewline
\rowcolor{lightgray} Precision analysis of specialized PDFs &  & Transverse-momentum dependent PDFs, nuclear, meson PDFs \tabularnewline
\hline 
\hline 
\multicolumn{3}{c}{\vspace{6pt} \textbf{NEW TASKS in the HL-LHC ERA} } \tabularnewline
\rowcolor{lightgray} Obtain complete \NNLOgen and \NNNLOgen predictions for PDF-sensitive processes & Improve models for correlated systematic errors & Find ways to constrain large-x PDFs without relying on nuclear targets\tabularnewline
Develop and benchmark fast \NNLOgen interfaces & Estimate \NNLOgen/\NNNLOgen theory uncertainties & New methods to combine PDF ensembles, estimate PDF uncertainties,
deliver PDFs for applications\tabularnewline
\hline 
\end{tabular}
\end{table}

\subsection{Non-perturbative QCD}
\label{sec:NPQCD}

\vspace{-1\baselineskip} \parbold{Parton distribution functions in the nucleon} \label{sec:PDFs}
  An overwhelming number of theoretical predictions for hadron colliders require PDFs \cite{Harland-Lang:2014zoa,Dulat:2015mca,Abramowicz:2015mha,Accardi:2016qay,Alekhin:2017kpj,Ball:2017nwa,Hou:2019efy,Bailey:2019yze,Bailey:2020ooq,Ball:2021leu,ATLAS:2021vod}, the non-perturbative functions quantifying probabilities for finding quarks and gluons in hadrons in high-energy scattering processes. PDFs contribute to precise measurements of the QCD coupling constant, heavy-quark masses, weak boson mass, and electroweak flavor-mixing parameters. PDFs often introduce the dominant source of uncertainty in collider experiments, such as in the CDF II high-statistics measurement of the $W$-boson mass \cite{CDF:2022hxs}. Reducing these uncertainties requires continuous benchmarking and improvements of the theoretical framework~\cite{PDF4LHCWorkingGroup:2022cjn,Accardi:2016ndt}.
The accurate knowledge of PDFs is also critical in searches for BSM interactions. 

Table~\ref{table:PDFupdates2021} illustrates the progress made on the PDF determinations since the previous Snowmass Summer Study in 2013 \cite{Campbell:2013qaa}. Details are presented in the Snowmass PDF whitepaper \cite{Amoroso:2022eow}. The \NNLOgen QCD accuracy has become the standard for the modern nucleon PDFs, with \NNNLOgen being on the horizon within the next decade. In addition, some PDF sets for precision physics include photon PDFs and QCD resummations. 
Current PDF predictions for parton luminosities agree within uncertainties at invariant masses $30 \lesssim m_X \lesssim 10^3$ GeV, relevant e.g. for Higgs and gauge-boson production, but in the gluon sector (gluon-gluon and gluon-quark parton luminosities), differences are seen at large masses \cite{Amoroso:2022eow}. These differences are a consequence of both methodology and data sets included in PDF fits. The available PDF ensembles account for a combination of experimental, theoretical, and methodological uncertainties  \cite{Kovarik:2019xvh} in different ways, and, as a result,  the provided PDFs can differ. The PDF-dependent cross sections can differ as well by the amounts exceeding the missing \NNNLOgen contributions. 

The bottom part of Table~\ref{table:PDFupdates2021} lists new tasks for the PDF analysis that emerge in the era of precision QCD.
While the most precise \NNLOgen or even \NNNLOgen theoretical cross sections should be preferably used, the accuracy of the theoretical predictions in the fits also depends on other factors. For complex \NNLOgen/\NNNLOgen calculations, their fast approximate implementations (such as fast \NNLOgen interfaces) must be developed. Propagation of correlated systematic errors into PDFs is a challenging task that requires combined efforts of experimentalists and theorists. Control of uncertainties requires, in particular, to either fit the experiments that are minimally affected by the unknown factors (for example, to include cross sections only on proton, rather than on nuclear targets), or to estimate the associated uncertainties from these factors directly in the fit. The PDF uncertainties must representatively reflect all PDF behaviors compatible with the fitted data \cite{Courtoy:2022ocu}. PDFs must be developed for a wide range of users in a format that optimizes for accuracy, versatility, and speed across a broad range of applications --- a highly non-trivial task.

\begin{figure}[t!]
    \centering
    \includegraphics[width=0.52\textwidth]{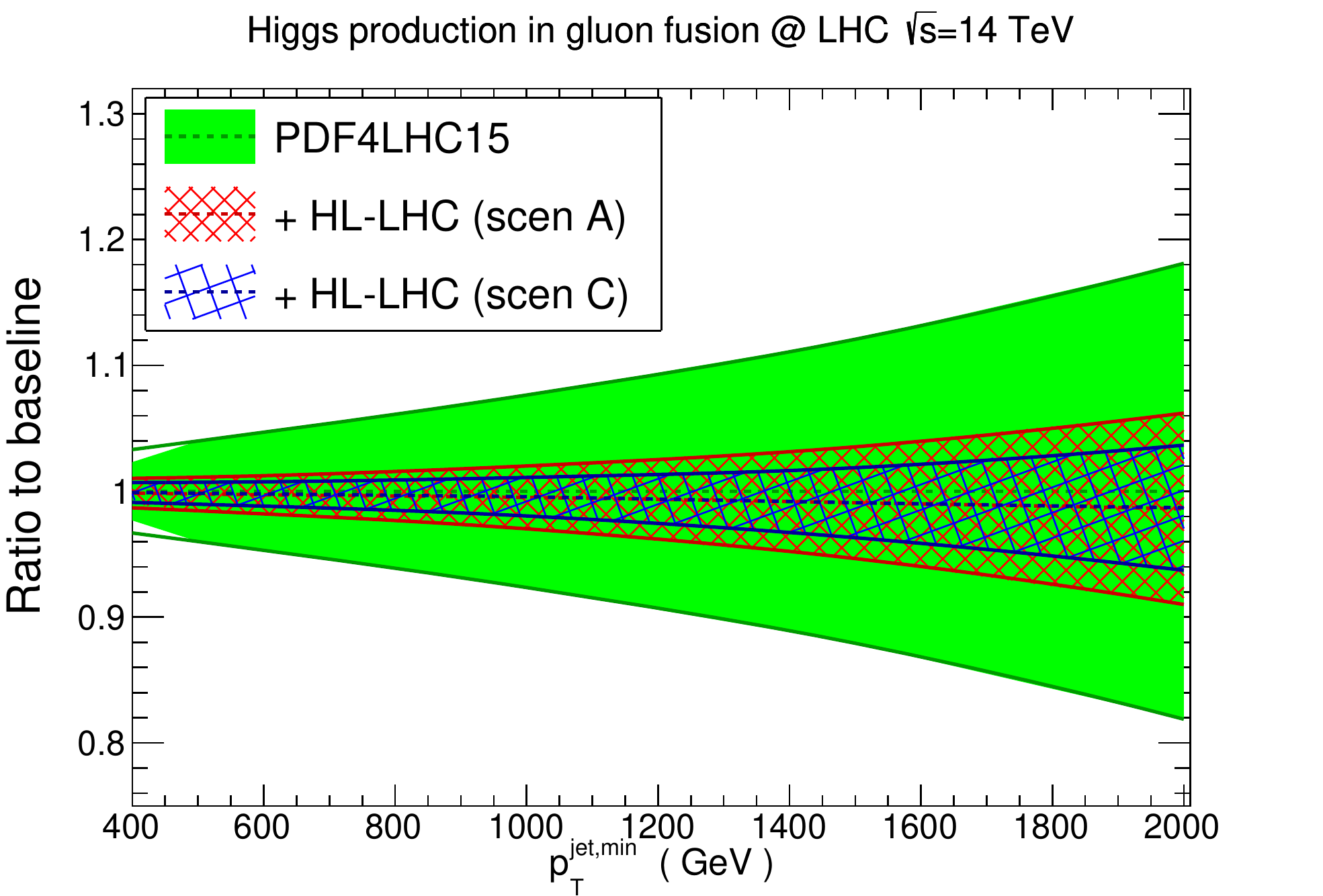}
    \includegraphics[width=0.47\textwidth]{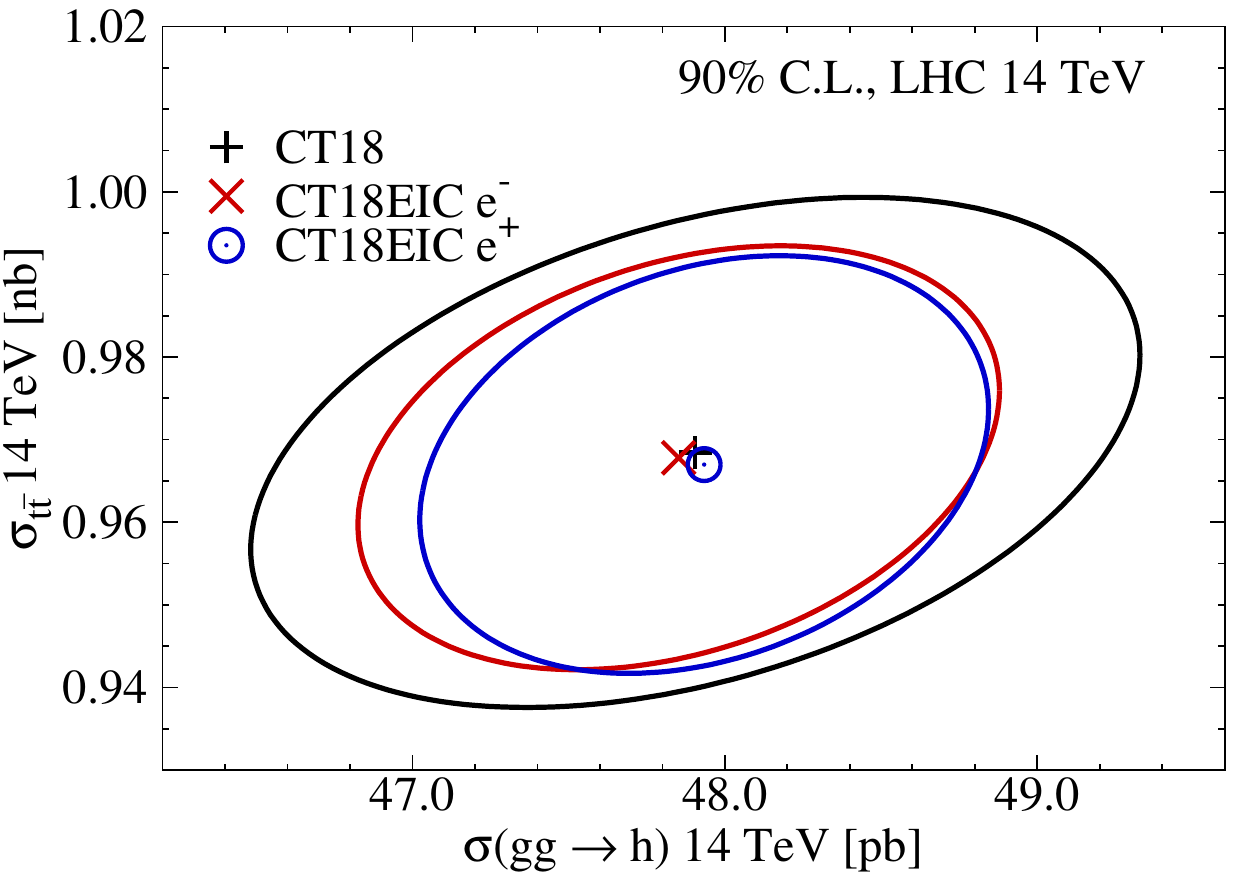}
    \caption{Examples of projections for PDF uncertainties in the HL-LHC era. Left: Uncertainties for \NNLOgen Higgs production via gluon fusion at $\sqrt{s} = 14$ TeV obtained with published PDF4LHC15 \NNLOgen PDFs  \cite{Butterworth:2015oua} (green band) and after additional constraints are imposed on these PDFs using simulated HL-LHC data in two scenarios (red and blue bands) \cite{AbdulKhalek:2018rok}. 
    Right: 90\% CL uncertainty ellipses for \NNLOgen predictions for $gg\to h$ and $t\bar t$ production at the LHC with $\sqrt{s}=14$~TeV obtained using CT18 \NNLOgen PDFs \cite{Hou:2019efy} and after imposing simulated constraints from inclusive DIS at the EIC \cite{AbdulKhalek:2021gbh}. 
    \label{fig:HiggsPDFerrors}
    }
\end{figure}

Recent studies \cite{AbdulKhalek:2018rok,AbdulKhalek:2021gbh} provide projections using various techniques for the reduction of PDF uncertainties under anticipated near-future theoretical and experimental developments. As an illustration, the left panel of Fig.~\ref{fig:HiggsPDFerrors} compares the current PDF uncertainty for $gg\to h$ production and its reduction when simulated HL-LHC measurements are included in the conservative (scen A) and optimistic (scen C) scenarios, using PDF4LHC15 \NNLOgen PDFs \cite{Butterworth:2015oua} as the baseline. The right panel shows an analogous projection for the reduction of the PDF uncertainty on the SM Higgs and $t\bar t$ cross sections at the LHC upon including the simulated measurements in DIS at the EIC, this time using the CT18 \NNLOgen framework \cite{Hou:2019efy}. The ability of the LHC measurements to reduce the PDF uncertainty critically depends on the control of systematic effects. Deep inelastic scattering and hadroproduction at the EIC will constrain the PDFs for the LHC high-mass BSM searches most directly and without systematic or new-physics factors relevant at the LHC.

\begin{figure}[htbp]
\includegraphics[width=0.50\textwidth]{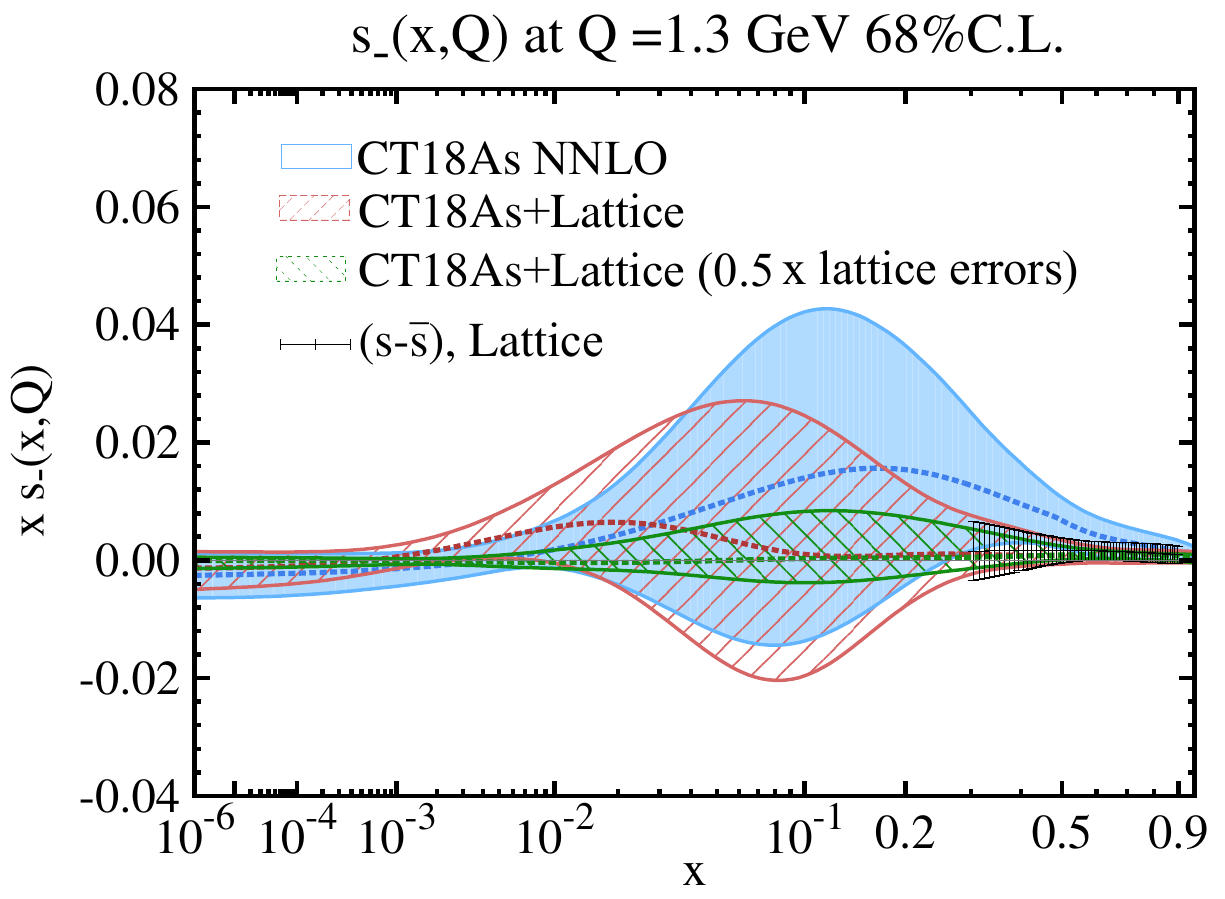}
\hspace{1.truecm}
\includegraphics[width=0.35\textwidth]{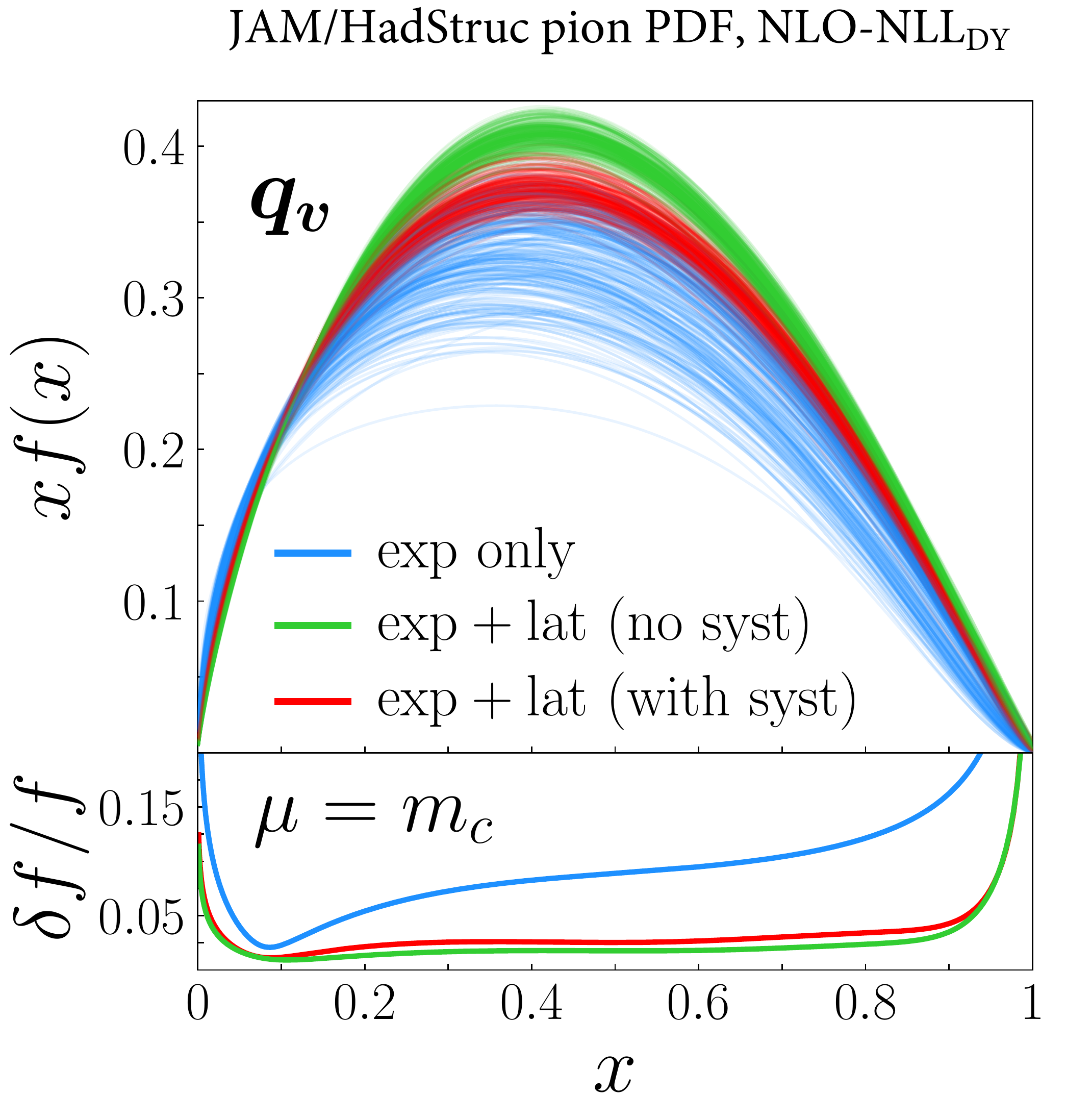}
\caption{
Left: Impact of constraints from lattice QCD (black dashed area) on the difference between strange quark and antiquark PDFs in a CT18As \NNLOgen fit \cite{Hou:2022sdf}. The red (green) error bands correspond to the current (reduced by 50\%) lattice QCD errors. Right: determination of a quark PDF in a pion using a combination of experimental and lattice QCD data, and including threshold resummation \cite{Barry:2022itu}. 
}
\label{fig:CT18Aas_Lat_Predicted} 
\end{figure}

\parbold{Predicting hadron structure in lattice QCD} \label{sec:LatticeQCD}
As lattice QCD techniques advance in computations of PDFs from first principles, unpolarized phenomenological PDFs in the nucleon serve as important benchmarks for testing lattice QCD methods \cite{Lin:2017snn,Constantinou:2020hdm}. Namely, precisely determined phenomenological PDFs in the nucleon serve as a reference to validate lattice and non-perturbative QCD calculations. The combination of the observation-driven PDF analysis and lattice QCD is thus especially promising and drives related studies of three-dimensional structure of baryons and mesons, including dependence on transverse momentum and spin. Figure~\ref{fig:CT18Aas_Lat_Predicted} (left) shows the impact of lattice QCD calculations on a quantity affecting precision measurements at hadron colliders --- the difference between the strange quark and antiquark PDFs. Such novel calculations can significantly constrain quantities that are difficult to assess with conventional PDF estimates. Figure~\ref{fig:CT18Aas_Lat_Predicted} (right) illustrates that recent lattice QCD calculations are now able to predict quark PDFs. Lattice QCD is most powerful in predicting various QCD charges and distributions of partons carrying 10\% of the hadron's energy or more. 

 A Snowmass whitepaper \cite{Constantinou:2022yye} summarizes rapid advances in lattice QCD calculations of PDFs and other QCD functions. New experiments and facilities will pursue exploration of the three-dimensional structure
described by transverse-momentum--dependent distributions (TMDs) as well as generalized parton distributions (GPDs) -- hybrid momentum- and coordinate-space distributions  that bridge the conventional form factors and collinear PDFs. These experiments will match the ongoing theoretical advancements that open doors to many previously unattainable predictions, from the $x$ dependence of collinear nucleon PDFs to TMDs \cite{Musch:2010ka,Musch:2011er,Engelhardt:2015xja,Yoon:2016dyh,Yoon:2017qzo} and related functions \cite{Shanahan:2020zxr,Zhang:2020dbb,Schlemmer:2021aij,Li:2021wvl,Shanahan:2021tst}, GPDs \cite{Chen:2019lcm, Alexandrou:2020zbe, Lin:2020rxa,Lin:2021brq}, and higher-twist terms -- a progress that was not envisioned as possible during the 2013 Snowmass study.

There remain challenges to be overcome in lattice calculations, such as reducing the noise-to-signal ratio, extrapolating to the physical pion mass, and increasing hadronic boosts to suppress systematic uncertainties. Computational resources place significant limitations on the achievable precision, as sufficiently large and fine lattices are necessary to suppress finite-size and higher-twist contaminating contributions. New ideas can bypass these limitations. With sufficient support, lattice QCD can fill in the gaps where the experiments are difficult or not yet available, improve the precision of global fits, and provide better SM inputs to aid new-physics searches across several HEP frontiers.

\parbold{Hadronization and fragmentation functions}
\label{sec:hadronization}
\NOTEAT{This section is rather verbose. I think we can keep a similar level of content with fewer words}\NOTEPN{Done.}
Hadronization is a process of formation of detected final-state hadrons from partons. Governed by non-perturbative dynamics, hadronization cannot be calculated analytically and 
is elusive in lattice calculations. Accurate description of hadronization is, however, 
absolutely indispensable for predicting production of hadronic states.

Fragmentation functions (FFs) -- the probabilities for producing hadrons from fragementing partons -- 
are instrumental for LHC experiments, as well as for extracting the spin-(in)dependent nucleon structure~\cite{Anselmino:2020vlp} at Belle~II and the EIC. The  Belle~II program will emphasize investigation of full multidimensional dependency of FFs with complex final states, such as dihadrons or polarized hyperons, in order to map out their factorization universality properties and kinematic dependencies.
Tagging on such final-state degrees of freedom allows  more targeted access to the hadron structure in semi-inclusive deep inelastic scattering (SIDIS) experiments, e.g. at JLab and the EIC \cite{Burkardt:2008ps,CLAS:2020igs,Hayward:2021psm,Courtoy:2022kca}, 
where TMD PDFs and FFs will be the primary means to investigate the mechanism of hadronization in a 3D-picture~\cite{AbdulKhalek:2022erw}.

Models of hadronization for the full event are necessary to predict background and signal processes in various BSM searches. 
Currently, modeling of backgrounds originating from light-quark fragmentation is mainly performed by MC event generators. Tuning those generators to a precision needed for discovery requires a model for correlated production of multiple hadrons that can only be verified with clean $e^+e^-$ hadroproduction data. 
High-luminosity data of this kind from
Belle~II, combined with data at other energies from LEP and SLD,  will provide a large lever arm in energy to confidently extrapolate the hadronization model from low to LHC energies and beyond. 

Where neither the single-hadron FFs nor the MC event generators are sufficient to model multi-particle fragmentation, semi-analytical approaches to account for particle correlations gain
more recognition. There have been significant recent efforts to define many-particle correlation observables that may obey factorization theorems, be sensitive to hadronization dynamics, can be interpreted within hadronization models (e.g. a QCD string model), and might be computed from QCD first principles in the future. Such correlation observables are already a focus at the LHC \cite{ATLAS:2020bbn}. 

Accurate knowledge of FFs in $e^+e^-$ collisions (especially, the  gluon FF ~\cite{dEnterria:2013sgr}) is of utmost importance for having an accurate ``QCD vacuum'' baseline to quantify ``QCD medium'' modifications of hadronization in proton-nucleus and nucleus-nucleus collisions \cite{Albino:2008aa,Accardi:2009qv}. 
Looking toward the future, an FCC-ee would offer a clean radiation environment 
to systematically explore color reconnection, hadronization, and multiparticle correlations (in spin, color, space, momenta) --- the non-perturbative final-state effects that
currently rely on phenomenological MC modeling and may limit the ultimate accuracy  at hadron-hadron colliders. 
In the $pp$ case, the color reconnection effects may constitute 20--40\% of the ultimately achievable uncertainty in the measurement of the top mass \cite{Argyropoulos:2014zoa} and can also impact limits in CP-violation searches in $ H \to W^+ W^-$ hadronic decays.
At the FCC-ee, $e^+e^- \to  W^+W^- \to q_1\bar{q}_2 q_3\bar{q}_4$ and $e^+e^- \to t\overline{t}$ processes, with the top quarks decaying and hadronizing closely to one another, would directly probe the color reconnection and interactions between hadronic strings~\cite{Khoze:1994fu,Christiansen:2015yca,Proceedings:2017ocd,Abada:2019lih}.

\subsection{Forward Physics \label{sec:ForwardQCD}}
\vspace{-1\baselineskip} The LHC experiments opened access to a wide range of forward and diffractive processes, driving advances in relevant QCD theory, such as charting the gluon at very low $x$, revealing dynamics at high partonic densities, and testing MC models for forward hadron production. Understanding small-$x$ dynamics in $pp$ collisions, already important at the (HL-)LHC \cite{Cepeda:2019klc,Azzi:2019yne}, is crucial for any future higher-energy $pp$ collider such as FCC-hh~\cite{FCC:2018vvp,FCC:2018byv,Mangano:2016jyj,Rojo:2016kwu}, where standard electroweak processes, e.g. $W$ and $Z$ production, are dominated by low--$x$ dynamics, and an accurate calculation of the Higgs production cross section requires accounting for BFKL resummation \cite{Fadin:1975cb,Lipatov:1976zz,Kuraev:1977fs,Balitsky:1978ic} or partonic saturation \cite{Golec-Biernat:1998zce}.

\parbold{Diffraction}
Some configurations
of final states with high forward multiplicities, as well as those with the absence of energy in the forward region (so-called rapidity gaps), in elastic, diffractive, and central exclusive processes originate from purely non-perturbative reactions, while others can be explained in terms of multi-parton chains or extensions of perturbative QCD such as the BFKL formalism. 
These processes are interesting for the exploration of electroweak and BSM physics. Understanding the elastic cross section and diffraction better, and probing models for Odderon~\cite{Lukaszuk:1973nt,Martynov:2018sga,Breakstone:1985pe, Erhan:1984mv,UA4:1986cgb,UA4:1985oqn,Nagy:1978iw} and Pomeron exchange, will be key studies at the HL-LHC, EIC, and any future hadron collider.
Further progress in this fundamental area requires the combination of  experimental measurements, including at the EIC and FPF, and theoretical work. The FPF also allows exploration of BFKL evolution and gluon saturation.

\begin{figure}[ht]
\centering
\includegraphics[width=0.48\textwidth]{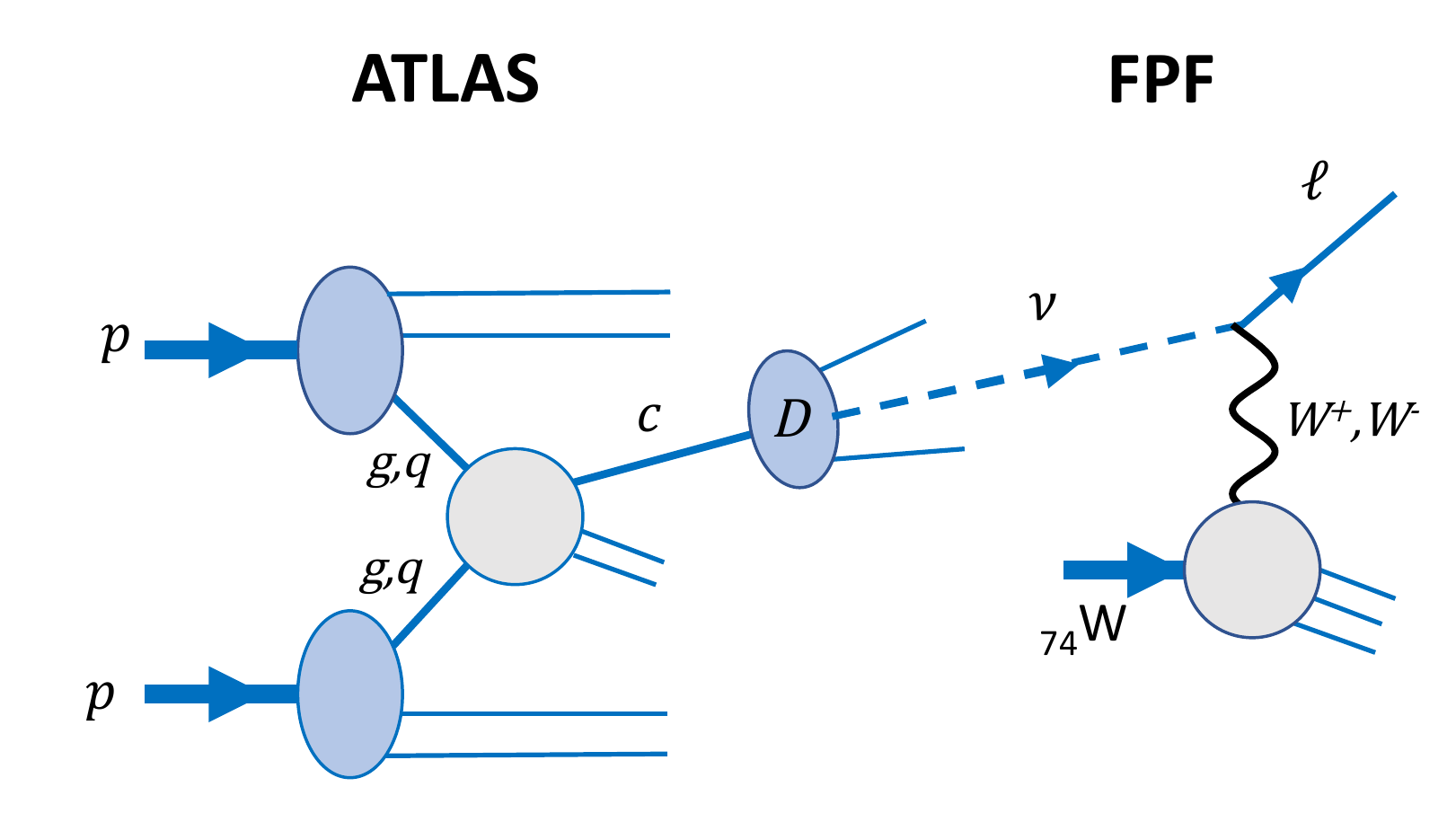}\quad
\includegraphics[width=0.48\textwidth]{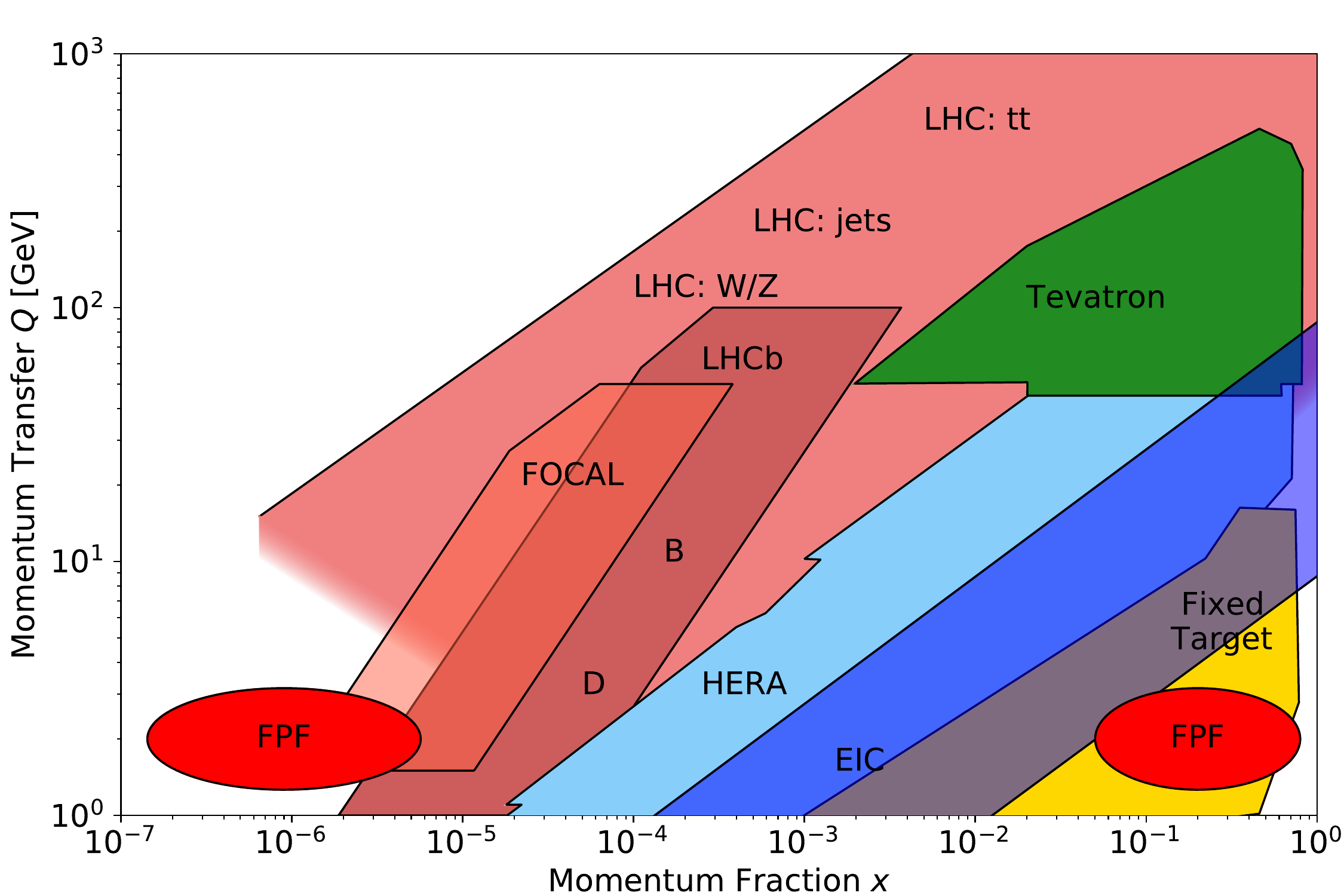}
\caption{Left: The production and detection processes for forward $D$-meson production at the HL-LHC followed by their decay into neutrinos falling within the FPF acceptance. Right: The $(x,Q)$ regions (red ovals) that can be accessed at the FPF via this process. 
}
\label{fig:FPF_kin}
\end{figure}

\parbold{Physics opportunities at the CERN Forward Physics Facility} \label{sec:FPF}
Given its unique configuration, the FPF \cite{Anchordoqui:2021ghd,Feng:2022inv}
 would extend the coverage of the LHC measurements (notably, the LHCb) at small $x$ by almost two orders of magnitude at low $Q$, reaching down to $x\simeq 10^{-7}$ (Fig.~\ref{fig:FPF_kin}). 
In its proposed main configuration, the FPF will detect far-forward neutrinos, produced from charm meson decays in a main LHC detector, by DIS on a tungsten target. Therefore, FPF measurements would provide a bridge between the physics program at the HL-LHC and that of a higher-energy $pp$ collider. 
Successful interpretation of FPF measurements will require a coordinated program including forward production at LHCb~\cite{LHCb:2013xam, LHCb:2015swx, LHCb:2016ikn,LHCb:2021stx}, large-$x$ charge current DIS at EIC~\cite{AbdulKhalek:2021gbh}, and small-$x$ scattering at the HL-LHC and future DIS facilities such as the Muon-Ion Collider~\cite{Acosta:2022ejc} and LHeC~\cite{LHeC:2020van}. In turn, with future experiments designed to detect neutrino interactions with energies between several hundreds of GeV and a few TeV, an energy range that has not been precisely probed for any neutrino flavor, the FPF  will significantly extend accelerator cross section measurements and will provide improved predictions for key astroparticle physics processes, such as ultra-high energy neutrino-nucleus and cosmic ray interaction cross sections.


Figure~\ref{fig:FPF_kin} also demonstrates that the FPF will be sensitive to very high--$x$ kinematics and in particular the intrinsic charm component of the proton~\cite{Brodsky:2015fna}. While charm production in $pp$ collisions is dominated  by gluon--gluon scattering, in the presence of a non--perturbative charm PDF in the proton, the charm-gluon initial state may be dominant for forward $D$-meson production.
FPF measurements, as part of a broader physics program including LHCb and the EIC, would provide complementary handles on high-$x$ intrinsic charm. At the small-$x$ end, FPF observations would reduce the currently large uncertainties on the expected flux of prompt neutrinos arising from the decays of charm mesons produced in cosmic-ray collisions in the atmosphere~\cite{Bertone:2018dse,Garcia:2020jwr,Gauld:2015kvh, Garzelli:2016xmx}.
These represent an important background for astrophysical neutrinos at neutrino telescopes such as IceCube~\cite{IceCube:2020wum} and KM3NET~\cite{Adrian-Martinez_2016}.

{\bf Neutrino-induced deep inelastic scattering (DIS)} mediated by charged current (CC), such as that at the FPF, provides essential flavor separation in 
global PDF fits~\cite{Gao:2017yyd} that is not achieved solely in neutral-current (NC) DIS. Hence FPF data can complement other planned experiments, such as the EIC. The coverage for CC DIS on nuclei at the FPF in Fig.~\ref{fig:FPF_kin} broadly overlaps with that for charged-lepton CC DIS at the EIC~\cite{AbdulKhalek:2021gbh,Khalek:2021ulf}.
Analogous information from previous neutrino-induced DIS measurements on nuclear targets (NuTeV~\cite{NuTeV:2005wsg}, NOMAD~\cite{NOMAD:2013hbk}, CCFR~\cite{NuTeV:2001dfo}, CHORUS~\cite{CHORUS:1997wxi}) plays the key role in global PDF fits of nucleon and nuclear PDFs (with the two related via nuclear corrections). 
Inclusive CC DIS and, especially, semi-inclusive charm production in CC DIS are the primary channels to probe the PDFs for strange quarks and antiquarks. The strange-quark PDF offers insights about the non-perturbative proton structure~\cite{Chang:2014jba}, while it also drives much of the PDF uncertainty in $W$ boson mass measurements at the LHC~\cite{Nadolsky:2008zw}. On the experimental side, determination of the (anti-)strangeness PDF has been a hot topic, as the different experimental results somewhat favor different shapes for $s-$quark PDFs~\cite{Alekhin:2014sya, Alekhin:2017olj,Hou:2019efy,Faura:2020oom}. The elevated PDF uncertainty from fitting such inconsistent experiments propagates into various pQCD predictions~\cite{Bevilacqua:2021ovq,Amoroso:2022eow,dEnterria:2022hzv}.


\subsection{Heavy Ions}
\label{sec:HeavyIons}

\vspace{-1\baselineskip} The relativistic heavy-ion (HI) program prioritizes the studies of the quark-gluon plasma (QGP), the partonic structure of nuclei, collectivity in small collision systems~\cite{CMS:2010ifv,CMS:2012qk,ATLAS:2012cix,ALICE:2012eyl,LHCb:2015coe}, and nuclear electromagnetic (EM) interactions~\cite{ATLAS:2017fur,CMS:2018erd}.
The HI studies have been a valuable part of the LHC physics program, with a variety of cross-cutting connections to the other areas of LHC physics summarized in Sec.~\ref{sec:CrossCuttingQCD}. 
Reference~\cite{Dainese:2703572} presents a
detailed plan for the HI measurements 
at the HL-LHC. 
These measurements will greatly benefit from planned detector upgrades for ALICE, ATLAS, and CMS. ALICE FoCal, a forward calorimeter that will allow us to probe parton distributions down to lower Bjorken-$x$ values than are previously accessible~\cite{ALICE:2020mso}. The increased pseudorapidity acceptance of charged particle tracking will be a boon to bulk particle detection. The upgraded Zero Degree Calorimeters (ZDCs)~\cite{ATLAS-ZDC-LHCC, CMS-ZDC-TDR} will improve triggering and identification for ultraperipheral collisions (see Sec.~\ref{sec:CrossCuttingQCD}). The addition of time-of-flight particle identification capability in the CMS timing detector~\cite{CMS:2667167} will allow differentiation among low-momentum pions, kaons, protons, and charmed states. The planned major upgrade of the ALICE detector for HL-LHC Run 5 (ALICE 3~\cite{ALICE:2803563}) will enable far-reaching studies of the QGP properties. The HL-LHC HI program is synergistic with the HI programs at RHIC and the EIC. The sPHENIX detector~\cite{PHENIX:2015siv,sPHENIX:2017lqb,FTR-18-025} at RHIC will start HI data-taking in 2023 with the aim to collect quality heavy-flavor meson, quarkonium, and jet data in Au-Au collisions.

\parbold{Hard probes}
High momentum-transfer interactions between partons in the nuclear medium serve as short-distance probes of the QGP. They can reveal the impact of QGP on color charges of fast-moving partons or slow-moving heavy quarks, commonly resulting in the attenuation of the jets~\cite{ATLAS:2010isq,CMS:2011iwn,CMS:2012ulu,CMS:2012ytf,CMS:2017ehl,ATLAS:2018dgb,CMS:2017eqd,ATLAS:2012tjt,ATLAS:2014ipv,CMS:2016uxf,ATLAS:2018gwx,CMS:2021vui,ALICE:2015mjv,CMS:2015hkr}, or modifications of the jet substructure~\cite{CMS:2013lhm,CMS:2014jjt,ATLAS:2014dtd,CMS:2017qlm,CMS:2018jco,ATLAS:2020wmg,ATLAS:2018bvp,CMS:2018mqn,CMS:2018fof,ALICE:2019ykw,ALICE:2018dxf,CMS:2020plq,CMS:2021otx}, referred to as jet quenching. 
ALICE, ATLAS, and CMS measurements will  significantly reduce statistical and systematic uncertainties in key measurements of medium modification of light or heavy quark jets produced in association with $\gamma$, $Z$, or $\mathrm{D^{0}}$~\cite{FTR-18-025, ATL-PHYS-PUB-2018-019}. 
Studies of medium dependence of jet quenching at the HL-LHC would be facilitated by collecting large $pp$ samples of events with the same signatures and low pileup. By comparing the LHC and RHIC data, we aim to constrain the temperature dependence of the transport coefficients of QGP.

The main HL-LHC experiments will also increase capabilities for detection of heavy-flavor (HF) mesons~\cite{CMS:2017uoy, CMS:2018eso,ALICE:2018lyv,CMS:2017qjw,ALICE:2015vxz,ALICE:2021mgk,CMS:2018bwt,CMS:2017exb,ATLAS:2018hqe} and quarkonia~\cite{CMS:2011all,CMS:2012bms,CMS:2012gvv, CMS:2016rpc,ATLAS:2010xzb,ALICE:2012jsl,CMS:2014vjg,ALICE:2015jrl}. Heavy quarks, $c$ and $b$,  are unique, slow-moving QGP probes produced during the early stages of the hard collision. In the medium, they radiate less energy than light flavors due to the dead-cone effect. Observations of HF mesons elucidate mechanisms for their thermalization, Debye color screening and recombination \cite{Kopeliovich:2014una,Aronson:2017ymv,Du:2017qkv} inside the QGP. 
Measurements of the $p_{\mathrm{T}}$ dependence of the quarkonium nuclear modification factors ($R_{\mathrm{AA}}$)  will discern whether high-$p_{\mathrm{T}}$ quarkonium formation is driven by the Debye screening mechanism or by energy loss  in the medium \cite{FTR-18-024}. Improved elliptic flow measurements of charm mesons in p-Pb~\cite{FTR-18-026} and of HF decay muons~\cite{ATL-PHYS-PUB-2018-020} and $\Upsilon(1S)$ ~\cite{FTR-18-024} in Pb-Pb collisions will provide insights both on the degree of thermalization of HF quarks at low $p_{\mathrm{T}}$ and on the recombination of bottomonia from deconfined $b$ quarks in the QGP. 
Production of strange $B$ mesons and charm baryons in $pp$ and Pb-Pb collisions~\cite{FTR-18-024} will investigate the interplay between the predicted enhancement of strange quark production and the quenching mechanism for beauty quarks, and on the contribution of recombination of HF quarks with lighter quarks to hadronization in HI collisions. Precise measurements of beauty mesons in p-Pb collisions~\cite{FTR-18-024} will help to establish the relative contributions of hadronization and nuclear-matter effects. Interpretation of these HL-LHC observations will be facilitated by the sPHENIX measurements of HF hadron spectra, particle multiplicities, and azimuthal anisotropy at lower energy, allowing especially to strongly constrain the HF diffusion coefficient and its temperature dependence.

\parbold{Hadronic structure}
Abundant production of light (anti-)nuclei at ALICE opens promising opportunities. As in the cases of light nuclei and  charmonium, the statistical hadronization or coalescence assumption can render a unique insight into the structure (e.g.\, tetraquark or molecular state) of exotic hadrons, such as $X(3872)$ studied by LHCb in high-multiplicity $pp$ collisions~\cite{LHCb:2020sey} and by CMS in Pb+Pb collisions~\cite{CMS:2021znk}. Those initial measurements will be followed up with high-statistics data in LHC Runs 3 and 4. In LHC Run 5, the ALICE 3 detector would precisely measure multi-charm baryons, expanding the hadronization studies in Runs 3 and 4. ALICE 3 would be a perfect tool to study the formation of light nuclei, hyper-nuclei, super-nuclei, and properties of exotic states, e.g. $X(3872)$ and the newly discovered $T_{cc}^+$.

{\bf Collectivity of small and large systems} will be studied with high-statistics samples of pp, pPb, and PbPb events at the HL-LHC. The pivotal upgrades of ATLAS and CMS trackers will extend observations of charged particles to a wide pseudorapidity range ($|\eta|<4$). For small systems, we expect a crucial improvement in the determination of the created system's size by measuring the Hanbury-Brown-Twiss (HBT) radii~\cite{ATL-PHYS-PUB-2018-020}. Azimuthally-dependent femtoscopy will measure the spatial ellipticity of the medium at freeze-out, e.g., to find the normalized second-order Fourier component of the transverse HBT radius as a function of the flow's magnitude. The extended $\eta$ acceptance in Run 4 will refine the characterization of the rapidity dependence of factorization breaking. Better  measurements of the forward-backward multiplicity correlation and multi-particle cumulants will deepen our understanding of the early-stage medium fluctuations~\cite{ATL-PHYS-PUB-2018-020}.

\begin{figure}[b]
    \centering
    \includegraphics[width=0.7\textwidth]{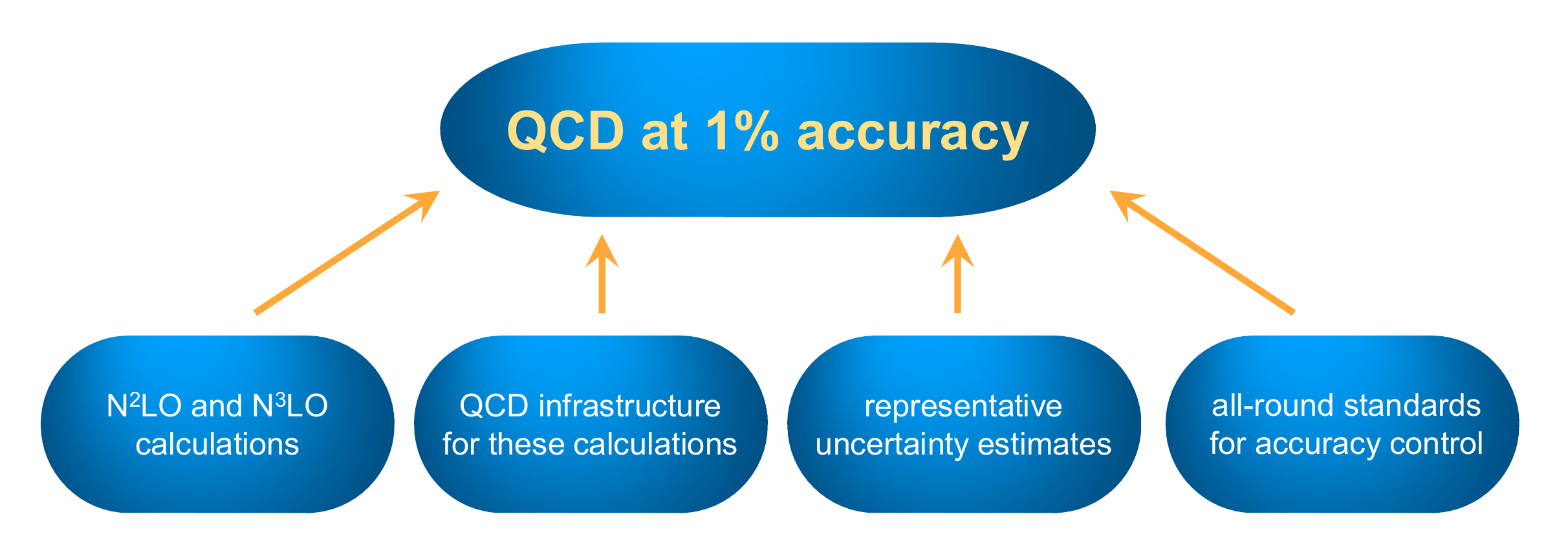}   
    \caption{\label{fig:QCDatOnePercent} Prerequisites for achieving percent-level accuracy in QCD calculations.}
\end{figure}

\subsection{Cross-Cutting QCD}
\label{sec:CrossCuttingQCD}
\vspace{-1\baselineskip} Successes across many areas depend on future developments at the intersections of QCD and other domains. To take advantage of precise perturbative QCD calculations, commensurate advances must be achieved in accuracy of long-distance QCD and electroweak contributions,  event generation, error analysis. These tasks require collaborations of experimentalists and theorists, model-builders and QCD experts, and, more broadly, support for the {\it QCD infrastructure} that adapts theoretical tools for experimental analyses according to standard protocols. This subsection presents examples of such cross-cutting issues from the QCD physics report~\cite{Begel:2022kwp}.

\parbold{Comprehensive uncertainty estimates}
Measurements of the $W$-boson mass and other LHC precision observables are affected by many systematic uncertainties arising from both the experimental and theoretical sides. NNLO PDFs used in many analyses are sensitive to modeling of experimental systematics that is not under full control in the PDF fits \cite{Amoroso:2022eow}. 
Accuracy of PDFs can be increased if experimental groups publish more complete models of systematic uncertainties  \cite{Cranmer:2021urp}.
To estimate these uncertainties, high-order calculations for complex final states must be adapted to be fast and practical \cite{Carli:2010rw,Kluge:2006xs,Bern:2013zja}. New types of complexity issues emerge in comparisons of multiparametric models with many parameters to very large data samples expected at the LHC Run 3 and HL-LHC. Undetected biases due to non-representative exploration of contributing systematic factors must be watched for, as has been recently demonstrated on an example of a PDF global fit \cite{Courtoy:2022ocu}. Figure~\ref{fig:QCDatOnePercent} illustrates that elevating the accuracy of QCD calculations to one percent requires both individual precise theoretical calculations as well as accurate supporting theoretical infrastructure that would allow to explore exhaustively the relevant systematic factors. Reaching this target also requires agreed-upon standards and practices for accuracy control at all stages of the analyses. 

\parbold{QCD in new physics searches and SMEFT fits}
The energy reach of many BSM searches at the HL-LHC depends on the interplay between
precision calculations of matrix elements and PDF analyses. At masses above a few TeV, production of BSM particles
requires large momentum fractions $x$, where PDF
uncertainties remain large due to a mix of influences from low statistics,
nuclear corrections, higher-twist contributions, intrinsic heavy-quark
components. Either forward particle production at the LHC or, often
more cleanly, DIS at the EIC can constrain PDFs in the large-$x$
region relevant for the TeV mass range. 

\begin{figure}[tb]
    \centering
   \includegraphics[width=0.45\textwidth]{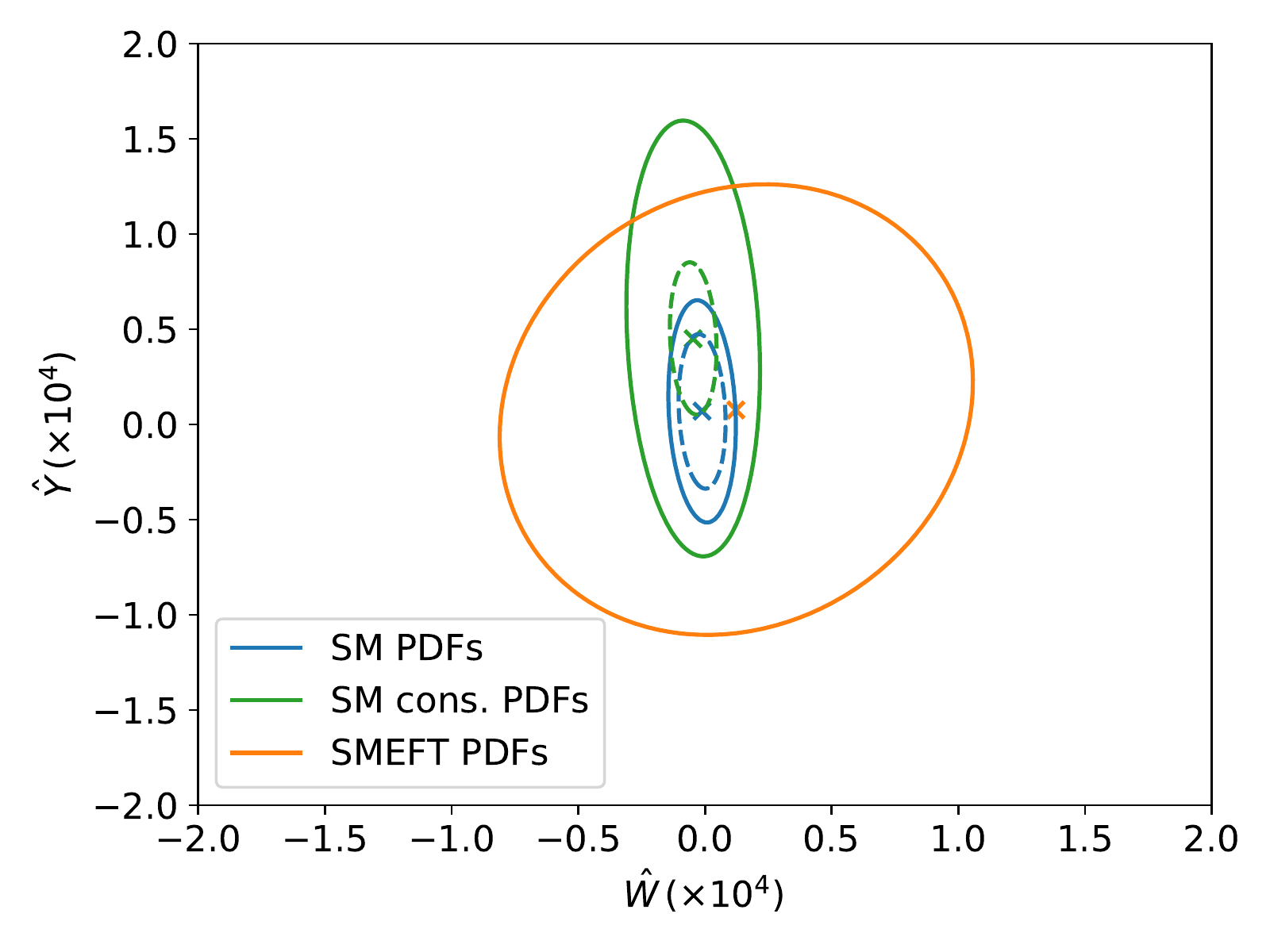}
    \includegraphics[width=0.45\textwidth]{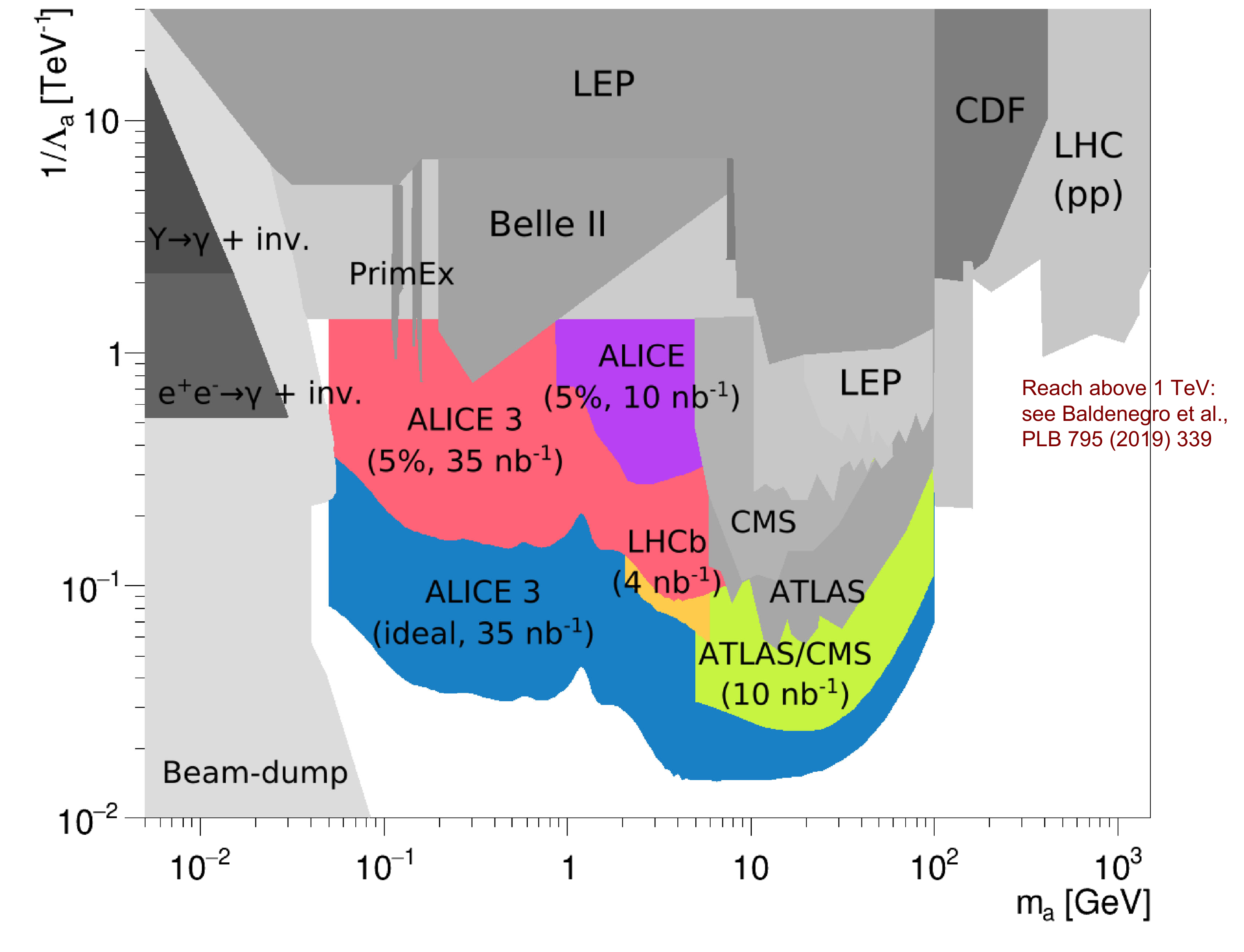}
    \caption{\label{fig:SMEFT_PDFs} Left: The 95\% confidence level bounds on the plane of the Wilson coefficients considered in Ref.~\cite{Greljo:2021kvv} obtained using either fixed SM PDFs (blue) or conservative SM PDFs that do not include high-energy data (green). PDF uncertainties are included in the solid lines and not included in the dashed lines. Results are compared to those obtained in a simultaneous fit of SMEFT and PDFs, when the PDFs are allowed to vary when varying the values of the Wilson coefficients (orange).
    Right: Compilation of exclusion ALPs limits obtained by different studies~\cite{Dainese:2703572, Bauer:2017ris,dEnterria:2022sut}, with limits from $\gamma\gamma$ scattering at ATLAS~\cite{ATL-PHYS-PUB-2018-018}, CMS~\cite{CMS-FSQ-16-012}, the projected performance with ALICE 3 detector~\cite{Adamova:2019vkf}, and exclusive $\gamma \gamma$ production with tagged forward protons at ATLAS and CMS-TOTEM \cite{Baldenegro:2018hng,Baldenegro:2019whq}.
    \label{fig:ALP-HI}}
\end{figure}

 Searches for deviations from SM examined in the language of EFT (e.g. SMEFT analyses) is an
 active research area (see Sec.~\ref{sec:gsmeft} and EF04 report~\cite{Belloni:2022due}). Although the proton structure parametrized by PDFs is intrinsically a low-energy
 input and should in principle be separable from the imprints of SMEFT
 operators, the complexity of the LHC environment might intertwine
 them \cite{Greljo:2021kvv,Carrazza:2019sec,ZEUS:2019cou,CMS:2021yzl,Liu:2022plj,Iranipour:2022iak}.
 As illustrated in the left Fig.~\ref{fig:SMEFT_PDFs}, fitting the
 SMEFT coefficients and PDFs together generally results in different
 constraints than when the fixed PDFs are used.
 Constraints on either the PDFs or EFT operators in low-energy
 experiments, such as the
 EIC,
 where some or all new physics contributions are
 absent, can be crucial for disentangling the SM/BSM degeneracies at
 the (HL-)LHC, especially for spin-dependent EFT operators \cite{Boughezal:2020uwq,Boughezal:2021kla}. 

\label{sec:ef:qcd:gammagamma}
{\bf Ultraperipheral ion collisions (UPCs)}
are electromagnetic interactions of relativistic heavy ions, occurring when the nuclei pass by with impact parameter $b>2R_A$, where $R_A$ is the nuclear radius \cite{Bertulani:2005ru,Contreras:2015dqa}. UPCs are energy-frontier for collisions associated with photons since these photon fields around ions interact with the opposing nucleus, in a photonuclear interaction, or with each other, in a $\gamma\gamma$ collision. 
Quantitative predictions for exclusive processes and/or small momentum fractions probed in the UPCs require coordinated advancements in perturbative QCD, nuclear, and electroweak theory, serving as an example of cross-cutting connections. In addition, the UPCs can be used to search for BSM phenomena. 
ALICE, ATLAS, and CMS have a suite of planned measurements of photonuclear reactions with upgraded detectors. LHCb is well positioned for exclusive production studies in UPCs in the forward direction. Exclusive photonuclear production of vector mesons, jets, and heavy quarks at the highest center-of-mass energies yields a tool to study nuclear parton densities (especially the gluon density) at momentum fractions down to $x \gtrsim 10^{-6}$, where saturation may be most pronounced~\cite{}. The expected performance at HL-LHC and the future plan have been detailed in Refs.~\cite{Citron:2018lsq,Klein:2020nvu}. In the $\gamma \gamma$ mode with tagging on two intact initial hadrons, the LHC can reach unprecedented sensitivities to quartic anomalous couplings, such as $\gamma \gamma \gamma \gamma$, $\gamma \gamma WW$, $\gamma ZZ$, $\gamma \gamma \gamma Z$, $\gamma \gamma t \bar{t}$ only to quote a few~\cite{Baldenegro:2022kaa,Baldenegro:2017aen,Fichet:2014uka,Fichet:2013gsa,Fichet:2016pvq,Fichet:2015vvy,Chapon:2009hh,Kepka:2008yx,Baldenegro:2019whq}. The final states in $\gamma \gamma$ interactions at the LHC are remarkably clean (like at LEP), resulting in sensitivities to quartic anomalous couplings and to the production of axion-like particles (ALPs) at high masses that is better by 2-3 orders of magnitude than in usual LHC searches. Figure~\ref{fig:ALP-HI} (right) illustrates the sensitivity to ALPs in $pp$ and heavy-ion interactions at the LHC, with the coupling plotted as a function of the particle's mass~\cite{dEnterria:2022sut}. Finally, the production of an electron and muon pairs via $\gamma\gamma$ interactions are of great interest since they give access to $\tau$ pairs, which are sensitive to BSM physics, including $\tau$ dipole moment, lepton compositeness or supersymmetry~\cite{Buhler:2022knp}. 

\parbold{Detectors and QCD theory for hadron colliders} \label{sec:FutureDetectors}
Detectors for a possible 100~TeV hadron-hadron collider must implement
innovative designs to achieve the necessary precision when measuring
the SM processes, while also precisely reconstructing multi-TeV
physics objects. They should be able to seek massive hyperboosted
objects inside highly collimated hadronic jets and to deal with a
factor-of-five larger pileup than at the HL-LHC. In theoretical
predictions, a predictive BFKL-like QCD formalism will be
necessary to quantify parton scattering at momentum fractions as low as
$10^{-7}$. Electroweak gauge bosons  $W$ and $Z$, leptons, and top
quarks will be copiously produced in a mix of QCD and EW interactions
and will need to be included into the PDFs together with quarks and gluons
\cite{Han:2020uid}. Detailed studies of hadronic and
electromagnetic showers  will be needed in the next few decades
to achieve the best predictions at future hadron collider energy and luminosity.


\section{The physics beyond the Standard Model}
\label{sec:BSM}

There are abundant reasons why physics beyond the SM of particle physics is likely and, in some cases, unavoidable.  
Such reasons are connected to the fundamental questions that lie at the core of the energy frontier program, as we discussed in Sec.~\ref{sec:EF-bigquestions},  answering which is among the highest priorities of particle physics.
Current and future experiments at the energy frontier offer unique capabilities to explore many of these questions.

In this context, energy frontier explorations follow three main leads:
\begin{enumerate}
\item 
Studies of phenomena that have been observed but for which a fundamental explanation is still lacking. These include:
\begin{itemize}
 \item What is the fundamental composition of Dark Matter?
 \item What is the additional source of CP violation needed to explain the matter-antimatter asymmetry observed in the universe? 
 \item  
 Possible observations of BSM physics referred to broadly as {\it Anomalies}.
\end{itemize}
\item
Following guiding principles forming the basis of the successful stories behind the SM and, more generally, of modern theoretical physics. These may offer us insight on where the theoretical framework is ``hinting'' for a more complete description of Nature, such as:
\begin{itemize}
 \item Naturalness.
 \item The structure of flavor.
\end{itemize}
\item 
Search for possible new phenomena that might not fit in the simplest theoretical extensions of the SM and for which, 
as history has shown many times, particle physics should maintain a wide open view for. This includes addressing questions like:
\begin{itemize}
\item Are there new interactions or new particles around or above the electroweak scale? 
\item Is lepton universality violated? 
\item Are there long-lived or feebly-interacting particles which have evaded traditional BSM searches?
\item Broadly speaking, how can we reduce biases in our searches and conduct them in a more model-independent way?
\end{itemize}
\end{enumerate}

Two main theoretical approaches in exploring BSM physics can be commonly identified. 
The first consists in seeking self-consistent theories that aim to address the questions above and can significantly boost our understanding of the fundamental laws of Nature. 
These well-motivated models of BSM physics, such as SUSY and Composite Models, are self-consistent to high-energy scales, and are excellent test cases for exploring possible experimental signatures and their interrelation. 
Looking beyond these prominent models, the landscape of possible experimental and theoretically-motivated models and signatures is very large. In this approach,  well-defined but incomplete theories extend specific areas without the expectation of full self-consistency. These {\it simplified} models or {\it portal} models are in some cases simplifications of complete theories. It is not practical nor useful 
to try to be exhaustive in projecting the scientific output of projects targeting all such models.
Instead, we focus on a representative set of models and signatures that are deeply connected with the fundamental questions above and
represent a wide range of physics that can be explored at the energy frontier.
Such an approach has the advantage of providing a manageable framework where different experimental results can be easily compared and, eventually, mapped into the parameter space of complete theories. However, the drawback is that those have intrinsically a larger degree of arbitrariness and should be viewed as simpler guiding frameworks for the more general exploration of BSM physics.

In this section, we summarize and chose a few representative benchmark models and scenarios from the BSM Topical Group report~\cite{Bose:2022obr}. These benchmarks include DM-driven considerations, as well as exciting recent development on long-lived particles.  We discuss their implications for the current and future collider programs.

\subsection{Composite Higgs Models}
\label{BSM:compositehiggs}

Determining whether the Higgs boson is an elementary or composite particle remains a fundamental open question.  The idea of a composite Higgs boson is attractive because it avoids the theoretical challenges associated with explaining the relatively small mass of a fundamental scalar particle. 
A composite Higgs boson requires a new strong gauge interaction whose coupling becomes strong above the TeV scale and binds together new elementary constituents. These constituents inevitably form not only the Higgs boson, but also many other bound states, much like the structure seen with QCD dynamics. In such models, the Higgs boson is the lightest bound state similar to the pion of QCD, protected by an approximate global symmetry, and observing other heavier resonances above the Higgs boson mass at the compositeness scale would be a tell-tale sign of Higgs compositeness. Current searches generically constrain the lowest lying resonances to be heavier than the TeV scale with lower mass limits in the range of 1-3 TeV for resonances with spin 1/2, 1, and 2. In addition to the direct production of new resonances, Higgs compositeness would also cause deviations in the couplings of the Higgs boson to gauge bosons and the (composite) top quark. These deviations are inversely proportional to the scale of compositeness and therefore require precision measurements for detection.

The phenomenology of Composite Higgs models is mainly governed by two parameters: the mass~(compositeness) scale $m_*$ and the coupling $g_*$, which sets the scale of the couplings in the EFT Lagrangian.  The strongly interacting model is expected to have $g_*>1$ couplings, while unitarity requires $g_* < 4\pi$. The Wilson Coefficients, defined in Ref.~\cite{deBlas:2019rxi}, can be all parameterized in terms of this mass scale and coupling, modulo order 1 factors. Different colliders have complementary sensitivities to the various operators ($C_\phi$,  $C_{2w}$, $C_w$); these are summarized in the BSM Topical Group report~\cite{Bose:2022obr}.
\begin{figure}[!ht]
  \begin{center}
    \includegraphics[width=0.55\textwidth]{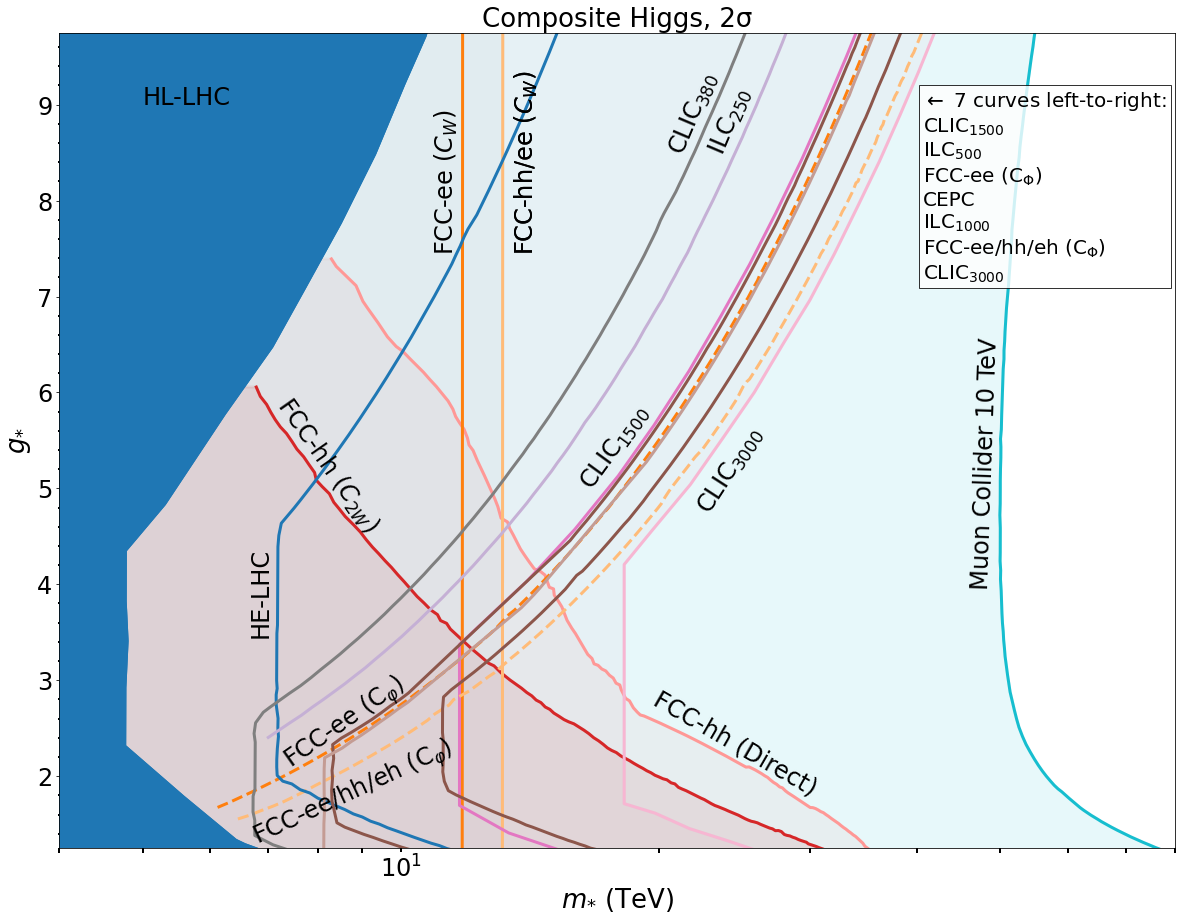} 
  \end{center}
  \caption{Exclusion (2-$\sigma$) sensitivity projections for future colliders. Plot based on Refs.~\cite{Ellis:2691414, Chen:2022msz}.}
    \label{fig:ef08_compHiggs}
\end{figure}
Sensitivity to a toy Composite Higgs model from several future colliders is shown in Fig.~\ref{fig:ef08_compHiggs}. Curves from HL-LHC, FCC-ee/eh/hh, and CLIC are taken from Ref.~\cite{Strategy:2019vxc}. Each Wilson coefficient has a different mass-dependence of the coupling sensitivity. Sensitivity to $C_\phi$ arises primarily through precision measurements of Higgs couplings; sensitivity to $C_{2w}$ arises from measurements of high energy Drell-Yan events; and sensitivity to $C_w$ more broadly comes from electroweak precision fits.  Also shown is the direct search sensitivity for a triplet vector $\rho$ resonance at FCC-hh.  The sensitivity from the 10 TeV Muon Collider is taken from studies of the tree level process $\mu^+ \mu^- \rightarrow hh\nu\bar{\nu}$~\cite{Chen:2022msz}, which provides good sensitivity for $C_{w}$ and $C_{2w}$, but not $C_\phi$.  Sensitivity to $C_\phi$ from Higgs coupling measurements at a Muon Collider~\cite{Buttazzo:2020uzc} are expected to be competitive, but are not shown here. We can also see the complementarity between direct resonance searches and the precision measurements on the SMEFT operators in this figure. This implies that if future discoveries point to signals of composite models, we will be able to use a whole class of operators to interpret other direct searches and pin down the underlying theory.

\subsection{SUSY Models}\label{sec:susy}

Supersymmety (SUSY) is a symmetry that extends the SM
fields by adding a set of partner fields with the same Yukawa couplings and gauge quantum
numbers but different spins. An extended Higgs sector is also required
for SUSY. The motivations for this symmetry include that it results
in the unification of gauge and Yukawa couplings at high energies,
radiative effects directly lead to electroweak symmetry breaking, and in
some versions it naturally contains a DM
candidate. Furthermore SUSY appears in low-energy realizations of
grand unified theories and superstrings, that allow for a consistent
quantization of gravity. SUSY, however, cannot be an unbroken symmetry
of nature, because particles with the same mass but different spin are
not observed. Instead SUSY is assumed to be broken by a set of soft
SUSY-breaking terms. These terms govern the masses of the predicted
SUSY partner particles.

There are many specific models within the SUSY framework. The
sensitivity studies presented here focus on $R$-parity conserving
decays in the minimal supersymmetric standard model (MSSM). In the
MSSM, there is a lightest supersymmetric particle (LSP) which is
protected from decay by $R$-parity and therefore contains a dark
matter candidate. There are many other models including those that
violate $R$-parity in different ways, and the next-to-MSSM (NMSSM) which includes an
additional singlet to dynamically generate the dimensionful $\mu$-term.

\begin{figure}[!ht]
  \begin{center}
    \includegraphics[width=0.75\textwidth]{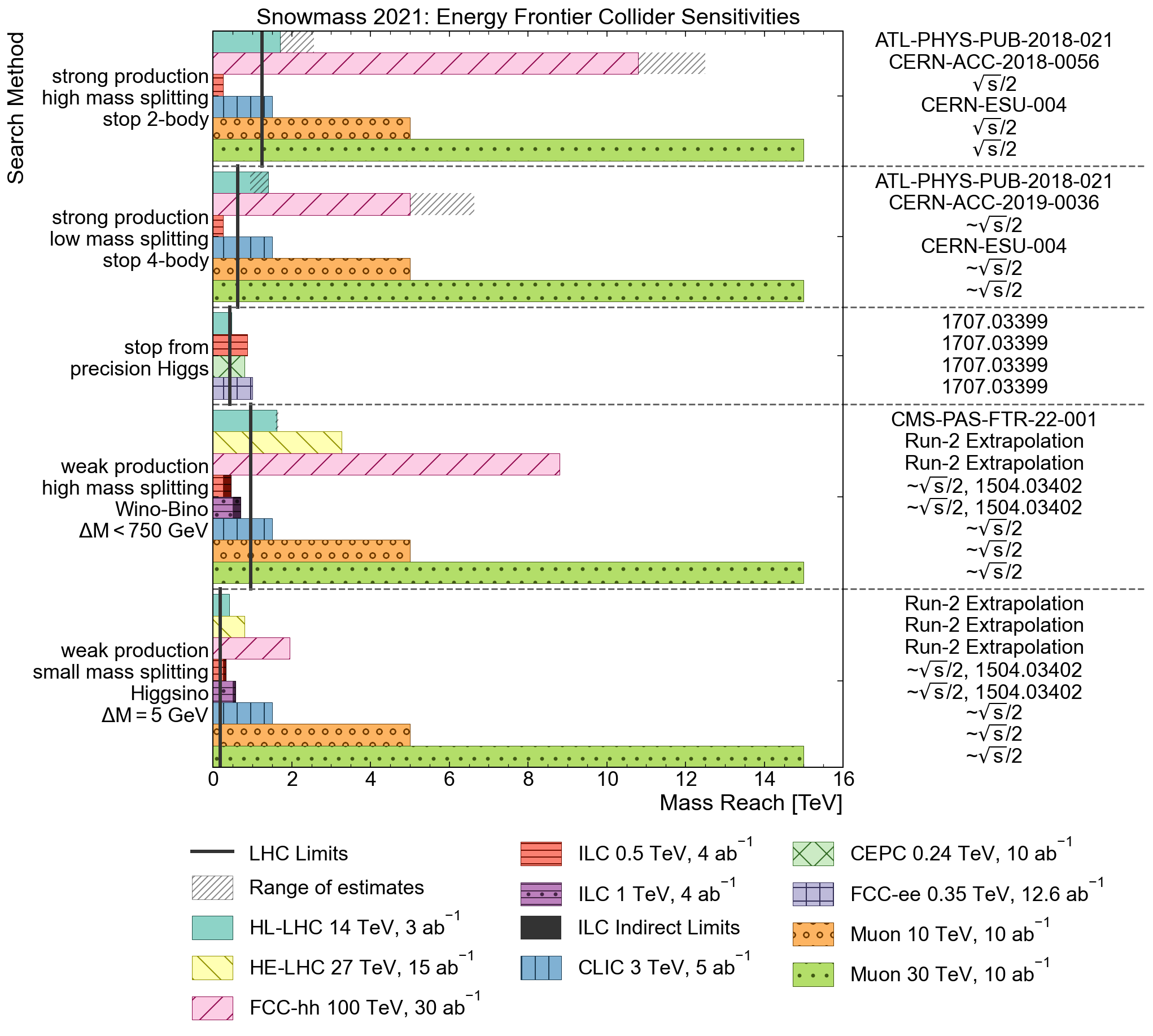} 
  \end{center}
  \caption{Comparison of 95\% exclusion SUSY sensitivities at different colliders for a representative set of scenarios, including small and large mass splittings for stop squarks, which are strongly produced, a large mass splitting Wino-Bino model, and a small mass splitting Higgsino model. The limits come from a combination of dedicated studies and extrapolations based on the collider reach program \cite{ColliderReach}. 
  The hashed gray band indicates the range of estimates in the case where both a dedicated study and Run-2 extrapolation are available.
 Current expected limits from the LHC are shown as vertical lines.
For the ILC limits (also relevant for other $e^+e^-$ colliders, not shown) there are indirect constraints from precision $e^+e^- \rightarrow f \bar{f}$ measurements \cite{Harigaya:2015yaa}}
    \label{fig:susy_summary}
\end{figure}

The Higgs-boson mass in the MSSM is strongly constrained but receives
logarithmic corrections due to stop-squark loops. To achieve the
observed Higgs mass, these loops need to be above order 1 TeV which is
the scale just being reached by the LHC (see Fig. \ref{fig:susy_summary}).
Conversely, for a small mixing between the two stop squarks and $\tan\beta \gg 1$,
the stop-quarks mass can be at most 5-10 TeV \cite{Draper:2016pys}.

This wide variety of phenomena and the fact that it is widely studied
make the MSSM a good context to make comparison plots between
different collider scenarios. Figure \ref{fig:susy_summary} shows the comparative 
sensitivity in the MSSM for a representative set of key points in the
model space. It includes large mass splittings for stop squarks, which are strongly produced, and two example weakly-produced scenarios. The first weak
production example is a classic Wino-Bino model with large mass splitting,
and the second is a Higgsino model with a small mass splitting motivated by naturalness considerations. The relevance of these plots goes beyond
the SUSY context. The relative sensitivity to weak and strong, large
and small mass-splittings are representative of what
sensitivity might be observed in other models with new states. The
studies focus on $R$-parity conserving SUSY where there is a
stable lightest-supersymmetric state that is only weakly
interacting. This is a challenging scenario, particularly for hadron
colliders which have pile-up effects, a range of parton-collision
energies, and reduced resolution and information about the momentum
conservation in the beam direction, however there are some $R$-parity-violating and other scenarios
which are at least as challenging and remain significantly unconstrained by the LHC.
The plot show the 95\% exclusion
limits. For discovery, the sensitivity at a hadron collider would be lower,
while at a lepton collider it would be quite similar.

The range of possible SUSY models is vast. Even within the MSSM, there
are many parameters and the complex interplay can lead to different
signatures. One way to understand this complex space is to construct
a Monte Carlo scan over the parameter space. For this purpose, a good candidate is the pMSSM,
which reduces the 120-parameter MSSM space to 19 free parameters, specified at the EW scale, based on assumptions
related to current experimental constraints (including those from flavor, CP violation, and EW symmetry
breaking) rather than details of the SUSY breaking mechanism. Then with
the scan points, the masses of particles, the relevant couplings, and impacts
on precision measurements, rare processes, and cosmology can be studied.

\begin{figure}[!ht]
\begin{center}
\includegraphics[width=0.49\hsize]{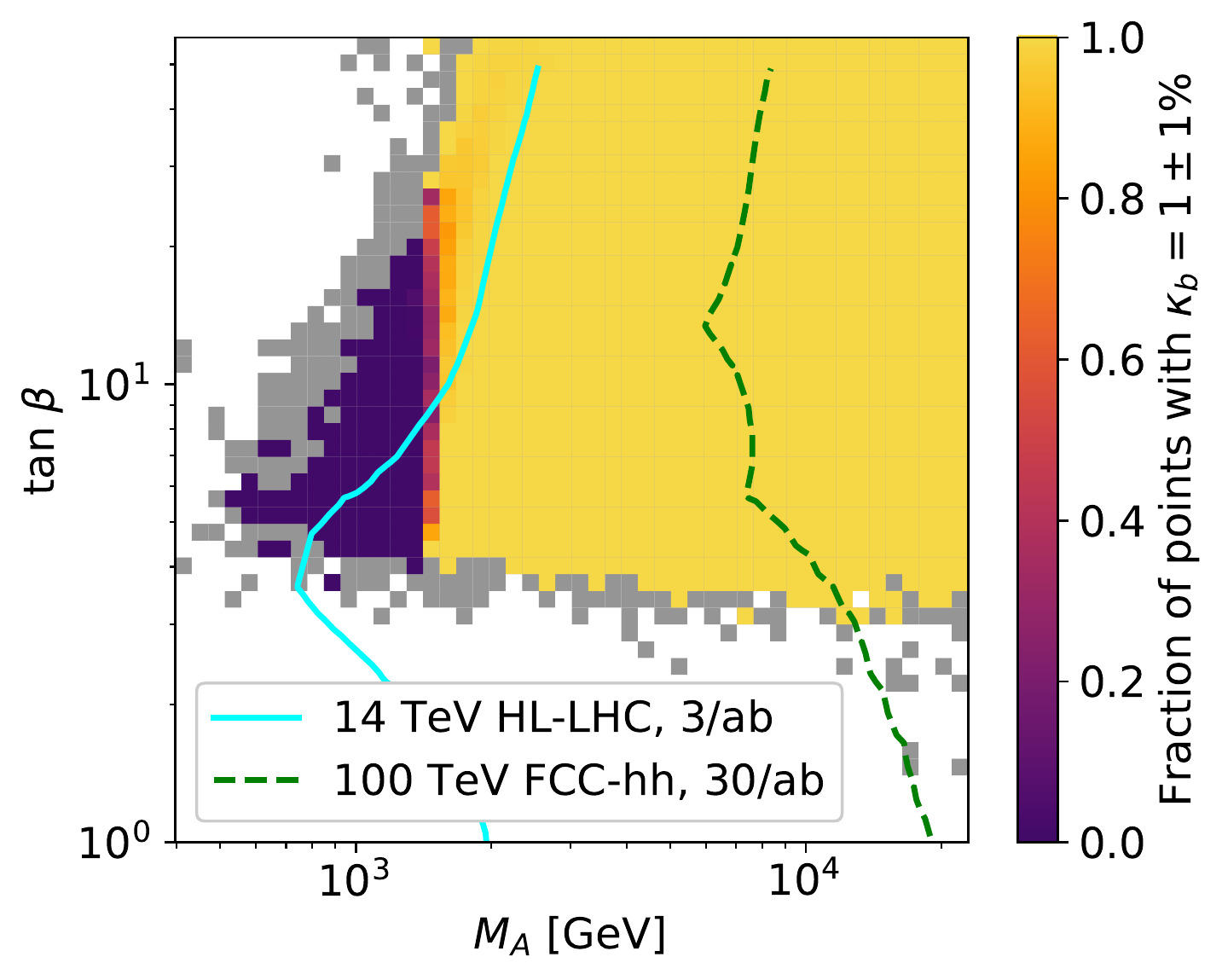}
\end{center}
\caption{The fraction of pMSSM scan points with $\kappa_{b}$ within $1\%$ of the SM expectation of unity as a function of $\tan\beta$ and $M_A$. The range of $1\%$ is chosen to approximately reflect the 95\% CL corresponding to the 0.48\% precision on $\kappa_b$ expected from the FCC-ee/eh/hh combination~\cite{deBlas:2019rxi}.  Expected 95\% CL exclusions from HL-LHC~\cite{Craig:2016ygr, ATLAS:2022hsp} and FCC-hh~\cite{Craig:2016ygr} are overlaid for reference. White bins include no scan points generated by the Markov chain Monte Carlo (McMC) procedure. Gray bins include scan points generated by the McMC, but rejected at a later step because of lack of consistency with current precision measurements and direct searches.}
\label{fig:pmssm}
\end{figure}

Figure \ref{fig:pmssm} shows the dependence of the $h(125)\rightarrow b\bar{b}$
branching fraction on the mass of the psuedo-scalar Higgs $m_A$ and
$\tan\beta$, the ratio of the up and down vacuum expectation values~\cite{pMSSM_2207_05103}. The branching fraction
is reported in terms of the coupling modifier $\kappa_b$ (ratio to the SM coupling). The plot shows the
fraction of pMSSM scan points with $\kappa_{b}$ within $1\%$ of the SM
expectation of unity, where the range of $1\%$ is chosen to
approximately reflect the 95\% CL corresponding to the 0.48\%
precision on $\kappa_b$ expected from a combination of precision
measurements at FCC-ee, FCC-eh, and FCC-hh~\cite{deBlas:2019rxi}.
Expected 95\% CL exclusions from direct searches for pseudoscalar
Higgs boson ($A$) at the HL-LHC and FCC-hh are overlaid for reference;
points to the left of the lines are excluded.  Exclusions at low
$\tan\beta$ are obtained from studies of $A\rightarrow
bb/tt$~\cite{Craig:2016ygr}, and those at high $\tan\beta$ come from
projections for $A\rightarrow \tau^{+}\tau^{-}$~\cite{Craig:2016ygr,
ATLAS:2022hsp}.  As is evident in the plot, direct searches for $A$ at
the HL-LHC are expected to provide better sensitivity to the MSSM than
the highest precision measurements of $\kappa_b$, which shows the
strongest MSSM-related deviation of any Higgs coupling parameter.

\subsection{New Bosons, Heavy Resonances, and New Fermions}

Direct searches for new states beyond specific models provide vital information in our explorations for new physics. The new states could appear as heavy resonances, such as new bosons and new fermions, well-motivated from the model-building perspectives. In this section, we chose the case of new bosons as an example. Various representative examples of studies for current and future facilities are discussed in the BSM Topical Group report~\cite{Bose:2022obr}. 
New heavy vector bosons are often regarded as the standard candle for BSM searches. The canonical example is a $Z'$ boson, which is a neutral vector particle coupling to a SM fermion and antifermion. From the phenomenological perspective, $Z'$ searches are generally characterized by the production coupling, the decay coupling, and the resonance mass, where the decay coupling is typically traded for the branching fraction to the desired final state. A coupling vs.~mass framework for $Z'$ searches~\cite{Dobrescu:2013cmh, Dobrescu:2021vak} helps distill the $Z'$ resonance signal from disparate ultraviolet models into the minimal new physics parameter space relevant for resonance searches at colliders.  This framework also enables direct comparison of experimental reach across different collider proposals, including a comparison of $e^+ e^-$, $pp$, and $\mu^+ \mu^-$ colliders as well as other collider options. 

\begin{figure}[!ht]
\begin{center}
\includegraphics[width=0.50\hsize]{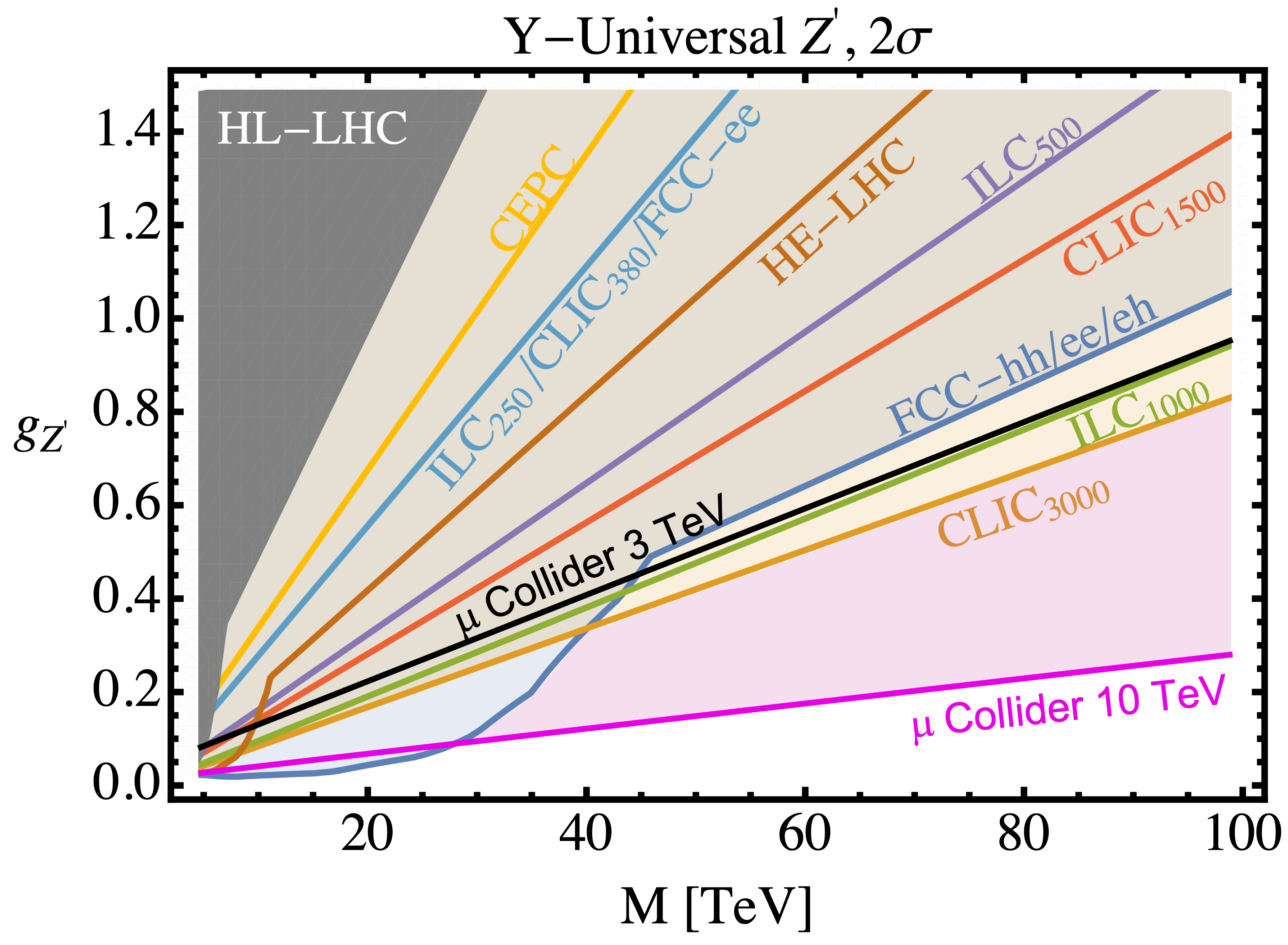}
\end{center}
\caption{Coupling versus mass reach at 95\% CL for $e^+e^-$ colliders (CEPC, ILC, CLIC and FCC-ee), $pp$ colliders (HL-LHC, HE-LHC and FCC-hh), an electron-proton collider (FCC-eh)~\cite{Ellis:2691414}, and the Muon Collider~\cite{Aime:2022flm}.}
\label{fig:UnivZp}
\end{figure}

Two specific $Z^{\prime}$ models studied in the many Snowmass 2021 contributions include the universal $Z^{\prime}$ model and the Sequential Standard Model (SSM). The universal $Z^{\prime}$ model features a $Z^{\prime}$ boson with unit charges for all SM fermions, hence its universal designation.  The sequential standard model (SSM) $Z^{\prime}$ boson follows the same coupling pattern of the SM $Z$ boson, and is the benchmark model most commonly used by experimental searches. Figure~\ref{fig:UnivZp} compares the sensitivity to a universal $Z^{\prime}$ at different colliders~\cite{Aime:2022flm, Ellis:2691414}. Included in the figure are new Snowmass 2021 results for the Muon Collider: it shows that a Muon Collider at $\sqrt{s}=3$ TeV is competitive with other colliders, with sensitivity nearly identical to ILC at $\sqrt{s}=1$ TeV. A Muon Collider at $\sqrt{s}=10$ TeV has the highest mass reach for a universal $Z^{\prime}$ with large couplings $g_{Z^{\prime}}$, probing masses $M_{Z^{\prime}}>100$ TeV, although only indirectly. A Muon Collider at $\sqrt{s}=10$ TeV is sensitive to smaller couplings than the other colliders, with the exception of FCC-hh, which has the highest sensitivity from direct searches within the mass region $M_{Z^{\prime}}<28$ TeV but can directly probe up to about 45~TeV. In this respect the FCC-hh and a 10-TeV Muon Collider will play a very complementary role.  Lepton colliders have an edge in sensitivity when the boson is so heavy that only indirect effects can be measured, arising from the fact that in the signal kinematic distributions the lepton collider experiments benefit from relatively smaller systematic uncertainties.
We can also see in this figure the complementarity between direct resonance searches and precision measurements of the SMEFT operators. 
Direct resonance searches allows us to go to small couplings within accessible energies at lepton colliders. Series of chiral determination of the BSM interference effects can also enable us to extract the new resonance's interaction structure~\cite{Han:2013mra,ILCInternationalDevelopmentTeam:2022izu,deBlas:2022ofj}.

All the different Snowmass contributions related to this topic can be organized into a summary table, see Table~\ref{tab:ZpTable}, to enable an illustrative comparison between the various $Z^{\prime}$ models and current and possible collider scenarios.  To enable the comparison and focus on the mass reach of the different colliders, we adopted the $g_Z^{\prime} = 0.2$ coupling parameter for the universal $Z^\prime$ model, since it roughly aligns with the mass reach for the SSM $Z^{\prime}$ model in the resonance channels studied. 

At  first glance, this table shows the obvious correlation that higher center of mass collider energy reaches higher values in $Z^\prime$ mass, where the orders of magnitude spanned in collider energy pay off in orders of magnitude in $Z^{\prime}$ mass reach.  This is justified since the resonance signal is assured when the $Z^{\prime}$ boson is within the kinematic reach of the collider.  Moreover, for a given operating point of a collider, we see that the two $Z^{\prime}$ model benchmarks have very comparable results, which reflects the fact that the underlying charge assignments of SM fermions to the $Z^{\prime}$ currents only differ by $\mathcal{O}(1)$ factors, and so these results would be broadly applicable in other models where $Z^{\prime}$ bosons couple to all SM fermions, such as in gauged $B - L$ models.  For more fermion-specific models, such as $L_{\mu} - L_{\tau}$ or gauged baryon number, which are equally relevant to the model benchmarks shown in Table~\ref{tab:ZpTable}, the distinction between the different colliders becomes dramatically more important since the $Z^{\prime}$ resonance would be produced via a tree-level coupling in some colliders while only produced via a kinetic mixing coupling or a loop-induced coupling in others.  As a first estimate, the corresponding reach for a point of comparison to Table~\ref{tab:ZpTable} would then adopt a coupling suppressed by a loop factor when the model does not couple to the initial partons at tree-level.

In Table~\ref{tab:ZpTable} the relationship between the $Z^\prime$ mass reach at 95\% CL and the mass reach at 5$\sigma$ depends on the collider type and final state. The two sensitivities are roughly equal for dilepton final states at $pp$ colliders, because the $Z^\prime$ peak is beyond the highest masses of the dilepton continuum background from electroweak production via Drell-Yan, a convincing and background-free exclusion or discovery. For dijet final states at $pp$ colliders, the direct searches for a $Z^\prime$ dijet mass bump has a 95\% CL mass reach that is roughly 20-30\% larger than the $5\sigma$ mass reach, because here the continuum background is larger from strong production of dijets via QCD. Finally, lepton colliders search within the kinematic distributions of fermion pairs for the indirect effects of a $Z^{\prime}$, with huge backgrounds at di-fermion masses significantly lower than the $Z^{\prime}$ pole mass, resulting in a 95\% CL mass reach that is roughly 60-100\% larger than the $5\sigma$ mass reach. Therefore, Table~\ref{tab:ZpTable} illustrates both the power of lepton colliders for indirect discovery of new physics, and the subsequent necessity of a higher energy to directly produce and confirm that new physics. 
\begin{table}[!ht]
\begin{center}
\begin{tabular}{|c|c|c|c|c|c|c|c|} \hline
\textbf{Machine} & \textbf{Type} & $\mathbf{\sqrt{s}}$ & $\mathbf{\int L dt}$ & \textbf{Source} & $\mathbf{Z^\prime}$ \textbf{Model} & $\mathbf{5\sigma}$ & \textbf{95\% CL} \\
        &      & (TeV)      & (ab$^{-1}$) &        &                  & (TeV)     & \textbf{(TeV)}   \\ \hline \hline
        & & & & RH~\cite{Harris:2022kls} & $Z^\prime_{SSM}\rightarrow$ dijet & 4.2 & \textbf{5.2} \\ \cline{5-8}
HL-LHC  & $pp$ & 14 & 3 & ATLAS~\cite{ATL-PHYS-PUB-2018-044} &  $Z^\prime_{SSM}\rightarrow l^+l^-$ & 6.4 & \textbf{6.5} \\ \cline{5-8}
    & & & & CMS~\cite{CMS-PAS-FTR-21-005} & $Z^\prime_{SSM}\rightarrow l^+l^-$ &
-- & \textbf{6.8} \\ \cline{5-8}
    & & & & EPPSU~\cite{Ellis:2691414} & 
   $Z^\prime_{Univ}(g_{Z^\prime}= 0.2)$ & -- & \textbf{6} \\ \hline
 ILC250, CLIC380 & $e^+e^-$ & 0.25 & 2 & 
 ILC~\cite{Suehara:2022pwv} & $Z^\prime_{SSM}\rightarrow f^+f^-$ & 4.9 & \textbf{7.7} \\ \cline{5-8}
or FCC-ee & & & & EPPSU~\cite{Ellis:2691414} &
    $Z^\prime_{Univ}(g_{Z^\prime}= 0.2)$ & -- & \textbf{7} \\ \hline
HE-LHC & $pp$ & 27 & 15 & EPPSU~\cite{Ellis:2691414} & $Z^\prime_{Univ}(g_{Z^\prime}= 0.2)$ & -- & \textbf{11} \\ \cline{5-8}
 & & & & ATLAS~\cite{ATL-PHYS-PUB-2018-044} & $Z^\prime_{SSM}\rightarrow e^+e^-$ & 12.8 & \textbf{12.8} \\ \hline
 ILC & $e^+e^-$ & 0.5 & 4 & ILC~\cite{Suehara:2022pwv} & $Z^\prime_{SSM}\rightarrow f^+f^-$ & 8.3 & \textbf{13} \\ \cline{5-8}
  & & & & EPPSU~\cite{Ellis:2691414} & $Z^\prime_{Univ}(g_{Z^\prime}= 0.2)$ & -- & \textbf{13} \\ \hline
CLIC & $e^+e^-$ & 1.5 & 2.5 & EPPSU~\cite{Ellis:2691414} & $Z^\prime_{Univ}(g_{Z^\prime}= 0.2)$ & -- & \textbf{19} \\ \hline
Muon Collider & $\mu^+\mu^-$ & 3 & 1 & IMCC~\cite{Aime:2022flm} &
$Z^\prime_{Univ}(g_{Z^\prime}= 0.2)$ & 
10 & \textbf{20} \\ \hline
ILC & $e^+e^-$ & 1 & 8 & ILC~\cite{Suehara:2022pwv} & $Z^\prime_{SSM}\rightarrow f^+f^-$ & 14 & 
\textbf{22} \\  \cline{5-8}
 & & & & EPPSU~\cite{Ellis:2691414} & $Z^\prime_{Univ}(g_{Z^\prime}= 0.2)$ & -- & \textbf{21} \\ \hline
 CLIC & $e^+e^-$ & 3 & 5 & EPPSU~\cite{Ellis:2691414} & $Z^\prime_{Univ}(g_{Z^\prime}= 0.2)$ & -- & \textbf{24} \\ \hline
 & & & & RH~\cite{Harris:2022kls} & $Z^\prime_{SSM}\rightarrow$ dijet & 25 & \textbf{32} \\ \cline{5-8}
FCC-hh & $pp$ & 100 & 30 & EPPSU~\cite{Ellis:2691414} & $Z^\prime_{Univ}(g_{Z^\prime}= 0.2)$ & -- & \textbf{35} \\ \cline{5-8}
 & & & & EPPSU~\cite{Helsens:2019bfw} &
$Z^\prime_{SSM}\rightarrow l^+l^-$ & 43 & 
\textbf{43} \\ \hline
Muon Collider & $\mu^+\mu^-$ & 10 & 10 & IMCC~\cite{Aime:2022flm} &
$Z^\prime_{Univ}(g_{Z^\prime}= 0.2)$ & 42 & \textbf{70} \\ \hline
\end{tabular}
\end{center}
\caption{List the operating point and mass reach, for 5$\sigma$ discovery and 95\% CL exclusion, of the SSM $Z^{\prime}$ model taken from Refs.~\cite{Harris:2022kls,Helsens:2019bfw,ATL-PHYS-PUB-2018-044,CMS-PAS-FTR-21-005,Suehara:2022pwv}, and the mass reach of the universal $Z^{\prime}$ model with a coupling $g_{Z^{\prime}}=0.2$ from Refs.~\cite{Aime:2022flm,Ellis:2691414}, determined from Fig.~\ref{fig:UnivZp}. }
\label{tab:ZpTable}
\end{table}

\begin{figure}[htb]
\begin{center}
\includegraphics[width=\hsize]{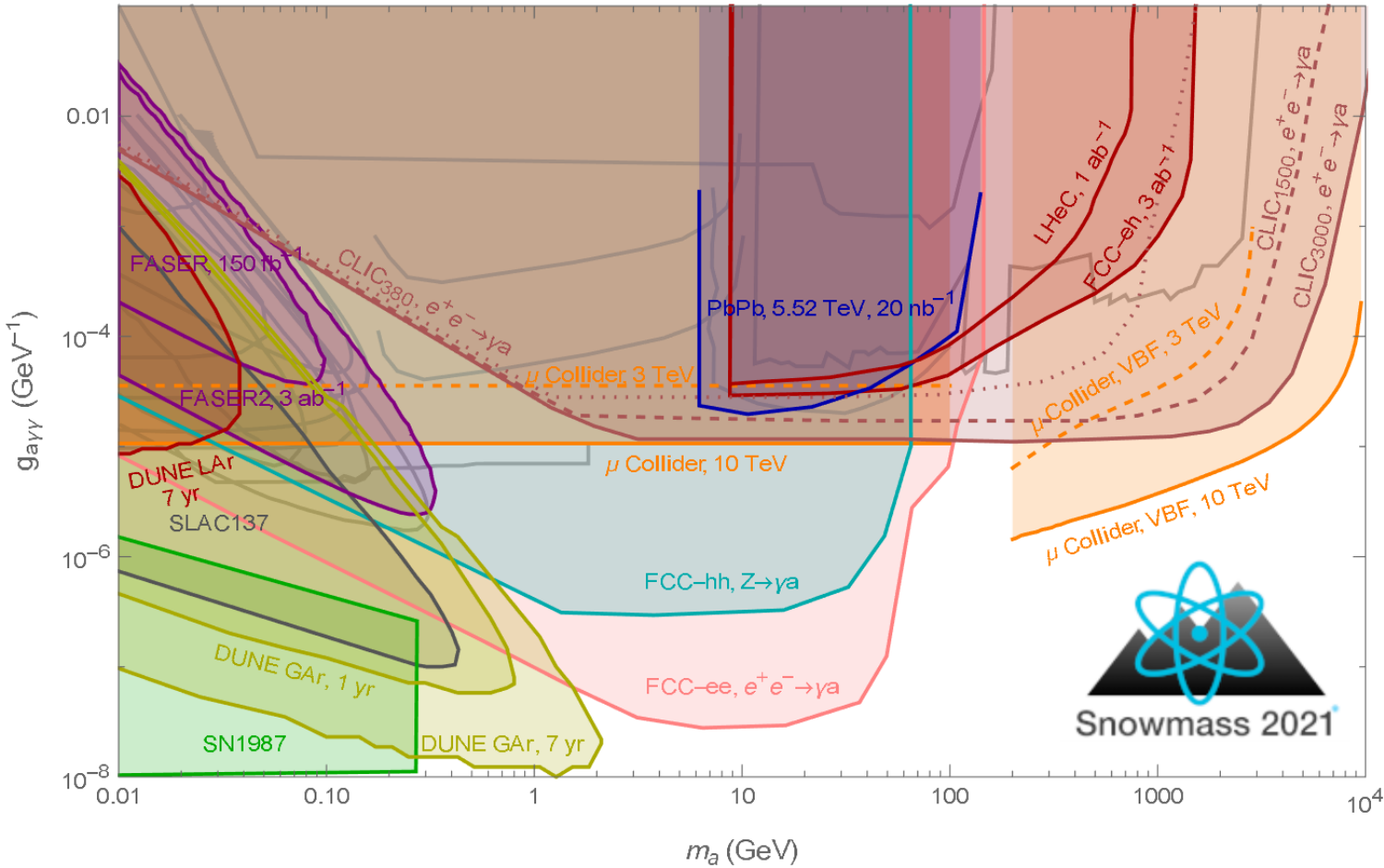}
\end{center}
\caption{ The ALP coupling in the diphoton channel $g_{a\gamma\gamma}$ versus 95\% CL mass reach is shown for multiple  colliders~\cite{Bernardi:2022hny,Strategy:2019vxc}, including Snowmass 2021 studies on the Muon Collider~\cite{MuonCollider:2022xlm,Han:2022mzp} (orange) at $\sqrt{s}=3$ TeV (dashed) and 10 TeV (solid) as well as from the DUNE near detector~\cite{Brdar:2020dpr} with liquid Argon technology (dark red) and gaseous Argon technology (dark yellow).}
\label{fig:ALPs-dijetres}
\end{figure}

Searches for light but very weakly coupled new particles are motivated by a variety of new physics scenarios. 
A prime example is the ALP. While ALPs have a rich phenomenology, including prompt and long-lived signatures, one of the main phenomenological target at experiments is the ALP coupling to two photons, allowing a smooth transition between traditional QCD axion and ALP parameters. Figure~\ref{fig:ALPs-dijetres} shows the results of Snowmass 2021 studies on the sensitivity to ALP, including the case of the Muon Collider~\cite{MuonCollider:2022xlm, Han:2022mzp} compared with other colliders~\cite{Bernardi:2022hny, Strategy:2019vxc} in the $m_a$ vs.~$g_{a \gamma \gamma}$ plane. For ALP decays to diphotons, a 10 TeV Muon Collider is the most sensitive to high ALP masses $m_a>200$ GeV, from vector boson fusion production processes ($VV\rightarrow a\rightarrow \gamma \gamma$). FCC-ee has the best sensitivity in the medium mass range $1 \lesssim m_a < 100$~GeV, from the associated production process ($Z\rightarrow \gamma a$), thanks to the potentially very large integrated luminosity expected at the $Z$ pole in circular $e^+e^-$ colliders. In the more near term, the best collider limits on ALPs coupling to photons over the range $m_a \approx 0.1-100$~GeV will be set by exploiting photon-photon collisions in ultraperipheral interactions of heavy-ions during the HL-LHC phase \cite{Bruce:2018yzs,dEnterria:2022sut}. It is worth noting that the ALP is typically expected to have non-suppressed coupling to gluons, in particular in its connection to the Strong CP puzzle of QCD~\cite{Hook:2019qoh,DiLuzio:2020wdo}. Having gluonic couplings changes the considerations for the search channels and the performance at different facilities appreciably (see recent phenomenological studies~\cite{Kelly:2020dda,Agrawal:2021dbo,Feng:2022inv}). 

The sensitivity to dijet resonances at $pp$ colliders was explored during Snowmass 2021 as discussed in Refs.~\cite{Harris:2022kls,Bernardi:2022hny,Guler:2022czi}. The process, $pp \rightarrow X \rightarrow 2\mbox{ jets}$, is an essential benchmark of discovery capability of $pp$ colliders and is sensitive to a variety of models of new physics at the highest mass scales.  The sensitivity to a dijet resonance is mainly determined by its cross section. The study considered strongly produced models, those with large production cross sections, that include scalar diquarks, colorons and excited quarks. At the highest resonance masses these strongly produced models can only be observed at a $pp$ collider, as lepton colliders can only produce diquarks and excited quarks in pairs at significantly lower masses. Also considered are weakly produced models, with production cross sections that are roughly two orders of magnitude smaller, that include $W^{\prime}$s, $Z^{\prime}$s and Randall-Sundrum gravitons, which can also be observed at lepton colliders as previously discussed.

\subsection{Long Lived Particles}

Particles with long lifetime arise in many generic BSM models. The space of signatures for these long-lived particles (LLP) signatures is very rich and complicated, ranging from exotic-looking tracks to heavy stable charged particles to various types of displaced objects (e.g. vertices, jets, leptons). Here we highlight two examples. More benchmark cases, results, and discussions can be found in the BSM Topical Group report~\cite{Bose:2022obr}.

The first example is that of LLPs that are electrically charged and can be produced by many different models. In the case of one particular signature, if the charged LLP decays within the detector, the LLP could produce a disappearing track signature if it decays to neutral and/or very soft particles that cannot be reconstructed. Disappearing tracks are particularly motivated in models of SUSY and dark matter.

\begin{figure}[!ht]
\begin{center}
\includegraphics[width=0.6\hsize]{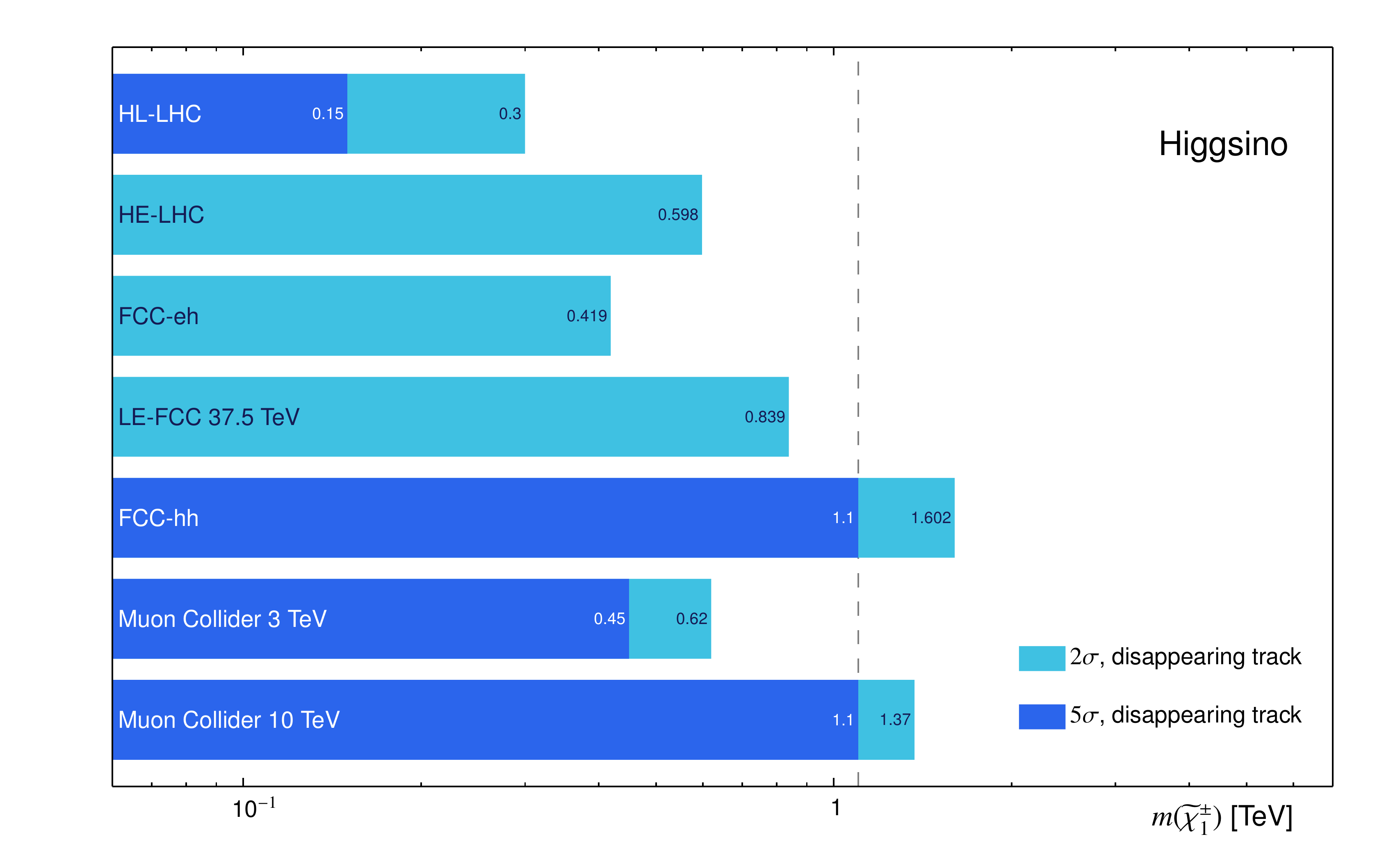}
\end{center}
\caption{Overview of the sensitivity to the pure Higgsino, assuming its natural mass splitting, for various future colliders. Figure adapted from Ref.~\cite{Capdevilla:2021fmj}.}
\label{fig:disappearingTracks2}
\end{figure}

 Figure \ref{fig:disappearingTracks2} shows the projected reach of disappearing track signatures at the HL-LHC~\cite{ATLAS:2022hsp}, HE-LHC~\cite{Han:2018wus}, LE-FCC~\cite{Ellis:2691414}, FCC-ee~\cite{Ellis:2691414}, CEPC~\cite{Ellis:2691414}, CLIC~\cite{Klamka:2790402}, ILC~\cite{PardodeVera:2020zlr}, FCC-eh \cite{Curtin:2017bxr}, FCC-hh~\cite{Saito:2019rtg}, and several high energy Muon Colliders~\cite{Capdevilla:2021fmj}, assuming a pure Higgsino with its natural mass splitting. Further discussion on these constraints and their implications for dark-matter can be found in the section on dark matter. The sensitivities are driven by many factors, and in particular, the proximity of the tracking system to the interaction points and low pile-up environment could help enhance them.

Many new physics models with an hidden sector could lead to long-lived signatures and displaced vertices. Here we use a simplified heavy neutral lepton (HNL) model, motivated by the neutrino mass model building and the seesaw mechanism, as an example to show the different coverage of displaced signatures in the current and future experiments. For more details on different models and search coverage, see the discussion in the BSM Topical Group report~\cite{Bose:2022obr}.
Extensions to the SM that account for neutrinos masses typically incorporate heavy neutrinos that are ``sterile'' with very small mixings to SM neutrinos, and they have masses much larger than the eV scale. Neutral leptons with masses on the MeV scale or higher are referred to as \emph{heavy} neutral leptons (HNLs).

\begin{figure}[htb]
    \centering
    \includegraphics[width=0.75\textwidth]{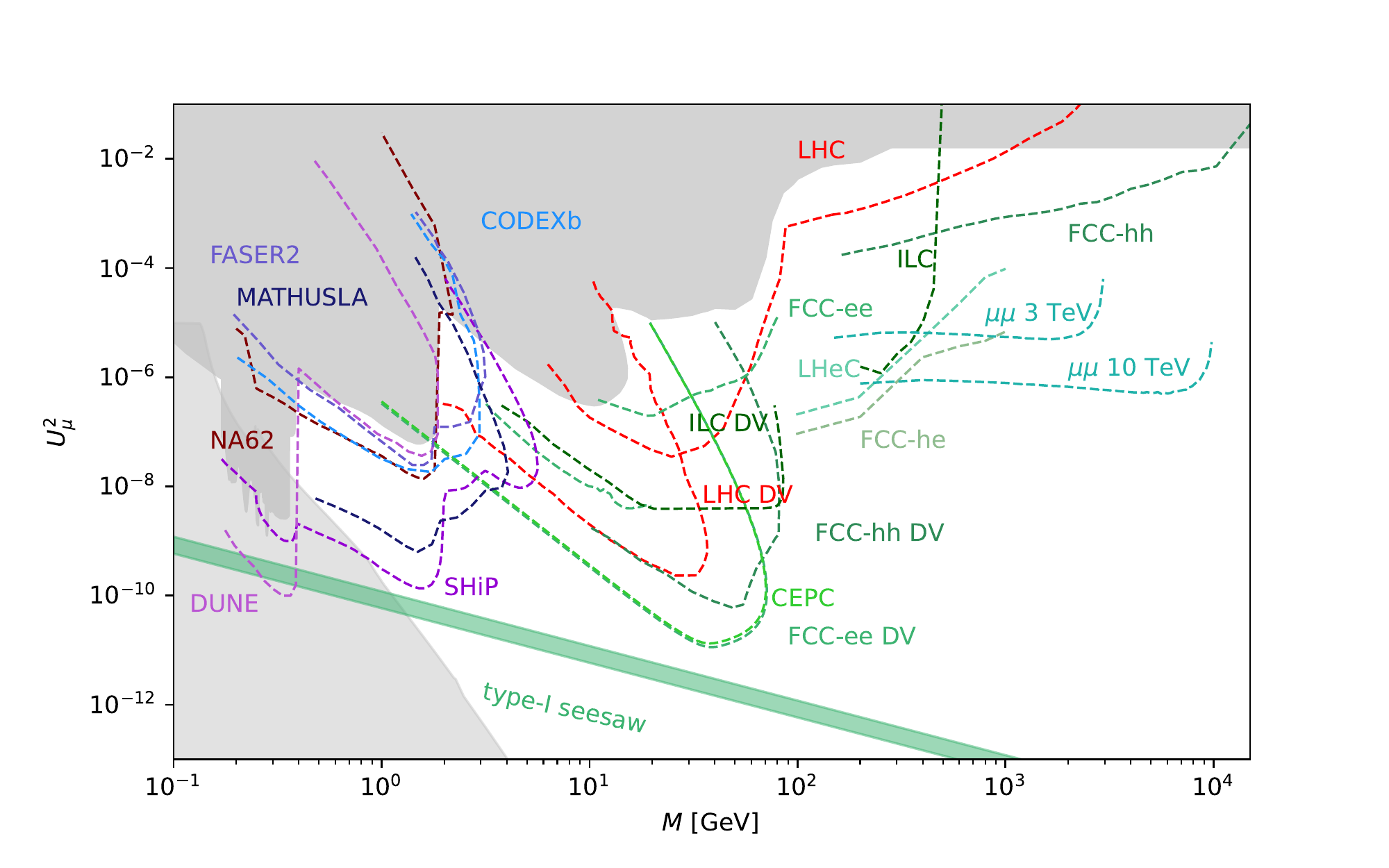}
    \caption{
    Constraints and future sensitivities for HNLs with mass $M$ and mixing $U_\mu^2$ with muon neutrinos (summed over three HNL flavours). \emph{Medium gray:} Constraints on the mixing of HNLs from past experiments~\cite{CHARM:1985nku,Abela:1981nf,Yamazaki:1984sj,E949:2014gsn,Bernardi:1987ek,NuTeV:1999kej,Vaitaitis:2000vc,CMS:2018iaf,DELPHI:1996qcc,ATLAS:2019kpx,CMS:2022fut}.
 \emph{Colourful lines:}
Estimated sensitivities of the main HL-LHC detectors~\cite{Izaguirre:2015pga,Liu:2019ayx,Drewes:2019fou}) and NA62~\cite{Drewes:2018gkc},
with the sensitivities of selected planned or proposed experiments~\cite{Ballett:2019bgd,FASER:2018eoc,SHiP:2018xqw,Gorbunov:2020rjx,Curtin:2018mvb,Aielli:2019ivi}, as well proposed future colliders~\cite{Alimena:2022hfr,Shen:2022ffi,BayNielsen:2017yws,Antusch:2016ejd,Pascoli:2018heg,Mekala:2022cmm,Antusch:2016vyf,Antusch:2019eiz,ZhenLiuTeamHNLMuC,TaoLiuTeamHNLMuC}.
\emph{Green band:} Indicative lower bound on the total HNL mixing $U_e^2+U_\mu^2+U_\tau^2$ from the requirement to explain the light neutrino oscillation data \cite{Esteban:2020cvm} when varying the lightest neutrino mass and marginalising over light neutrino mass orderings.
\emph{Light gray:} Lower bound on $U_\mu^2$
 from BBN~\cite{Sabti:2020yrt,Boyarsky:2020dzc}.
 More details can be found in Refs.~\cite{Bose:2022obr,Abdullahi:2022jlv}
    }
    \label{fig:HNLtype1disp}
\end{figure}

In the timescale of HL-LHC and beyond, many proposed experiments could offer discovery potential for Type-1 Seesaw HNLs, particularly in the low-mass/small-coupling region where long-lived searches will be required. Figure~\ref{fig:HNLtype1disp}, adapted from~\cite{Alimena:2022hfr, HNL_PresFut}, shows the expected reach of experiments such as FASER2, MATHUSLA, CODEXb, and future colliders. 
To guide the eye, the “type-I seesaw” line indicates the approximate parametric scaling associated with a simplified model with just a single neutrino flavor. Realistic three-generation models can populate the experimentally accessible regions in Fig.~\ref{fig:HNLtype1disp}.


\subsection{Dark Matter}
\label{sub:darkMatter}

The existence of dark matter (DM) is some of the most concrete evidence for particle physics beyond the Standard Model. 
However, little is known about DM beyond its gravitational effects. 
One of the central questions of particle and astroparticle physics is ``\textit{What is the nature of dark matter and how does it interact with ordinary matter?}".

Observing non-gravitational interactions of DM would bring us closer to answering this question.
Many theoretical hypotheses foresee interactions between the DM and the SM to explain the measured relic dark matter density in the universe. 
Such DM-SM interactions are also key to terrestrial searches for dark matter.
Since DM-SM interactions are generally feeble (as a consequence of DM's \textit{darkness}), DM produced in SM-particle collisions would escape detection at collider experiments. 
Nevertheless, these invisible particles can be discovered when collisions appear to fail to conserve transverse momentum, leading to missing transverse momentum in a hermetic detector. 

While finding the connection between new invisible particles at colliders and astrophysical DM requires an observation of DM in astrophysics experiments\footnote{The search for dark matter must be conducted in synergy between different Frontiers, using multiple probes and assumptions. This is discussed in the Snowmass 2021 cross-frontier report on DM complementarity~\cite{Boveia:2022syt}.}, to verify the new particle's cosmological connection,
collider experiments have unique abilities to study DM-SM interactions and the properties of DM in detail, including discovering new high energy particles that mediate the interactions.
The historical discovery of other invisible particles -- neutrinos -- offers an example of such synergy: an observation of neutron decays at low energies required high energy colliders for the study of neutrino interactions and the discovery and study of the weak force mediators.

Following this example, here we focus on DM search benchmarks where the SM-DM interaction is mediated by either an existing or a new particle, which in turn decays into invisible (DM) particles and can also decay back into SM particles allowing for a discovery in visible final states. 
Collider experiments can also investigate theories of complex dark sectors and the wealth of particles they hypothesize.


\subsubsection{Testing the traditional WIMP paradigm}

Among the possible theories of particle DM, WIMPs have been the most prominent target of experiments.
These are also theories to which colliders are especially well-suited.
In the traditional WIMP scenario, DM consists of a single particle of mass between roughly 1 GeV and 100 TeV.
As a thermal relic, its mass and coupling to the SM allow them to reach a common thermal equilibrium in the early universe and control the present abundance of dark matter, largely independent of the universe's initial conditions and evolutionary history.
WIMP candidates feature in many theories with connections to other EW-scale new physics, including supersymmetry, the archetype for the WIMP idea~\cite{Jungman:1995df}.

Since the 2013 Snowmass report, experiments at LHC Run 2, as well as at underground facilities and with astrophysical observations, have made significant progress in ruling out some possible WIMP parameters. Nevertheless, others remain a compelling target for DM searches at colliders and beyond (see, among numerous others, Refs.~\cite{Leane:2018kjk,GAMBIT:2021rlp,Arcadi:2017kky,Boddy:2022knd,Blanco:2019hah,Bottaro:2021snn,Smirnov:2019ngs}, for example).

Broadly speaking, WIMPs can be classified according to the way in which the DM particle coupled with the Standard Model.
Here we discuss some of the most widely-studied model categories. 

\paragraph{Minimal WIMPs.}

One simple, predictive WIMP scenario identifies the DM particle as the lightest member of an EW multiplet.
Examples are the Higgsino (a Dirac fermion doublet) and the Wino (a Majorana fermion triplet) in supersymmetry.
More general cases have also been considered \cite{Cirelli:2005uq,Cirelli:2009uv}.
In these models, the interaction strengths are fixed by the SM gauging couplings, and the only free parameter, the mass of the dark matter particle, $m_{\chi}$, is chosen to give the observed thermal relic abundance~\cite{Planck:2018vyg}.
These ``thermal'' targets are typically in the TeV range~\cite{Belotsky:2005dk,Hisano:2006nn,Cirelli:2007xd,An:2016gad,Mitridate:2017izz,DelNobile:2015bqo} and it is critical to probe them at future high-energy colliders.

\begin{figure}[!ht]
\centering
\includegraphics[height=1.8in]{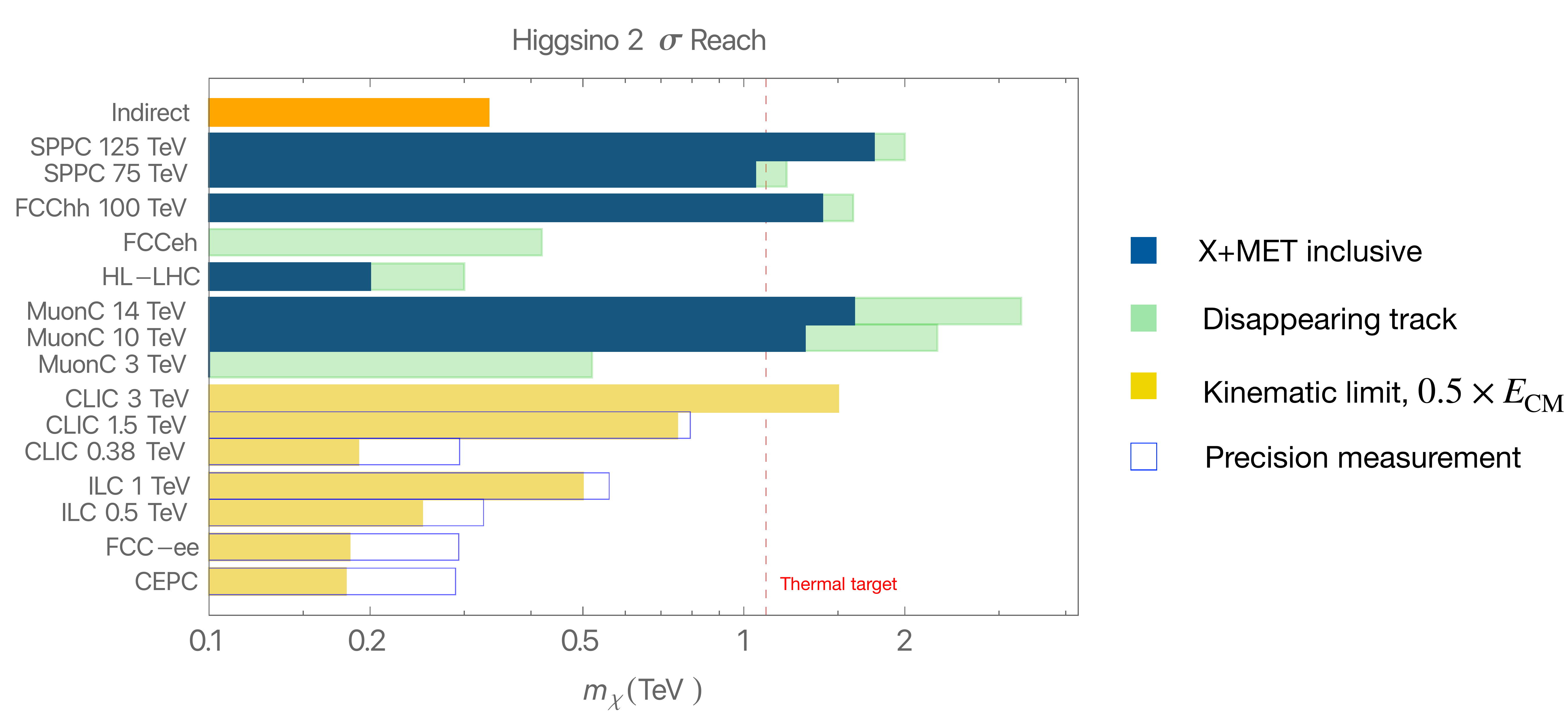}
\includegraphics[height=1.8in]{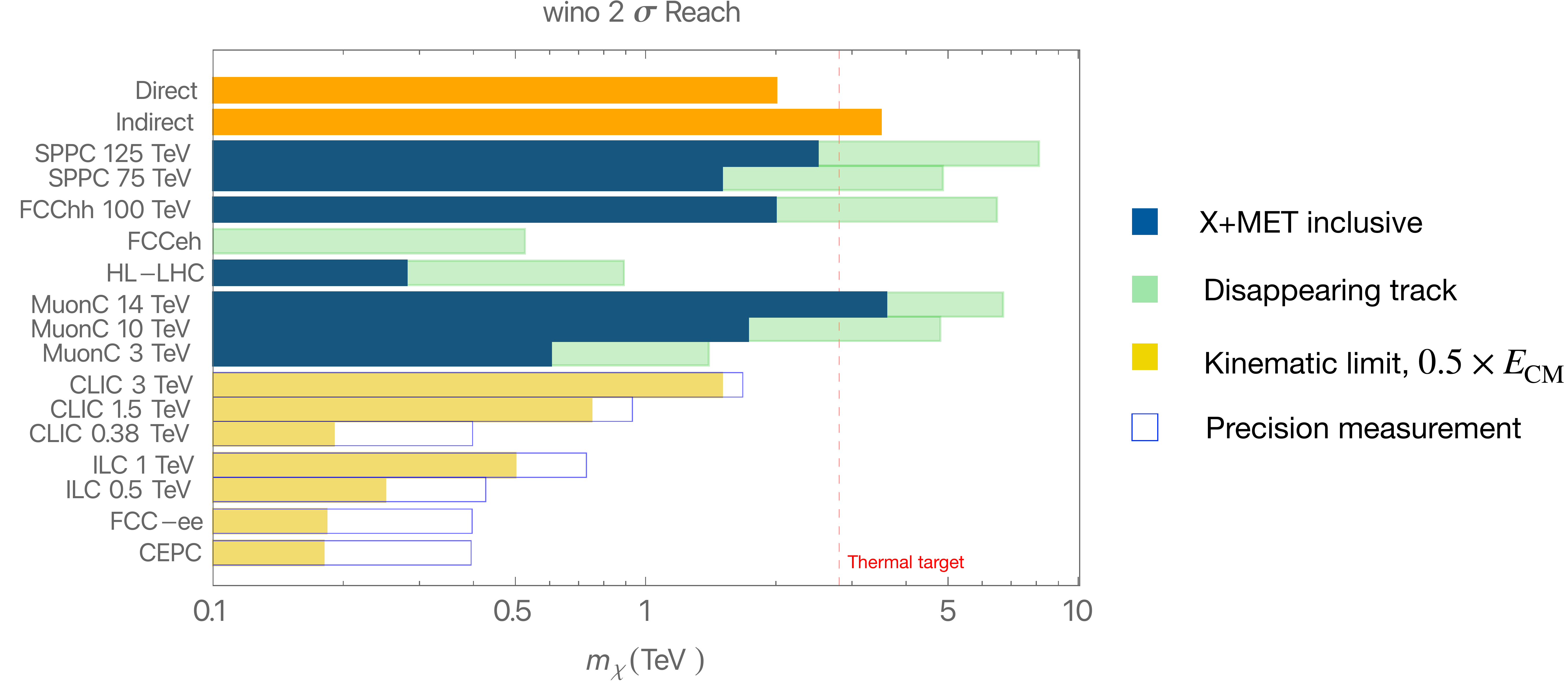}
    \caption{A summary of the reach of future colliders for simple WIMPs from four search strategies, as indicated in the legend. For comparison, the reaches of the direct and indirection detectors are also included (orange bars at top). For lepton colliders where a detailed study is not available, the kinematic limit $m_\chi = 0.5 \times E_{\rm CM}$ is used to indicate potential reach; Muon-collider studies suggest this is likely to be an overestimate. Hadron-collider projections are from \cite{Saito:2019rtg,CidVidal:2018eel}, while lepton-collider projections are from projections in \cite{Han:2020uak,Capdevilla:2021fmj,Han:2022ubw}.}
    \label{fig:WIMPSummary}
\end{figure}

Figure~\ref{fig:WIMPSummary} summarizes projections for $2 \sigma$ exclusions from future colliders for Higgsino and Wino DM.
New studies of the reach of a high energy Muon Collider \cite{Han:2020uak,Han:2022ubw,Capdevilla:2021fmj,Bottaro:2021srh,Bottaro:2021snn,Bottaro:2022one,Black:2022qlg} were developed within Snowmass 2021 in addition to the studies within the Briefing Book for the European Strategy for Particle Physics Update~\cite{Strategy:2019vxc}.

At hadron colliders, the dominant signature of WIMP DM is often (but not always) large missing transverse momentum recoiling against SM particles (e.g. jets)~\cite{Low:2014cba,Han:2018wus}.
At lepton colliders, other signatures involving SM EW gauge bosons and leptons come into play~\cite{Han:2020uak,Bottaro:2021snn,Han:2022ubw,Bottaro:2022one,Black:2022qlg}. 
These signals are relatively insensitive to the splitting between masses of the EW multiplet and, hence, more robust against variations beyond the minimal scenario.
The loop-induced mass splitting among the components of the EW multiplet also results in a disappearing track signature which can enhance the reach but is more sensitive to the mass splitting and detector backgrounds. 

The basic lesson from Fig.~\ref{fig:WIMPSummary} is that high energy colliders, such as a hadron collider with $E_{\rm CM} \simeq 100 $ TeV or a Muon Collider with $E_{\rm CM} \simeq 10 $ TeV, can definitively test these scenarios.
High energy $e^+ e^-$ colliders, with energies up to 3 TeV, can cover lower-mass regions.

\paragraph{Higgs mediation.}

DM could also couple to the SM via \textit{portals}, which is a direct coupling via gauge-invariant operators. 
The Higgs boson provides a prime example: as a spin-0 particle, this `Higgs portal' allows a renormalizable coupling with the DM that can have a sizable effects on SM Higgs properties.
Searches at colliders are powerful probes of the Higgs portal.
For example, DM production would enhance tiny rate of invisible decays of the Higgs predicted by the SM, provided the DM mass is less than half the Higgs mass.
Precision measurements of the Higgs couplings, another main objectives of a future collider, would also contribute to probe the Higgs portal scenario.
Future prospects for the Higgs portal were studied in the European Strategy physics Briefing Book~\cite{Strategy:2019vxc} and are discussed in the BSM Topical Group report~\cite{Bose:2022obr}.

Models involving a larger extension of the scalar sector can also be probed with Higgs measurements and BSM Higgs searches. 
Example of such extensions are the Inert Doublet Model, where an extra scalar doublet provides a DM candidate, and a 2HDM where an additional pseudoscalar has direct couplings to DM. 
The HL-LHC and lepton colliders are expected to be sensitive to large parts of the parameter space of these models, as can be seen in studies targeting specific signatures \cite{ATLAS:2022hsp, CMS:2022sfl,Kalinowski:2022fot}.

\paragraph{BSM mediation.}
Another category of models involves DM interacting with the SM only via one or more new bosons.
Although this category encompasses a huge set of possible models, it is likely that only a few new particles will be relevant in the early phase of a discovery.
The relevant phenomenology can be captured by simplified models of collider DM production.

The simplified models used to design present LHC searches, described in Ref.~\cite{Abercrombie:2015wmb}, are inspired by the vector and scalar bosons of the SM. 
They involve a single fermionic DM species and a mediator particle, either a vector ($Z^\prime$) with pure vector or axial-vector couplings to SM and DM fermions, or a scalar or pseudoscalar boson with Yukawa couplings to the SM and DM fermions.
This involves 4 to 6 free parameters such as the masses of the DM and the mediator, and their couplings to the SM and DM.

Contributions to Snowmass 2021 have significantly extended the projections made for these models by the recent European Strategy Update Briefing Book~\cite{Strategy:2019vxc}. 
Prior results, originally focusing on relatively large coupling values appropriate for LHC Run 2~\cite{Boveia:2016mrp,Albert:2017onk}, have now been extended to arbitrary coupling values~\cite{Albert:2022xla}. 
This provides a better picture of how future experiments would improve upon current searches and complement direct-detection and indirect-detection experiments, especially when the dark matter mass is less than a few GeV. 
The new studies of coupling dependence are especially useful in understanding the effect of different coupling values assumed for the signal at colliders~\cite{Boveia:2022jox}, while direct detection only depends on a specific combination of the couplings and the mass of the mediator. 


\begin{figure}[htb!]
\centering
\subfloat[]{\label{subfig:gqscan-dmLight}\includegraphics[width=0.32\textwidth]{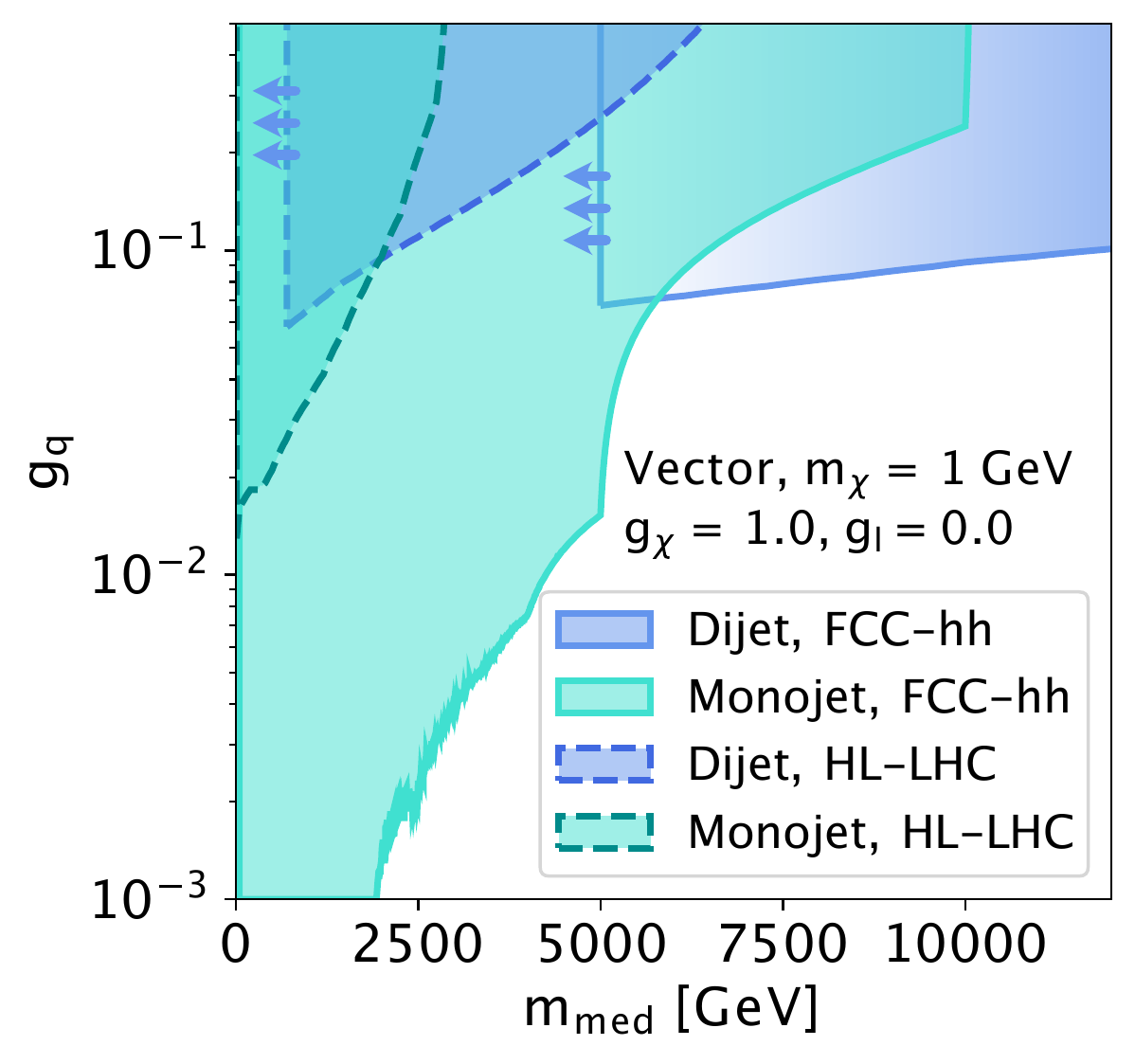}}
\subfloat[]{\label{subfig:gchiscan-dmLight}\includegraphics[width=0.32\textwidth]{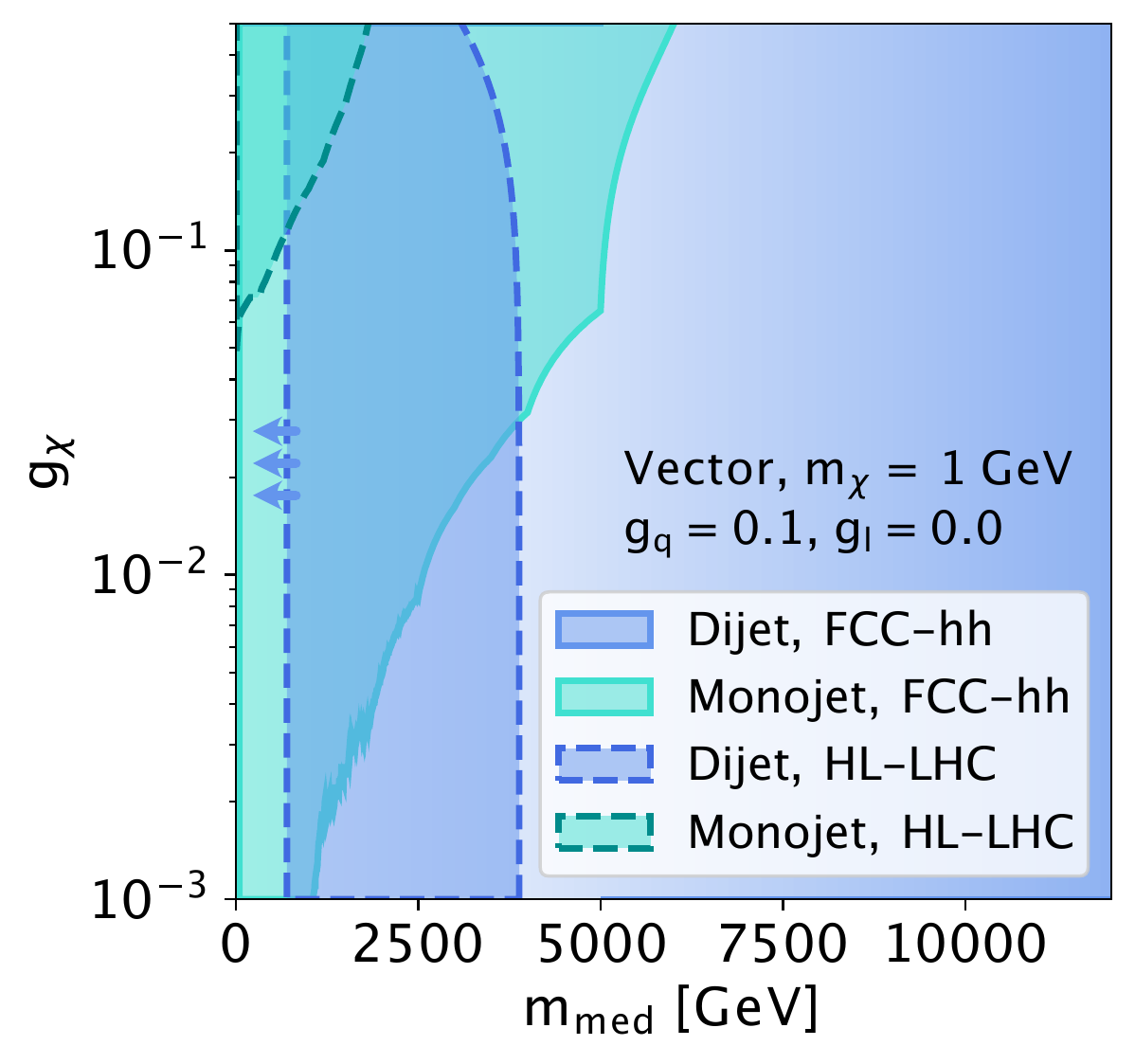}}
\subfloat[]{\label{subfig:glscan-dmLight}\includegraphics[width=0.32\textwidth]{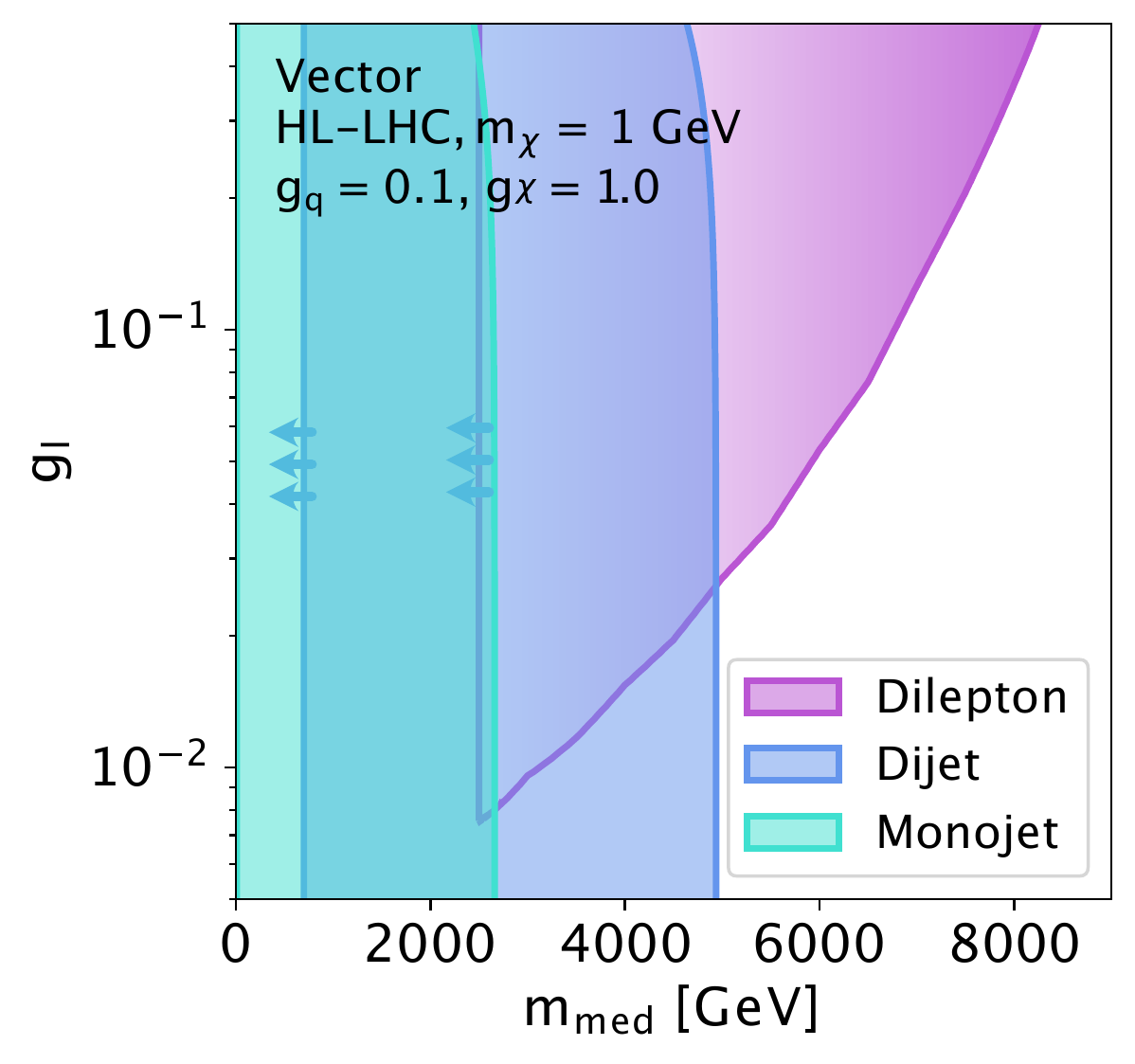}}
\\
\subfloat[]{\label{subfig:couplingscan-dd-monojet}\includegraphics[width=0.48\textwidth]{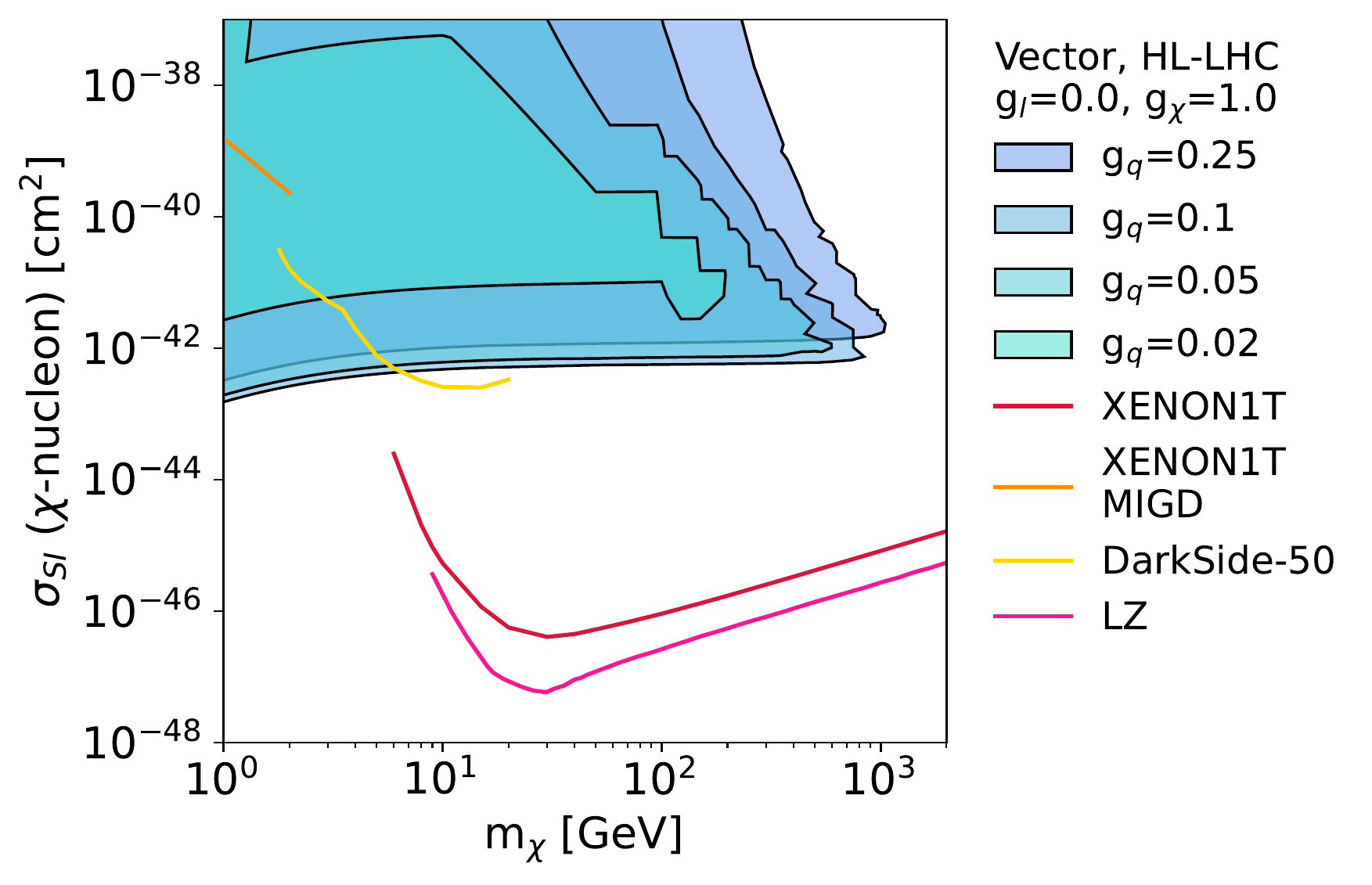}}

\caption{\textbf{Top}: Projected exclusion limits on the couplings $g_q$ \protect\subref{subfig:gqscan-dmLight}, $g_\chi$ \protect\subref{subfig:gchiscan-dmLight}, and $g_l$ \protect\subref{subfig:glscan-dmLight} for a vector mediator at the HL-LHC and FCC-hh (for quark and DM couplings only). The result is shown as a function of the mediator mass $m_{med}$; the mass of the DM candidate is fixed to 1 GeV in all cases. The coupling on the $y$ axis is varied while the other two couplings are fixed: in~\protect\subref{subfig:gqscan-dmLight}, $g_\chi$=1.0 and $g_l$=0.0; in~\protect\subref{subfig:gchiscan-dmLight}, $g_q$=0.1 and $g_l$=0.0; and in~\protect\subref{subfig:glscan-dmLight}, $g_q$=0.25 and $g_\chi$=1.0. 
The arrows in the lower edge of the contours indicate that other searches for lower mass mediators that are normally performed at colliders could be sensitive to these models, but are not shown because the inputs received focused on the highest mediator masses only.
It is worth noting that the lower bounds for both HL-LHC and FCC-hh for mediator (resonance) searches shown in the figure can be significantly improved by using non-standard data taking techniques such as data scouting/trigger-level analysis described in Section \ref{sec:computational_resources} and in Refs. \cite{CMS:2016ltu,ATLAS:2018qto} specifically for dijet resonances.
\textbf{Bottom}: Example of the effects on the HL-LHC exclusion limits in $\sigma_{\mathrm{SI}}$ for the monojet search \protect\subref{subfig:couplingscan-dd-monojet} when varying the quark coupling, $g_q$. The dark matter coupling is held fixed to $g_{\mathrm{DM}}=1$ and there is no coupling to leptons. Limits from existing direct detection experiments~\cite{DarkSide:2018bpj,XENON:2018voc,XENON:2019zpr,LZ:2022ufs} are shown for context (a recent re-analysis of the DarkSide-50 data~\cite{DarkSide-50:2022qzh} has subsequently become available).}
\label{fig:couplinglimits-hl-lhc-allanalyses}
\end{figure}

The top panels of Fig.~\ref{fig:couplinglimits-hl-lhc-allanalyses} show the projections for future hadron colliders for couplings to the SM fermions and DM fermion. These are derived from the projections for missing-momentum signatures of the mediator (e.g. jet plus missing transverse momentum, or \textit{mono-jet}) as well as searches for the visible decays (e.g. dijet and dilepton resonances). 
Observing both visible and invisible mediator decays would provide complementary views of DM-SM interactions. 
Hadron colliders with a higher center-of-mass energy can reach higher mediator and dark matter masses. 
Searches for invisible particles should be able to constrain models with much smaller couplings than current searches, especially at mediator masses below the TeV. 
These results can also be recast in terms of dark-photon benchmark models used by the Rare Processes and Precision Measurements Frontier~\cite{Artuso:2022ouk}, as discussed in the next Section. 

The reach of future colliders for these simplified models is dependent on the interaction strength between the mediator and the SM, especially the quark coupling of the mediator, which governs its production. 
It is expected that lepton colliders will also strongly constrain models with lepton couplings, especially in the case of polarized beams~\cite{Kalinowski:2021vgb}.  

In order to understand when direct and indirect detection experiments can confirm that collider signals are connected to DM, the European Strategy Briefing Book compared collider and direct/indirect detection experiments, highlighting where a simultaneous discovery would be possible.

Following those studies, the searches shown in the top row panels of Fig.~\ref{fig:couplinglimits-hl-lhc-allanalyses} can be reparameterized as limits on the DM-nucleon cross section for direct detection searches, as shown in Fig. \ref{subfig:gchiscan-dmLight}. 
As expected from Refs.~\cite{Boveia:2016mrp,Albert:2017onk}, as future collider searches reach smaller SM-DM and SM-SM couplings, their coverage in the DM-nucleon cross section plane improves.
When the coupling sensitivity limit approaches, collider projections gradually disappear. 

The sensitivity of collider searches also depends on the ratio of the DM to mediator mass. 
Searches at colliders are most sensitive when the mediator can be produced directly from SM particle collision, and when the mediator is much heavier than the DM particle. 
However, a strength of collider experiments is that constraints on the dark interactions for these models can also be obtained when the DM is too heavy to be produced directly. 
In BSM mediation (as well as EFT) models, different mass hierarchy hypotheses are plausible, so different mass ratios as well as different couplings should be tested.
We refer to the following section for a specific choice of a benchmark for a vector mediator, motivated by thermal relic DM. 

The lessons from these plots remain similar to Ref.~\cite{Strategy:2019vxc}. 
When particle DM is discovered in a direct- or indirect-detection search, Fig.~\ref{subfig:gchiscan-dmLight} illustrate -- in a necessarily model-dependent fashion for the specific simplified model considered -- when collider searches for invisible particles would also be sensitive to production of the mediator. 
In roughly these regions, both types of searches would have complementary discovery potential, as discovery at a direct-detection experiment would be combined with further study of the type and properties of the DM-SM interaction at a hadron collider.

These figures also indicate where collider searches for invisible particles would add unique sensitivity to the coverage from the other DM experiments.
Nevertheless, even in these regions, other experiments would be essential to confirm that the invisible particles are indeed associated with galactic dark matter.

\subsubsection{Beyond WIMP dark matter}

As the astrophysical evidence for dark matter does not point to a particular mass scale for DM particles, there are many other intriguing DM hypotheses beyond the WIMP that can be probed at colliders and neighboring facilities.
These viable possibilities include extending the thermal relic paradigm to lighter DM (below 1 GeV), provided it has feeble couplings to SM, or more complex dark sectors signatures that can also include long-lived particles.
Here, we will highlight some representative beyond-WIMP DM models and the prospects for their discovery at future colliders and complementary facilities. 

\paragraph{Portal models for DM: dark photon and dark Higgs.}

The experimental characteristics of feebly-coupled, low-mass thermal relic DM particles can be represented by a small set of relevant interactions, mediated by new \textit{portal} particles, akin to the simplified models mentioned above.
One such portal particle is a new, low-mass vector boson, commonly called \textit{dark photon}, that couples to DM and to the SM via a new electromagnetic-like interaction (kinetic mixing).
A dark photon benchmark model has been adopted by the Rare Processes and Precision Measurements Frontier~\cite{Artuso:2022ouk} to compare the sensitivity of accelerator experiments to thermal DM below a GeV. 
Colliders can also probe such a dark photon in a complementary way.

\begin{figure}[!ht]
\centering
\subfloat[]{\label{subfig:darkPhoton}\includegraphics[width=0.55\textwidth]{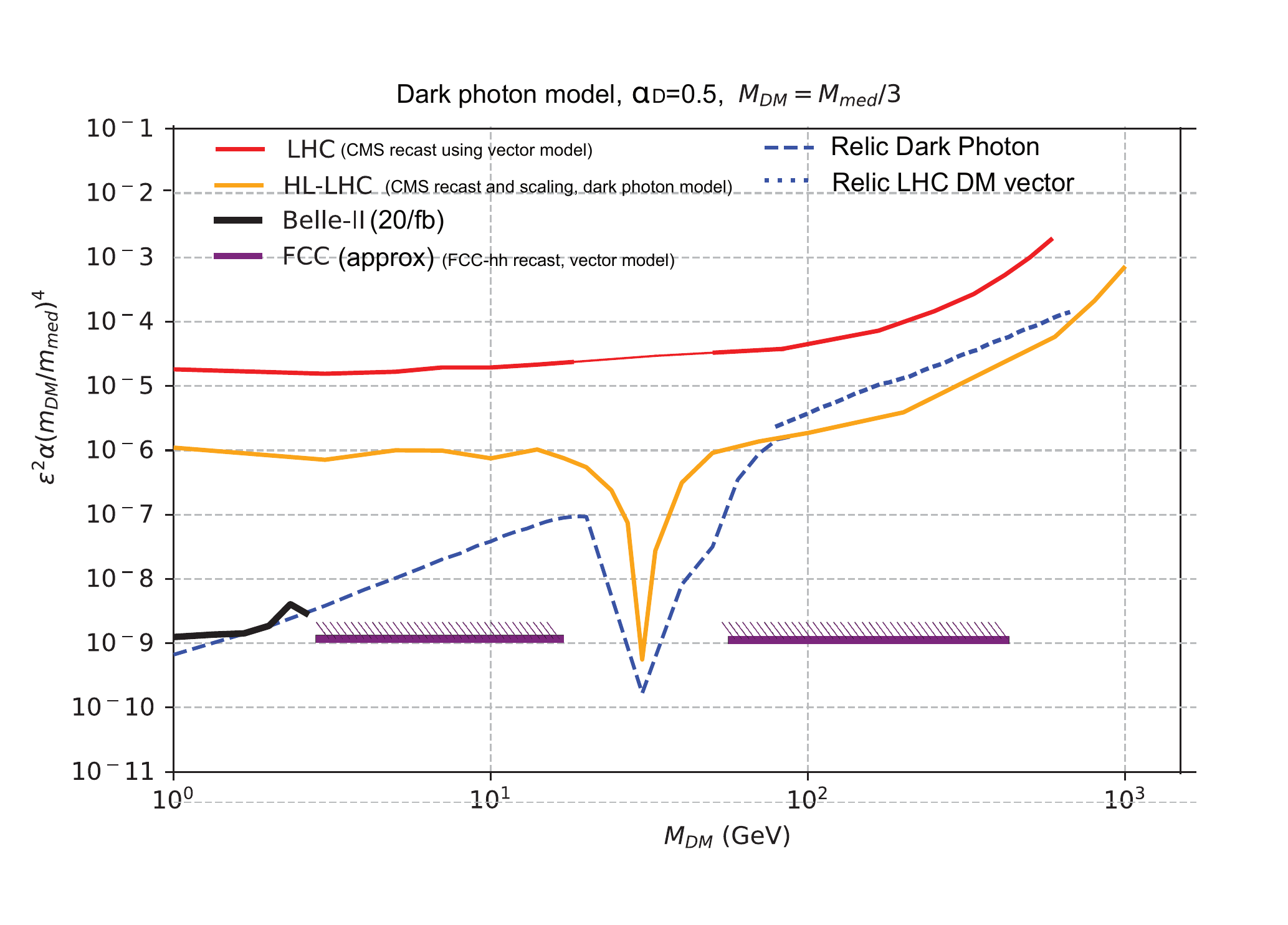}}
\subfloat[]{\label{subfig:flavorfulScalarResonances}\includegraphics[width=0.45\textwidth]{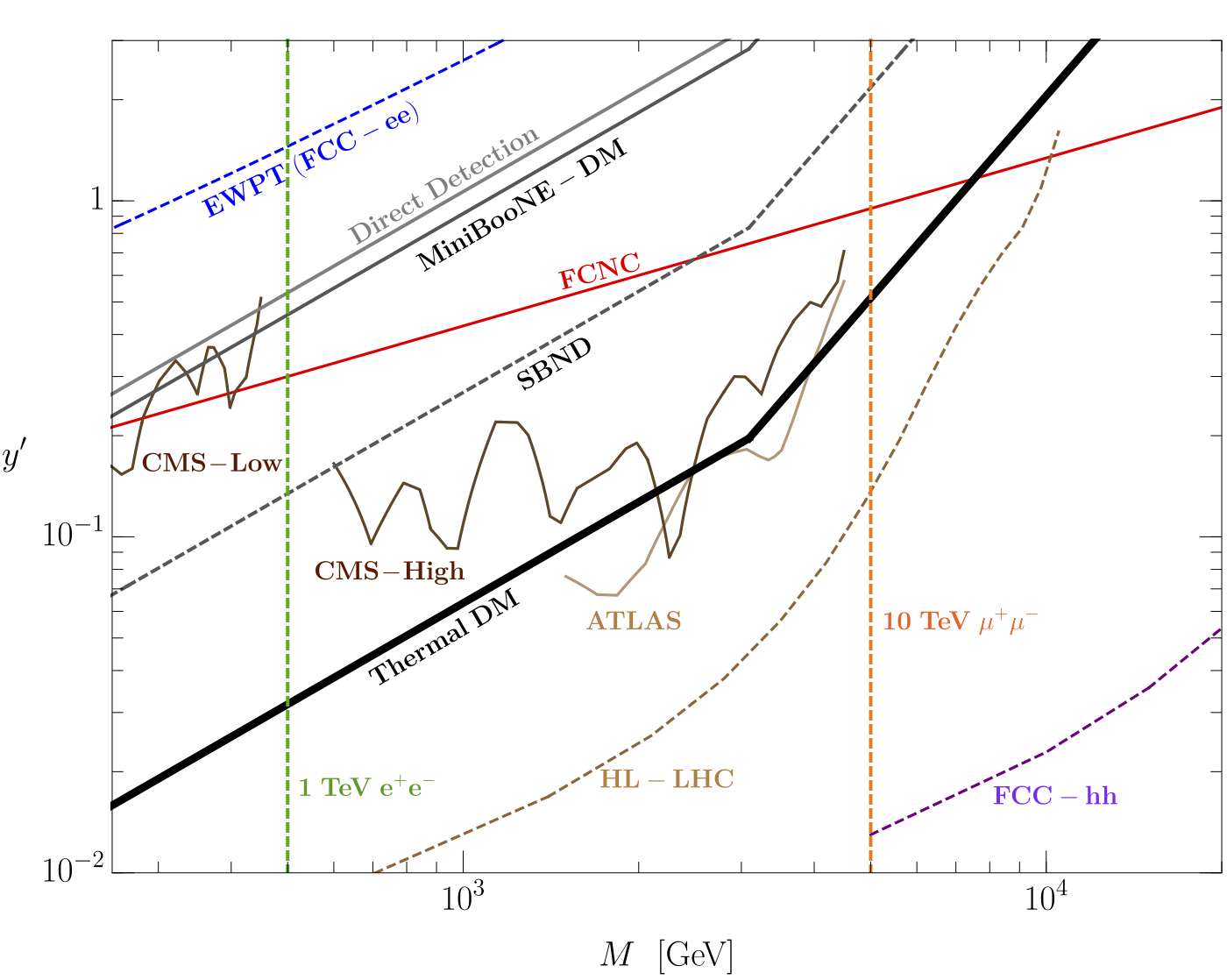}}
\caption{\protect\subref{subfig:darkPhoton} Comparison of the expected constraints for the dark photon model, starting from the reinterpretation of the results \cite{CMS:2021far} in terms of the simplified vector mediator model (\textit{LHC}), and for HL-LHC, FCC-hh and Belle-II 20/fb \cite{Belle-II:2022cgf,Belle-II:2018jsg} in terms of the dark photon model and \protect\subref{subfig:flavorfulScalarResonances} existing constraints and future prospects for the up-specific scalar mediator $S$ in the renormalizable completion described in \cite{Batell:2018fqo, Batell:2021xsi}. Each figure also depicts the parameter combinations that would reproduce a thermal relic.}
\label{fig:darkPhotonandflavorfulScalarResonances}
\end{figure}

The left panel of Fig.~\ref{fig:darkPhotonandflavorfulScalarResonances} reinterprets present and future collider searches for vector-mediated DM in terms of the parameters of this dark-photon benchmark, for a pseudo-Dirac DM particle with a mass below a TeV. 
It shows that the HL-LHC dataset is needed to reach the thermal relic milestone for DM masses above roughly 100 GeV, while FCC-hh is needed to cover the remaining parameter space\footnote{This model assumes that the DM particle constitutes the entire astrophysical DM abundance. For the thermal relic in all such simple models, this is not a strict requirement if other processes not present in minimal model enhance or deplete the DM abundance.}.
High-energy collider experiments are therefore needed, together with accelerator and B-factory experiments, to explore the full mass range for DM in this scenario. 

Although not immediately obvious from the formulation of minimal benchmarks~\cite{Batell:2018fqo, Batell:2021xsi, Rizzo:2022qan}, many portal models require both low and high mass particles to be self-consistent. 
New TeV-scale particles at loop-level are needed to generate the kinetic mixing interaction that characterizes dark-photon models.
These particles present discovery opportunities at colliders such as HL-LHC or the FCC complex, and could be inferred from precision measurements at lepton colliders \cite{Rizzo:2022qan}. 
For instance, if a light scalar singlet (a dark Higgs) decaying to light DM couples predominantly to one SM fermion family, then new heavy states are required to keep the model self-consistent at higher energy scales~\cite{Batell:2018fqo, Batell:2021xsi, Egana-Ugrinovic:2019wzj, Egana-Ugrinovic:2019dqu}.
In this example, the Higgs-boson mass and the DM mass are both of the order of a GeV, while the other particles needed to complete the model have masses of the order of a TeV. 
Figure \ref{subfig:flavorfulScalarResonances} shows estimates for current and future collider searches for these new particles.

Similar considerations are generically applicable to different types of portal models, encouraging complementary probes for the lower-energy phenomenology of this model and higher-energy particles required for the completion. 
Light portal particles (e.g. light dark Higgs, dark photon) with feeble couplings can be discovered at Rare and Precision Frontier experiments~\cite{Artuso:2022ouk}, while the higher-energy particles can be discovered at future hadron or Muon Colliders.
A corresponding discovery in low-threshold Cosmic Frontier experiments~\cite{Chou:2022luk} would determine the cosmological nature of the dark matter particle. 

\paragraph{More complex dark sectors.}

Taken further, dark sectors could reach or exceed the complexity of the Standard Model, for example in the class of \textit{Hidden Valley} models \cite{Strassler:2006im} where a portal particle feebly couples ordinary matter to a more complex dark spectrum.
Dark sector models can include a large number of states with different masses and lifetimes, where a viable DM candidate arises from the uncharged states of the dark sector~\cite{Bernreuther:2019pfb,Cline:2016nab,Mies:2020mzw,Dienes:2022zbh}. 
These lead to qualitatively different signals with respect to WIMP dark matter. 

Alternate mechanisms to obtain the DM abundance as well as new forces can point to new dark matter candidates and parameter regions.
In particular, new QCD-like confining forces within the dark sector (a `dark QCD'~\cite{Bai:2013xga}) can give rise to stable dark hadrons that can be viable DM candidates.
These dark hadrons can be embedded in \textit{dark showers} that could be discovered at present and future colliders \cite{Albouy:2022cin}.
A dark-sector--confining phase transition can also lead to new viable masses for dark matter \cite{Asadi:2021yml}.
As for the other models, connecting collider signals to cosmology requires other experiments\footnote{Additional information on the dark sector spectrum can come from the lattice community, see e.g. \cite{Kulkarni:2022bvh}}.

As these models offer myriad ways to connect colliders to cosmology via the dynamics of a complicated dark sector, future efforts to explore such a connection are needed to identify regions of parameter space that are cosmologically interesting, and to help contextualise future discoveries in multiple Frontiers.

\paragraph{Dark sector discoveries at facilities co-located with EF colliders: the FPF example.}

The signatures generated by feebly-coupled models often include particles with long lifetimes, as discussed earlier. 
Covering these signatures at colliders requires both general-purpose and highly-optimized experiments.
In particular, portal models and beyond-WIMP models are targets for dedicated facilities and experiments designed to use existing high-energy beams and supplemented by information recorded at traditional collider experiments. 
Collisions of high energy beams produce a large, forward flux of secondary particles, including light DM of dark-sector particles, that are best investigated by dedicated experiments in facilities downstream from the standard detectors. 

An example of such a facility is the FPF, proposed to be installed in the long shutdown between the LHC Run 3 and the start of the HL-LHC~\cite{Anchordoqui:2021ghd,Feng:2022inv}. 
In particular, the proposed FLArE \cite{Batell:2021blf} experiment would be sensitive to thermal-relic DM mediated by a dark photon, and experiments such as FASER2 \cite{FASER:2018eoc} and FORMOSA \cite{Foroughi-Abari:2020qar} would target a number of DM and dark-sector scenarios. 
The opportunities for DM discoveries (as well as of neutrino measurements relevant to cosmic DM experiments) offered by the FPF would further enhance the HL-LHC DM and dark-sector search program, especially since development and construction can be performed in parallel.

Future colliders offer opportunities for similar compelling experiments for DM physics.
Smaller-scale experiments can be optimized to cover a number of beyond-WIMP scenarios that complement and enhance what can be discovered at the larger, general-purpose experiments, while exploiting the same colliding beams. 

\subsubsection{Dark matter complementarity}

The search for DM has become a major science driver across all experimental frontiers (CF, EF, RF and NF) and the cross-cutting AF, IF, UF and TF frontiers. Discovering its microscopic physics requires collaboration within and across all of these communities, and a variety of experiments at different scales. The first step toward a unified strategy, understanding how the efforts of each community complement the others, is discussed in a separate community whitepaper, the Snowmass 2021 Dark Matter Complementarity report~\cite{Boveia:2022syt}.
This document, developed by the combined effort of these communities, sketches opportunities for cooperation, lays out a road map for a coherent strategy and presents case studies for discovery.\\



\section{The Energy Frontier vision}~\label{sec:EFvision}

The EF community has proposed several opportunities for pursuing its scientific goals, among them the most prominent ones are Higgs-boson factories and multi-TeV colliders at the Energy Frontier. These projects have the potential to be truly transformative as they will push the boundaries of our knowledge by testing the limits of the SM and directly discovering new physics beyond the SM.
\footnote{The EF vision, as outlined in the following, has been formulated from the input received directly from the Energy Frontier community during the Snowmass process, including EF wide meetings and workshops, Topical Group meetings, and Agor\'{a} events on future colliders. Full documentation of these activities can be found on the Snowmass 2021 EF wiki page~\cite{SnowmassEFwiki}.}

The EF aims to facilitate US leadership in an innovative, comprehensive, and international program of collider physics. The timescales to fully realize the EF vision extend to the end of this century, and the ultimate goals can only be realized if our actions foster a vibrant, diverse, and intellectually rich US EF community. Maintaining and strengthening such a community is only possible if our plans reflect the aspirations of and provide a rich set of opportunities for Early Career physicists.

During the coming decade it is essential to complete the highest priority recommendation of the last community planning exercise (P5) and to fully realize the scientific potential of the HL-LHC by collecting and analyzing at least 3 ab$^{-1}$ of integrated luminosity. 
The precision electroweak measurements possible at an $e^+e^-$ collider (Higgs factory) would greatly extend and complement the scientific results provided by the HL-LHC. The EF endorses making the investments now to enable US leadership in a Higgs factory and start the construction for a Higgs factory in parallel with HL-LHC operations. To realize this goal, the US EF community needs to expand the R\&D, in collaboration with the international community on the detector and accelerator technologies which will be required for a Higgs factory. In addition, the global HEP community should consider opening a dialogue for a US-based site for a future $e^+e^-$ collider.

The next step in our exploration of the fundamental properties of matter requires the exploration of the multi-TeV energy scale. Any deviations observed at HL-LHC or an $e^+e^-$ Higgs factory would strongly motivate such a program. Two possible and potentially complementary paths forward appear most promising to develop the capability to explore the energy-scale frontier: 1) a 100 TeV (or higher) hadron collider, and 2) a high-energy Muon Collider. 
The community proposes the US (in collaboration with international partners) embark on a R\&D program addressing high-priority critical aspects of accelerators to reach these high energies, and to develop the detector technologies needed to withstand the complex backgrounds and high radiation environments envisioned for these two types of future colliders. 

{\bf Thus, the energy frontier believes that it is essential to complete the HL-LHC program, to support construction of a Higgs factory, and to ensure the long-term viability of the field by developing a multi-TeV energy frontier facility such as a Muon Collider or a hadron collider.} 

A key role in the success of the US EF at the HL-LHC, at future $e^+e^-$ colliders (Higgs factories) as well as at future multi-TeV colliders is played by the Theory Frontier: model building, precision calculations and simulations are necessary for precision measurements and searches of new physics. The theory community must be adequately funded to support the success of the EF community as a whole. In addition, the EF community thrives on collaborating with other frontiers within HEP, and has relied on cross-fertilization opportunities available via the interdependence with various fields and  opportunities, see Sec.~\ref{sec:opportunities}.

It is essential for the US and the global HEP community to develop an integrated plan for future colliders to pursue the reach of our ultimate goal of uncovering new particles, new forces, and unveiling more fundamental laws of nature.

\subsection{The immediate-future Energy Frontier collider}

The immediate future is the HL-LHC. The physics case for this program rests on its ability (1) to extend the direct search for new elementary particles, (2) to measure the couplings of the Higgs boson at a level that is sensitive to corrections from physics beyond the SM, (3) to demonstrate the presence of a self-coupling of the Higgs boson, (4) to measure the couplings of the top quark at a level that is sensitive to corrections from physics  beyond the SM, and (5) to extend our understanding of QCD and strong interactions by improving the precision of the corresponding relevant measurements. 

The results from the HL-LHC set the basis for any vision of the future of the EF program. The EF scientific program at the LHC is thriving and Run~3 will triple the current integrated luminosity. The HL-LHC is scheduled to start operation in 2029 and to increase the integrated luminosity by another factor of 5 over the following decade. The HL-LHC program spans a very wide range of physics topics, where some legacy results are expected to remain the most sensitive for a long period of time after the end of the HL-LHC data-taking. The unprecedented data set accumulated at the HL-LHC will allow the study of SM phenomena that remain elusive so far because of their small rates, and to extend the reach of searches for new processes beyond the SM. These data will boost the potential of the experiments to make direct discoveries which may revolutionize the human understanding of nature. These results rely both on innovative detector technology, and a major theoretical undertaking in reducing the expected systematics.

The study of the Higgs-boson self-interaction is one of the primary goals of the HL-LHC due to its role in cosmological theories, involving, for example, the vacuum stability. Higgs-boson pair production is a flagship measurement, with a projected evidence of a di-Higgs signature at the 4$\sigma$ level, by the end of HL-LHC running. Among all the Higgs-boson self-interaction terms, the trilinear self-interaction is the only one in the reach of the HL-LHC and it is parametrized by the coupling strength, which can be measured with a sensitivity of $50\%$.

Studying the properties of the Higgs boson is a key mission of the HL-LHC.  Uncertainties in the signal strength modifier and coupling measurements in the main production and decay channels will reduce from their current levels of 10-50\%  to less than 5\%, moving Higgs-boson measurements into the regime of precision physics, and allowing for spotting deviations from the SM.
Rare processes such as Higgs-boson decays to $c\overline{c}$, which are challenging, and currently at the level of about 7.6 times the SM cross section, will become accessible at the HL-LHC. The sensitivity to the H $\rightarrow$ invisible branching ratio would reduce from the current $\sim$20\% to a few percent, approaching the SM prediction of 0.1\%. 
For the heavy Higgs bosons predicted by BSM theories with extended scalar sectors, the reach would increase to masses up to 1 TeV. 

The HL-LHC will also extend the sensitivity in direct searches for BSM particles.  For example, in SUSY, the reach for gluinos and squarks will increase by up to 1-2 TeV, while chargino, neutralino and slepton reach will increase by up to 0.5-1 TeV.  This will allow to make a more conclusive statement on the naturalness hypothesis.  The reach for new resonances decaying to SM particles will extend on average by 2-3 TeV. Moreoever, HL-LHC will provide the ability of studying resonance decays to lighter BSM particles, such as in the case of $Z^\prime$ decays to charginos, which are barely accessible at the LHC, due to typically small branching ratios.  Another set of dedicated searches will look for dark matter, typically in final states with invisible dark matter and visible mono-X and increase the current sensitivity. Sensitivity to long-lived particles will be especially enhanced at the HL-LHC due to improved and innovative detectors.  The reach for Higgsinos via a disappearing track search will increase to masses up to 350 GeV. 
Sensitivity to long lived neutralinos decaying to photon and graviton 
is expected to increase 
to masses of 700 GeV, improving the reach in short $c\tau$ and at high masses.  
Additionaly, dedicated displaced muon reconstruction will improve cross section reach for smuons or dark photons by 1-2 orders of magnitude for different values of lifetime with respect to Run 3.

Expected measurements of QCD interactions, e.g. in jet, photon as well as in $W$ and $Z$ boson productions, will considerably improve the understanding of PDFs at low and high momentum fraction $x$, which are critical for carrying out the rich physics program of precision Higgs-boson measurements as well as BSM searches.
Heavy Ion studies at the HL-LHC will include measurements of parton densities in broad kinematic range and search for saturation, precision measurement of macroscopic long wavelength QGP properties, developing a unified picture of collectivity across colliding systems, assessing microscopic parton dynamics underlying QGP properties, and performing precision QED and BSM physics, for example in ultraperipheral collisions.

The international collaborations at the LHC recognize the importance of the Snowmass 2021 process to the HEP community in the US and beyond. Continued strong US participation is in particular critical and essential to the success of the HL-LHC physics program, including the Phase-2 detector upgrades, the HL-LHC data-taking operations, and the physics analyses based on the HL-LHC dataset. 

Additionally, auxiliary experiments and facilities are proposed to take advantage in far forward kinematic regions. 
Forward physics facilities allow to further extend the breadth of the HL-LHC physics: they can study regions of parameter phase space for BSM, for example in LLPs and DM searches, that would otherwise remain uncovered, and can perform novel QCD and neutrino measurements in the very forward region. 

We note that cross domain collaboration and studies with the EIC measurements (under the purview of the DOE nuclear physics) identified the synergies and complementarities in many physics studies, experimental technologies, analysis techniques, and theoretical tools.

The US HEP community is heavily involved in ATLAS, CMS, LHCb, and ALICE, and it contributes to aspects of the LHC accelerator infrastructure. More than half of the US HEP community is involved in the LHC. Over the last years, US institutions have graduated about 100 PhDs/year based on research carried out with LHC data. During the last decade, ATLAS and CMS have together published more than 2000 scientific papers in peer reviewed journals. This has had a significant impact on the advancement of the field, and is an unprecedented achievement compared to any previous collider experiment.
In addition, the PhDs apply the valuable skills they learn during their research to many domains of science and industry, which is an essential positive economic return of the LHC program. The vibrant scientific program of the HL-LHC will continue this tradition and provide excellent training environment to  the next generation of students and postdoctoral researchers. Given the broad portfolio of HL-LHC, similar number of student cohorts, i.e.  about 100 PhDs/year from the US are expected to benefit from the HL-LHC program.

{\bf In conclusion, our highest immediate priority accelerator and project is the HL-LHC, the successful completion of the detector upgrades, operations of the detectors at the HL-LHC, data taking and analysis, including the construction of auxiliary experiments that extend the reach of HL-LHC in kinematic regions uncovered by the detector upgrades.}

At the same time, the time scales for realizing what comes next require also an effort to advance preparations for the next collider of the intermediate future during this time frame. This is reflected in the discussion on needs and timelines in Sec.~\ref{sec:exec-summary}.

\subsection{The intermediate-future Energy Frontier collider}

The intermediate future is the an $e^+e^-$ Higgs factory, either based on a linear or a circular collider.  The physics case for this program rest on its ability to (1) measure the couplings of the Higgs boson to the sub-percent level and discern the pattern of modifications from beyond the Standard Model, (2) search for exotic Higgs decays due to a "Higgs portal" to a hidden sector of forces,  (3) measure the parameters of the Standard Model, including the $Z$, $W$, top, and Higgs-boson masses to very high precision, and to provide stringent tests of couplings in the electroweak sector, (4) measure the electroweak couplings of the top quark at a level that can clearly reveal corrections from beyond the Standard Model, and (5) perform precision measurements of QCD phenomena which are testing-grounds of QFT in both perturbative and non-perturbative regimes, and provide complementary information relevant to cosmology and BSM physics.

{\bf The $e^+e^-$  colliders are the vehicle that will enable a high-precision physics program in the EW sector by increasing the precision of SM measurements.
The physics case for an $e^+e^-$ Higgs factory is compelling and the program is possible essentially with current technology.
The various proposed facilities have a strong core of common physics goals that underscores the importance of realizing at least one such collider somewhere in the world.  A timely implementation of a Higgs factory is important, as there is  considerable US support for initiatives that can be achieved on a time scale relevant for early career physicists. }

Circular $e^+e^-$ colliders will be implemented in stages as electroweak, flavor, QCD, Higgs-boson, and top-quark factories by spanning the energy range from the $Z$ pole (and below) up to the top-quark pair threshold and beyond. The highlights of such colliders are the highest luminosities at the $Z$ pole, $WW$ threshold, and Zh energies (the “intensity frontier”); the exquisite beam energy calibration at the $Z$ pole and $WW$ threshold; the possibility of center-of-mass energy monochromatization at $\sqrt{s} = m_h$; the compatibility with four interaction points; and the cleanest experimental environment. Circular $e^+e^-$ colliders are therefore excellent Higgs factories: they produce over a million Higgs bosons in three years at 240 GeV center-of-mass energy. They provide the most precise determination of the Higgs-boson coupling to the $Z$ boson, and of the Higgs-boson width (and mass), and they could provide the opportunity for the discovery of the Higgs-boson self-coupling and for a first measurement of the electron Yukawa coupling. These colliders are also much more than Higgs factories. At the $Z$ pole, the Tera-$Z$ factory, with several trillions $Z$ produced, offers opportunities for a comprehensive set of electroweak measurements with the best prospects for precision, such as $Z$-boson mass and width (10’s of keV), effective weak mixing angle (few 10$^{-6}$), a direct determination of the electromagnetic coupling constant, which allow sensitivity to mass scales up to 70 TeV to be reached. In addition, a Tera-$Z$ factory has  comprehensive programs for QCD physics, e.g. the most precise measurement of the strong coupling constant, flavor and rare decay physics, e.g. the search for lepton-flavor violation, as well as direct searches for heavy neutral leptons, axion-like particles, and other feebly coupled dark matter particle candidates. Collisions at the $WW$ threshold will allow the most precise determination of the $W$ mass (300 keV) and width (1 MeV), while running at the top-pair threshold, will provide the best prospect for the top mass measurement.

Linear $e^+e^-$ colliders are primarily aimed at precision measurement of the Higgs-boson properties with the aim to potentially flag violations of the SM. The linear $e^+e^-$ colliders will also run at center-of-mass energies covering the production thresholds of the $Z$ boson, $WW$ pairs, $Zh$ pairs, $\rm{t\bar{t}}$ pairs, and $tH$ and Higgs-boson pair production. The center-of-mass energy can be chosen flexibly depending on new discoveries at the LHC, or elsewhere. The $e^+e^-$ linear collider will use polarized electron and positron beams to enhance signal reactions and allow the measurement of helicity-dependent observables, multiplying the physics output per unit of luminosity. Beam polarization also enables the suppression of backgrounds and provides cross-checks for the control of systematic errors. In electroweak measurements, beam polarization gives direct access to $Z$ left-right asymmetries, which are very sensitive precision probes. For example, in a dedicated run at the $Z$ pole, it allows a measurement of $\sin^2\theta_w$ at the level of $10^{-5}$, comparable to the TeraZ capability. The nominal $e^+e^-$ collider physics program begins with running at a center of mass energy of 250 GeV. At this energy the total cross section and the branching ratios for all Higgs decays can be determined, including decays to invisible final states. It provides the potential to search for exotic Higgs decays. These measurements will improve our knowledge of Higgs-boson couplings by a large margin over HL-LHC results. A global fit using an EFT framework allows for a determination of the $hbb$ couplings to 1\%, the $hWW$ and $hZZ$ to 0.7\%, and all other important Higgs boson couplings to close to 1\%. These levels of precision are sufficient to be sensitive to new physics beyond the reach of direct searches at the LHC. The second step in the linear $e^+e^-$ collider program would be an energy upgrade to $\sim$600 GeV. This would improve the precision of all measurements from the 250 GeV running by a factor two. Beyond the improved precision of the Higgs-boson couplings, the couplings of the top quark can be explored and it is possible to search for pair production of new particles produced in EW interactions that are difficult to discover at LHC. At $\sqrt{s}$ of 500 GeV, the Higgs pair production can be observed and the Higgs-boson trilinear coupling can be measured with a precision of 22\%. 

As the physics case for a Higgs factory is strong, the US should participate in global efforts to construct an $e^+e^-$ collider. Planning the center-of-mass energy of this collider is important, and planned upgrades of the center-of-mass energy will provide access to a broader spectrum of Standard Model physics, including top-quark physics, which is an important component of the energy frontier physics program. The accelerator R\&D should aim at developing sustainable facilities for Higgs research at a reasonable cost. 
To enable the realization of a Higgs factory in the shortest possible timescale, a targeted program on detector R\&D for Higgs factories should be supported with an immediate start and with initial emphasis on studies that are applicable across proposed facilities. 
The critical detector R\&D areas have been identified as 4D tracking and vertexing, i.e. providing precision time and spatial measurements in a single detector unit, low-mass detectors, e.g. monolithic detectors that embed the electronics in the sensing elements, wireless data transmission technologies, implementation of advanced AI/ML algorithms in on-detector electronics, dual calorimetry among others, see Sec.~\ref{app:det}. 

{\bf The Energy Frontier also supports the possibility of a Higgs factory in the US. Given global uncertainties, consideration should also be given to the timely realization of a possible domestic Higgs factory, in case none of the currently proposed global options are realized.}

\subsection{The long-term-future Energy Frontier collider}

In the long-term future we envision a collider that probes the multi-TeV energy scales, i.e. up to about hundred TeV center-of-mass energies. 
Its physics case rests on its ability to  (1) produce the fundamental particles that generate the mechanism of electroweak symmetry breaking, (2) produce particles with flavor-dependent couplings to quarks and leptons, (3) search for thermal dark matter particles into the region of strong coupling in the dark sector, and (4) in general explore the unknown at the highest possible energy scale.

A 10-TeV scale Muon Collider with sufficient integrated luminosity provides an indirect energy reach similar or better to that of a 100 TeV proton-proton collider (e.g. FCC-hh, SppC), although its direct reach is limited to 10 TeV.
The similarity between high energy lepton colliders (effectively $W-W$ colliders) and hadron circular colliders (effectively gluon-gluon colliders) is outstanding.
Studies indicate that both the muon and hadron colliders have
similar reach and can significantly constrain scenarios motivated by the naturalness principle. The 100 TeV
hadron collider will have an  advantage when it comes to searching for colored states, while the Muon Collider naturally has better access to EW states. 
Multi-TeV muon  colliders will  have the benefit of excellent signal to background ratios, taking advantage of the vector boson fusion production processes.  
As the multi-TeV colliders are planned for after the Higgs factory, they will benefit from the precision studies of the Higgs-boson properties in understanding the possible scale of new physics.  One of the key measurements from the multi-TeV colliders is the one of the Higgs self-coupling to a  precision of a few percent, and the scanning of the Higgs potential.

This program to enable new physics insight into higher scales is currently limited by technological readiness. In addition to the most prominent projects that have been proposed, i.e. hadron and Muon Colliders, other auxiliary proposals include high-energy $e^+e^-$ colliders using plasma wakefield or structure advanced acceleration.  All of these proposals require substantial accelerator R\&D.

A vigorous R\&D program into accelerator and detector technologies will be critical to position the US and international community at the forefront of this research on a long term. This R\&D program must specifically enable instrumentation research that goes beyond current projects, so that the detector technology will be available to make use of these high-energy colliders once they can be built. The critical detector R\&D 
areas have been identified as 4-dimensional tracking with precision timing detectors,  small area silicon pixels in a high radiation environment. Besides that, particle flow calorimeters with hybrid segmented crystal, and fiber readout may offer a better alternative to silicon, see Sec.~\ref{app:det}. 

With the start of the Snowmass 2021 studies, there has been a surge in the interest of the US and the international community for the  Muon Collider option because of advances in technology and analysis methodologies. There is synergy with the formation of the International Muon Collider Collaboration hosted at CERN. About a third of the contributed papers in the EF are on Muon Colliders. Since the last Snowmass study in 2013, there has been substantial progress in understanding the physics case, the detector requirements, as well as developing novel techniques to address the major beam induced backgrounds.

{\bf The investment in detector and collider R\&D for hadron and muon facilities, and planning for discussion of siting options for Muon Colliders have to start now, and run in parallel with the HL-LHC and any $e^+e^-$  precision electroweak program. Enabling this future, also requires strong input from every area of the theoretical community to understand the discovery potential of such colliders. }


\subsection{Opportunity for US as a site for a future Energy Frontier Collider}

Our vision for the EF can only be realized as a worldwide program, and CERN as host of the LHC has been the focus of EF activities for the past couple of decades. In order for scientists from all over the world to buy into the program, the program has to consider siting future accelerators anywhere in the world.
The US community has to continue to work with the international community on detector designs and develop extensive R\&D programs, and the funding agencies (DOE and NSF) should vigorously fund such programs (as currently the US is severely lagging behind).

The US community has expressed a renewed ambition to bring back EF collider physics to the US soil, while maintaining its international collaborative partnerships and obligations, for example with CERN. The international community also realizes that a vibrant and concurrent program in the US in EF collider physics is beneficial for the whole field, as it was when Tevatron was operated simultaneously as LEP.

{\bf The US EF community proposes to develop plans to site an $e^+e^-$  collider in the US. 
A Muon Collider remains a highly appealing option for the US, and is complementary to a Higgs factory. 
For example, some options which are considered as attractive opportunities for building a domestic EF collider program are: 
\begin{itemize}
\item A US-sited linear $e^+e^-$ (ILC/CCC) Collider
\item Hosting a 10 TeV range Muon Collider
\item Exploring other $e^+e^-$ collider options to fully utilize the Fermilab site
\end{itemize}
}

Planning to proceed in multiple parallel prongs may allow us to better adapt to international contingencies and eventually build the next collider sooner. Such a strategy will also help develop a robust long term plan for the global HEP community.  Therefore, there are requests to assess the cost of siting a linear $e^+e^-$ Higgs factory collider option, and a multi-TeV Muon Collider in the US. In addition, to realize a successful US $e^+e^-$ linear collider program, cost reduction options, and targeted accelerator R\&D, e.g. for the Cool Copper Collider (CCC), is very important in the near term.

Some of the options listed above capitalize on the existing facilities, and on expertise in key areas of accelerator and detector R\&D at Fermilab. Among other sites, Fermilab is proposed as an ideal one for a Muon Collider with a center-of-mass energy reach at the desirable 10-TeV scale. The synergy with the existing/planned accelerator complex and the neutrino physics program at Fermilab is an additional stimulus for such investment of effort. 
A roadmap of the accelerator R\&D timeline, see Ref.~\cite{Adolphsen:2022ibf} and Sec.~\ref{sec:EFColliderTimelines}, indicates that a 3 TeV Muon Collider is possible by 2045, though technically limited. A set of Muon Collider design options, with one of the siting options being at Fermilab, should be considered as an integral part of a global discussion for siting and selecting an international Muon Collider. 

A goal should be to prepare a document summarizing design for the Fermilab-sited Muon Collider in time for the next Snowmass. 
Investment in this effort will reinvigorate the US high-energy collider community and enable much needed global progress towards possible discoveries at the next energy frontier.

The proposals and R\&D efforts to address the innovative detector developments for Higgs factories are  well underway globally and many challenges are resolved. Bold “new” projects such as a linear $e^+e^-$ CCC, and a Muon Collider will offer the next generation some challenges to rise to. It will inspire more young people from the US to join HEP and in the long term help with strengthening the vibrancy of the field. 
Realizing our ultimate goal will require significant funding and government support. The community feels that there is potential to raise funds and obtain government buy-in for a future collider project located in the US. However, funding is not all that is needed. We also need a future program that continues to inspire the next generation of high energy physicists, and one that entices the next generation of graduate students to choose high energy physics as their field.

\section{Acknowledgments}

We acknowledge the input from the EF community-at-large in Spring/Summer 2022  (included verbatim in Ref.~\cite{SnowmassCommunityInput}) and in particular Alain Blondel, James Brau, Joel Butler, Sekhar Chivukula, Dmitri Denisov, Sarah Eno, Stephen Gourlay, Paul Grannis, Christophe Grojean, Tao Han, Patrick Janot, Andy Lankford, Lucie Linssen, Michael Peskin, Chris Rogers, Vladimir Shiltsev, Jianchun Wang for their help in  developing and formulating the vision of the Energy Frontier.  We also thank the convenors of the e$^+$e$^-$ Collider Forum, Maria Chamizo Llatas, Sridhara Dasu, Ulrich Heintz,
Emilio Nanni, John Power and Stephen Wagner, as well as  the convenors of the Muon Collider Forum, Derun Li, Diktys Stratakis, Kevin Black, Sergo Jindariani, Fabio Maltoni and Patrick Meade. Last but not least, we thank the Department of Energy for supporting the March 2022 EF workshop.  

\appendix
\chapter{EF Appendix}
\label{sec:Appendices}
This appendix presents the requirements, needs and challenges for the detector developments for future collider physics, including detection technologies, software and computing, event simulations and analysis techniques as well prospects for Monte Carlo event generators. In addition it discusses opportunities and challenges that the EF community will encounter in future EF projects, e.g. attraction and mentoring of young minds, diversity and inclusion, needed investments etc. The appendix ends presenting the timeline and expected costs of the major EF projects that were discussed in Snowmass 2021.

\section{Detectors, reconstruction, simulation and data analysis}
\label{app:det}

Enabling particle detection techniques, data computation and analysis methods, as well as precision calculations, event generators, detector and accelerator simulations are critical for any future project at the energy frontier.

Different collider projects have different requirements on detector layout and performance. For example, $e^+e^-$ colliders require detectors with high granularity and low material budget, while hadron colliders have additional requirements on radiation tolerance for detectors to be placed close to the beam pipe.
The detector design directly impacts the computational resources and affects the data analysis methodology. 
Needs for software and computing resources should be accounted for at the start of the detector design stage to inform an optimal design strategy and ensure the best possible physics performance of reconstruction algorithms.
Requirements for computational resources are set by detector design parameters such as number of readout channels, luminosity, trigger system, data compression as well as the experimental computing model and the analysis methods. 
The sophistication of artificial intelligence and machine learning (AI/ML) techniques is growing ever more, and collider physics should continue to play a pioneering role in this expanding field of research by capturing and leading such developments. As we experienced at the LHC experiments, AI/ML techniques have become necessary tools to perform precision measurements and expand the reach of searches for new physics. We expect AI/ML techniques will play even greater roles at the HL-LHC and future colliders The direction of current research suggests that such techniques will be employed in future experiments not only in offline data analyses but also in online data selection and reconstruction, making their accuracy and reliability even more critical~\cite{Shanahan:2022fzy, CompF:sm2021}. 

Most measurements at high-energy colliders require the understanding of radiative corrections on observables. A close collaboration between theory and experiments is mandatory to design observables that have a reduced exposure to high-order effects and exploit detector capabilities as well as the latest developments in precision calculations and Monte Carlo simulations. 
A fundamental tool to design detectors, develop software and perform data analysis in all experiments is the Monte Carlo simulation of final state particles entering the detector. Essential software packages need to be supported to produce Monte Carlo samples, including physics generators and the Geant4 detector simulation toolkit. The computational needs for such simulations strongly depend on the size of the MC samples, the complexity of the detectors, the precision and accuracy needed for the physics models embedded in these tools, as well as the data formats. For future collider feasibility and design studies, accelerator modeling is also of fundamental importance~\cite{https://doi.org/10.48550/arxiv.2209.08177, CompF:sm2021}.

Some of the most relevant aspects of needs for detector, method and tool development and research for the coming decade will be discussed in this subsection.

\subsection*{Detectors}

Particle detectors are key to address science challenges and their development is based on our understanding of fundamental laws of physics.
Therefore, there is a “virtuous cycle” between fundamental laws of nature and enabling detector concepts and techniques. Such virtuous cycle must remain strong and unbroken in order to lead to greater physics discoveries. 
In this context, detectors in high-energy physics face a huge variety of operating conditions and employ technologies that are often deeply intertwined with developments in industry. The environmental credentials of detectors are also increasingly in the spotlight.

At the EF, one can distinguish two major drivers for detector R$\&$D: detector upgrades towards future hadron and Muon Colliders and development of advanced technologies for the future $\ee$ machines. 

The detectors for the next $\ee$ Higgs factory must provide excellent precision and efficiency for all basic signatures, i.e. electrons, photons, muon and tau leptons, hadronic jets, and missing energy over an extensive range of momenta. The tracking resolutions should enable high-precision reconstruction of the recoil mass in the $e^+e^- \rightarrow Zh$ process for instance. These inherently very accurate physics requirements ask for integrated concepts with unprecedented precision, minimal power consumption and ultra-light structures. 
The demands for high resolution (granularity) and low material budget on one hand, and low power consumption on the other hand, exceed significantly what is the state-of-the-art today. Such leaps in performance can only be achieved by entering new technological territory in detector R$\&$D. To quote just some examples, several new concepts for silicon sensor integration, such as monolithic devices, are being pursued for pixel vertex detectors, new micro-pattern gas amplification detectors (MPGD) are explored for tracking and muon systems, and the particle-flow approach to calorimetry promises to deliver unprecedented jet energy resolution~\cite{Barbeau:2022muf}. 

The Basic Research Needs for High Energy Physics Detector Research \& Development document (BRN in short)~\cite{osti_1659761} compiles a list of requirements for transformative and innovative technologies at the next generation of energy frontier experiments focused on precision Higgs and SM physics, and searches for BSM phenomena, such as (1) low-mass, highly-granular tracking detectors and (2) highly-granular calorimeters, both with high-precision timing capabilities.

Emerging novel vertex and tracking detector technologies are the vital backbone for the success at a future electron-positron machines. These will operate in an environment with high (continuous or bunched) beam currents, a minimum distance from beam axis of about 20 mm, a requirement of $< 5~\mu m$ single point resolution, high granularity ($< 30 \times 30~\mu m^2$), power dissipation ($<50~mW/cm^2$), low mass ($\sim 0.1~\%$ of $X_0$, or 100 $\mu m$ Si-equivalent per layer).
The clear challenge is unprecedented spatial resolution, to be achieved with ultra-small pixels, and thus extremely low material budget. Very thin detector assemblies are mandatory, while providing high stiffness and stability. 
The tracking resolutions should enable high-precision reconstruction of the recoil mass in the $e^+e^- \rightarrow Zh$ process, as shown in Table~\ref{tab:physics-req}, and allow very efficient $b$ and $c$ tagging and tau-lepton identification through the reconstruction of secondary vertices.
 The detector solenoid magnetic field must be limited to 2 T 
 for the Tera-$Z$ operation at future circular electron-positron colliders, to avoid a blow up of the vertical beam emittance and a resulting loss of luminosity. The 2 T magnetic field limit is not a significant problem since the momentum scale of the produced partons is typically distributed around 50 GeV and does not exceed 182.5 GeV. 

\begin{table}[!ht]
\begin{center}
 \caption{Physics goals and detector requirements~\cite{osti_1659761,MuonCollider:2022ded}.}
\begin{tabular}[c]{|c|c|c|l|}
\hline
Initial state	&	Physics goal	& Detector & 	Requirement	\\
\hline
\hline
$e^+e^-$ & $h$ZZ sub-\% & Tracker & $\sigma_{p_T}/p_{T}$=0.2\% for $p_T <$ 100 GeV \\ 
&&& $\sigma_{p_T}/p_{T}^2=2 \cdot 10^{-5}/$ GeV for $p_T >$ 100 GeV \\ 
&& Calorimeter & 4\% particle flow jet resolution\\
&&& EM cells 0.5$\times$0.5 cm$^2$, HAD cells 1$\times$1 cm$^2$\\ 
&&& EM $\sigma_E/E=10\%/\sqrt{E}\oplus 1\%$\\ 
&&& shower timing resolution 10 ps\\ 
& $hb\bb/h c\bar{c}$ & Tracker & $\sigma_{r\phi} = 5\oplus 15 ( p\sin\theta^{\frac{3}{2}})^{-1}\mu$m\\ 
&&& 5$\mu$m single hit resolution\\ 
\hline
pp-100 TeV  & Higgs  & Tracker & $\sigma_{p_T}/p_{T}$=0.5\% for $p_T <$ 100 GeV \\ 
&&& $\sigma_{p_T}/p_{T}^2=2 \cdot 10^{-5}/$ GeV for $p_T >$ 100 GeV \\ 
&&& $ 300 $ MGy and $\approx$ 10$^18$  n$_{eq}$/cm$^{2}$\\ 

&& Calorimeter & 4\% particle flow jet resolution\\
&&& EM cells 0.5$\times$0.5 cm$^2$, HAD cells 1$\times$1 cm$^2$\\ 
&&& EM $\sigma_E/E=10\%/\sqrt{E}\oplus 1\%$\\ 
&&& shower timing resolution 5 ps\\ 
&&& 4 MGy / 5 GGy  and  $\approx$ 10$^16$/10$^18$ n$_{eq}$/cm$^{2}$ central/forward \\
\hline
$\mu$  & Higgs \& LLP & Tracker & 30 ps timing resolution and 0.01 rad angular resolution \\
&&& 5$\mu$m single hit resolution\\ 
\hline
\end{tabular}
\label{tab:physics-req}
\end{center}
\end{table}

Gaseous and semiconductor detectors are the two main types of tracking detectors; other ones include fiber-based or transition radiation tracking devices. While gaseous detectors offer sizeable low-mass volumes and many measurements require an excellent pattern recognition and ultimate dE/dx measurement (e.g. cluster counting technique can be exploited instead of the charge-analog information), the silicon-based approach offers the most accurate single point resolution. 
The breakthrough technology is expected to come from monolithic devices incorporating complex readout architectures in CMOS foundries.
To minimize material budget, new technologies, like stitching, will allow developing a new generation of large-size CMOS MAPS with an area up to the full wafer size~\cite{Hoeferkamp:2022qwg,MAPS}. 
In addition to large-size sensors, it may also be useful to bend thin (50~$\mu m$) sensors to make cylindrical assemblies. 
Fast picosecond-time sensors based on Low-Gain Avalanche Detectors (LGAD)~\cite{LGAD} and 3D-devices can be also exploited. Aiming for an excellent position and timing resolution ($\sim$~10~ps and $\sim$~10~$\mu m$) with GHz counting capabilities to perform 4D tracking, LGAD-based technologies represent a very attractive option for PID and TOF applications.
For future applications at the EF, alternative technologies include beyond state-of-the-art interconnection technologies, such as 3D vertical integration, through-silicon-vias (TSV), or micro bump-bonding, which, while retaining the advantages of separate and optimized fabrication processes for sensor and electronics would allow fine pitch interconnects of multiple chips.

The "particle flow" (PFlow) concept, originally developed for the electron-positron Linear Collider (LC), aims at measuring the energy of all the particles in a jet, exploiting track information for the charged particles, ECAL for prompt photons, and HCAL to capture the neutral hadrons. Due to this, PFlow has led to calorimeter designs, as part of a complex system of inter-connected detectors rather than as a stand-alone device. 
Present and future challenges in calorimeters are closely linked to all aspects of ultimate exploitation of the PFlow technique and the dual-readout calorimetry~\cite{Pezzotti:2022ndj} approach, developed by the RD52, DREAM, IDEA and CalVision collaborations.
Silicon photomultipliers have seen a rapid progress in the last decade, becoming the standard solution for scintillator-based devices, but also 
enabling substantial improvement in dual-readout calorimetry, based on scintillating and clear plastic fibers embedded in absorber structures.
Ultra-fast timing in calorimetry can be also used to resolve the development of hadron showers, by separating their electromagnetic and hadronic components, and therefore simplifying the implementation of the particle flow algorithm. In general, space–time tracking could be used in  many physics analysis at LHC – Higgs, BSM searches for long-lived particles, by measuring precisely the time-of-flight between their production and decay, and/or in assigning beauty and charm hadrons to their correct primary vertex. An ultimate concept is to develop 4D real-time tracking system for a fast trigger decision and to exploit 5D imaging reconstruction approach, if space-point, picosecond-time and energy information are available at each point along the track.

An emerging effort during Snowmass has focused on strange-quark tagging~\cite{Albert:2022mpk}. Particle identification at high momenta could further boost strange tagging capabilities at future $\ee$ machines as well as the analysis sensitivity in constraining the available phase space for new physics. Gaseous Ring Imaging Cerenkov system (RICH) detector can be capable of $\pi$/K separation up to 25 GeV. A preliminary study based on ILD geometry at ILC, shows that in a compact RICH with a radial extension of 25 cm, the Cherenkov angle resolution can be maintained at the level of $\sim$5 mrad in magnetic fields up to 5 T. This leads to a discrimination power of 3$\sigma$ between kaons and pions up to momenta of approximately 25 GeV. Further simulation studies and system optimization are needed to evaluate globally the impact of a RICH system on object reconstruction, such as PFlow jets, and on other physics benchmarks, when used in conjunction with silicon tracking detectors.

Future hadron collider experiments have an additional requirement of radiation hardness. The radiation levels close to the beam pipe will require technologies that can withstand radiation levels that are an order of magnitude higher than those expected at the HL-LHC. Such demanding requirements urge a long term R\&D on radiation-tolerant technologies, especially for tracking systems.
In addition, both hadron and Muon Collider experiments need to use fast-timing detectors to suppress backgrounds.
More specifically, a factor-of-five larger pileup at  a 100 TeV hadron collider than at the HL-LHC is expected to pose stringent requirements on the detector
design~\cite{Aleksa:2019pvl}: future hadron colliders need timing of the order of 5-10 ps per track to suppress pileup and correctly assign tracks to vertices.
At a multi-TeV Muon Collider to reject a good fraction of beam induced background, accurate timing information with a resolution of 30 ps is assumed to be available in the vertex detectors~\cite{MuonCollider:2022ded}.

At future hadron colliders, detector capabilities to reconstruct highly boosted objects are fairly
challenging (for instance, the average $Z$ boson from $ZZ$ production
would shower mostly within a single LHC calorimeter cell).  
This challenge is accentuated by so-called ``hyper-boosted'' jets,
whose decay products are collimated into areas the size of single
calorimeter cells. Holistic detector designs that integrate tracking,
timing, and energy measurements are needed to mitigate for these
conditions~\cite{Larkoski:2015yqa,Chang:2013rca,Elder:2018mcr,
Spannowsky:2015eba,ATLAS-CONF-2016-035,Gouskos:2642475}.  Hadronic and electromagnetic shower
components up to several TeV need to be simulated, where
extrapolations to these  high energies come with large
uncertainties. Differences in the hadronic shower  simulation models
in Geant4~\cite{GEANT4:2002zbu} have been reported for pions in the
energy range 2--10~GeV~\cite{CALICE:2019vza}. Detailed studies of
hadronic showers  will be needed in the next few decades to achieve
the best possible precision in QCD measurements at future
colliders.

While research has always required state-of-the-art instrumentation in trigger and data acquisition systems, the demands for the next generation of hadron colliders are the increasingly large local intelligence, integration of advanced electronics and data transmission functionalities (e.g. using FPGA).
Another important trend is the progressive replacement of the complex multi-stage trigger systems by a new architecture with a single-level hardware trigger and a large farm of computers to make the final online selection and to reduce Level-1 trigger rate to O(kHz) for permanent storage.
A technical challenge is also associated to run both online trigger algorithms and offline reconstruction on heterogeneous computing platforms, including GPUs and FPGAs, for which code portability to multiple computing platforms is a challenge to be resolved. This illustrates the need to bridge the gap between computing needs and available resources, at least in an immediate future, as well as the trend towards moving more complex algorithmic processing into the online systems. 
Modern technologies allow the integration of significant intelligence at the sensor level and many different R$\&$D lines are being explored, like local hit clustering for strip and pixel detectors, local energy summing for calorimeters, local track-segment finders.
The use of advanced ML algorithms, such as neural networks (NN), boosted decision trees (BDT) and many others, is a long-standing tradition in particle physics since 1990’s and has been already key enabler for discoveries (e.g. single-top production at Tevatron). Bringing the modern algorithmic advances from the field of ML from offline applications to online operations and trigger systems is another major challenge.

As the high-energy particle physics community and particularly the Snowmass community begins to design future detectors, it is important to keep the many, varied LLP signatures in mind, lest we design new detectors that are biased against them. For example, overly-aggressive filtering can introduce biases that limit the acceptance for displaced tracks~\cite{Jindariani:2022gxj}. At the same time, we can develop technologies, such as dedicated trigger algorithms~\cite{Alimena:2021mdu}, displaced tracking algorithms~\cite{Kotwal:2019zia}, and timing detectors~\cite{CMS:2667167,Chekanov:2020xco}, to explicitly reconstruct and identify LLPs. Careful studies of beam-induced backgrounds will be necessary to reduce and/or quantify these background contributions without removing possible LLP signals. Other important factors to consider for LLPs include the time between collisions and how that interplays with the detector readout, as well as the size of the beamspot, the amount of pileup and the material budget of the detector areas closest to the interaction point. \\
Different geometry choices that provide similar hermeticity for prompt particles can differ drastically in their ability to reconstruct particles that do not originate from the interaction point. In particular, high granularity at large radii enables better reconstruction efficiency of displaced tracks and vertices, and helps to distinguish them from beam-induced and non-collision backgrounds. \\
A high volume, (partially) shielded subdetector system like the current muon systems at LHC experiments would therefore play an important role in searches for LLPs at future hadron colliders. For a future $e^+e^-$ collider, on the other hand, the background yields are expected to be much lower and it could be beneficial to invest the equivalent amount of space into a larger inner detector, and restrict the muon system to the minimum required for muon identification. Finally, Muon Colliders~\cite{MuonCollider:2022xlm} come with a new set of challenges for LLP searches, as their detectors are bombarded from both sides with ultrahigh energy electrons/positrons from the in-flight decay of the muon beam \cite{Jindariani:2022gxj,Ally:2022rgk}. It is difficult to shield the detectors from this qualitatively new beam background, but over $99\%$ background rejection can be achieved by making use of timing and angular measurements from paired layers~\cite{Ally:2022rgk}. Whether simultaneously a good signal efficiency for LLPs can be maintained needs to be studied further.

Future collider detectors will also face a large number of diverse engineering challenges, in the areas of system integration, power distribution, cooling, mechanical support structures, and production techniques. Within the field of particle physics, technologies developed under generic R$\&$D studies or with the aim to address experiment-oriented challenges at future colliders provide a boost in innovation and novel designs that often suit the needs of the Intensity or Cosmic Frontiers, i.e. neutrino or astroparticle physics.

\subsection*{Monte Carlo Event Generators}
\label{sec:mcsimulations}
Nearly all high-energy experiments rely on the detailed modeling of multi-particle final states through 
Monte Carlo simulations~\cite{Campbell:2022qmc}. A particular strength of general-purpose simulation tools 
derives from the factorization of physics effects at different energy scales, making their underlying 
physics models universal. 
Uncertainties on experimental measurements are often dominated by effects 
associated with event modeling. These uncertainties arise from the underlying physics models 
and theory, the truncation of perturbative expansions, the parametrization or modeling of 
non-perturbative QCD effects, the tuning of model parameters, and the fundamental 
parameters of the theory. Addressing and reducing the uncertainties is crucial to meet 
the precision targets in current and future measurements. 

The experimental facilities discussed in this report span a wide range of energies, 
beam particles, targets (collider vs.\ fixed target), and detected final states. 
Each experiment may require some dedicated theory input to the simulation, such as 
high-precision QED calculations for Tera-$Z$ or an electroweak parton shower for a Muon Collider.
Other aspects, such as parton-to-hadron fragmentation or hadronic transport models can be similar
for many facilities, enabling the modular assembly of (parts of) a generator from existing codes 
when targeting a new facility. In this manner, previously gained knowledge and experience 
can be transferred, and a more comprehensive understanding of the physics models is made possible
by allowing them to be tested against a wealth of data.
These cross-cutting topics in event generation have been identified as a particular
opportunity for theoretical developments in a broad HEP program~\cite{Campbell:2022qmc}.

The extraction of SM parameters at the HL-LHC will depend on the precision of perturbative QCD and EW
calculations, both fixed order and resummed, and on their faithful implementation in particle-level MC simulations. 
The results of some analyses will however also be limited by the number of Monte Carlo events 
that can be generated, and computing efficiency will play a crucial role. 
Future highest-energy colliders, including a potential Muon Collider, will likely require electroweak effects
to be treated on the same footing as QCD and QED effects.

The Forward Physics Facility at the LHC will leverage the intense beam of neutrinos, 
and possibly undiscovered particles, in the far-forward direction. These measurements 
will require an improved description of forward heavy flavor -- particularly charm -- production,
neutrino scattering in the TeV range, and hadronization inside nuclear matter, 
including uncertainty quantification~\cite{Feng:2022inv}.

Future lepton colliders would provide permille level measurements of Higgs-boson couplings, and will reach even higher precision in the measurement of the $W$-boson mass (one part in $10^{-6}$) and the top-quark mass (one part in $10^{-4}$). The unprecedented experimental precision will require event generators 
to cover a much wider range of processes than at previous facilities, both in the Standard Model and beyond. 
In addition, predictions for the signal processes must be made with extreme precision, involving
QED up to fourth and EW corrections up to second order. Some of the related methodology is available 
from the LEP era, while other components will need to be developed from scratch.

With the next generation neutrino experiments not being limited by statistical uncertainties,
and all running and planned experiments using nuclear targets, one of the leading systematic uncertainties
to their measurements arises from the modeling of neutrino-nucleus interactions.
This requires the use of state-of-the-art nuclear-structure and -reaction theory calculations.
While not a traditional topic for general-purpose high-energy event generators, there is strong 
overlap with topics relevant to simulating high-energy neutrino DIS in the FPF detectors and at IceCube.
This is expected to become an area of active development and cross-collaboration between frontiers.

The EIC  will use highly polarized beams and high luminosity to probe the spatial and spin structure
of nucleons and nuclei. Simulating spin-dependent interactions of this type at high precision is 
currently not possible with standard event generators and requires the development of new tools 
at the interface between particle and nuclear physics. It is expected that measurements at the LHC 
can greatly benefit from these developments~\cite{MC4EICReport,AbdulKhalek:2022erw}. Various experiments also require the understanding of heavy-ion collisions and nuclear dynamics 
at high energies as well as intricate heavy-flavor effects.

In addition to the physics aspects, there are similar computational aspects, 
such as interfaces to external tools, handling of tuning and systematics, 
data preservation, the need for improved computing efficiency, 
and connections to artificial intelligence and machine learning~\cite{Butter:2022rso}. The following sub-section will present the computational requirements at future experiments. In the context of preparing to fulfill the needs for MC samples in near-future and far-future EF programs, it should be highlighted that the MC event generation processes, especially at hadron colliders when high-order QCD corrections are included, are very CPU-intensive, and the Geant4-based detector simulation applications are among the largest consumers of computing resources at the HL-LHC. Therefore, it is important to develop new and faster computational techniques for MC event generation and support detector simulation applications to maintain and improve the accuracy of physics modeling at future collider experiments~\cite{https://doi.org/10.48550/arxiv.2209.08177}.

\subsection*{Computational requirements}
\label{sec:computational_resources}
Experiments require computational resources during their design, operation, and data analysis phases. Experiments must generate and simulate collision events and other backgrounds, reconstruct events, optimize their design, trigger on collisions, reconstruct events, calibrate the experiment, and analyze the reconstructed data to extract physics.

Software trigger systems are ubiquitous for hadron collider detectors, which face significant data reduction challenges even before recording events to long-term storage. Such triggers are large computing farms that must execute a pared-down reconstruction of high multiples of the eventual recorded event rate in real time, and hence typically constitute very powerful computing sites on their own which can be repurposed when collisions are not being recorded. The challenges of offline reconstruction and software triggering go hand-in-hand.

There are many physical resources that are needed --- long-term storage, both ``hot'' and ``cold'' (today represented by disk and tape); computers, both traditional CPU and accelerators like GPUs; and network bandwidth. Given the speedy evolution of computing, it is hard to predict what mixtures of available technologies will be optimal on the timescale of new energy frontier experiments. Nevertheless the scale of the computing problems posed by proposed facilities is roughly indicated by the data volume of the experiments. Estimates are summarized in Table~\ref{tab:computing-resources}. FCC-hh is the only facility with offline data sizes exceeding those of HL-LHC by more than an order of magnitude.

\begin{table}[!ht]
\begin{center}
 \caption{Computational resources expected at future Energy Frontier colliders.}
\begin{tabular}[c]{|c|c|c|c|}
\hline
Collider Scenario	&	Event size	&	Event rate	&	Data/year \\
\hline
\hline
HL-LHC general purpose expt & 4.4 MB & 10 kHz  &	0.6 EB	\\ 
FCC-ee $Z$-pole, one expt & 1 MB & 100 kHz & 2 EB\\
CEPC 240 GeV, one expt & 20 MB & 2 Hz & 260 PB\\
ILD 500 GeV & 178 MB & 5 Hz & 14 PB\\
CLIC 3 TeV, 1 expt & 88 MB & 50 Hz & 110 PB\\
Muon Collider, 1 expt & 50 MB & 2 kHz & 2 EB \\
FCC-hh, 1 expt & 50 MB & 10 kHz & 10 EB \\
\hline
\end{tabular}
\label{tab:computing-resources}
\end{center}
\end{table}

MC simulations, including event generation, detector simulation and event reconstruction, generally constitute the majority of offline computing and hot storage use by experiments. For example, ATLAS projections for HL-LHC anticipate $\approx 70$\% of CPU and $\approx 60$\% of disk use to arise from MC simulation \cite{ATLAS:2020pnm,cmshllhcprojections}. More specifically, the CPU consumption in ATLAS is dominated by the detector simulation that is driven by the complex geometry of the calorimeter and the event generation is the second-leading contribution, whereas in CMS the event reconstruction is the dominant contribution while the detector simulation and event generation constitute the second and third contributions respectively.
In general the proportion of resources for MC simulations is expected to be even higher for lepton colliders, due to the more democratic cross sections of relevant processes. 
Therefore efforts to maintain or increase the physics precision and accuracy of MC event generators and Geant4 physics models while speeding up the code utilizing computing hardware accelerator and machine learning techniques and specialized hardware should be aggressively pursued, given their major impact in our ability to extract physics from near-future EF programs and future  facilities.

The optimal way to provision the required resources will likely evolve over time. Trigger farms will still need to be located in physical proximity to experiments due to latency requirements, but offline processing may take additional advantage of resources that are shared with other sciences (such as supercomputing centers) or which are provided by industry (cloud resources). The design of experiment computing architectures will be influenced by the cost structure and technology availability imposed by such use.

The high energy physics community is already active in cross-collaboration forums such as the HEP Software Foundation~\cite{HSFPhysicsEventGeneratorWG:2020gxw,HSFPhysicsEventGeneratorWG:2021xti,CompF:sm2021} to find solutions that can meet the challenges brought forward by the amount of data delivered by HL-LHC and future colliders. 

LHC experiments are also putting in place non-traditional analysis workflows and computing architectures~\cite{Acosta:2022sax} in order to exploit the physics potential of the data discarded by their triggers. 
For example, real-time analysis (Data Scouting in CMS, Turbo Stream in LHCb, and Trigger-level analysis in ATLAS) move part of the data reduction (reconstruction and calibration) into the trigger system, allowing data to be recorded with significantly lower trigger thresholds with negligible increase in bandwidth. 
These workflows can also exploit upgrades to the processing capabilities of trigger systems designed to tackle more challenging data taking conditions; for example, LHCb and CMS have begun to employ GPUs in their software trigger in Run-2 to parallelize problems such as particle tracking.

\subsection*{Artificial Intelligence and Machine Learning}

Artificial intelligence (AI) and ML have come to pervade particle physics. Particle identification and event classification in data analysis and software triggering routinely use ML models. There are many proposals to extend the use of AI/ML in other realms which are just beginning to be explored, and we mention a few examples below, and for a comprehensive review we refer to the Snowmass 2021 Computational Frontier report~\cite{CompF:sm2021}.

The ability to find unexpected deviations from the Standard Model --- and not just classify signals from specifically-targeted processes --- has been a goal in the field for a long time and is important for fully exploiting the datasets collected by new facilities. Efforts are underway using a number of techniques to try to find such events, using semi-supervised and unsupervised learning methods. Such methods could be implemented at the data analysis level or even earlier in the trigger system. 
Similar anomaly detection techniques can also improve effective operation of experiments by rapidly identifying periods of bad data. Algorithms that can be used for identifying outliers can also learn the optimal way to compress data with tolerable losses in fidelity, and studies are in progress to assess how these can be employed for HL-LHC and future colliders. 

Work is in progress in the US and internationally to provide tools to transform ML models into FPGA code. This opens the door to deploying ML in hardware triggers, improving signal discrimination for experiments (especially at hadron colliders) with strong bottlenecks at this level.
Closely related to this area of research is the development and testing of a computing model for optimal utilization of modern computing hardware platforms. More specifically, the development of portable code that runs on multiple computing hardware platforms, both locally and remotely, is a necessary step towards utilizing the available computing resources when running ML algorithms~\cite{Shanahan:2022fzy}.

Generative ML techniques hold the potential to accelerate event generation and simulation. By learning accurate approximations to a full physics model that can be executed much faster, generative ML can increase the amount of Monte Carlo that can be produced with limited computing resources, improving the optimization of analyses and final statistical uncertainties~\cite{Butter:2022rso}.

The core technology of model training (optimizing a function of many parameters relative to some objective) motivates the idea of ``differentiable programming,'' in which more generic user code, written in an appropriate framework, can be rapidly optimized in a similar fashion if analytic derivatives are available. If deployed at scale, this opens the possibility of end-to-end optimization of data analysis and other computations, as well as more sophisticated methods of handling systematic errors.

Detector design is also an optimization problem. By linking together appropriate methods, it may be possible to do a practical end-to-end optimization of a detector from scratch using target physics measurements as the benchmark, rather than intermediate figures of merit for each subdetector (such as momentum or energy resolution). The simultaneous global optimization of detector designs could allow improved physics performance for reduced cost.

From a theoretical perspective, there are cases where it is possible to implement machine learning models that respect important symmetries by construction and are potentially capable of being mapped onto first-principles theory; such approaches may provide acceleration for otherwise computationally-difficult problems, or rigorous interpretability. The use of neural networks for tackling inverse problems (determining regions of theory space that are compatible with observation) is another promising direction at this interface~\cite{Butter:2022rso}.

As it is the case with other data selection and data processing tools used in high energy physics, cross-collaboration and cross-field activities should be encouraged. 
Challenges such as having the shortest possible time-to-insight on large amounts of data, and extracting small signals from large backgrounds, are not unique to high energy physics. 
Developing and discussing solutions for similar problems in different fields can be mutually beneficial. 
Computer science and industry are routinely developing ML algorithms, tools and hardware beyond the state-of-the-art, and their involvement in high energy physics projects can accelerate the field's development.

\subsection*{Analysis reinterpretation, preservation and open data}
\label{sec:reinterpretationAndPreservation}

Preservation of experimental data, simulation, software and analysis products is necessary to ensure a long-term return on investment in the EF (and in general in other Frontiers). FAIR principles (Findability, Accessibility, Interoperability, and Reuse of digital assets)~\cite{FairPrinciples} should be used as guidance not only for data but also for research software. This approach allows for reproducible scientific results, that in turn can facilitate and enable new analyses and can be easily reinterpreted under different theoretical hypotheses. 

In order to make this possible, the Computational Frontier recommends that: 

\begin{itemize}
    \item current and future experiments have a strategy and sufficient resources to preserve data and analysis capabilities, including beyond their lifetimes;
    \item substantial investments are made to support efforts on FAIR data and analysis preservation. This also includes theoretical inputs, statistical analysis, AI models, metadata, archives etc. The careers of those working on these topics should be equally supported. 
\end{itemize}

International coordination with current and future particle and astroparticle physics consortia such as the European Science Cluster of Astronomy and Particle Physics ESFRI research infrastructures~\cite{ProjectEscape} (including the HL-LHC) can help adopting these recommendations, given that multiple funding agencies worldwide require data and analysis preservation, as well as open data and research products. 

In terms of making experimental data publicly accessible (Open Data), future experiments at the EF can discuss future policies together with other communities (e.g. Cosmic Frontier) where opening the data is common practice, in order to understand how to balance the needs for recognition of the scientists producing the data with the growing requirements to make the data available, as well as the challenges that opening data entails in terms of physics analyses that are not done under the careful scrutiny of the collaborations. 
It would be desirable for future EF experiments to design Open Data policies starting from the beginning of their life cycle, also taking inspiration from the experience and examples within the CERN experiments~\cite{CernOpenData}.


\section{Enabling the Energy Frontier research}~\label{sec:EFCollab}

This section attempts to address from the point of view of the EF community several broad questions that pertain the whole HEP community, such as opportunities and challenges in future collaborations, how to build a cohesive and diverse community, and expected needed investments to prepare for major future experiments.

\subsection*{Collaborations}
\label{sec:opportunities}
\textit{What opportunities exist for cross-frontier, cross-disciplinary, or international collaboration and cooperation in the coming decade to enhance our ability to address the issues identified (including training or mentorship) ? How do these collaborations affect the timescales or resources needed for these activities ?}

The EF community in the US is fully integrated in and interdependent with the broad international community of particle physicists, which includes experimentalists, theorists, accelerator and detector physicists in the various domains that pertain to HEP as well as nuclear physics. The cross-fertilisation between different domains of particle physics is a strength of the EF and will continue to offer opportunities in the future, thus has to be nurtured and supported. 

\paragraph{Interdependence between Frontiers.}

Given the strong dependence of the EF on collider technology, there is a unique opportunity for collaborations with the Accelerator Frontier. We have seen growing interest in the EF community for the development of collider technologies that enable the targeted physics research. 

There is a clear interdependence between the communities of experimentalists and theorists. Such interdependence offers an opportunity for the EF and Theory Frontier to work hand-in-hand on physics studies for the next generation of collider physics experiments.

At the LHC we already experience a strong cross-fertilization between the EF and the Rare Processes and Precision Measurements Frontier. Such cooperation will be even more useful and necessary in the HL-LHC and at future colliders to tackle experimental and theoretical challenges, for example on detector solutions as well as analysis and computational methods and techniques. On the latter, the cooperation with the Instrumentation and the Computational Frontiers is instrumental for the development and eventual use of transformative technologies that will enable the realization of novel detectors we well as readout, reconstruction and simulation techniques for future collider experiments.

The EF must keep working in concert with the Theory, Rare Processes and Precision Measurements, Cosmic, and Neutrino Frontiers to discover particle dark matter and probe its interactions with the SM at future collider experiments. Sections~\ref{sub:BSMPhysics} and \ref{sub:darkMatter} provide examples of how searches for invisible particles with future experiments at the energy frontier are a vital complement to the efforts in the other frontiers. Collider experiments offer unique possibilities for new physics discoveries as well as the ability to connect signals of astrophysical DM observed at the other frontiers with measurements of the fundamental interaction(s) that are responsible for such signals.
To highlight this complementarity that is central to understanding the nature of DM, representatives from the different frontiers developed the Snowmass 2021 Dark Matter Complementarity report~\cite{Boveia:2022syt}.

\paragraph{Interdependence between US and international communities.}

The US community in the EF is fully embedded in the international community. Mutual support and exchanges of ideas, expertise and future plans is advisable and inevitable. The growth of one member of the community is the growth of the whole community. Constructive and multi-directional communication within the broad international community is vital to keep the field vibrant and make concrete scientific progress.

\paragraph{Cross-fertilization between Energy Frontier and other domains of physics.}

The planned construction of EIC in the USA within the next decade offers a unique opportunity for strengthening the central role of the US scientific community in the field of particle physics, seen as a whole. The consideration of EIC, within the EF Topical Groups for this Snowmass study, has been instrumental in fostering opportunities for cross-fertilization between the two fields of HEP and Nuclear Physics in terms of physics, experimental technologies, data analysis techniques and methods, as well as theoretical and phenomenological tools. The two (international) communities are well integrated, beyond the artificial and too-often restricting boundaries set by funding agencies. 

Emerging technologies and detection techniques in other realms of physics, e.g. quantum computing and detectors, novel materials, new detection techniques and methods, including AI/ML etc. will offer new opportunities for enhancing our experimental capabilities in ways that can be difficult to predict with our current knowledge. 

\subsection*{Building a diverse Energy Frontier community}
\textit{How can we ensure that the US particle physics community is vibrant, inclusive, diverse, and capable of addressing the scientific questions identified, and of fulfilling our obligations to society during this period ?}

A vibrant, inclusive, diverse and capable scientific community are necessary for any success of the US scientific community in the EF. These can be achieved by innovation as well as empowerment and training of the next generation of leaders in the EF.

\paragraph{Continue to innovate and empower.}

A vibrant EF community requires a large base of well motivated early career scientists with a vision for the future of the field. Such motivations and vision are acquired by training, mentorship, and empowerment. The complexity of modern experiments and the dispersion of large collaborations too often prevent young scientists from developing a vision and solid foundation in different sectors, as they are often confined in specific niches of their research. It is important for young scientists  to participate in different and diverse collaborations, similar to the variety of scientific methods and techniques they engage in.

In large collaborations such as those at the LHC, the contribution to specific research is often hidden by large author lists in papers or conference presentations given by people who did not directly participate to the presented studies. In addition, too often students do not have an opportunity to present their analyses at a conference during the course of their studies due to the complex decision making process of the speaker committees. The main vehicle for an early career physicist to get visibility inside and outside the collaboration is by appointments in managerial roles such as subgroup or group convenorship, which has become a coveted (instead of longed-for) role by young physicists and used as a trampoline for a permanent job. Such strict hierarchy and opaqueness of large collaboration often causes lack of ownership of projects and physics studies by early career physicists. A more transparent and merit-based empowerment of early career scientists will help the community grow in the long term.

Inclusiveness and diversity are at the core of the scientific endeavor, which is motivated solely by the goal of advancing fundamental scientific knowledge with a potential societal impact. 


By pursuing our goals of fundamental knowledge at the energy frontier we will inevitably continue to push the boundaries of innovation in several areas of research and we will fulfill our obligations towards society, not only with augmented knowledge but also with concrete technological advances.

\paragraph{Training the next generation of scientists.}

The EF experiments are the largest scientific experiments in the world and offer young researchers unique exposure to open research within diverse international collaborations. Junior scientists engaged in particle physics research receive hands-on training in quantitative, computational, engineering domains as a part of their work on big collider data.  The EF community has been pioneering new technologies and methods (both theoretical and experimental) that have found applications in a variety of scientific fields that greatly benefited society as a whole. The Energy Frontier has a long-standing record of training young scientists. Keeping this tradition is of paramount importance as it is necessary to keep the field thriving. There is a unique opportunity to engage early career scientists with challenging new problems on physics prospects for future colliders, detector design, software development, analysis and computational techniques. The vast range of challenges that we face at future colliders are the perfect training ground for future scientists.

\subsection*{Investments essential to the progress of the Energy Frontier }
\textit{What investments need to be made during 2025-2035 for the continuing scientific, technical, or community progress identified by your frontier in the decades beyond, on what timescales can these be implemented, and what resources would be required ?}

The realization of the EF scientific program depends on transformative advances in several critical areas of research, such as innovation in collider and detector technologies, use and development of cutting-edge computational and data acquisition techniques as well as novel theoretical ideas and accurate calculations.
These areas of research need significant and immediate support and investment in order to accomplish the energy frontier programmatic goals and have a competitive scientific community in the next decade.  The resources to be allocated will have to be commensurate to the vast range and amount of scientific output that is expected by those experiments. Such a scientific output is not only measured in terms of scientific articles and size of the community, but also in terms of the societal impact that can be considered as a return of investment, such as training of young professionals, dissemination of scientific knowledge, development of new technologies and methods, etc.  History has demonstrated that the activities of the Energy Frontier community have always had large societal return of investment, therefore the resources allocated to such enterprises should be commensurate to their expected return. 

\begin{itemize}
    \item \textbf{Collider R\&D:} progress in the energy frontier is dependent on progress in collider technologies. It is vital for future energy frontier experiments to secure support for collider R\&D that enables transformative changes in accelerator and magnet technologies in the next decade, for electron, hadron, and Muon Colliders.
    \item \textbf{Detector Technology R\&D:} as the 2020 BRN~\cite{osti_1659761} report identified, there is an opportunity to capture advances in detector technologies, e.g. for low-mass and high granularity detectors, to use at future electron colliders. In parallel there is the need at future hadron and Muon Colliders for innovative technologies that can withstand an unprecedented level of harsh environmental conditions.
    \item \textbf{Computing Resources:} computing is a fast-evolving field, mostly driven by industry. However, HEP has historically pioneered techniques, e.g. ML, and has developed its specific solutions to its specific computational and data acquisition requirements. While it is difficult to predict advances in such a dynamic field in the next decade,  we surely know that we have an opportunity to capture and lead the progress in computing for the next generation of accelerator-based experiments. For example, while lepton colliders do not provide the same computing challenges as hadron colliders, the sheer amount of channels and information to be analyzed requires the use of cutting-edge computational resources for all proposed collider experiments. Not to be forgotten is the ever growing computational need in theoretical calculations and simulations.
    \item \textbf{Theoretical Physics:} Advances in theoretical calculation and modeling for new physics are expected to maintain an important role in the unveiling of new opportunities for experimental measurements and  searches for new physics beyond the Standard Model. As the experimental measurements become more precise, theoretical calculations need to become more accurate. In the past decade we have seen transformative advances, often called 'revolutions', in theoretical calculations and Monte Carlo simulations that allowed to achieve unprecedented levels of precision in measurements that are dominated by theoretical or modeling uncertainties, and  extended the reach of searches at the HL-LHC. Similarly, we have a unique opportunity to capitalize on the expected progress in the next decade. For example, at $e^+ e^-$ colliders, where extremely high precision is expected to be reached in a broad range of experimental measurements, high-order calculations as well as fast and accurate Monte Carlo simulations are needed to match that precision.
    In another direction, as shown in a number of examples in this report as well as in the report of the Theory Frontier~\cite{Craig:2022cef}, theorists have discovered new ideas in QFT and model building which have led to new observables and clarified the meaning of others in the quest for new physics.  For example in Higgs-boson physics, contributions from theory have  been crucial to build a precision physics program and have led to qualitative changes in studying the possibility of an electroweak phase transition, new ideas for investigating the relation between the Higgs boson  and the flavor dynamics of fermions, and EFT techniques that have all broadened the experimental program. Therefore, continued investment in theory is crucial to the success of the EF.
    
\end{itemize}

\newpage
\section{Timelines and Costs of Proposed Future Colliders }~\label{sec:EFColliderTimelines}

\begin{figure}[!ht]
\begin{center}
\includegraphics[width=0.98\hsize]{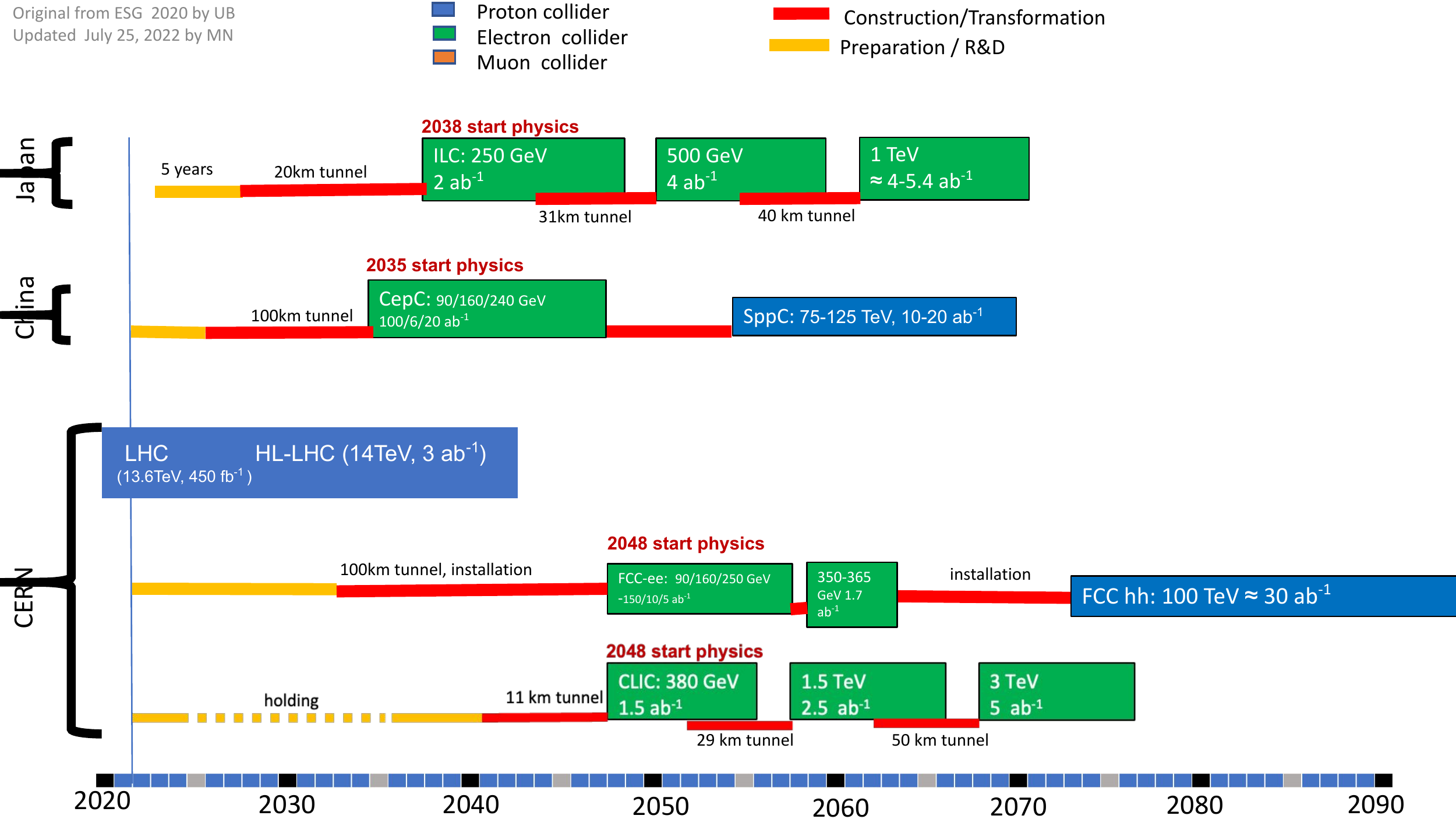}

\end{center}
\caption{Projected timelines for R\&D, construction, and physics operations for some of the leading proposed future collider options.}
\label{fig:futureCollOne}
\end{figure}

The projected timelines for R\&D, construction, and physics operations for some of the leading proposed future collider options, such as ILC sited in Japan, CepC in China, LHC, HL-LHC, CLIC and the integrated FCC option considered during Snowmass 2021, are shown in Fig.~\ref{fig:futureCollOne}. The yellow bar shows the time needed for preparation and R\&D, prior to construction or transformation that are represented by the red bar. The green and blue boxes represent electron and proton colliders respectively, together with the operating center of mass energy and the integrated luminosity. 

\begin{figure}[!ht]
\begin{center}
\includegraphics[width=0.98\hsize]{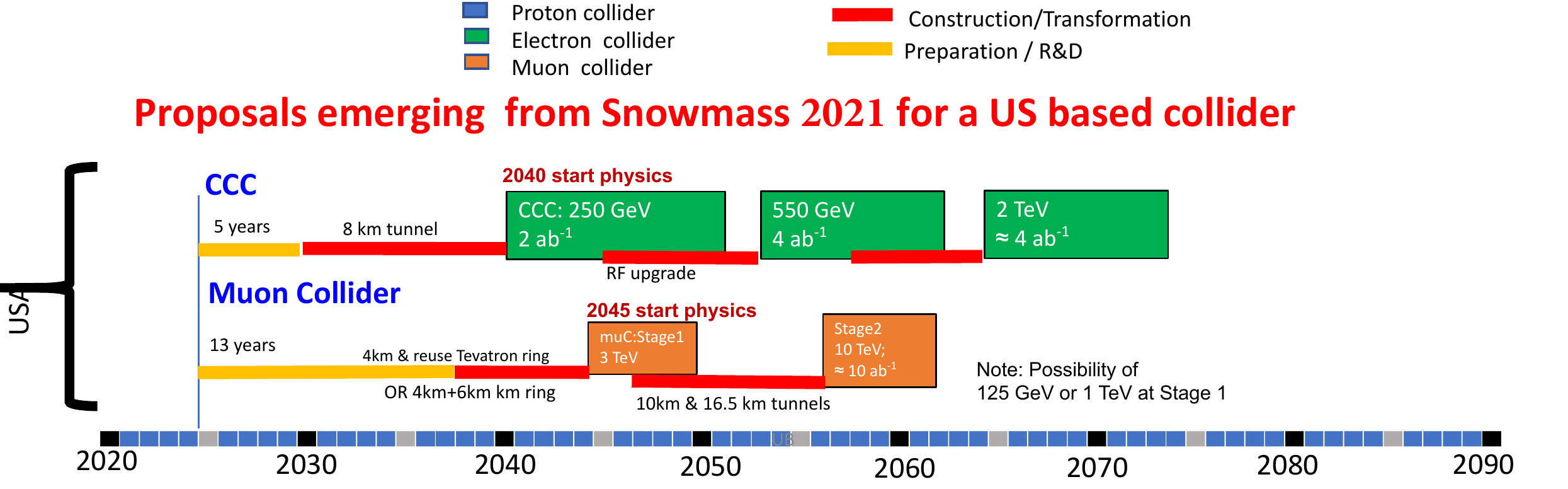}

\end{center}
\caption{Approximate timelines for proposals for ILC/CCC and Muon Collier emerging from Snowmass 2021 for a US based collider option.}
\label{fig:futureCollTwo}
\end{figure}

Timelines for proposals for ILC/CCC and Muon Collier emerging from Snowmass 2021 for a US based collider option are shown in Fig.~\ref{fig:futureCollTwo}. It is important to note that the timelines shown are technologically limited, and some challenges need to be sorted out, for example, successful R\&D and feasibility demonstrations for CCC and Muon Collider are essential at a short timescale. Additionally, evaluation progress in the international context and discussion on international cost sharing is a top priority.

Cost estimates, construction start and end dates for the large future collider projects are summarized in Table~\ref{tab:costCollider}. Costs are taken from the collider Implementation Task Force (ITF) report~\cite{Roser:2022sht}  by the Accelerator Frontier. Please refer to the ITF document for explanations on how they were estimated and associated caveats.

\begin{table}[!ht]
\begin{center}
 \caption{Cost estimates, construction start and end dates for the large future collider projects.}
 \label{tab:costCollider}
\begin{tabular}[c]{||l|c|c|c||}
\hline \hline
Project	&\multicolumn{3}{c||}{Construction} \\	
	&	Start date (yr)	&	End date (yr)	& Cost B\$
\\ \hline
{\bf Higgs Factories}	&		&		&		\\
CepC	&	2026	&	2035	&	12-18	\\
CCC (higgs Fac)	&	2030	&	2040	&	7-12	\\
ILC (higgs Fac)	&	2028	&	2038	&	7-12	\\
CLIC	&	2041	&	2048	&	7-12	\\
FCC-ee	&	2033	&	2048	&	12-18	\\ \hline
{\bf Multi-TeV Colliders}	&		&		&		\\
Muon Collider (3 TeV)	&	2038	&	2045	&	7-12	\\
Muon Collider (10 TeV)	&	2042	&	2052	&	12-18	\\
SppC	&	2043	&	2055	&	30-80	\\
HE CCC	&	2055	&	2065	&	12-18	\\
HE CLIC (3 TeV)	&	2062	&	2068	&	18-30	\\
FCC-hh	&	2063	&	2074	&	30-50	\\ \hline\hline
\end{tabular}
\end{center}
\end{table}

The R\&D cost estimates for medium-scale projects and near-term suggested contributions to be considered by U.S. EF community (and proponents of the options) are listed in Table~\ref{tab:RDCollider}. In the spirit of Snowmass, such numbers are to be taken as an approximate scale of investments. 

\begin{table}[!ht]
\begin{center}
 \caption{Suggested R\&D investment to be considered by the U.S. EF community with approximate start and end dates.}
 \label{tab:RDCollider}
\begin{tabular}[c]{||l|c|c|c||}
\hline \hline
Project	&\multicolumn{3}{c||}{R\&D} \\	
	&	Start date (yr)	&	End date (yr)	& Cost M\$
\\ \hline
Higgs Factory detector R\&D	&	now	&	2035	&	~100-150	\\
CCC higgs factory	&	2024	&	2028	&	~100	\\
CCC High Energy	&	2045	&	2050	&	~200	\\
Muon Collider  (1-3 TeV)	&	now	&	2040	&	~300	\\
Muon Collider (10 TeV)	&	2040	&	2047	&	~200	\\
 \hline\hline
\end{tabular}
\end{center}
\end{table}



%
%

\bibliographystyle{Energy/utphys.bst}
\bibliography{Energy/EnergyINSPIRE.bib, Energy/EnergyCUSTOM.bib}